\documentclass[11pt,letterpaper]{report}          % for default format

\usepackage{pstcol}
\usepackage{units}

\usepackage{epsfig}

\usepackage{varioref}
\usepackage{graphicx}
\usepackage[intlimits]{amsmath}
\usepackage{amssymb}

\usepackage{dcolumn}
\usepackage{tabularx}
\usepackage{ltxtable}

\usepackage[normalem]{ulem}

\usepackage{thesis}

\newcolumntype{d}[1]{D{.}{.}{#1}}

\newcount\aghour
\newcount\agminute
\aghour = \time
\divide\aghour by 60 
\agminute = \time
\begingroup
\multiply\aghour by 60
\global\advance\agminute by -\aghour
\endgroup

\input{macros.texinc}

\newif\ifagsdouble
\input{linespacing.texinc}

\renewcommand{\contentsname}{Table of Contents}

\title{A precision measurement of the muon decay parameter delta}
\author{Andrei Gaponenko}
\degree{Doctor of Philosophy} % or "Doctor of Philosophy", this string appears
\dept{Physics}  % Write Computing Science or Civil Engineering.
\permanentaddress{1518--1450 Chestnut St\\ Vancouver, BC, V6J 3K3\\ Canada}

\examiners{Prof. Andrzej Czarnecki~(Co-Supervisor),Dr. David R.~Gill~(Co-Supervisor),Prof. Douglas M. Gingrich,Prof. Kim~H.~Chow,Prof. Alex Brown,Prof. Douglas Bryman~(External~Examiner)} %Names of people on you committee
\convocationseason{Spring}   % or "Spring", again string appears in thesis.
\dedication{To Nate Rodning}  % Optional, both this and frontpiece
\begin{document}

\admin        % use when you wish to produce the whole works, ie.,
              % approval page, release page ect.
\begin{abstract}   % enviorment for abstract.

% The limit is 350 words for PhD.

The muon decay parameter delta characterizes momentum dependence of
the parity-violating muon decay asymmetry.  A new measurement of delta
has been performed using the first physics data recorded by the TWIST
experiment at TRIUMF.  The obtained value,
$\delta=0.74964\pm0.00066\,\text{(stat.)}\pm0.00112\,\text{(syst.)}$,
is consistent with the Standard Model expectation $\delta=3/4$.  This
is the first determination of $\delta$ performed using a blind
analysis technique.  Combined with other data, the measurement sets
new model-independent limits on effective right-handed couplings of
the muon.  Improved limits on the product of another muon decay
parameter, $\xi$, and the muon polarization in pion decay, $\Pmu$, are
obtained in the form: $0.9960<P_{\mu}\xi\le\xi<1.0040$, at 90\%
confidence level.  Implications for left-right symmetric models are
discussed.

% Physics, but not only.  Mention the learning experience for TWIST
% and optimization for future measurements.  See slides from the LBL talk.

\end{abstract}

%%%%%%%%%%%%%%%%%%%%%%%%%%%%%%%%%%%%%%%%%%%%%%%%%%%%%%%%%%%%%%%%
\begin{acknowledgements} 
{\ifagsdouble\doublespacing\else\relax\fi

% Acknowledgements.

%\baselineskip=0.9\baselineskip

I am very grateful to my first TWIST supervisor, Nate Rodning
(deceased), for his support and care during the early stage of my
program.  Nate's lead in the construction of the detector was an
important factor in the success of the experiment.

I want to thank my co-supervisor Dave Gill, who was my first contact
in everyday work on the experiment, and whose help in preparation of
this thesis was invaluable.  I am thankful to my co-supervisor Andrzej
Czarnecki for the many useful discussions.  I would like to thank Doug
Gingrich, a member of my supervisory committee, for his help in
resolving some administrative issues.

I would like to thank every member of the examining committee for
their review of this work.

%% TWIST
 
I am grateful to the entire TWIST collaboration for their efforts on
the experiment, and for the productive friendly atmosphere we all
could enjoy.  
I want to particularly thank the people I mostly interacted with:
Dick Mischke, whose relentless efforts ensured a timely
completion of the measurement,
Glen Marshall, who taught me many practical things about doing
particle physics experiments,
Rene\'e Poutissou,
who was the first person I started working with at TRIUMF,
Konstantin Olchanski,
Peter Gumplinger,
Vladimir Selivanov, 
Art Olin,
Maher Quraan,
and fellow graduate students
Jim Musser, Blair Jamieson, and Rob MacDonald.
I particularly benefited from the many insightful discussions with
Carl Gagliardi.  From him I first learned that generating a
Monte-Carlo with a non positive definite theoretical spectrum was
possible, an important ingredient in the fitting technique we used.

Some people outside of TWIST also contributed to this work.
I am thankful to
Andrej Arbuzov, who made available to us his computer code 
for calculation of radiative corrections,
Mihail Chizhov, for the discussions of the proposed tensor interaction,
Dan Melconian, who
provided me with the data used to plot the nuclear beta decay curve on
Fig.~\ref{fig:lrs-manifest},
and a University of Alberta graduate
Lara De Nardo, from whom I got a LaTeX template for this thesis.

I am grateful to Jean-Michel Poutissou for sharing his wisdom in
our discussions of careers in science and academia.

I wish to express my deepest gratitude to my parents
Nikolay Gaponenko and Lyudmila Gaponenko,
my sister Lyuba and my brother Pavel, for their enduring and
whole-hearted support.  I recall that when I was about six years old,
my father always had time to discuss my latest drawings
of yet another \textsl{perpetuum mobile}.  This was the beginning of my
physics studies.

I am very grateful to my wife Elena for bearing with me 
during all the years I was preoccupied with the experiment.

%----------------
\vspace{2ex}
An important contribution to the success of this work was the 
continuing financial support from the Alberta Ingenuity Fund,
which allowed me to direct all my efforts toward this research.

}\end{acknowledgements}

%%%%%%%%%%%%%%%%%%%%%%%%%%%%%%%%%%%%%%%%%%%%%%%%%%%%%%%%%%%%%%%%
% Now the table of contents etc.

{
%\thispagestyle{fancy}
%\ifagsdouble\doublespacing\else\relax\fi
\tableofcontents

}

{
\ifagsdouble\doublespacing\else\relax\fi
\listoftables  % if you have any

\listoffigures % if you have any
               % minimal support for list of plates and symbols (Optional)
}

\begin{listofsymbols}
% % You are responsible for formating this page
% This is where you can define variables and such, assuming you remember
% what they actually mean.

%% Thesis regulations: if thesis is signle-spaced, there must be
%% double spasing _between_the_entries_ here.
\begingroup
\ifagsdouble\doublespacing\else\relax\fi

{\noindent\Large\bf Acronyms}

\begin{description}
\item[CDF] Collider Detector at Fermilab, the name of an experiment.

\item[CKM] The Cabibbo-Kobayashi-Maskawa quark mixing matrix.

\item[CL] Confidence level.

\item[D0] The name of an experiment at Tevatron.

\item[DC] Drift chambers.

\item[FWHM] Full width at half maximum.

\item[GARFIELD] A program for detailed simulation of gaseous detectors.

\item[GEANT] 1) A framework for writing simulation programs for
  particle physics.\\  2) The same as MC.

\item[MC] Monte-Carlo simulation.

\item[MOFIA] \TWIST{} track fitting program.

\item[PC] Miltiwire proportional chambers.

\item[PDG] Particle Data Group.

\item[TCAP] Time of a capacitor probe signal relative to the trigger
  time.  Used in a time-of-flight cut, chapter
  \ref{chapter:reconstruction}.  The capacitor probe detects passage
  of a bunch of protons in the primary beam.

\item[TDC] Time to digit converter. An electronic circuit providing a
  digital readout of arrival time of a signal.

\item[TRIUMF] Tri-University Meson Facility, a cyclotron lab.

\item[TWIST] TRIUMF Weak Interaction Symmetry Test, the name of an experiment.

\end{description}

% $p_t$ \\[1ex]
% 
% $p_z$ \\[1ex]

\endgroup
\end{listofsymbols}

               % Now a set up command any you are off
%\doublespacing % Optional; default is \singlespacing; you can also use
               %                \onehalfspacing or \truedoublespacing

\ifagsdouble\doublespacing\else\relax\fi

\bodyoftext
%  ... your magnificient thesis ... 
%  hopefully more than two lines! Use standard Latex sectioning commands
%  like \chapter ect. End with the bibliography

%%%%%%%%%%%%%%%%%%%%%%%%%%%%%%%%%%%%%%%%%%%%%%%%%%%%%%%%%%%%%%%%
% AG: coordinates of the corners of the text area on the page:
%
%     (108,720) --------------------------- (503, 720)
%         |                                     |
%         |                                     |
%         |                                     |
%         |                                     |
%         .                                     .
%         .                                     .
%         .                                     .
%         |                                     |
%         |                                     |
%         |                                     |
%      (108,73) ---------------------------- (503, 73)
%
% ==> width = 396, height = 648

%%%%%%%%%%%%%%%%%%%%%%%%%%%%%%%%%%%%%%%%%%%%%%%%%%%%%%%%%%%%%%%%

\chapter{\label{chapter:introduction}Introduction}
%\chapter{Introduction}

%\thispagestyle{fancy}

\begingroup
\def\be{{\bar{e}}}
\def\bu{{\bar{\nu}}}
\def\zetag{\zeta_g}

\def\vudL{|V_{ud}^L|}
\def\vudR{|V_{ud}^R|}

% SM incomplete. Search for new physics. High energy---direct observation,
% or precision measurements---virtual effects.
% Complimentary to collider and nuclear beta decay measurements.

The Standard Model of particle physics
\cite{glashow:1961tr,weinberg:1967tq,salam:1969sm} has been very
successful in describing available experimental data.  The only
observed deviation from the original theory is the recent observation
of neutrino oscillations, but that can be easily accommodated in the
model's framework and does not lead to a conceptual change.  Despite
the many successes, the theory is generally believed to be
incomplete\vc{, see e.g. Section~VII in~\cite{Gaillard:1998ui}}.
Numerous extensions of the Standard Model have been
proposed, and experimental searches for New Physics are ongoing.  The
experimental efforts explore two complementary approaches.  One type
of experiment aims at direct observation of new particles.  These
searches require the energy of the collision to be high enough to produce the
supposed heavy particle being looked for, and their reach is limited
by the capabilities of the accelerator.  The other direction of such research
exploits contributions of hypothetical particles to known processes
through virtual (loop) effects.  These experiments can be done at low
energies.  The mass-scale reach of this kind of search is limited by
the precision of the measurement and by the theoretical precision of
the calculation of the ``known'' processes.

Muon decay $\mu\to{}e\nu\bar{\nu}$, studied by \TWIST{} (TRIUMF Weak
Interaction Symmetry Test experiment), is one of a few processes in
particle physics that can be unambiguously calculated with high
accuracy in the framework of a theoretical model.  The purely leptonic
nature of the decay eliminates many uncertainties due to the internal
structure of the particles.  The strong interaction, which at present
can not be accurately evaluated from first principles, enters only
through higher order radiative corrections.  The fractional hadronic
contribution to the energy spectrum can be estimated as
$0.07\,(\alpha/\pi)^2\approx0.4\times10^{-6}$
\cite{davydychev:2000ee}, so any related uncertainty is negligible for
the current state of the field.  On the other hand muons are easy to
produce in large quantities at an accelerator.  That means high
experimental statistics is affordable, and precision experiments can
be done to test theoretical predictions.  Thus the decay of a muon is
an ideal low energy process with which to investigate the 
%
%vb: space-time
%
\vc{Lorentz}
structure of the weak interaction.

%%%%%%%%%%%%%%%%%%%%%%%%%%%%%%%%%%%%%%%%%%%%%%%%%%%%%%%%%%%%%%%%
\section{The 4-fermion interaction formalism}

An approach useful in searches for new physics is to start with a very
general description of the process, then try to limit the possibilities.
The most general, local, derivative-free, Lorentz-invariant and
lepton-number conserving four fermion interaction was introduced
by Michel~\cite{michel:1949qe}.  Using the notation of \cite{fetscher:1986uj},
which represents particles by fields of definite chirality 
\cite{scheck:1977yg,scheck:hfp,mursula:1984zb}, the
interaction \vc{matrix element} can be written in a ``helicity projection form'' as
\begin{eqnarray}
M & =& \frac{4\,G_F}{\sqrt2}
  \sum_{\begin{array}{c}
	\gamma=S,V,T\\
	\epsilon,\mu = R,L
  	\end{array}
	}
  {
     g^\gamma_{\epsilon\mu}
     \langle{}\be_\epsilon|\Gamma^\gamma|(\nu_e)_n\rangle{}
     \langle{}(\bar{\nu}_\mu)_m|\Gamma_\gamma|\mu_\mu\rangle{}.
  }
\label{eq:hpf}
\end{eqnarray}
Here $G_F$ is the Fermi coupling constant, while $\gamma$
labels scalar, vector or tensor type of interaction:
\begin{equation*}
\Gamma^S = 1, \qquad
\Gamma^V = \gamma^\alpha, \qquad
\Gamma^T = \frac{1}{\sqrt{2}}\,\sigma^{\alpha\beta}.
%% %vb:\Gamma^T = \frac{1}{2\sqrt{2}} \left(
%% \Gamma^T = \frac{\vc{i}}{2\sqrt{2}} \left(
%%     \gamma_{\alpha}\gamma_{\beta}-\gamma_{\beta}\gamma_{\alpha}
%% \right)
\end{equation*}
In the last equation
$\gamma^{\alpha}$ are the Dirac gamma matrices,
and $\sigma^{\alpha\beta}=\frac{i}{2}(\gamma^\alpha\gamma^\beta-\gamma^\beta\gamma^\alpha)$.
The indices $\epsilon$ and $\mu$ indicate the chirality (handedness)
of the spinors of the charged leptons:
\begin{equation*}
\psi_{R,L}=\frac12\,(1\pm\gamma_5)\,\psi
\end{equation*}
The chiralities $n$ and $m$ of the ${\nu}_e$ and $\bar{\nu}_\mu$
spinors, respectively, are uniquely determined for given $\gamma$,
$\epsilon$ and $\mu$.  The tensor term in~\eqref{eq:hpf} requires
special attention.  Due to the identity
$\gamma_5\,\sigma_{\alpha\beta}=\frac{i}{2}\,\epsilon_{\alpha\beta\lambda\rho}\,\sigma^{\lambda\rho}$,
the coupling constants $g^T_{RR}=g^T_{LL}=0$.  So the general
interaction~\eqref{eq:hpf} is defined by 10 complex parameters.  Since
a common phase does not matter, the interaction is fully described by
19 real independent coupling constants.  The usual convention is to
absorb the overall strength of the interaction into $G_F$, and
normalize $g^\gamma_{\epsilon\mu}$ \cite{fetscher-gerber-singapore}
as:
\begin{equation}
%\begin{aligned}
\begin{matrix}
{} & \frac14|g^S_{RR}|^2 & + & \frac14|g^S_{RL}|^2 
  &+ & \frac14|g^S_{LR}|^2 &+& \frac14|g^S_{LL}|^2
\\[6pt]
 + & |g^V_{RR}|^2 &+& |g^V_{RL}|^2
 & +& |g^V_{LR}|^2 &+& |g^V_{LL}|^2 
\\[6pt]
 && + &3|g^T_{RL}|^2 
 & + &3|g^T_{LR}|^2  
&&& = & 1.
\end{matrix}
%\end{aligned}
\label{eq:gmunu-norm}
\end{equation}

Using \eqref{eq:hpf} as a starting point, it is straightforward to
calculate the differential rate, in energy and angle, of positrons
emitted in muon
decay~\cite{bouchiat-michel:1957,kinoshita-sirlin:1957,fetscher-gerber-pdg:2004}:
\begin{equation}
\frac{d^2\Gamma}{dx\,d\costh} = \frac{m_{\mu}}{4\pi^3}\,
W_{e\mu}^4 G_F^2 \sqrt{x^2-x_0^2}
\:\big\{
F_{\text{IS}}(x) + P_{\mu}\costh F_{\text{AS}}(x)
\big\}
\label{eq:theoretical-spectrum-2d}
\end{equation}
with
\begin{align}
F_{\text{IS}}(x) &= x\,(1-x) + \frac29\rho\,(4x^2-3x-x_0^2) 
+ \eta\,x_0 \,(1-x)
+ F_{\text{IS}}^{\text{RC}}(x)
, 
\label{eq:theoretical-spectrum-fis}
\\[6pt]
F_{\text{AS}}(x) &= \frac13\xi\,\sqrt{x^2-x_0^2}\,
\left[
  1-x + \frac23\delta
  \left(
  4x-3 + \left(\sqrt{1-x_0^2}-1\right)
  \right)
\right]
+ F_{\text{AS}}^{\text{RC}}(x)
.
\label{eq:theoretical-spectrum-fas}
\end{align}
Here $m_{\mu}$ is the muon mass, $W_{e\mu}=(m_{\mu}^2+m_e^2)/2m_{\mu}$
is the maximum energy of the emitted positron, $x=E_e/W_{e\mu}$ is the
reduced positron energy, $\theta$ is the angle
%
%v1: between the muon spin and the positron momentum, and 
%
between the positron momentum and an arbitrary direction $\vec{z}$, 
$x_0=m_e/W_{e\mu}$ is the dimensionless electron mass,
$-1\le{}P_{\mu}\le1$ is the muon polarization with respect to $\vec{z}$, and
$F_{\text{IS}}^{\text{RC}}(x)$ and $F_{\text{AS}}^{\text{RC}}(x)$ are
radiative corrections.  The muon decay parameters $\rho$, $\eta$,
$\xi$, $\delta$, are real numbers expressed through bilinear
combinations of the coupling constants $g^\gamma_{\epsilon\mu}$,
and the indices IS and AS label the isotropic and anisotropic terms.

%   In particular,
% \begin{align}
% \xi\delta & = aaa\\
% \xi & = bbb
% \end{align}

%
% A 100\% parity violation is postulated, and only left-handed fermionic
% fields interact weakly.  This means that the SM corresponds to a
% $g^V_{LL}$ interaction, with all other couplings being zero.
% (The ``V-A'' interaction.)
%
In the Standard Model muon decay is mediated by a $W$ vector boson.
It is postulated that only left-handed fermionic fields interact
weakly, that is, the degree of parity violation is 100\% (``V-A''
interaction).  This means that the SM corresponds to only $g^V_{LL}$
being non zero, and leads to the following values of decay parameters
\begin{equation}
\rho =\frac34,\qquad\eta=0,\qquad\xi=1,\qquad\delta=\frac34.
\label{eq:sm-michelpars}
\end{equation}
Many extensions of the Standard Model give rise to other couplings
that modify~\eqref{eq:sm-michelpars}.

The contact interaction \eqref{eq:hpf} is not renormalizable, so only
a tree level result can be obtained in a consistent way for the most
general case.  A calculation of radiative corrections requires either
restricting the interaction to a V,A type, or specifying an underlying
model leading to the effective interaction \eqref{eq:hpf}.
Radiative corrections to the spectrum are significant
\cite{behrends-finkelstein-sirlin:1956,kinoshita-sirlin:1957,berman:1958}
and have been calculated under different assumptions by many authors.
Very detailed results computed within the Standard Model are available
\cite{arbuzov:2001ui,aaa-leadinglog:2002,arbuzov:2002cn,arbuzov:2002rp,arbuzov:2003ce}.

%%%%%%%%%%%%%%%%%%%%%%%%%%%%%%%%%%%%%%%%%%%%%%%%%%%%%%%%%%%%%%%%
\section{How general is the ``general'' matrix element?}

This section%
\footnote{%
  Material in this section is based on my term write-up for the
  Quantum Field Theory---II course at U of Alberta (2000).
}%
 discusses assumptions underlying \eqref{eq:hpf}.

First of all note that \eqref{eq:hpf} assumes that all the four
particles involved in muon decay are fermions.  The two neutral
particles, which are not observed in modern experiments measuring the
muon decay spectrum, may be of a different nature.  For example
supersymmetric theories predict the decay of a muon into an electron and
two light scalar sneutrinos mediated by a wino.  Such a decay cannot
be described accurately by a parameterization of the four-fermion
interaction~\cite{fetscher:1986uj}.

To understand what other assumptions are implied by \eqref{eq:hpf},
let us consider the $S$-matrix element of the
$\mu\rightarrow e\nu\bar{\nu}$ decay.
There are four particles involved, fully described by 16 kinematic
variables, the components of the four-momenta. These variables obey 4
relations $p_i^2=m_i^2$.  
The conservation laws corresponding to 10 generators of the
Poincar\'e group impose 10 additional constraints.
Thus \vc{$16-4-10=2$} independent invariants can be constructed from the four
4-momenta~\cite{landafshitz-IV:1982}.  For example $s=(p_\mu -
p_{\nu_\mu})^2$ and $t = (p_e - p_{\nu_e})^2$.  In the general case
the $S$-matrix element should have the form~\cite{landafshitz-IV:1982}
\begin{equation}
M_{fi}=\sum_n{f_n(s,t)\,F_n}
\label{eq:s-matrix}
\end{equation}
where the functions of the kinematic invariants, $f_n$, are called
invariant amplitudes, and the $F_n$ are invariants which depend
linearly on the wave amplitudes and 4-momenta of all the particles
concerned.  Equation \eqref{eq:s-matrix}~may be understood to include
all radiative corrections.

The spin part of \eqref{eq:s-matrix} is more general than
\eqref{eq:hpf}.  The linear momentum dependence of $F_n$ in some cases
can be absorbed into $f_n$ by applying the equation of motion, for
example:
\begin{equation}
p_2^\alpha\, 
  \langle{}\bar{\psi}_1|\gamma_\alpha|\psi_2\rangle{}\langle{}\bar{\psi}_3|\psi_4\rangle{}
=
  m_2
  \langle{}\bar{\psi}_1|\psi_2\rangle{}\langle{}\bar{\psi}_3|\psi_4\rangle{}
\label{eq:dirac-reduction}
\end{equation}
But one can imagine a term for which this reduction will not work.  For
example, replace $p_2\rightarrow p_3$ in~\eqref{eq:dirac-reduction}.
So, ``derivative-free'' is an assumption.  This conclusion seems to
contradict a statement in \cite{greiner-muller}, p.5, that ``in the
interaction term the $4\times4$ differential operators can be reduced
to (constant) $4\times4$ matrices.''  However, a contact interaction is
assumed in~\cite{greiner-muller} but not in the $S$-matrix approach.
So for the spin part of the matrix element the assumption of
``locality'' of the effective interaction makes it also
``derivative-free''.  Extensions of the four-fermion interaction that
allow for a linear momentum dependence of the spin part have been 
proposed \cite{Poblaguev:1990tv,Chizhov:1994db}.

The invariant amplitudes $f_n(s,t)$ in \eqref{eq:hpf} are just
constants.  To build $f_n$ we need a mass parameter, \vc{$M$, so that
$f_n(s,t)=\tilde{f}_n(\tilde{\vphantom{t}s},\tilde{t})$, where $\tilde{f}_n$ is a
function of dimensionless variables $\tilde{\vphantom{t}s}=s/M^2$,
$\tilde{t}=t/M^2$.  Taylor} expansion of the amplitude should look
like
\vc{\begin{equation}
%   f_n(s,t) = f_n(0,0) + t/M^2 +\ldots
   f_n(s,t) = f_n(0,0) + \frac{\p{\tilde{f}_n}}{\p\tilde{t}}\,\frac{t}{M^2}+\frac{\p{\tilde{f}_n}}{\p\tilde{\vphantom{t}s}}\,\frac{s}{M^2}+\ldots
\end{equation}}
In the context of a gauge theory $M$ is the mass of an intermediate
boson, e.g. $M=M_W\gg m_\mu$, and replacement of $f_n(s,t)$ by a constant
can be justified.  Again, this assumption is approximately equivalent
to the assumption of a contact interaction: a heavy mediator means a
short-range force.

The interaction term~\eqref{eq:hpf} contains also the assumption that
lepton number is conserved.  However, this assumption is not
essential.  Langacker and London~\cite{langacker:1988cm} have shown
that a Hamiltonian allowing both lepton flavor violation and total
lepton number violation still leads to the same decay
spectrum~\eqref{eq:theoretical-spectrum-2d}.  Moreover, there is a
one-to-one correspondence between combinations of coupling constants
of the lepton-number non-conserving Hamiltonian and coupling constants
in~\eqref{eq:hpf}.

%%%%%%%%%%%%%%%%%%%%%%%%%%%%%%%%%%%%%%%%%%%%%%%%%%%%%%%%%%%%%%%%
\section{Tests for new physics\\ with the muon decay parameters}

Early measurements of the muon decay spectrum helped to establish the
current theory of the electroweak interaction.  The much more precise
experimental data available today are still in good agreement with the
Standard Model.  The best measurement of the muon decay parameter
$\delta$ before \TWIST{} was \cite{Balke:1988delta}
$\delta=0.7486\pm0.0026\text{(stat)}\pm0.0028\text{(sys)}$.  In the
rest of the chapter we will consider some constraints on new physics
that can be imposed by a more precise measurement of muon decay
parameters.  We will concentrate on parameters $\delta$, a measurement
of which constitutes the subject of this thesis, and $\xi$, which was
constrained by the presented measurement of $\delta$.  At the end of
this section we briefly mention some models where $\delta$ differs
from the SM value of $3/4$, which have been excluded by other
experiments.

\subsection{Model-independent search for right-handed interactions}

By re-ordering \eqref{eq:gmunu-norm} we can write the 
fractions of decays where the muon interacts $\mu$-handedly 
and the positron $\epsilon$-handedly as \cite{fetscher:1986uj}
\begin{align}
Q_{RR} &= \frac14|g^S_{RR}|^2 + |g^V_{RR}|^2 \label{eq:qrr}\\
Q_{LR} &= \frac14|g^S_{LR}|^2 + |g^V_{LR}|^2 + 3|g^T_{LR}|^2   \\
Q_{RL} &= \frac14|g^S_{RL}|^2 + |g^V_{RL}|^2 + 3|g^T_{RL}|^2 \\
Q_{LL} &= \frac14|g^S_{LL}|^2 + |g^V_{LL}|^2 \label{eq:qll}
\end{align}
The fraction of muons decaying through a right-handed interaction
$Q^{\mu}_{R}=Q_{RR}+Q_{LR}$ can be expressed through Michel parameters
$\xi$ and $\delta$~\cite{fetscher-gerber-pdg:2004}:
\begin{equation}
Q^{\mu}_{R}=\frac12\,\left\{1 + \frac13\,\xi-\frac{16}{9}\,\xi\delta \right\},
\label{eq:qmur}
\end{equation}
and thus is measurable by \TWIST{}.  The non-negative quantity
$Q^{\mu}_{R}$ is exactly zero in the Standard Model.  Any deviation
from zero would indicate that the right-handed muon component
participates in the decay process through either a scalar, or vector,
or tensor, interaction.

\subsection{\label{section:lrs}Left-right symmetric models}

In the Standard Model the charged weak current is purely $V-A$.  A
natural assumption is that the $V+A$ current is suppressed, but not
exactly zero \cite{lipmanov}.  Left-right symmetric models
\cite{Pati:1973rp,Pati:1974yy,Mohapatra:1974hk,Mohapatra:1974gc,Senjanovic:1975rk}
extend the electroweak gauge group to include at least $SU(2)_{R}$ and
refer to a spontaneous symmetry breaking mechanism to explain parity
violation. A general $SU(2)_{L}\times{}SU(2)_{R}\times{}U(1)$ case is
considered in \cite{Herczeg:1985cx}.  The charged gauge boson fields
are mixed:
\begin{align}
W_L & = 
%\hphantom{e^{i\omega}(}
\cos\zeta\,W_1 + \sin\zeta\,W_2,\\
W_R & = e^{i\omega}(-\sin\zeta\,W_1 + \cos\zeta\,W_2),
\end{align}
where $W_L$, $W_R$ are the interaction eigenstates, $W_1$, $W_2$, are
the mass eigenstates, $\zeta$ is a mixing angle, and $\omega$ is a
CP-violating phase.  

The $W_R$ boson can contribute to muon decay only if the right-handed
neutrinos are light enough, so that the process is kinematically
allowed.  
% 
% For the case $m_{\nu_R}\lesssim1\ldots10\text{ MeV/c^2}$ 
% there are stringent limits $M_R\gtrsim{}\OO(1\text{ TeV})$ from 
% cosmological abundance of ${}^4He$
% \cite{Steigman:1979xp,Cyburt:2004yc}, so muon decay mostly 
% contributes to constraining models with intermediate mass 
% Dirac neutrinos \cite{Langacker:1989xa}.
% 
The muon decay parameters affected in left-right symmetric models are
$\rho$ and $\xi$.  Note that the spectrum shape
\eqref{eq:theoretical-spectrum-2d}--\eqref{eq:theoretical-spectrum-fas}
depends on the combination $P_\mu\xi$, not on $\xi$ itself.  The
polarization of muons from charged pion decay, $P_\mu$, in left-right
symmetric models is also different from unity.  Introducing the
notation
\begin{align}
t &= \frac{g_R^2\,m_1^2}{g_L^2\,m_2^2},\\[3pt]
t_\theta & =\frac{g_R^2\,m_1^2\,\vudR}{g_L^2\,m_2^2\,\vudL}, \\[3pt]
\zetag &= \frac{g_R}{g_L}\,\zeta
\end{align}
we can write \cite{Herczeg:1985cx}
\begin{align}
\rho &= \frac34\,(1-2\zetag^2), \label{eq:rls-rho}\\
\xi  &= 1-2\,(t^2+\zetag^2), \label{eq:rls-xi}\\
P_\mu &= 1-2\,t_\theta^2 -2\,\zetag^2-4\,t_\theta \zetag \cos(\alpha+\omega).
\label{eq:rls-pmu}
\end{align}
Here $g_L$, $g_R$ are the coupling constants, $m_1$, $m_2$ are the
masses of $W_1$ and $W_2$, $V_{ud}^{L,R}$ are the elements of the
left- and right-handed quark mixing matrices, and
$\alpha=\arg\{V_{ud}^R\}$ is a CP violating phase. ($V_{ud}^L$ is chosen to be real.)

It follows from \eqref{eq:rls-xi}--\eqref{eq:rls-pmu} that
a measurement of $P_\mu\xi$ constrains both the mass of the second
charged gauge boson and the mixing angle.

\subsection{\label{section:kappa}Non-local tensor interaction}

The ISTRA experiment observed a statistically significant deficit of
$\pi^{-}\to{}e^{-}\bar{\nu}\gamma$ events in the $E_\gamma>21$~MeV,
$E_e>70\text{~MeV}-0.8E_\gamma$ region \cite{Bolotov:1990yq}.  To explain it, a
new momentum transfer dependent tensor interaction has been suggested
\cite{Poblaguev:1990tv}.  This idea was discussed in the literature.
In particular, \cite{Quin:1993vh} pointed out possible difficulties
the hypothesis may have explaining nuclear beta decay data.  However
it could not be excluded \cite{Herczeg:1994ur}.  Recently another
experiment, PIBETA \cite{Frlez:2003pe}, also observed a deficit of
radiative pion decay events (using $\pi^{+}$) in a similar kinematic
region, renewing an interest in the problem.

The suggested tensor interaction is non-local (momentum transfer
dependent), and is not included in Eq.~\eqref{eq:hpf}.  A new coupling
constant, $g^T_{RR}$, needs to be introduced.  The new interaction
term, which should be added to \eqref{eq:hpf}, can be written as:
\begin{gather}
-\sqrt{2}\,G_F\,
g^T_{RR}\,
     \langle{}\be_R|\sigma_{\alpha\lambda}|\nu_e\rangle{}
     \,\frac{4q_\alpha q_\beta}{q^2}\,
     \langle{}\bar{\nu}_{\mu}|\sigma_{\beta\lambda}|\mu_R\rangle{}
     \label{eq:gtrr-term}
% \\
% - \sqrt{2}\,G_F\,
% g^T_{LL}\,
%      \langle{}\be_L|\sigma_{\alpha\lambda}|\nu_e\rangle{}
%      \,\frac{4q_\alpha q_\beta}{q^2}\,
%      \langle{}\bar{\nu}_\mu|\sigma_{\beta\lambda}|\mu_L\rangle{}.
\end{gather}
This term contains only left-handed neutrinos and can interfere with
the standard decay mode.  This interference leads to a higher
experimental sensitivity to that interaction.

There is a field theoretical extension of the Standard Model
\cite{Chizhov:1993gt}, which produces the effective tensor
interaction~\eqref{eq:gtrr-term}.  In \cite{Chizhov:2004hk} a
prediction for the spectrum of positrons from muon decay is made based
on the pion decay data.  It is shown that the muon decay parameter
$\delta$ is very sensitive to the new interaction
\begin{equation}
\delta\approx\frac34\,\left(1-6\,|g^T_{RR}|^2\right),
\label{eq:kappa-effect}
\end{equation}
and with the suggested value $g^T_{RR}\approx0.013$ almost a $10^{-3}$
deviation of $\delta$ from the Standard Model value can be expected.

\subsection{Historical models\label{sec:historical-models}}

The muon decay parameters have been discussed in the context of
supersymmetric theories with light sneutrinos
\cite{Buchmuller:1984gp,Balke:1988delta}.  However LEP data at the Z
pole \cite{Alexander:1991qw,Abreu:1991wj,Decamp:1991uy,Adriani:1993gk}
and above \cite{Heister:2002jc,Abdallah:2003xe} constrain
$m_{\tilde{\nu}}\gtrsim30\ldots94$~\text{GeV}, depending on the
assumptions.  Therefore muon decay with sneutrinos in the final state
is kinematically forbidden.

An explanation of the LSND anomaly suggested by Babu and Pakvasa
\cite{Babu:2002ic} involves a lepton number violation decay
$\mu^{+}\to{}e^{+}+\overline{\nu_e}+\overline{\nu_i}$
$(i=e,\mu,\tau)$.  Since the model requires
$\rho=\delta\approx0.7485$, it can be tested by \TWIST{}.  In 2003 the
KARMEN collaboration put a strict limit on the emission of
$\overline{\nu_e}$ from $\mu^+$ decay \cite{Armbruster:2003pq},
excluding the explanation at 90\% confidence level.

\endgroup

%%%%%%%%%%%%%%%%%%%%%%%%%%%%%%%%%%%%%%%%%%%%%%%%%%%%%%%%%%%%%%%%
\chapter{Experimental setup\label{chapter:experiment}}
\begin{figure}[tp]
\includegraphics[width=\textwidth]{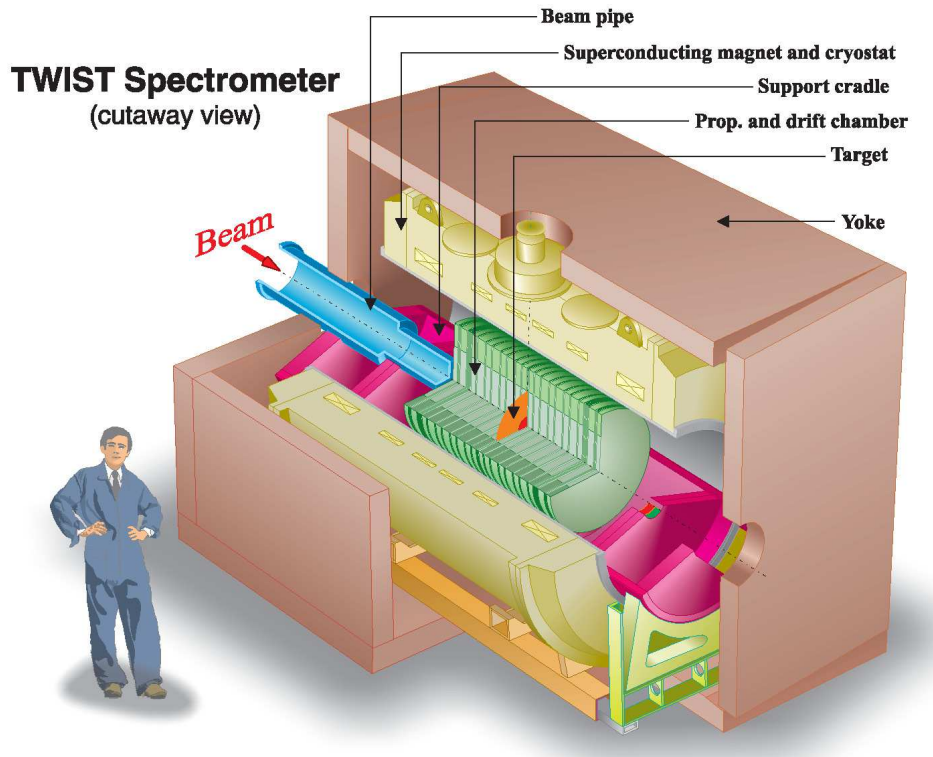}
\caption[\TWIST{} detector]{A drawing of the \TWIST{} detector
\cite{Henderson:2004zz}.%
\label{fig:twist-drawing}}
\end{figure}

The \TWIST{} experiment is designed to measure the spectrum of
positrons from muon decay in a wide range of energy and angle.  A
conceptual view of the spectrometer is shown on
Fig.~\ref{fig:twist-drawing}.  An important feature of the detector is
its planar geometry \cite{Khrutchinsky:1997}, which gives the
possibility to correct for the average energy loss of decay positrons
with high precision in a data-driven way (chapter
\ref{chapter:ecalib}).  The experiment uses a highly polarized surface
muon beam \cite{Pifer:1976ia} from the M13 secondary beam line at
TRIUMF.  The beam rate of about $2.5\times10^{3}$ muons per second is
low enough to typically have no more than one muon at a time in the
detector.  A muon is stopped in the center of a symmetric stack of
planar wire chambers and decays at rest.  A 2~T uniform magnetic field
preserves the direction of the spin of the stopped muons.  The decay
positron spirals in the magnetic field, leaving hits on the wires.
The hits are recorded by TDCs and analyzed offline to reconstruct the
trajectory of the particle and determine its energy and angle with
respect to the magnetic field.  A detailed description of the \TWIST{}
apparatus is given in \cite{Henderson:2004zz}.  The rest of this
chapter summarizes different aspects of the experimental setup.

% \begin{figure}[tbp]
% \includegraphics[width=\textwidth]{plots-unused/m13cut.eps}
% \caption{M13 beam line.%
% \label{fig:m13beamline}}
% \end{figure}

%%%%%%%%%%%%%%%%%%%%%%%%%%%%%%%%%%%%%%%%%%%%%%%%%%%%%%%%%%%%%%%%
\section{Muon beam\label{section:m13}}

\begin{figure}[tbp]
\includegraphics[width=\textwidth]{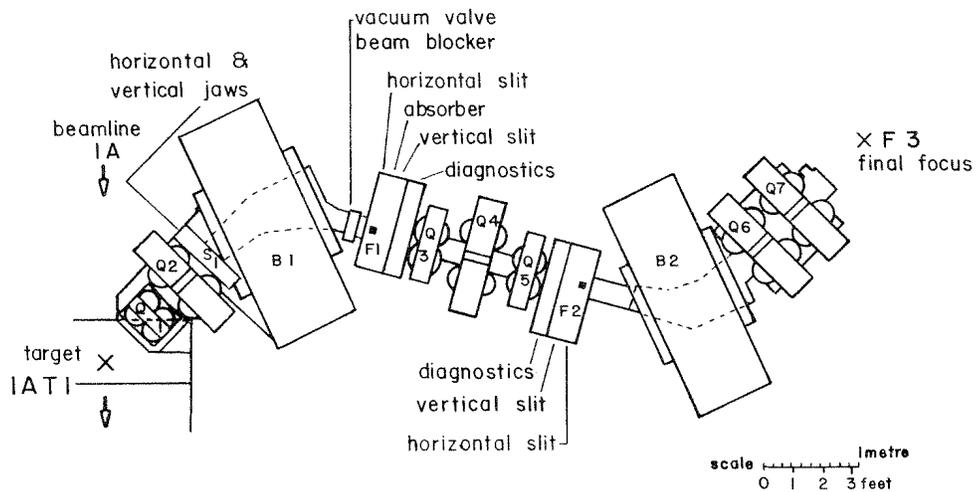}
\caption[M13 beam line]{M13 beam line layout \cite{Oram:1980nx}.  B1
and B2 are the dipole, and Q1--Q7 are the quadrupole magnets.  The
production target 1AT1 is seen by M13 at $135^{\circ}$ with respect to
the primary proton beam, the bends in B1 and B2 are $60^{\circ}$
each.%
\label{fig:m13beamline}}
\end{figure}
The \TWIST{} detector is installed in the M13 secondary beam line
\cite{Oram:1980nx} at the TRIUMF cyclotron.
Fig.~\ref{fig:m13beamline} shows the M13 layout.  The cyclotron
produces a 500~MeV quasi-continuous proton beam, with 4~ns proton
bunches striking a production target every 43~ns.  During the 2002
\TWIST{} data taking a beryllium production target was used.  Among
the particles produced when beam protons interact with the target are
positive pions.  The dominant decay mode $\pi^+\to\mu^+\nu_{\mu}$
results in the production of muons, which can be transported through
the M13 beam line to the experiment.

The process $\pi^+\to\mu^+\nu_{\mu}$ is a two body decay, thus the
momentum of the $\mu^+$ in the rest frame of the $\pi^+$ is fixed,
$p_\mu=29.79$~MeV/c.  The relationship between the muon spin and the
muon momentum is also predicted by theory.  In the Standard Model the
spin is antiparallel to the momentum.  This relationship may be
altered if the muon scatters in a material, since the Coulomb
scattering of the nonrelativistic muons changes the momentum direction
without influencing the spin.  To preserve high polarization of the
muon beam such interactions should be minimized.

The surface muon beam technique \cite{Pifer:1976ia} utilizes those
pions that stop in the production target, then decay at rest.  Passage
through a material causes muons to also lose their momentum, and that
loss, like the depolarization, is proportional to the amount of
material crossed.  By tuning the beamline to select muons that lost
only a limited amount of momentum the depolarization can be
controlled.  The muons accepted by the properly tuned beamline come
from pions decaying in a thin layer of material close to the surface
of the target, with some ``cloud'' muon contamination from pion decays
in flight.  The 43~ns time structure of the beam makes possible the
elimination of prompt particles produced at the time the protons hit
the target, which includes the ``cloud'' muons.  Since the life time
of $\pi^+$ is 26~ns, most of ``surface'' muons are emitted between the
proton pulses.  As discussed on page \pageref{page:pmu-for-delta}, the
measurement of $\delta$ requires high muon polarization, though a
knowledge of the precise value of the polarization is not important.
For the 2002 \TWIST{} physics data taking the M13 beam line was tuned
to the momentum 29.6~MeV/c, with a momentum \vc{acceptance} of 1.3\%, resulting
in a higher than 90\% muon polarization seen by the \TWIST{}
spectrometer.

At surface muon momenta the beam contains mostly positrons, muons, and
a small fraction of pions \cite{rpm-msc}.  The positron and pion beam
backgrounds are removed by the reconstruction software (chapter
\ref{chapter:reconstruction}).  Data were also taken at 120~MeV/c for
calibration purposes.  At this momentum the beam predominantly
contains pions.

% At surface muon momenta the flux of positrons in M13 with the Be
% target is about a factor of 30 higher than flux of the surface muons
% (\cite{rpm-msc}, pages 17 and 20).  There is also a small admixture of
% pions in the beam (less than 1\% of the surface muon flux,
% \cite{rpm-msc}).  These backgrounds are removed by the reconstruction
% software (chapter \ref{chapter:reconstruction}).

%%%%%%%%%%%%%%%%%%%%%%%%%%%%%%%%%%%%%%%%%%%%%%%%%%%%%%%%%%%%%%%%
\section{\TWIST{} detector}

\subsection{Wire chambers}

The \TWIST{} apparatus uses wire chambers as the primary source of
information.  Two types of chambers are employed in the detector:
drift chambers (DC), and proportional chambers (PC).  They are similar
in construction and use 15~$\mu$m sense wires.  Their cathodes are
made from nominally 6.35~$\mu$m thick doubly aluminized Mylar foils.
(Monte-Carlo simulation of the detector,
chapter~\ref{chapter:monte-carlo}, uses a measured mass density
instead of the thickness, and also accounts for effect of stretching
of the foils on their thickness.)  The pitch of the sense wires is 4~mm
for the DCs and 2~mm for the PCs. The cathode-to-cathode distance is
4~mm in all cases.  The PCs use a CF${}_4$/isobutane gas mixture, the
high drift velocity of which provides fast response.  One of the main
functions of PCs in the analysis is to resolve tracks of different
charged particles in time.  The DCs use dimethylether (DME) gas, which
gives high spatial resolution.

The wires are positioned at $45^\circ$ to the vertical direction to
reduce the gravitational sag, therefore instead of the $X$ and $Y$ the
chambers measure the $U$ and $V$ coordinates as defined in
Appendix~\ref{chapter:twist-coordinates}.  The wire chambers are
assembled into modules, each module having two or more wire planes.
The volume between the chambers is filled with a
helium(97\%)/nitrogen(3\%) mixture.
%
% To reduce the gravitationl sag the chambers are mounted in the
% detector so that the wires form $\pm45^{\circ}$ angles with the
% vertical direction.  UV coordinates...
%
%
\begin{figure}[tp]
\includegraphics[width=\textwidth]{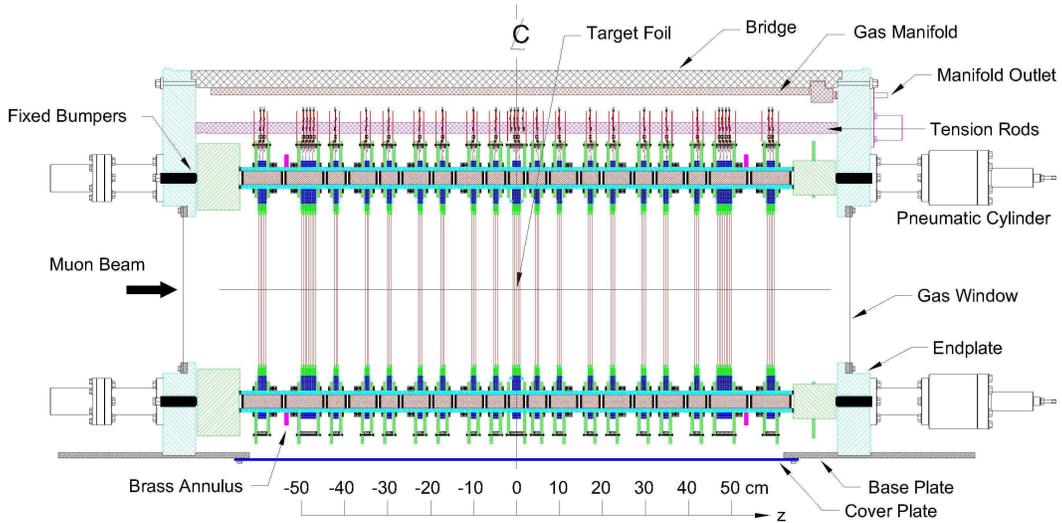}
\caption[Side view of the \TWIST{} cradle]{Side view of the \TWIST{}
cradle \cite{Henderson:2004zz}.%
\label{fig:stack}}
\end{figure}
A side view of the stack of wire chambers is shown on
Fig.~\ref{fig:stack}.  The order, as seen by an incoming muon, of the
modules is: PCs (4 planes), the ``dense stack'' (8 DCs), seven modules
of ``sparse stack'' containing a pair of DCs each, the target module.
The downstream (after the stopping target) arrangement mirrors the
upstream.  The DC and PC planes are numbered sequentially, with the
numbers increasing along the path of a muon.  DC~22 is the last drift
chamber and PC~6 is the last proportional chamber before the stopping
target.  

%
% Materials (from PDG_material_calc.sxc)
% 0.5*(17.375(tgt foil) + 9(tgt. graphite)) = 13.19 mg/cm^2 half TGT
% (for zero angle positrons).
% 
% 0.8cm*3.3333(CF4-ISO) + 3*0.825(foils) + 4cm*0.192(He-N2) = 5.91
% mg/cm^2 till the sense volume of the closest DC
%
% Total: 13.19 + 5.91 = 19.10 mg/cm^2 for 0 degrees
% 19.10 * sqrt(2) = 27 mg/cm^2 for 45 degrees.

The target module consists of 4 PC planes.  The central cathode foil
in this module also serves as the muon stopping target.  During the
2002 data taking this target was made from 125~$\mu$m Mylar, with
conductive graphite coating on both sides.  (A high purity aluminum
stopping target was not available at the time.)  Depolarizing interactions
in the target rendered 2002 data unsuitable for an improved
measurement of $P_{\mu}\xi$, but the extraction of $\delta$ does not
require a precise knowledge of the value of $P_{\mu}$ (page
\pageref{page:pmu-for-delta}).

An important advantage of the \TWIST{} detector is the small amount of
material in the tracking volume, leading to smaller effects from
scattering and energy loss and thus to smaller related uncertainties.
The thickness of one pair of DCs is only $1\cdot10^{-4}$ radiation
lengths.  Also, the positrons cross only about $25$~mg/cm${}^2$ of
material before entering the tracking volume at the first DC, compared
to $240$~mg/cm${}^2$ before the tracking volume in the previous
measurement \cite{Balke:1988delta}.

\begin{figure}[tp]
\centerline{\includegraphics[width=0.6\textwidth]{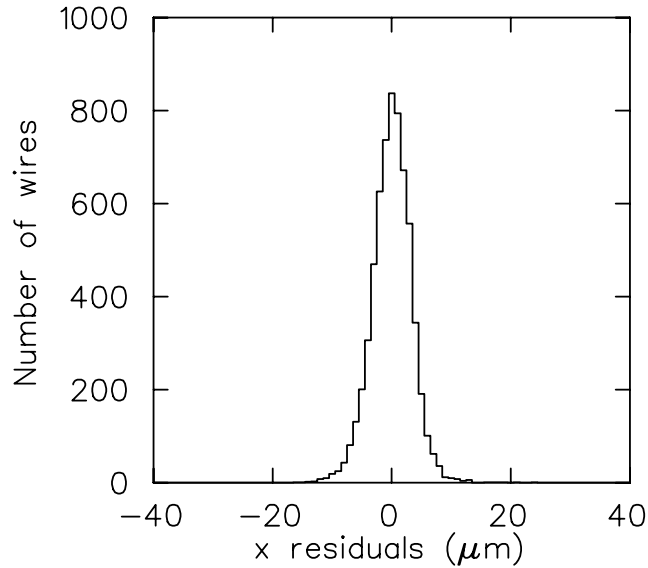}}
\caption[Wire positioning]{Deviation of wires from their nominal
  positions.  $\sigma=3.3$~$\mu$m \cite{Henderson:2004zz}.%
\label{fig:wire-positions}}
\end{figure}
A very high mechanical precision has been achieved in the production
of the detector.  The positions of the planes in the Z direction
(along the beam) are defined by precise Sitall ceramic spacers with a
negligible coefficient of thermal expansion.  The cumulative error on
the Z position is less than 50~$\mu$m over the whole length of the
detector \cite{Henderson:2004zz}.  The positions of wires within each
plane are accurate to a few microns, see
Fig.~\ref{fig:wire-positions}.  The relative alignment of different
wire planes within the detector has been accomplished using 120~MeV/c
pion tracks taken with the spectrometer magnet off.  The precision of
this alignment is 5~$\mu$m (translations) and
0.01~degrees (rotations) \cite{Henderson:2004zz}.

\subsection{Electronics and DAQ}

A 195~$\mu$m thick scintillator mounted between the beam pipe and the
upstream end of the \TWIST{} detector is used to provide a trigger
signal during physics data taking.  The scintillator is read out by
two PMTs and the trigger condition is a coincidence of the two
signals.  The thinness of the scintillator makes the system less
sensitive to beam positrons, while muons, which have a higher density
of energy loss at this momentum, produce a strong signal.  An
essential feature of the trigger is that it is unbiased: since the
decay positron is not used by the trigger system, the trigger
efficiency is not correlated with the muon decay parameters.

An electric signal from a DC or PC wire is fed to a pre-amplifier mounted
inside the wire chamber module.  The output of the pre-amplifier is
connected to a post-amplifier/discriminator in a CAMAC crate outside
of the detector.  The discriminator circuit provides a time over
threshold signal, which is recorded by a multihit TDC with 0.5~ns time
resolution.  Upon receiving a trigger signal the TDC analyzes its
internal buffer, and any activity from 6~$\mu$s before to 10~$\mu$s
after the scintillator hit is read out via FASTBUS by a PowerPC.  Time
of the leading edge, as well as the width (time over threshold) of the
signal are recorded for each of up to 8 hits per wire.  Data are sent
through an Ethernet connection to a dual 1GHz Pentium Linux computer
running a MIDAS \cite{midas} based data acquisition system
\cite{twist-daq}, which writes them to a disk buffer, then to SDLT
tapes.

%v1: The combined amplification of the electronic channel and of the drift
%v1: chamber cell give an effective threshold of 1.6 electrons collected
%v1: from a track to produce a hit \cite{Henderson:2004zz}.  The wire
%v1: chambers operated at a better than 99.9\% efficiency
%v1: \cite{Henderson:2004zz}, without a single dead or noisy channel during
%v1: the 2002 data taking.

The gas gain of the drift chambers combined with the electronic
amplification leads to an effective threshold of 1.6 electrons
collected from a track to produce a hit \cite{Henderson:2004zz}.  The
wire chambers operated at about 99.95\% efficiency
\cite{Henderson:2004zz}, without a single dead or noisy channel during
the 2002 data taking.

% Cross-talk??

In addition to the TDC data, the DAQ logs hundreds of ``slow control''
variables.  They include voltages and currents for individual wire
planes, gas flows through the chambers, proton beam current,
temperatures at numerous locations in the \TWIST{} detector, NMR
measured magnetic fields of the spectrometer and beamline magnets,
currents of the beam line magnets, atmospheric pressure, etc.

\subsection{Spectrometer magnet\label{section:magnet}}

\begin{figure}
\includegraphics[width=\textwidth]{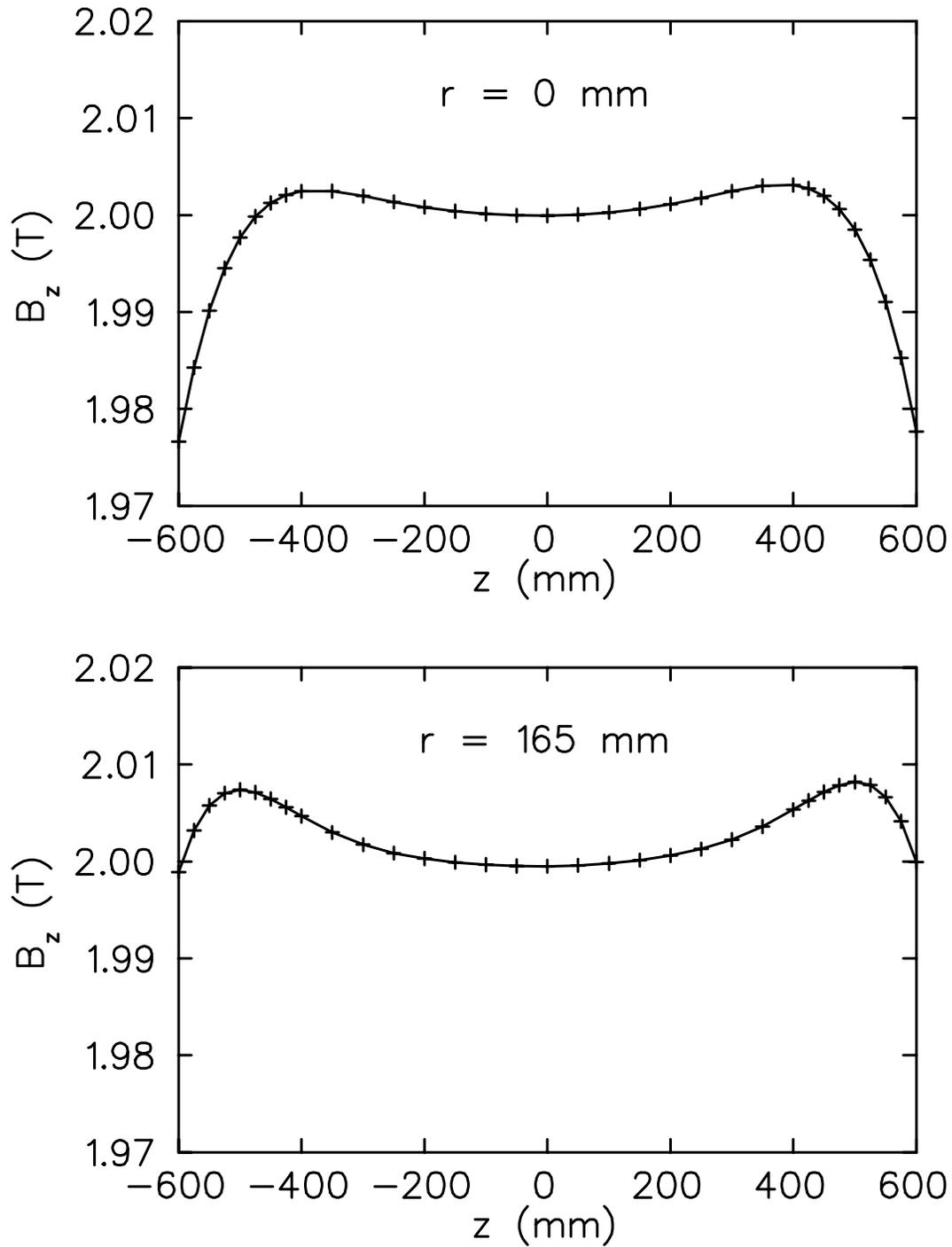}
\caption[Magnetic field map]{$B_z$ vs $z$ on the detector axis (top),
  and at the edge of the tracking volume (bottom).  Note the zoomed
  vertical scale.  The limits of the tracking volume in $z$, defined
  by the outermost DCs, are $\pm500$~mm.  The radius of the tracking
  volume is defined by the size of the wire planes.  Plots
  from~\cite{roberto-fieldmap}.%
\label{fig:fieldmap}}
\end{figure}

The stack of wire chambers is placed inside of a 2~T superconducting
solenoid.  The solenoid, together with the outside steel yoke,
produces highly uniform magnetic field in the tracking volume.  The
$B_z$ component of the field was mapped, Fig.~\ref{fig:fieldmap} shows
representative curves from the measurements.  
An OPERA-3d \cite{opera-3d} simulation model was tuned to the measured $B_z$.  The
OPERA-3d simulation produces the complete
$\vec{B}(\vec{\vphantom{B}r})$ field (as opposed to simply
$B_z(\vec{r})$) that is used by \TWIST{} Monte-Carlo and track
reconstruction software.  The simulated map reproduces the measured
$B_z$ to better than 3~Gauss in the tracking region, giving the
relative accuracy of 
$1.5\times10^{-4}$.

\subsection{Beam degraders}

To center the distribution of muon stopping position in the target,
there is the possibility to fine tune the amount of material in the
path of muons.  This capability is provided by a gas degrader, a
21.67~cm long volume installed between the vacuum window of the beam
pipe and the trigger scintillator.  The gas degrader contains a
He/CO${}_2$ mixture.  The fraction of CO${}_2$ can be varied from 0\%
to 100\%, affecting energy loss of muons in the degrader and their 
final stopping position.   

It is also possible to install a plastic film in the path of muons.
It was used to shift the stopping distribution to the upstream end of
the stack of wire chambers to acquire the Monte-Carlo verification
data (chapter~\ref{chapter:monte-carlo}).

%%%%%%%%%%%%%%%%%%%%%%%%%%%%%%%%%%%%%%%%%%%%%%%%%%%%%%%%%%%%%%%%

%%%%%%%%%%%%%%%%%%%%%%%%%%%%%%%%%%%%%%%%%%%%%%%%%%%%%%%%%%%%%%%%
\chapter{\TWIST{} data\label{chapter:twist-data}}
%\chapter{\TWIST{} data\label{chapter:twist-data}}

With a quasi-continuous beam from the TRIUMF cyclotron, the
definition of a ``data run'' is arbitrary.  In \TWIST{} the DAQ was
usually instructed to split data into files of about 1.9~GB each.  One
such file is a ``data run''.  A typical run contains
about $8.5\times10^5$ data events (triggers), and was acquired in
about 7~minutes for nominal surface muon beam.  Data quality was
monitored on per-run basis, and, if a problem was detected, a complete run
was excluded from the analysis.

A ``data set'' was defined as the amount of data required to achieve a
statistical precision of ${}\sim10^{-3}$ on the muon decay parameters.
While acquiring a data set, all controllable running conditions were
kept unchanged.  (However variations in e.g. atmospheric pressure
could still introduce differences between runs within a data set.)
Set A became significantly smaller than other sets because of an
off-line rejection of bad runs.  Table~\ref{table:twist-data}
summarizes data sets used for the extraction of $\delta$ and
systematic studies.  Note that set~A has much lower statistics than
other data sets.  This is why set~B was used in this work to quote typical
numbers or show example plots.
%
%================
\begin{table}[hbtp]
\begingroup
\begin{tabularx}{\textwidth}{|p{4.5em}|l|c|r|X|}
\hline {Set name}  
& Dates 
& \parbox{3.3em}{Number\\[-3pt]of runs}
& {\vrule height3.9ex width0pt
  depth3.3ex}\parbox{3.5em}{Fiducial\\[-3pt] events,\\[-3pt] millions} & Comment \\
%
% &\runnums{7967--8177}
\hline A   &Oct 8--9 & 165 & 7.9 & nominal \\
%
% &\runnums{11847--12168}
\hline B  & Nov 21--23 & 318 &15.9 & nominal\\
%
% set 1=1.96 &\runnums{13197--13536}
\hline 1.96~T &Dec 2--4 & 338 & 16.5& 1.96~T spectrometer field\\
%
% set 22 = 2.04 &\runnums{13700--13970}
\hline 2.04~T &Dec 7--9 & 240 & 12.7& 2.04~T spectrometer field \\
%
%  &\runnums{11165--12578}
\hline Cloud & Nov 6--28& 561 & 12.4& Cloud muon beam \\
%
%%%%%%%%%%%%%%%%%%%%%%%%%%%%%%%%%%%%%%%%%%%%%%%%%%%%%%%%%%%%%%%%
%
% set 7, \runnums{12759--12921}
\hline DS Al &Nov 29--30 & 160& 7.7& Outside materials systematic \\
% set 4, \runnums{7581--7957}
\hline Slightly Upstream & Oct 5--8&307 & 7.3& Stopping location systematic\\
% set 21, \runnums{8966--9374}
\hline Low rate & Oct 13--20& 338 & 17.9& Beam intensity systematic %Trigger rate
				%$1.1\times10^3$~\persec
\\
% set 11, \runnums{8542--8964}
\hline High rate & Oct 11--13& 341& 14.1&Beam intensity systematic %Trigger rate $4.7\times10^3$~\persec
\\
% set 16, \runnums{9377--9737}
\hline B2${}+{}$10G & Oct 20--21& 348& 15.4& Channel magnets systematic \\
\hline
\end{tabularx}
\caption[Data sets mentioned in the
  thesis]{\label{table:twist-data}Data sets mentioned in the thesis.
  Fiducial region is defined in chapter~\ref{chapter:reconstruction}.}
  All dates are in 2002.
\endgroup
\end{table}

%%%%%%%%%%%%%%%%%%%%%%%%%%%%%%%%%%%%%%%%%%%%%%%%%%%%%%%%%%%%%%%%
\chapter{Monte-Carlo simulation\label{chapter:monte-carlo}}
% \chapter{Monte-Carlo simulation\label{chapter:monte-carlo}}

\TWIST{} uses a very detailed Monte-Carlo simulation program (MC),
which is based on the GEANT \cite{geant3} package.  It produced
digitized output in the same format as the DAQ, except that additional
MC-specific information may be included.  Production of a large amount
of Monte-Carlo events, matching \TWIST{} data statistics
(chapter~\ref{chapter:twist-data}), was made possible by the use of
WestGrid computing facility~\cite{westgrid}.

The geometrical description of the detector contains all the
components of the hardware with which a muon or a decay positron could
possibly interact.  Each individual wire of the wire chambers is
implemented in the software.  The wire planes are offset and rotated
to their as-measured positions.  A map of the magnetic field (sections
\ref{section:magnet}), which extends to the outside of the yoke, is
used to propagate charged particles in the simulation.

The initial kinematics of an event contains a muon, and possibly other
muons and/or beam positrons.  The probability of having the pile-up
particles is determined by the specified muon and beam positron rates.
The positions and directions of flight of the muons are sampled from
experimentally measured distributions $\{x,dx/dz\}$ and $\{y,dy/dz\}$
and reproduce the observed position-angle correlations.  The beam
particles, muons and positrons, are started outside of the yoke, and
GEANT tracking propagates them through the fringe field of the
spectrometer magnet into the \TWIST{} detector.

In the \TWIST{} Monte-Carlo, unlike the standard GEANT3, the direction
of the muon spin is also tracked in the magnetic field.  The initial
spin direction is defined as antiparallel to the muon momentum.
Depolarizing interactions of the stopped muons are simulated as a step
function followed by an exponential relaxation.  Setting the initial
polarization to $-0.935$ and the time constant to $5.8\cdot10^{-5}$~s
\cite{blair-depolarization-rate} reproduces the behavior of $P_\mu(t)$
in data as it is observed for $t>1$~$\mu$s (this time limit is
discussed in chapter \ref{chapter:reconstruction}).
A mismatch in polarization between data and
Monte-Carlo would not bias the value of $\delta$ in the analysis, but
a large difference in the average $P_\mu$
%
% between the data and Monte-Carlo values 
%
could require generating larger ``derivative'' (chapter
\ref{chapter:fitting-all}) samples to achieve the same statistical
precision.

%
% Dave's mustop plots:
%
% https://twist.phys.ualberta.ca/forum/view.php?bn=twist_general&key=1052947164

The energy loss of a muon is simulated by GEANT, and determines the
stopping position of the particle.
The simulation of the muon stopping process has been validated using
special data runs with muons stopping in the middle of the upstream
part of the chamber stack.  Because each of the chamber modules is
much thinner than the stopping target, the muon stopping distribution
in these runs spreads out over several wire chambers, and the
distribution of the last (the most downstream) wire plane hit by muon
can be used to observe the shape of the stopping distribution.  It has
been shown \cite{dave-mustop-plots} that the simulation matches the
shape of the stopping distribution well, but a constant offset
equivalent to an about 86~\mum{} of additional plastic (Mylar) is
required in Monte-Carlo to match the mean stopping position of the
muons.
The peak of the stopping distribution within the target is not
directly observable in data, but the tails of the distribution are
still accessible through the last plane hit information.
The following procedure was used to determine the setting of the gas
degrader for the nominal data taking.  A histogram of the last muon
hit from Monte-Carlo, with the stopping distribution centered in the
target, was compared with similar histograms from data for different
settings of the gas degrader.  The setting corresponding to the best
match to the Monte-Carlo distribution was used for the data taking.
All nominal Monte-Carlo sets were generated with the muon stopping
distribution centered in the target.
The energy calibration procedure (chapter \ref{chapter:ecalib})
compensates for any remaining differences in the average muon stopping
position.

% Step size tuning?

A muon decay subroutine returns the energy and angle of the decay
positron with respect to the muon spin as dictated by the Michel
parameters input.  The theoretical decay spectrum includes full
$\OO(\alpha)$ radiative corrections with exact electron mass
dependence, as well as leading and next-to-leading logarithmic terms
of $\OO(\alpha^2)$, leading logarithmic terms of $\OO(\alpha^3)$,
corrections for soft pairs, virtual pairs, and an ad-hoc
exponentiation
\cite{arbuzov:2001ui,aaa-leadinglog:2002,arbuzov:2002cn,arbuzov:2002rp,arbuzov:2003ce}.
The actual implementation of the muon decay spectrum is separated from
the rest of the Monte-Carlo code to make possible a blind analysis, as
is explained in section \ref{section:blind-analysis}.

% KO's tuning of the clustering parameters:
%
% https://twist.phys.ualberta.ca/forum/view.php?site=twist&bn=twist_general&key=1053913884
%
% PC TDC time and width spectra:
%
% https://twist.phys.ualberta.ca/forum/view.php?site=twist&bn=twist_montecarlo&key=1087421585

The wire chamber response is simulated by randomly creating ionization
clusters along the path of a charged particle when it crosses a drift
cell, calculating the drift time of each cluster to the wire, and
simulating the overlap of different clusters to produce an
above-the-threshold signal.  The obtained time of the signal is
smeared to simulate electronics effects.  The parameters of the method
are derived from a detailed GARFIELD \cite{garfield} simulation study
\cite{selivanov-tn-81}, and the ``electronics'' smearing is tuned to
\TWIST{} data.  The critical piece of information affecting the
accuracy of track reconstruction (chapter
\ref{chapter:reconstruction}) is the time of the leading edge of DC
chamber signals. A good match of the simulated distribution to
\TWIST{} data has been demonstrated
\cite{olchansk-chamber-clustering}.

% : the average cluster spacing, the time duration of a
% (rectangular) cluster, the average number of clusters required to
% overlap to produce a detectable signal, and the time smearing,
% 
% are chosen depending on the wire chamber type and the particle type.
% 
% For the primary particle of interest, the decay
% positron, the parameters are constants 
% 
% The dependence of the cluster spacing on the particle energy is
% approximated by a constant 
% 
% Since the analysis (chapter \ref{chapter:reconstruction}) does not
% requre a reconstruction of the muon track, and since the drift
% properties of PC chambers are not used by the reconstruction, the most
% attention was paid to reproducing the signal that a positron creates
% in a DC cell.
% 

Full GEANT physics interactions are enabled, so such processes as
creation of delta-electrons, or conversion of a bremsstrahlung photon
into an electron-positron pair, may create additional hits on the
detector wires.  

%dave: The thresholds of 10~keV for muons and 20~keV for
%dave: positrons and gammas were used in MC production.  Test runs at
%dave: different thresholds showed no sensitivity to that settings.

% Sensitivity to MC energy thresholds:
%
% https://twist.phys.ualberta.ca/forum/view.php?bn=twist_physics&key=1082585671

% TDC ``gate'' and overflow?

%%%%%%%%%%%%%%%%%%%%%%%%%%%%%%%%%%%%%%%%%%%%%%%%%%%%%%%%%%%%%%%%
%\section{Accuracy of the simulation of positron interactions}

% An important consideration in a validation of a simulation is that the
% validation must not ``tune out'' the effect we are trying to measure.

Interactions of decay positrons in detector materials distort the
reconstructed spectrum and can lead to biases in the values of the
measured decay parameters.  \TWIST{} relies on Monte-Carlo to
compensate for these effects (chapter~\ref{chapter:fitting-all}).
Knowing the accuracy of the simulation of the interactions is
important to determine the corresponding systematic uncertainty
(chapter~\ref{chapter:systematics}). To verify a claim that in GEANT3
\begin{quote}
the cross-sections of the electromagnetic processes are well
reproduced (within a few percent) from 10 keV up to 100 GeV, both for
light (low Z) and for heavy materials
\end{quote}
(\cite{geant3}, section PHYS001), data runs with muons stopped in the
upstream end of the detector have been taken.  In these runs a decay
positron emitted in the downstream direction crosses the whole
spectrometer.  The upstream and downstream halves of the detector can
be used as two independent devices to measure the momentum and angle
of the positron before and after the central target, thus allowing an
extraction of the energy loss and angular scattering distributions
from data. These distributions can be compared with similar
distributions obtained from a corresponding Monte-Carlo simulation.
The Monte-Carlo validation data were taken in 2003, with a high purity
aluminum stopping target of a known thickness.  Thus the validation,
unlike the final result, is not affected by the uncertainty in the
thickness of the graphite coating on the Mylar stopping target used
for the main 2002 data sets.

%%%%%%%%%%%%%%%%%%%%%%%%%%%%%%%%%%%%%%%%%%%%%%%%%%%%%%%%%%%%%%%%

% Using Rob's analysis of 2003 data,
%       usdsdiffs_data_altgt_5cmtgtradius_040903.root
% and   usdsdiffs_mc_altgt_5cmtgtradius_040903.root
%
% root [0] .x plot_dpcosth.C
% N    data = 76844,       mc = 86036
% mean data = -0.126944,   mc = -0.122867, diff = -0.0040764 +- 0.00130904
% RMS  data = 0.268787,    mc = 0.257955

\begin{figure}
\begin{tabular}{c}
%\includegraphics[height=0.45\textheight,width=\textwidth,keepaspectratio=false]{plots/monte-carlo/eps/dpcosth.eps}
%\\
%\includegraphics[height=0.45\textheight,width=\textwidth,keepaspectratio=false]{plots/monte-carlo/eps/dpcosth_logy.eps}
%
%
\includegraphics[width=\textwidth]{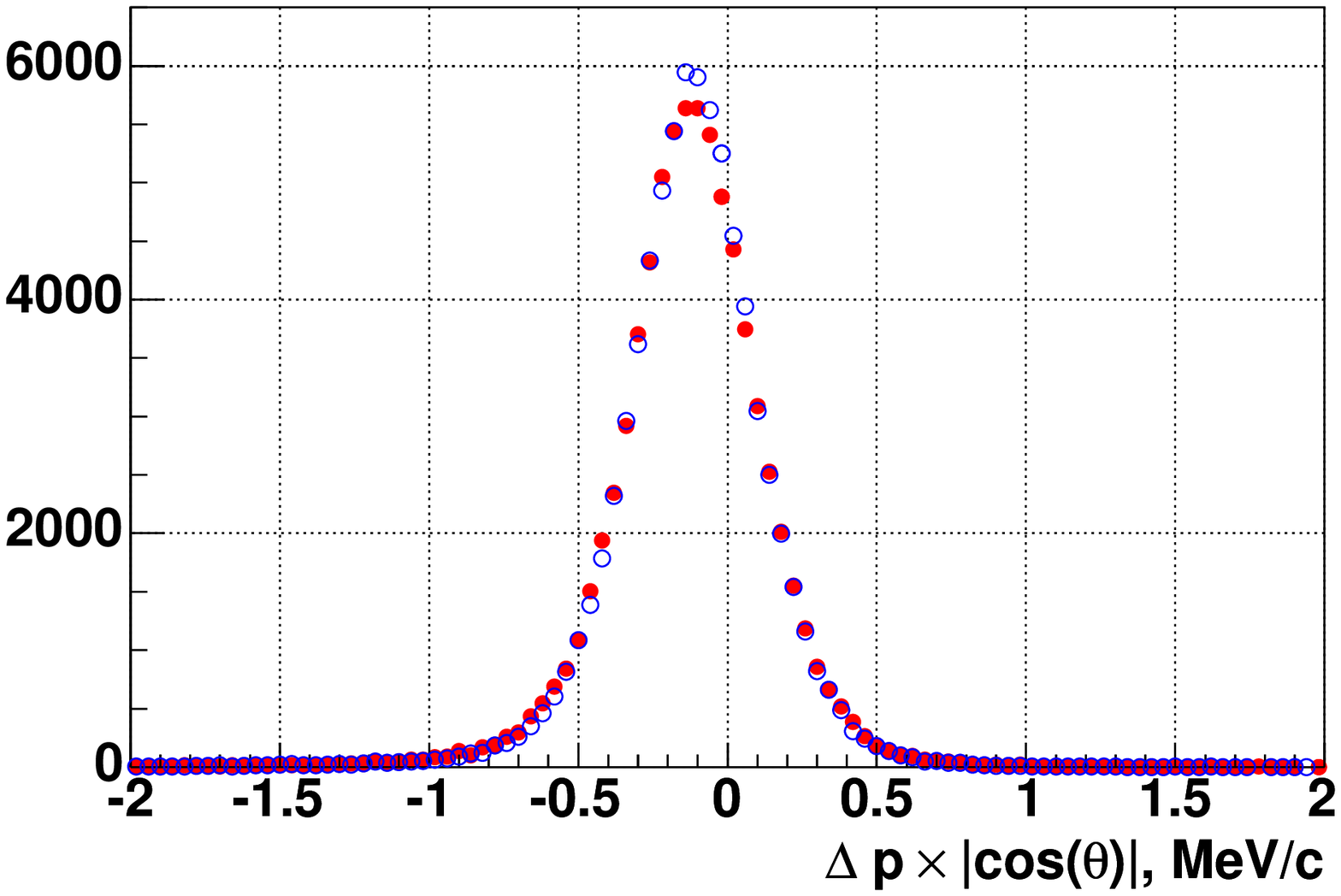}
\\
\includegraphics[width=\textwidth]{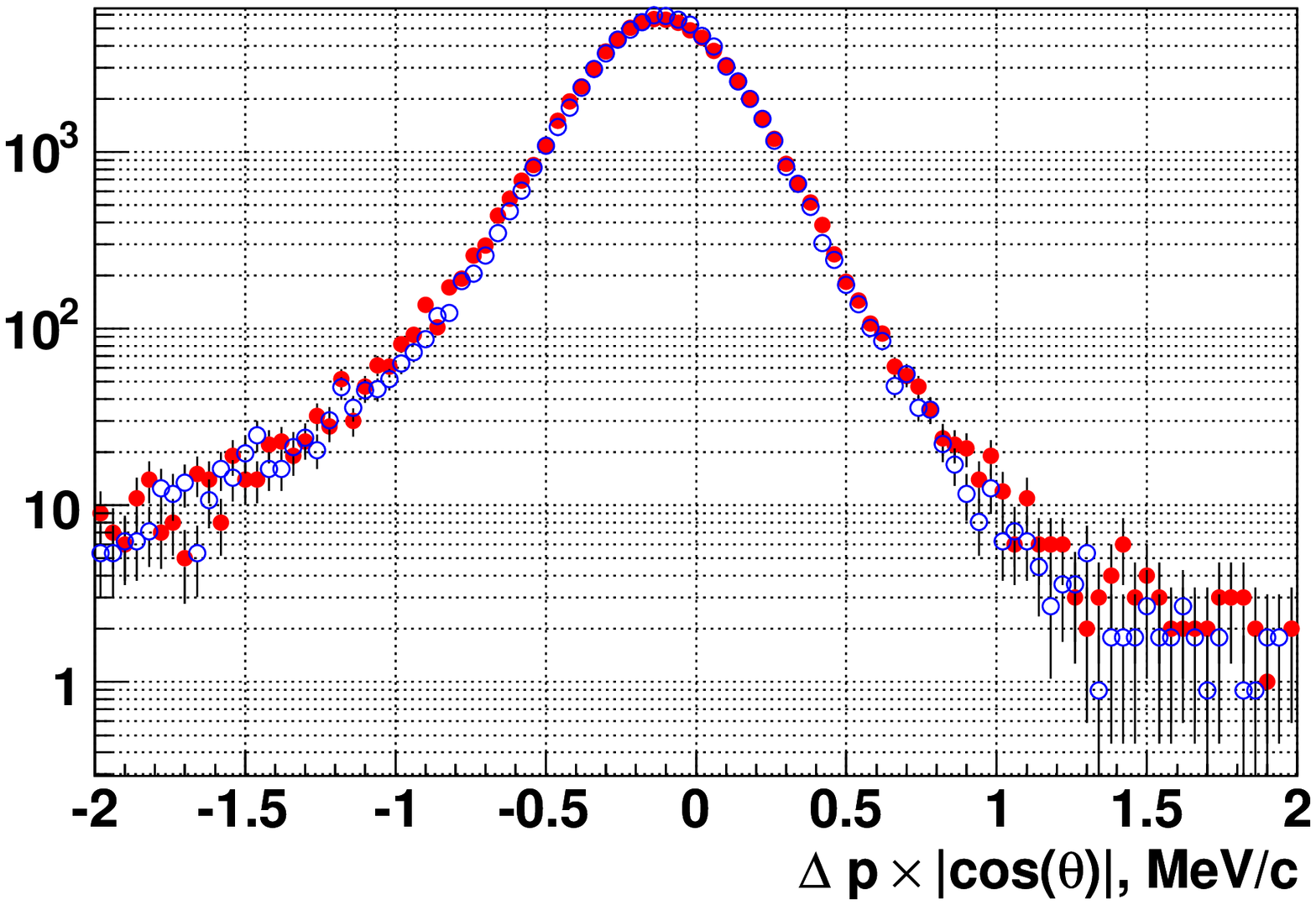}
\end{tabular}
\caption[Positron momentum loss in data and
  Monte-Carlo]{\label{fig:validation-dpcosth}Positron momentum loss
  in data (solid) and Monte-Carlo (empty markers). Top: linear
  vertical scale, bottom: logarithmic scale.  The Monte-Carlo
  histogram is normalized to data.  
  Mean values of the distributions are $-126.9$~keV/c (data) and
  $-122.9$~keV/c (MC), the difference is $4.1\pm1.3$~keV/c.
  The RMS is 0.269~keV for data, 0.258~keV for MC. (The RMS in this
  study does not represent \TWIST{} momentum resolution, see text.)
  Analysis by Rob MacDonald.  }
\end{figure}

% 
% root [2] .x plot_thdiff.C                                                                 
% N    data = 76844,       mc = 86036
% mean data = -0.00289897,         mc = -0.00189999, diff = -0.000998987 +- 8.18977e-05
% RMS  data = 0.0164523,   mc = 0.0165532
% Info in <TCanvas::Print>: eps file theta_diff.eps has been created

\begin{figure}
\begin{tabular}{c}
\includegraphics[width=\textwidth]{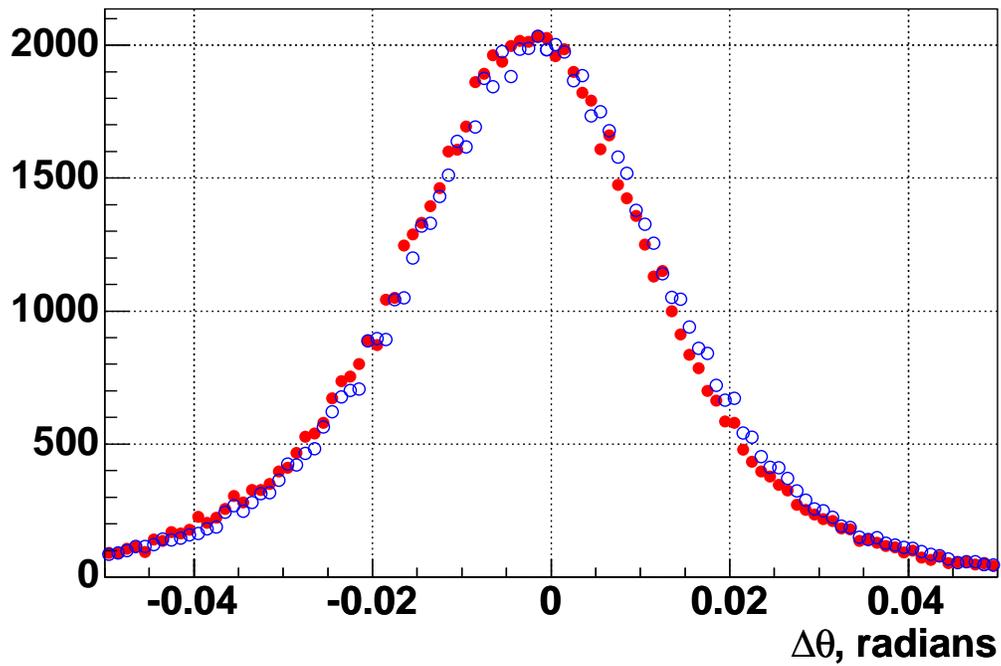}
\end{tabular}
\caption[Positron angle change in data and
  Monte-Carlo]{\label{fig:validation-thetadiff}Positron angle change
  in data (solid) and Monte-Carlo (empty markers). The Monte-Carlo
  histogram is normalized to data.  The mean angle change is
  $-2.9$~mrad for data, $-1.9$~mrad for MC.  The RMS is 16.4~mrad for
  data, 16.6~mrad for the Monte-Carlo.  (The RMS in this study does
  not represent \TWIST{} angular resolution, see text.) Analysis by
  Rob MacDonald.}
\end{figure}

%%%%%%%%%%%%%%%%%%%%%%%%%%%%%%%%%%%%%%%%%%%%%%%%%%%%%%%%%%%%%%%%

Fig.~\ref{fig:validation-dpcosth} shows an overlay of the momentum
loss distributions for data and Monte-Carlo.  There
$\Delta{p}=p_{\text{downstream}}-p_{\text{upstream}}$ is the
difference between the positron momenta reconstructed by the two
halves of the detector.  In \TWIST{} geometry momentum loss of a
positron is proportional to $1/|\costh|$ (see
chapter~\ref{chapter:ecalib}); a factor of $|\costh|$ is used in the
study to compensate for a possible difference in the angular
distributions of the emitted positrons between data and Monte-Carlo
while comparing the momentum losses.  The difference between the data
and the simulation does not exceed 5\% for both the average losses,
and the widths of the distributions.  To separately look at ``hard''
processes, an arbitrary bound of 1~MeV/c was defined.  The tails of
the distributions shown on Fig.~\ref{fig:validation-dpcosth} were
integrated from $-\infty$ to $-1.12$~MeV/c (i.e. to 1~MeV/c below the
average).  The discrepancy in the fraction of such ``hard scatter''
events was found to be 14\%.  \emph{Therefore the numbers we used to
estimate a systematic uncertainty due to the quality of GEANT
simulation of positron interactions are 14\% for hard interactions,
and 5\% for intermediate and soft interactions.}

The distribution of
$\Delta\theta=\theta_{\text{downstream}}-\theta_{\text{upstream}}$ is
shown on Fig.~\ref{fig:validation-thetadiff}.  The width of the
distributions agrees to about 1\%, but there is a noticeable shift in
the mean value of $\Delta{\theta}$.  The interpretation of the mean
value of this distribution is complicated.  Since there is more phase
space at higher angles due to the $\sin\theta$ factor, a naive
expectation is to observe a small positive shift in $\Delta\theta$.
This has been confirmed by a Monte-Carlo study \cite{andr-mean-theta},
which demonstrated a ${}\sim1$~mrad positive bias. An effect of a
non-uniformity of the magnetic field is negligible
\cite{andr-mean-theta}.  However the distribution of the
\emph{reconstructed} $\Delta\theta$ has a negative mean for both data
and Monte-Carlo.  Thus is should be attributed to biases in the track
reconstruction, which affect the two halves of a track in a different
way.  At least two causes for such biases are known.  In the GEANT
validation studies positrons originate in PC~4, and no information
from PC~1--3 is available for pattern recognition
(chapter~\ref{chapter:reconstruction}) for the upstream part of a
track, but full information is available for the downstream part.
Also, the track fitting always assumes that a positron originates in the
stopping target to determine the sign of a time-of-flight correction
to drift chamber hits.  Therefore the correction was applied with the
wrong sign for the upstream parts, and with the correct sign for the
downstream parts of tracks in the validation study.  Such biases are
common for the data and the simulation, and may shift the
distributions to negative $\Delta\theta$.  A possible explanation for
the difference in the central values between data Monte-Carlo is a
misalignment of the detector to the magnetic field.
Chapter~\ref{chapter:systematics} evaluates a systematic uncertainty
associated with such a misalignment.

It has to be noticed, that the widths of the distributions on
Fig.~\ref{fig:validation-dpcosth} and
Fig.~\ref{fig:validation-thetadiff} do not represent \TWIST{}
resolutions.  The lack of information from PC1--3, and the wrong
time-of-flight correction mentioned above worsen the tracking quality.
Moreover, a positron in these studies goes through twice as much
material as a positron from a muon stopped in the target.  Also, a
track is reconstructed in the two halves of the detector, so random
fitting errors contribute twice.

% The final 14% hard scatter systematics number:
% 
% https://twist.phys.ualberta.ca/forum/view.php?site=twist&bn=twist_physics&key=1094238154
%
% NB: the above uses Al target to get necessary statistics.

% 
% 
% \rrr{The accuracy of the hard scatter simulation: tail counts for dP more
% than 1MeV below the average.  2002 numbers: consistent, 2003: 14\%
% difference.}
% 
% 
% \label{fig:rob-style-plots}
% 
%\rrr{EC and cut histograms below may be looked at as MC verifications.}

%%%%%%%%%%%%%%%%%%%%%%%%%%%%%%%%%%%%%%%%%%%%%%%%%%%%%%%%%%%%%%%%
\chapter{Spectrum reconstruction\label{chapter:reconstruction}}
% \chapter{Spectrum reconstruction\label{chapter:reconstruction}}
\begingroup
\newcommand\evttype{Event Type}
\newcommand\dkwintime{Decay Window Time}

The primary purpose of the spectrum reconstruction is to produce a
2-dimensional spectrum of decay positrons in momentum and angle using
information available from the \TWIST{} detector.  

% The \TWIST{} method of extraction of the decay parameters
% (chapter~\ref{chapter:fitting-all}) relies on a comparison of
% reconstructed data and Monte-Carlo spectra, and effectively takes away
% any reconstruction biases that are common to data and the simulation.
% Therefore a goal in the development of the reconstruction codes was
% a minimization of the dependence of the result to various
% imperfections of the simulation, which miminizes the systematic error
% (chapter~\ref{chapter:systematics}).

Real and simulated data are analyzed by the same reconstruction code
in essentially the same way. The only non-trivial difference in
treating the two cases is the crosstalk removal, which is discussed
below.  Technically the analysis was done in two stages.  The first
stage, required a large amount of computations and was performed at
WestGrid~\cite{westgrid}.  It consisted of the following steps:
\begingroup
%\ifagsdouble\onehalfspacing\else\relax\fi
\begin{description}

\item[Crosstalk removal.]  A hit on a wire may induce a signal in a
  different channel through electronics crosstalk.  Most of crosstalk
  in \TWIST{} is induced by highly ionizing muons, therefore it does
  not affect decay positrons a microsecond later.

  The effect has been studied in detail in the hardware by pulsing a
  channel and observing the crosstalk signal on an oscilloscope.
  \label{page:crosstalk-algorithm}Crosstalk signals have
  characteristics allowing for their identification and removal by
  software \cite{lalancette-crosstalk}: there is a 5--65~ns delay from
  the generating pulse, and the width of a crosstalk signal
%
% (typically ${}\approx20$~ns)
%
  is much smaller than the width of a real signal.

  An algorithm utilizing these characteristics removes crosstalk hits
  from real data events before any other analysis is done.  Because
  crosstalk is not simulated by \TWIST{} Monte-Carlo, and also because
  the width of chamber signals in the simulation is not tuned to
  data, this algorithm is not invoked on Monte-Carlo events.

\item[Windowing.] Chamber hits are grouped in time.  Different groups
  correspond to tracks from different particles.

% Blair's postings on classification:
%
% https://twist.phys.ualberta.ca/forum/view.php?site=twist&bn=twist_software&key=1048842376
%
% +Corrections from
% https://twist.phys.ualberta.ca/forum/view.php?site=twist&bn=twist_software&key=1048884700
%
\item[Classification.]  For each time window, characteristics of its
  chamber hits are used to guess what process has occurred
  there~\cite{blair-classifications}.  For example, a decay positron
  hits wire chambers in only one half of the detector, while a beam
  positron crosses the whole detector.  Based on the number and type
  of time windows in an event, and the times of different windows, a
  classification code is assigned to the
  event~\cite{blair-classifications}.
  
% KO's FG postings:
% 
% Introduced multitrack ability, combinatorics:
% https://twist.phys.ualberta.ca/forum/view.php?site=twist&bn=twist_software&key=1081884201
%
% Final 2002 analysis:
% https://twist.phys.ualberta.ca/forum/view.php?site=twist&bn=twist_software&key=1086409076
%
% My posting on magic wavelength:
% 
% https://twist.phys.ualberta.ca/forum/view.php?site=twist&bn=twist_software&key=1050476690

\item[Pattern recognition.]  

  A cluster of consecutive wires hit in a plane determines only one
  transverse coordinate of a passing track.  Information from an
  orthogonal pair of wire planes can be used to determine a ``space
  point'' on the track.  

  For a time window, the pattern recognition uses space points formed
  by the hits belonging to the window to find parameters of possible
  helical tracks passing through the points.  It uses a combinatorial
  technique \cite{olchansk-firstguess-1,olchansk-firstguess-2} and can
  find more than one track per window.

% dt_geo.00038:
%
% Sparse:
%   
%    24     4.9998 0.0  -135.0        80
% 
%    26    10.1982 0.0  -135.0        80
% 
%    28    17.3965 0.0  -135.0        80
% 
%    30    22.5961 0.0  -135.0        80
% 
%    32    29.7944 0.0  -135.0        80
% 
%    34    34.9940 0.0  -135.0        80
% 
%    36    42.1922 0.0  -135.0        80
%
% The Z diffs are: ~5,7,5,7,5,7 cm
%
%
% 
% Dense:
% 
%    37    46.9921 0.0   135.0        80
%    38    47.3921 0.0  -135.0        80
%    39    47.7921 0.0   135.0        48
%    40    48.1921 0.0  -135.0        48
%    41    48.5921 0.0  -135.0        48
%    42    48.9921 0.0   135.0        48
%    43    49.3921 0.0  -135.0        48
%    44    49.7921 0.0   135.0        48

  A feature of the arrangement of the wire chamber modules in the
  detector, which can be seen on Fig.~\ref{fig:stack}, is that the
  distances between the corresponding planes of different pairs form a
  pattern: 5.2~cm, 7.2~cm, 5.2~cm, 7.2~cm, 5.2~cm, 7.2~cm, with only
  two independent distances, for the 7 modules closest to the stopping
  target on either side.  That means that a helix with the wavelength
  $L=5.2+7.2=12.4$~cm has only two distinct measurements of its
  transverse position, so its radius can not be reconstructed using
  the space
  points~\cite{andr-magic-wavelength-1,andr-magic-wavelength-2}. (In a
  projection along the detector axis, the 7 space points collapse into
  only 2 points on the circle---projection of the helix.)

  The ``dense stacks'' of wire planes at the outer edges of the
  detector, along with information from PC chambers, help to resolve
  the ambiguity. However the worsening of the quality of
  reconstruction of such tracks still prompted the introduction of a
  fiducial cut on $p_z$, described below, to stay away from the
  ``magic wavelength'' zone.

\item[Wire center fits.]

  Tracks are fit to the positions of hit wires in a time window. The
  ``narrow windows'' technique \cite{James:1982mg} is used.  To
  account for multiple scattering, kinks at the positions of sparse
  stack chamber modules are allowed \cite{Lutz:1987vs}.  The magnetic
  field map (section~\ref{section:magnet}) is used.  The resulting fit
  has sufficient precision to resolve most left-right ambiguities.
  
\item[Track fitting]
  
  The parameters of the tracks are refined using drift time
  information.  The necessary space-time relations are obtained from
  GARFIELD.  Kinks and magnetic field map are used.

\end{description}
\endgroup
%
%%================================================================
%% Set B
\begin{figure}[tbp]
\begin{tabular}{c}
\includegraphics[width=\textwidth]{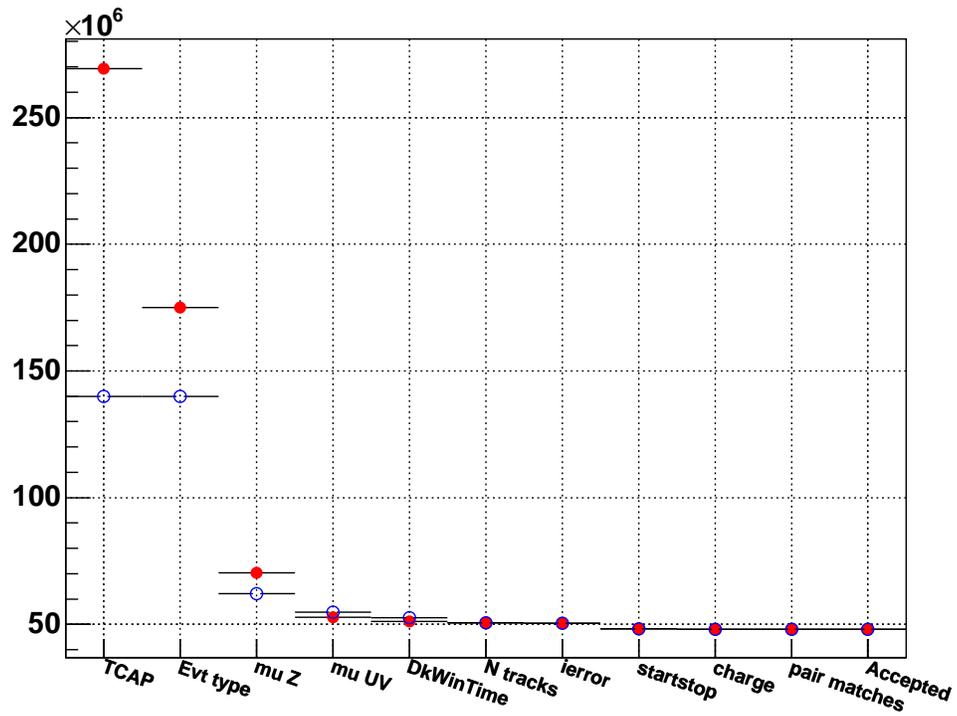}\\
\includegraphics[width=\textwidth]{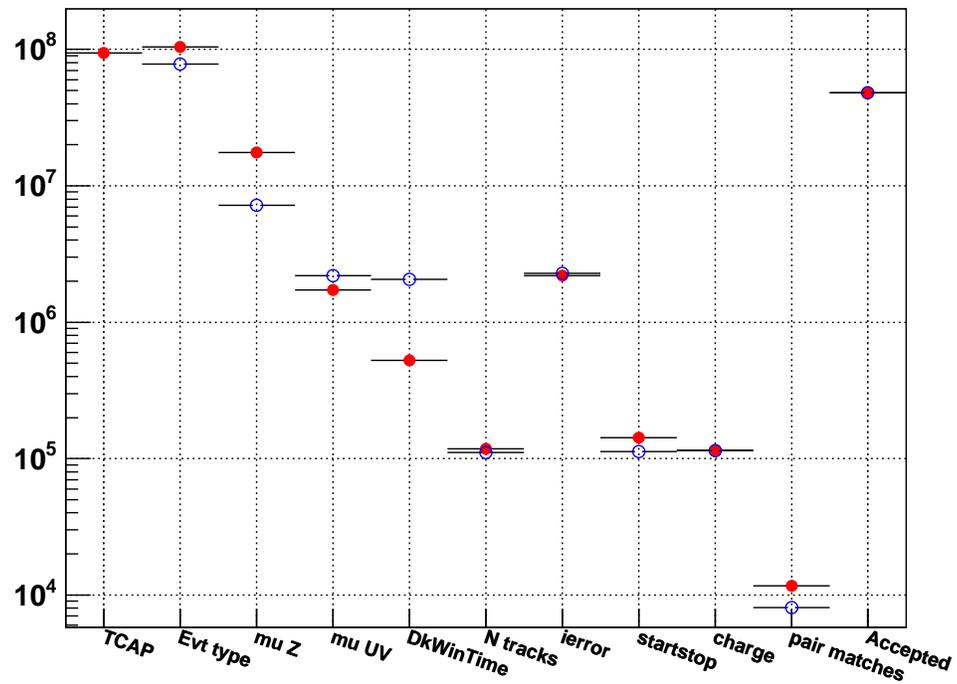}
% \framebox[\textwidth]{\vbox to0.4\textheight{\vfill}{\Large Events
%     before cut}}\\
% \framebox[\textwidth]{\vbox to0.4\textheight{\vfill}{\Large Events
%     rejected by cut, log scale}}
\end{tabular}
\caption[Cut passage]{\label{fig:rec-cuts}
  Top: each bin contains the number of events before the cut; bottom:
  the number of events rejected by cut.
  Solid markers: data, empty markers: MC.  Monte-Carlo is normalized
  to data by matching the number of accepted events.
}
\end{figure}

No cuts on events were imposed at the first stage, and a summary of
each event was written out in a ``ROOT tree'' format~\cite{root}.  The
output data were subsequently analyzed to select events and tracks to
be used in the final spectrum.  The following cuts, illustrated on
Fig.~\ref{fig:rec-cuts}, were used.
% 
%  Note that events where the muon decayed earlier than \rrr{about
% 1~\mus{}} after the trigger were classified as ``DC overlap'' events,
% and thus rejected by the \evttype{} cut which is before the
% \dkwintime{} cut.
%
\begingroup
%\ifagsdouble\onehalfspacing\else\relax\fi
\begin{description}

\item[TCAP.]  This is a cut on the time of flight of the trigger
  particle through the M13 beam line.  It selects muons from stopped
  pions that decayed in between the proton pulses from the cyclotron,
  and rejects ``cloud'' muons (section~\ref{section:m13}).  The
  purpose of the cut is to improve muon polarization.
  It also rejects triggers from beam pions and prompt positrons.

  The time of flight cut is only applied to data events.  The
  Monte-Carlo does not simulate the 43~ns time structure of the beam,
  and is instead produced using measured parameters of the surface
  muon beam with the TCAP cut applied.

\item[\evttype.] This cut uses the event classification code, and is a
  combination of several requirements.  The event must be triggered by
  a muon, and not a beam positron.  There must be only one identified
  muon.  A unique decay time window must be identified.

  The decay must happen at least 1.05~\mus{} after the trigger to
  ensure that DC hits with the longest drift time ${}\sim1$~\mus{}
  from the muon do not affect reconstruction of the decay positron.
  Since the muon life time is about 2.2~\mus{}, this requirement
  alone rejects 38\% of events.

  Pile-up beam positrons are allowed, but they must be well separated
  in time (at least 1.05~\mus{}) from both the muon and the decay
  positron to avoid an overlap of DC hits.  Events with DC overlap
  constitute another major fraction of all events rejected by the
  \evttype{} cut.  It is important to note that rejection of DC
  overlap events does not introduce a bias in the spectrum, because
  the probability of an overlap does not depend on the momentum and
  angle of the decay positron.

  On the other hand, beam particles within ${}\sim100$~ns of the decay
  can not be reliably separated by the classification code from such
  processes as creation of delta-electrons or backscattering of the
  decay positron.  Therefore all such ``PC overlap'' events are kept
  at this stage.

\item[Muon Z.]  Require that the event is consistent with the muon
  stopped in the central target,  i.e. require that the last chamber
  hit by the muon is PC~6.

\item[Muon radius.] Constrain muon stopping position in the transverse
  plane to $r<2.5$~cm.

\item[\dkwintime.]  Require the decay to occur before 9~\mus{} from
  the trigger, so that long drift time hits are not cut off by the
  10~\mus{} DAQ limit.

\item[N tracks] Require at least one track candidate, as defined by
  the pattern recognition, in the decay window.

% \item[zero weight]

\end{description}
\endgroup
All the cuts above are essentially event-level cuts.  An event may
contain more than one successfully reconstructed track.  There are
several reasons for multiple tracks to be found.  For example, due to
a hard scatter a decay track may be split into two helical parts,
before and after the scatter, by the pattern recognition.  That would
lead to two fit tracks for a single decay positron.  Or a decay
positron may backscatter off material outside, re-enter the tracking
volume, and produce a second track, which may cross the whole
detector.  Such events may be indistinguishable from a decay being
overlapped by a beam positron particle.  A genuine beam positron
overlap is yet another possibility to produce multiple tracks.

% My posting on a comparison of the tree summings:
% 
% https://twist.phys.ualberta.ca/forum/view.php?site=twist&bn=twist_physics&key=1093914516

% Number of events vs cut, relevant distributions before each cut for
% data and MC.

The challenge is to identify the ``correct'' track to be included in
the spectrum.  It is handled as following~\cite{andr-tree-summing}.  A
set of ``decay candidate tracks'' is created.  Initially it consists
of all tracks identified in the decay window by the pattern
recognition.  Then each of the candidates is subject to the cuts
below.  Failed candidates are eliminated from the set.  A
corresponding event-level cut is defined as the requirement that the
set of decay candidates is not empty after the track cuts.
\begingroup
%\ifagsdouble\onehalfspacing\else\relax\fi
\begin{description}

\item[ierror] Eliminate track candidates that did not produce a
  successful fit.

\item[startstop] Require that the ends of a fit track are on the
  ``correct side'' of the detector, as determined by the
  classification.  I.e. the track begins and ends in the upstream half
  for upstream decay type events, in the downstream half for
  downstream decay type events.  Tracks that cross the central target
  are always rejected here. 

\item[charge] The direction of the winding of the helix must be
  consistent with a positive particle that originated in the central
  target.

\item[pair matches] This piece of code essentially attempts to
  ``glue'' pieces of tracks that were split in the reconstruction
  because of a hard scatter~\cite{andr-tree-summing}.

  The cut is done in two stages.  First, for each track among the
  decay candidates, a set of ``anti-tracks'' is found as described
  below.  An ``anti-track'' for a decay candidate is a track in the
  decay window that, if successfully ``glued'' to the candidate, would
  indicate that the candidate track should not be used.

  Then the closest distance of approach\footnote{The distance used is
  the shortest distance between two tracks in the transverse plane,
  not in 3D.}  (CDA) is calculated between the candidate and each of
  its anti-tracks.  If CDA is less than 0.5~cm for at least one
  anti-track, the candidate is rejected: probably the anti-track and
  the candidate track belong to the same particle.  The decision is
  done on purely geometrical grounds: do the two tracks intersect?
  They can belong to the same particle only if they do.

  % The CDA distribution is shown on
  % Fig.~\ref{cda}, the Z of the closest approach of track pairs on
  % Fig.~\ref{pair_zmin}.

  \textbf{Selection of anti-tracks}

  The set of anti-tracks for a decay candidate is found through the
  following procedure:  
  \begin{enumerate}
  \item Start with all good fits in the decay window, but exclude the
    decay candidate itself, to form a set of anti-track candidates.

  \item Exclude all tracks which overlap with the decay candidate in
    Z.  If two tracks have hits in the same DC plane, they can't be
    from the same particle.  (Sharing of DC hits in not allowed in the
    track fitter. One DC hit could belong to no more than one fit
    track.)

  \item Exclude all tracks which are in the same half of the detector
    and farther from the target than the decay candidate.  
    \vc{We want to keep the decay candidate because it is
    closer to the target, and therefore provides more accurate
    information on the momentum and angle of the positron at the decay
    point, even if the anti-track belongs to the same particle.}

  \end{enumerate}
  % Fig.~\ref{anticuts} shows the mean number of anti-track candidates
  % satisfying the criteria above.

\end{description}
\endgroup
%
%%================================================================
%% Set B
\begin{figure}[tbp]
\begin{tabular}{c}
% Natural size:
%\includegraphics[width=\textwidth,height=0.726\textwidth,keepaspectratio=false]{plots/reconstruction/eps/extra_tracks_p.eps}
% To fit |pz| cut on this page and keep the next page unaffected
\includegraphics[width=\textwidth,height=0.66\textwidth,keepaspectratio=false]{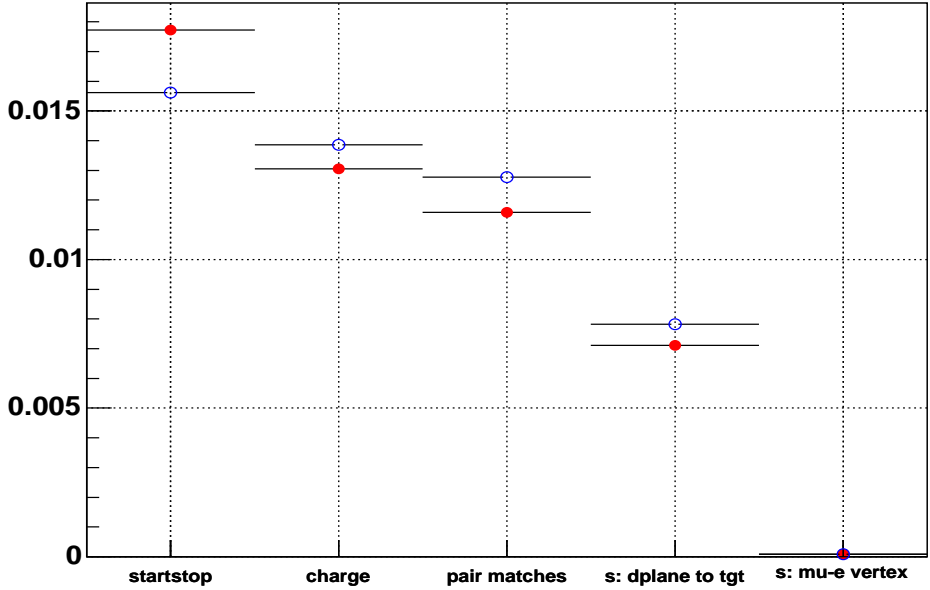}
\end{tabular}
\vspace*{-12pt}
\caption[Extra tracks]{\label{fig:num-extra-decay-candidates-profile}
  The mean number of extra decay track candidates before cut or
  select. Solid markers: data, empty markers: MC.
}
\end{figure}
%%================================================================
%
At this point an event is accepted and will produce an entry in the
final spectrum.  However there is still no guarantee that a unique
decay track has been identified.  In fact about 0.7\% of events still
contain more than one decay track candidate, as can be seen on
Fig.~\ref{fig:num-extra-decay-candidates-profile}.  The following
algorithm~\cite{andr-tree-summing} is used to decide among the
multiple candidates.  These ``selects'', unlike cuts, never reject all
candidates, but choose one.
\begingroup
%\ifagsdouble\onehalfspacing\else\relax\fi
\begin{description}

\item[dplane to target] Find the DC plane closest to the central
  target, which is used by a decay candidate.  Eliminate all
  candidates that do not use a hit from that plane, and therefore
  start farther from the target.

\item[mu-e vertex] The remaining ${}\lesssim10^{-4}$ fraction of
  multiple track cases is resolved by the proximity of extrapolation
  of the positron tracks to the muon stopping position.  Since the
  error on the two muon coordinates is significantly different because
  of the large scattering near its stopping position, an elliptical
  metric is used:
  $R_{\text{ellipse}}=(U_e-U_\mu)^2+(V_e-V_\mu)^2/\sigma_{vu}^2$, with
  $\sigma_{vu}=1.7$ found empirically by comparing RMS of
  muon-positron mismatch in the $U$ and $V$ directions.

  The code is guaranteed to select only one decay track per event.

\end{description}
\endgroup

The momentum and $\costh$ values of the selected tracks represent an
unbinned decay spectrum.  This ``raw'' spectrum was used to perform
the energy calibration procedure, described in
chapter~\ref{chapter:ecalib}.  Then calibration results were applied
to the unbinned decay spectrum (chapter~\ref{chapter:ecalib}), and the
corrected spectrum was filled into a 2-dimensional histogram in
momentum and $\costh$, which was further used to perform a fit and
extract values of the muon decay parameters
(chapter~\ref{chapter:fitting-all}).  An example of a reconstructed
data spectrum is shown in Fig.~\ref{fig:michel-data}.

\label{page:fiducial}The fiducial region used in the extraction of
$\delta$ is defined by the following constraints:
\begingroup
%\ifagsdouble\onehalfspacing\else\relax\fi
\newcommand{\bb}[1]{\boldsymbol{#1}}
\begin{description}
\item[$\bb{p}\bb{<}\bb{50}$~MeV/c] The shape of the reconstructed spectrum at the end
  point is defined by the detector resolution.  On the other hand, in
  the bulk of a smooth spectrum its distortion due to detector
  resolution is a second order effect.  Excluding the end point region
  from the decay parameter fits (chapter~\ref{chapter:data-fits})
  drastically reduces the sensitivity of the result to discrepancies
  between resolutions for real and simulated data.

  Another reason to stay away from the end point is to keep the decay
  parameter fits statistically independent from the energy calibration
  fits (chapter~\ref{chapter:ecalib}).

\item[$\bb{|p_z|>13.7}$~MeV/c] This cut eliminates tracks affected by the
  ``magic wavelength'' problem, which is discussed above in this chapter.

\item[$\bb{p_t<38.5}$~MeV/c] That requirement, together with the $r<2.5$~cm
  cut on the muon stopping position, insures that the decay positron
  track is radially contained within the instrumented region of the
  detector.

\item[$\bb{0.50<|\cos\theta|<0.84}$] Events at high angles (small
  $|\costh|$) are more affected by multiple scattering and momentum
  struggling, leading to worse resolution.  They are also more
  difficult to reconstruct.

  At small angles, it becomes difficult to determine the wavelength
  ($p_z$) of the tracks.  Reconstruction biases observed in this
  region lead to a deviation of the average energy loss
  $p_{\text{rec}}-p_{\text{MC}}$ from the $1/|\costh|$ behavior
  required for energy calibration (chapter~\ref{chapter:ecalib}).

  %\cite{andr-small-angle-biases}
  
\end{description}
\endgroup 
The shape of the fiducial region in the momentum and $\costh$ plane
can be seen in the upper left panel of
Figs.~\ref{fig:residuals-setA}--\ref{fig:residuals-cloud}.

% My posting on resolutions
%
% https://twist.phys.ualberta.ca/forum/view.php?site=twist&bn=twist_physics&key=1101774660
% --- the plots, and FWHMs.
%
% https://twist.phys.ualberta.ca/forum/view.php?site=twist&bn=twist_abstracts&key=1102037717
%
% Note: A Gaussian fit to the (rec-MC) distribution of cos(theta) gives
% results from
% 
% sigma=0.0059 (self-consistent fit within +-3 sigma)
% 
% to 
% 
% sigma=0.0043 (self-consistent fit within +-1 sigma).
% 
% The FWHM of the distribution is 0.01 (see my previous posting), that
% "corresponds" to sigma=0.01/2.3=0.0043.  The RMS may be 0.018,
% depending on how much tail you include.
% 
% Having the choice among those coarse numbers, we decided to quote
% 0.005 as the cos(theta) resolution.
% 

The average resolutions in the fiducial region are: 150~keV (FWHM) for
momentum, 0.015~rad (FWHM) for the $\theta$ angle, and 0.01 (FWHM) for
$\costh$.  Distributions of the differences of reconstructed and
Monte-Carlo values for events reconstructed within the fiducial are
shown on Fig.~\ref{fig:resolutions}.

% %%================================================================
% %% Set B
% \begin{figure}[tbp]
% \begin{tabular}{cc}
% \includegraphics[width=0.5\textwidth]{plots/reconstruction/eps/michel_data.eps} 
% & \framebox[0.5\textwidth]{\vbox to0.4\textheight{\vfill}{\Large P spectrum}}\\
% \framebox[0.5\textwidth]{\vbox to0.4\textheight{\vfill}{\Large $\costh$ spectrum}}
% & \framebox[0.5\textwidth]{\vbox to0.4\textheight{\vfill}{\Large Asymmetry(p)}}
% \end{tabular}
% \caption[Data spectrum from set B]{\label{fig:michel-data}
%   Reconstructed data spectrum from set~B.
% }
% \end{figure}

%%================================================================
%% Set B
\begin{figure}[tbp]
\begin{tabular}{c}
\raisebox{0.13\textheight}{{\large\textbf{a)}}}\includegraphics[width=0.95\textwidth]{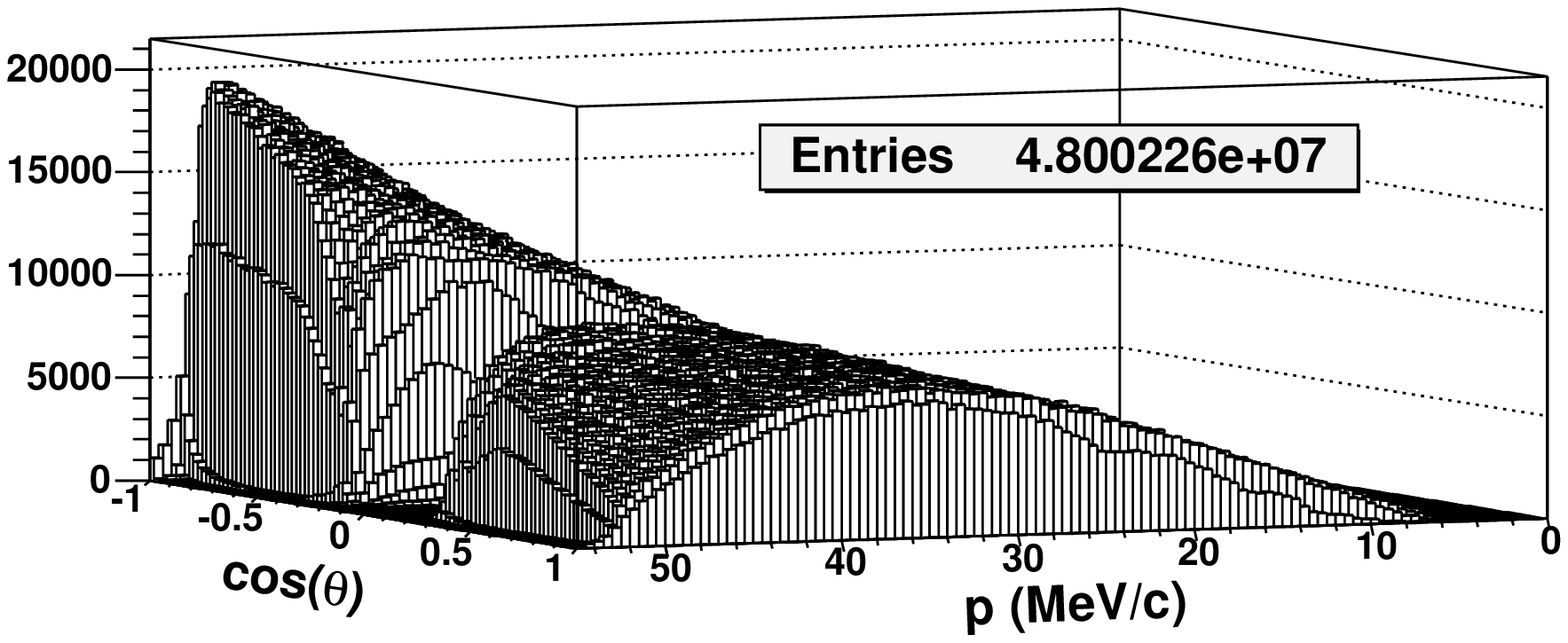} \\[-2mm]
\raisebox{0.13\textheight}{{\large\textbf{b)}}}\includegraphics[width=0.95\textwidth]{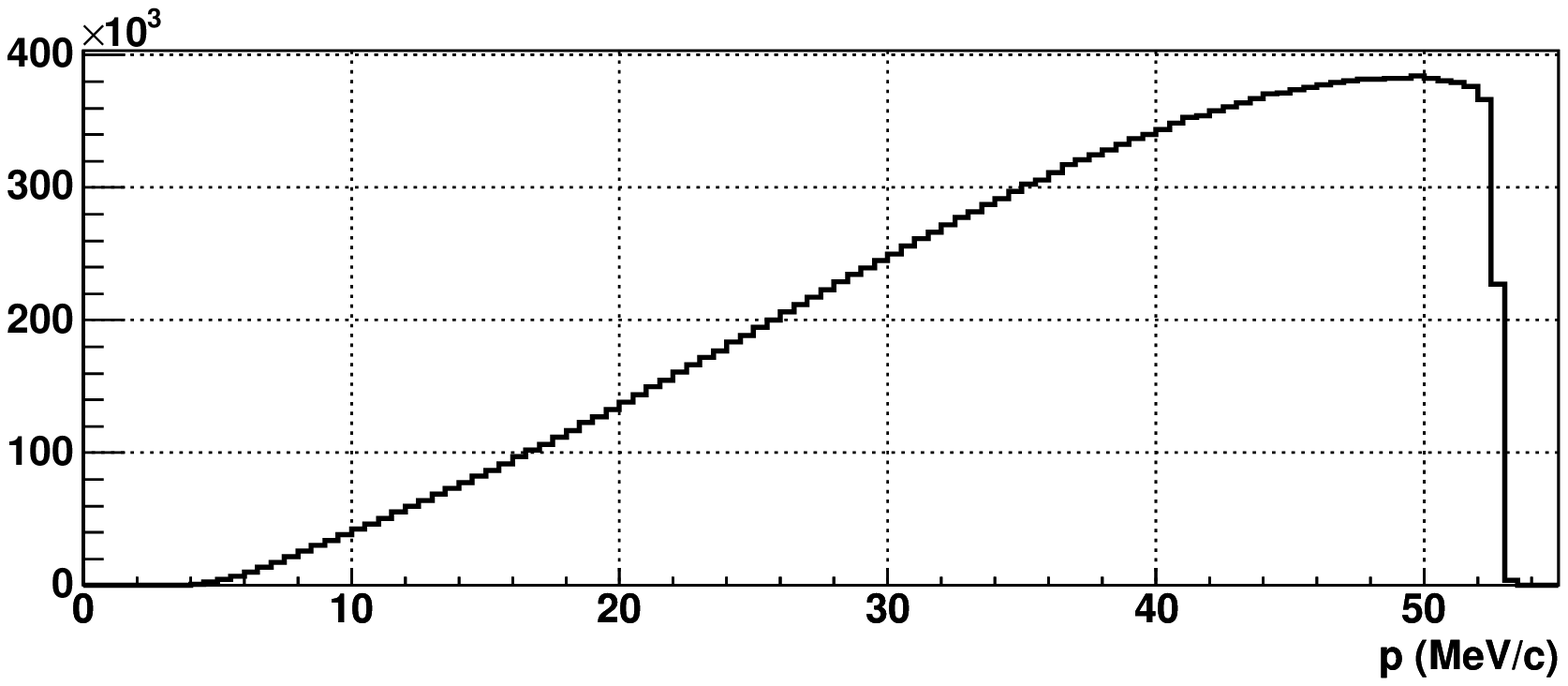} \\[-2mm]
\raisebox{0.13\textheight}{{\large\textbf{c)}}}\includegraphics[width=0.95\textwidth]{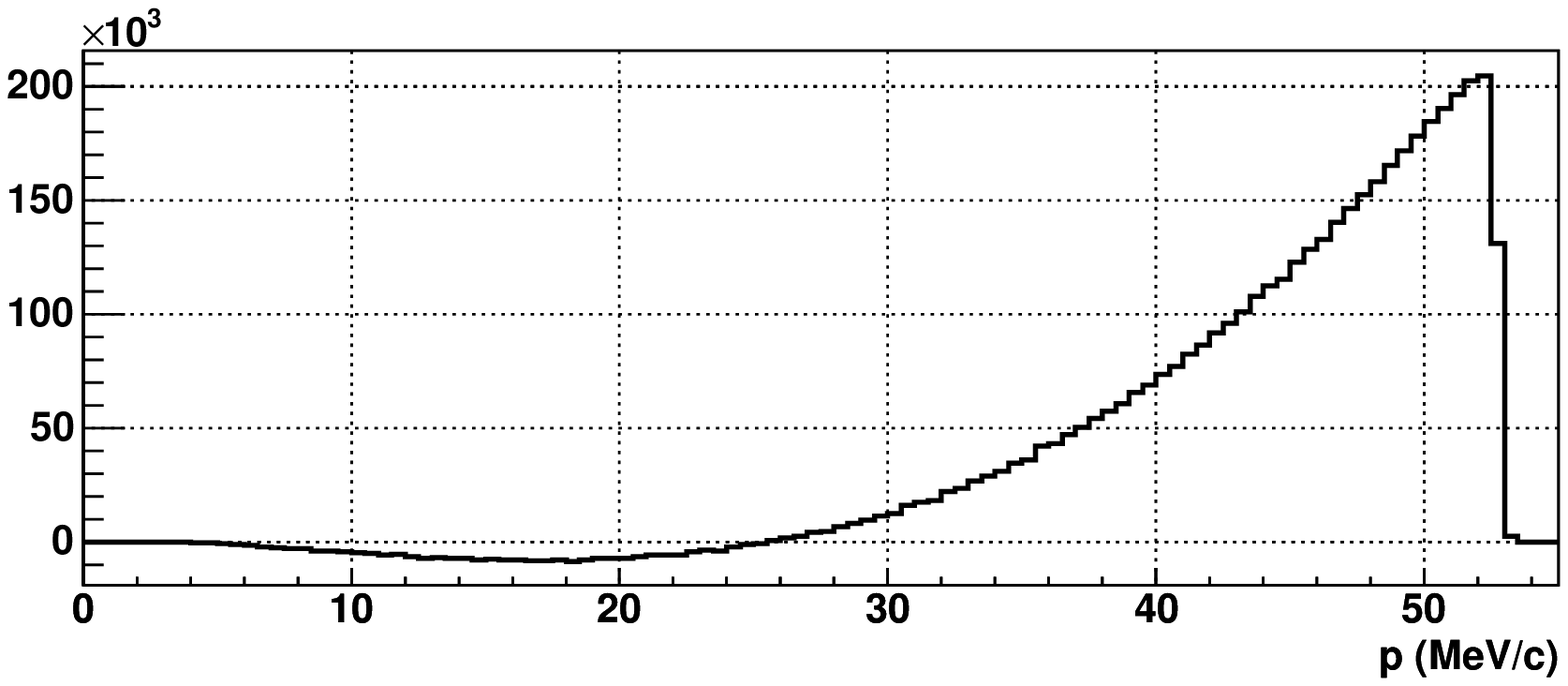} \\[-2mm]
\raisebox{0.13\textheight}{{\large\textbf{d)}}}\includegraphics[width=0.95\textwidth]{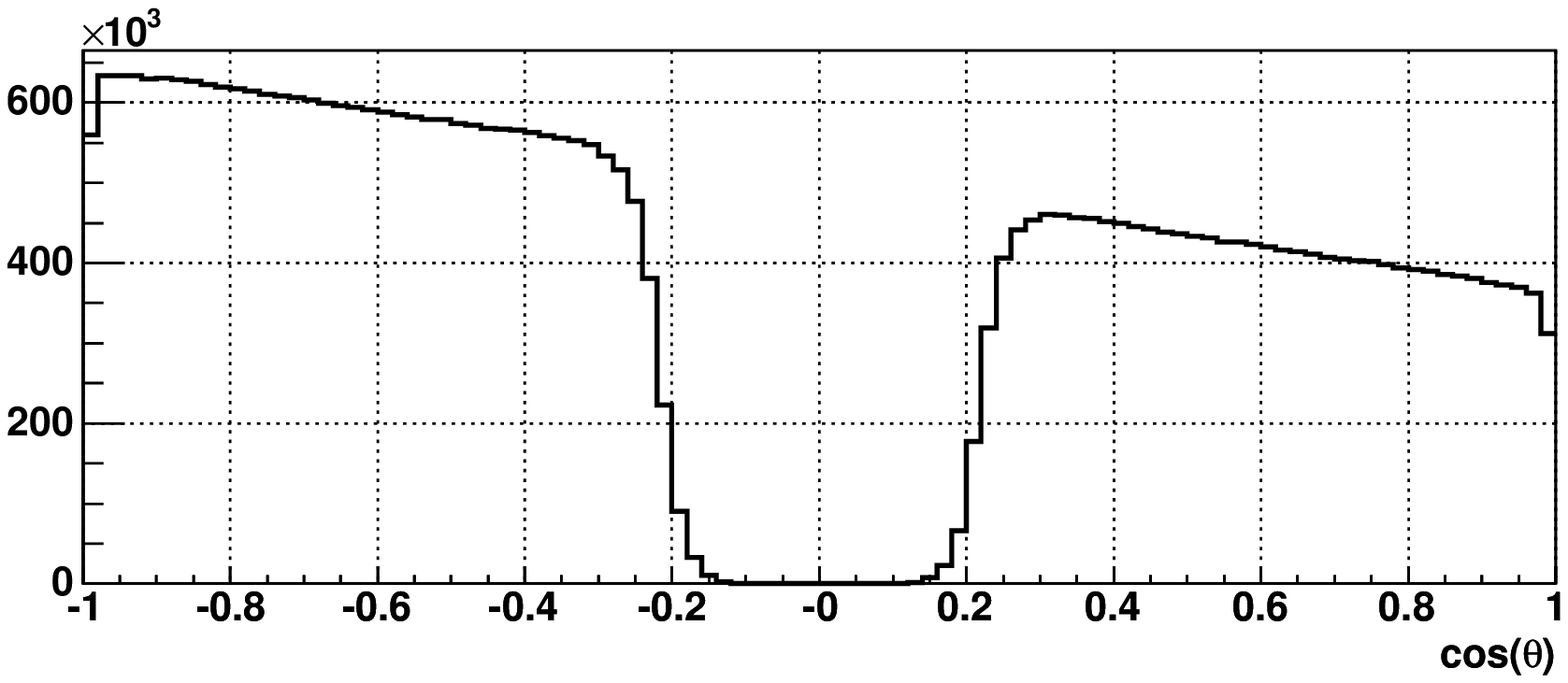}
%\framebox[\textwidth]{\vbox to0.23\textheight{\vfill}{\Large Asymmetry(p)}}
\end{tabular}
\vspace*{-5mm}
\caption[Data spectrum from set B]{\label{fig:michel-data}
  Reconstructed data spectrum from set~B. a) In the momentum-$\costh$ plane.
b) Momentum spectrum for $0.50<|\costh|<0.84$, proportional to
  $F_{\text{IS}}(p)$.
c) Difference of momentum spectra for $-0.84<\costh<-0.50$ and
  $0.50<\costh<0.84$, proportional to $F_{\text{AS}}(p)$
d) $\costh$ spectrum for $20<p<50$~MeV/c.
}
\end{figure}

%%================================================================
%% a subset of gen148, /home/andr/analysis/twt.20041127/datav.v2/gen148an1
\begin{figure}[tbp]
\begin{tabular}{cc}
\includegraphics[width=0.5\textwidth]{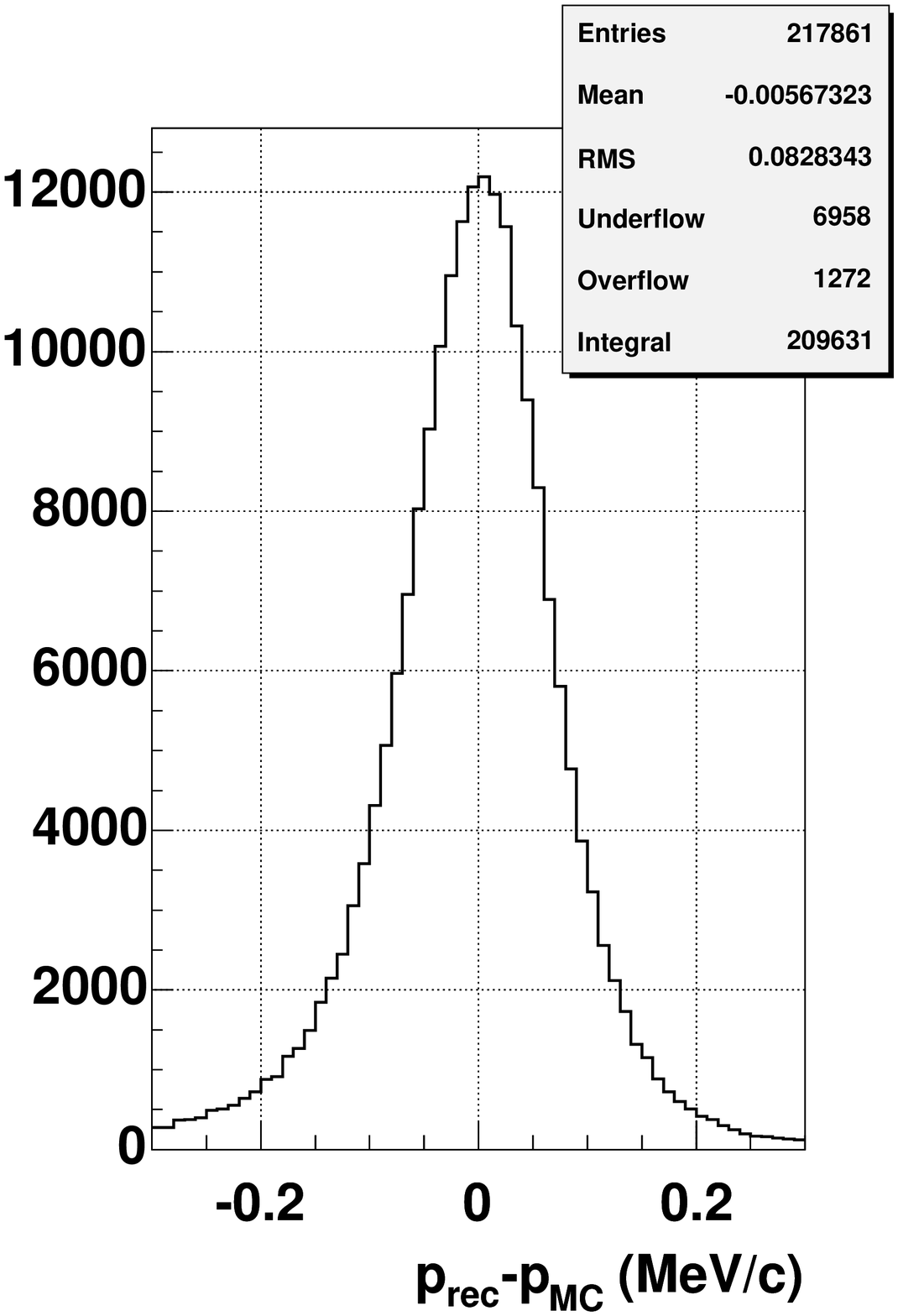}&
\includegraphics[width=0.5\textwidth]{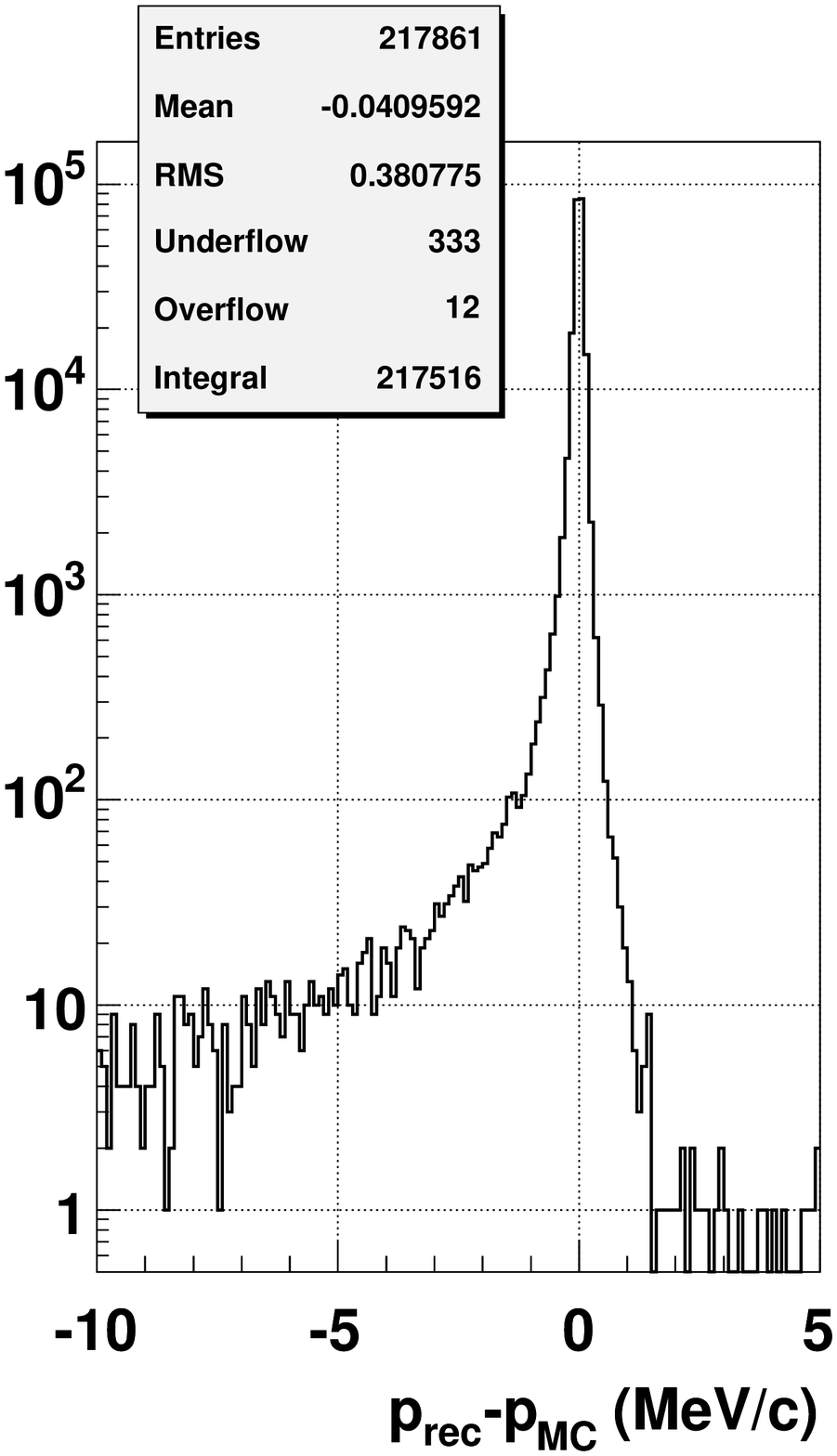}\\
\includegraphics[width=0.5\textwidth]{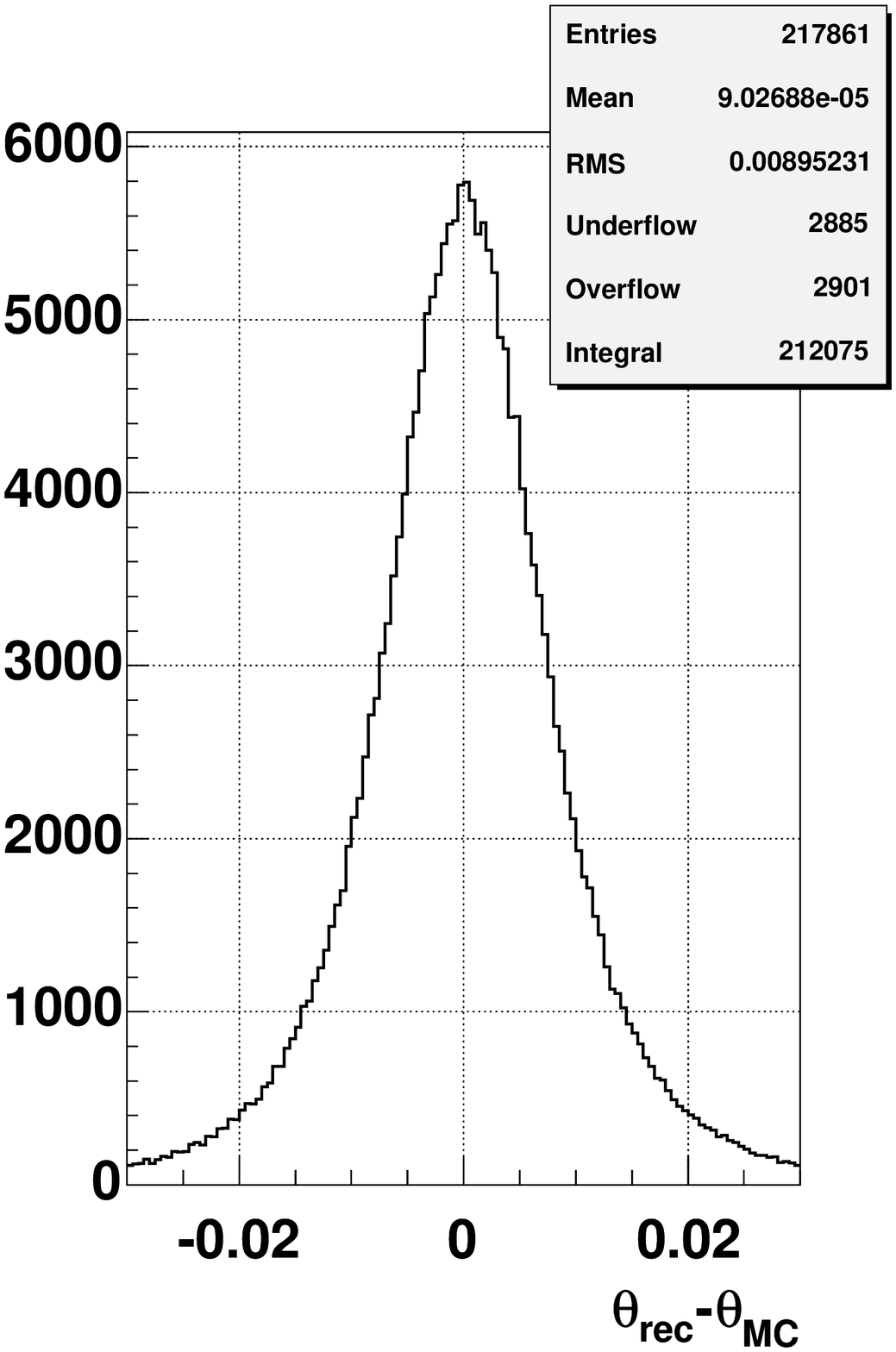}&
\includegraphics[width=0.5\textwidth]{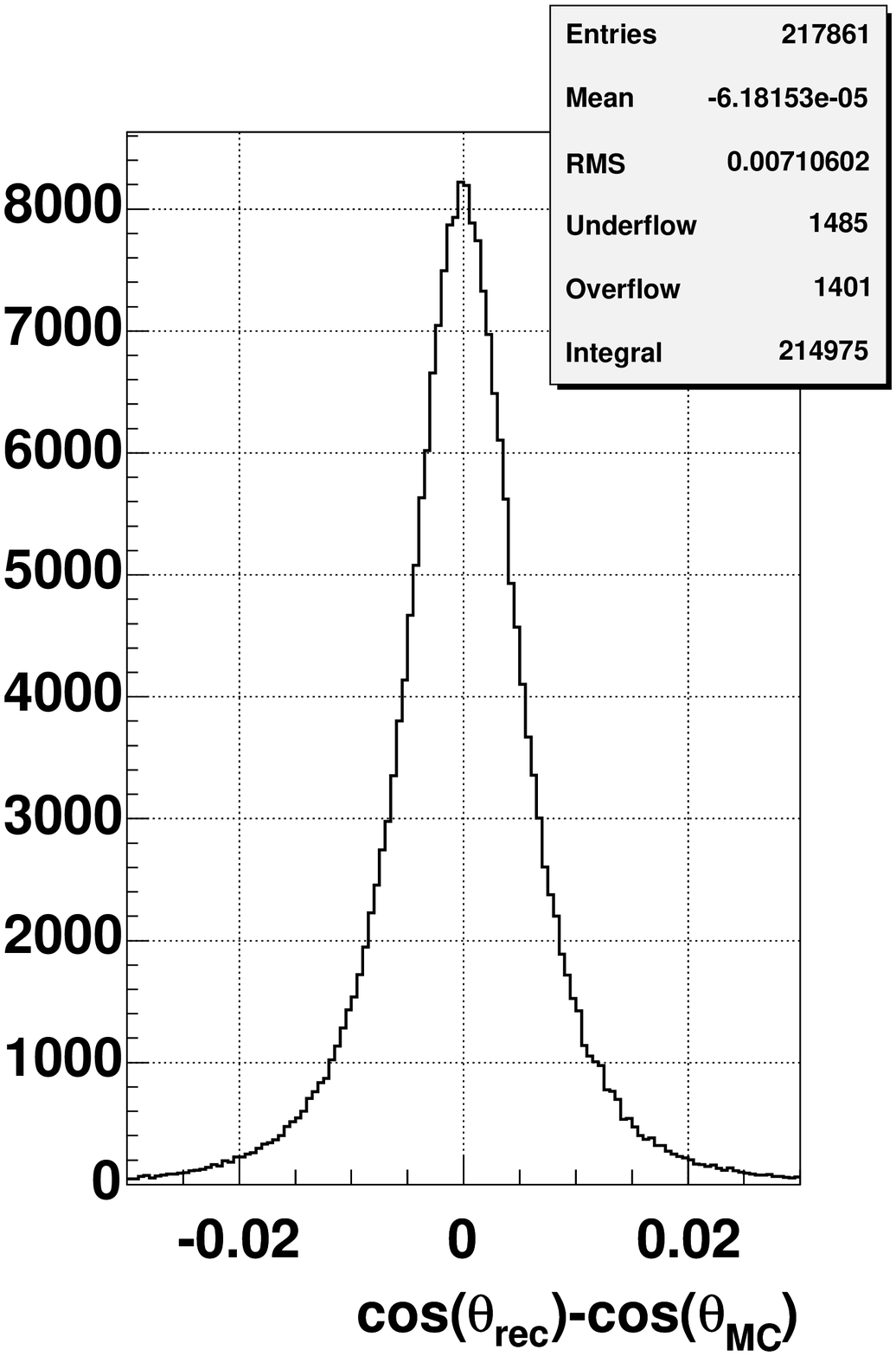}
\end{tabular}
\caption[Momentum and angle resolution]{\label{fig:resolutions}
  Distributions of differences of reconstructed and Monte-Carlo values.
  Top row: momentum, on different scales.  Bottom left: $\theta$, bottom
  right: $\costh$.  }
\end{figure}

\endgroup
%

%%%%%%%%%%%%%%%%%%%%%%%%%%%%%%%%%%%%%%%%%%%%%%%%%%%%%%%%%%%%%%%%
\chapter{Energy calibration\label{chapter:ecalib}}
% \chapter{Energy calibration\label{chapter:ecalib}}
\begingroup

The momentum of a decay positron measured by \TWIST{} is affected by
energy loss in the detector materials.  For a muon stopping
distribution that is not centered in the target the average momentum
shift is different for upstream and downstream decays, introducing an
asymmetry in the spectrum. Also the average energy loss may be
different in data and the simulation even for a centered stopping
distribution.
The momentum also scales with the ratio of the ``true'' magnetic field
seen by the particle to the field used by the reconstruction program.
The ratio is unity for the Monte-Carlo, but the field map is not a
perfect representation of the real field.
These effects may lead to different spectrum distortions for real and
simulated data, producing a bias in the extracted values of the muon
decay parameters.  The goal of the energy calibration procedure is to
correct the reconstructed spectra to compensate for these differences
between the data and the simulation.

The calibration is done on physics data, using the same reconstructed
spectrum as for the extraction of the decay parameters.  The
calibration point is provided by the sharp edge of the muon decay
spectrum at the upper kinematic limit.  Its position is determined by
the muon and the positron masses, and is therefore known.  

The planar geometry of the \TWIST{} spectrometer leads to an exact
$1/|\costh|$ dependence of the amount of material traversed by a decay
positron \cite{Khrutchinsky:1997}.  This dependence is obvious for the
flat stopping target, cathode foils, and the gas layers.  For the
cylindrical wires, that dependence still applies on average, because
the probability of hitting a wire changes as $1/|\costh|$.  In the
15--53~MeV/c \TWIST{} range of momenta, this translates into a linear
dependence of the average energy loss of the positron on $1/|\costh|$.
That known dependence provides a way to disentangle the effects of the
magnetic field scale, which is angle-independent, and the energy loss.
The position of the spectrum end point as a function of angle is given
by
\begin{equation}
  p^{\text{raw}}_{\text{edge}}(\theta) =
  \left(1+\frac{{\beta}}{p_0}\right)
  \left(p_0 -\frac{{\alpha}}{|\cos(\theta)|}\right).
  %, \quad   E_0\approx52.83\text{~MeV}
  \label{eq:ec-main}
\end{equation}
Here $p^{\text{raw}}_{\text{edge}}(\theta)$ is the end point position
of the ``raw'' reconstructed spectrum, $p_0\approx52.828\text{~MeV}$
is the kinematic spectrum limit, and the constants $\alpha$ and
$\beta$ describe the momentum loss and the field scale mismatch,
respectively.  To accommodate a non-centered muon stopping
distribution, the calibration procedure allows for different energy
loss parameters in the upstream and the downstream parts of the
detector, $\alpha=\au$ or $\alpha=\ad$.  The difference of the
upstream and downstream energy losses $\alphadiff=\au-\ad$ is
proportional to an offset of the muon stopping distribution from the
centered position, while the sum $\alphasum=\au+\ad$ does not depend
on the muon stopping position and is a measure of an effective
thickness of the detector.

A determination of the calibration parameters $\beta$, $\au$, $\ad$ is
not trivial.  The challenge is to determine the end point position,
$p^{\text{raw}}_{\text{edge}}(\theta)$, on reconstructed data, where
the sharp edge of the theoretical spectrum is smeared by the detector
resolution.  A model function describing the spectrum shape at the end
point is needed.  This complicated shape is a convolution of the muon
decay spectrum and the detector resolution, and a model function can
not be expected to perfectly describe the data distribution.
Therefore the result of a fit depends on the range of momenta used.  An
objective procedure establishing the range must be developed.  Since
the bias of a fit depends on the relative position of the fit range
with respect to the end point, that procedure should be adaptive and
produce the same relative position for different absolute positions of
the end point in the input data.  It also has to be noticed that
fitting in the end point region by definition involves transition from
high to low per bin statistics, and care must be exercised in handling
the statistical issues properly.

A straightforward approach of fitting
$p^{\text{raw}}_{\text{edge}}(\theta)$ independently for different
angles, with an adaptive choice of fit range, and then applying
Eq.~\eqref{eq:ec-main} to fit a straight line through the resulting
points, has been tried~\cite{andr-ecalib-tn,andr-old-ec-posting}.
Different end point model functions were used, some including the
effects of radiative corrections on the muon decay
spectrum~\cite{andr-ecalib-tn}.  An important result of these studies
is a demonstration that the end point indeed has a $1/|\costh|$
dependence.  However to stabilize the independent end point fits,
especially for small downstream angles where the statistics is low, a
large range of momenta ${}\approx2$~MeV/c needs to be used.  It is
more difficult to find a model function describing the data in a large
range.  Another issue is that the shape of the data distribution
becomes sensitive to the values of the muon decay parameters if a
large range of momenta is looked at, which is an undesirable effect.
Also, there is no single goodness of fit criteria in this approach.
The goodness of the final straight line fit does not include any
information about goodness of the individual end point fits.

%%================================================================
%% A sketch of the end point model function
\begin{figure}[tbp]
\begin{tabular}{c}
\includegraphics[width=\textwidth]{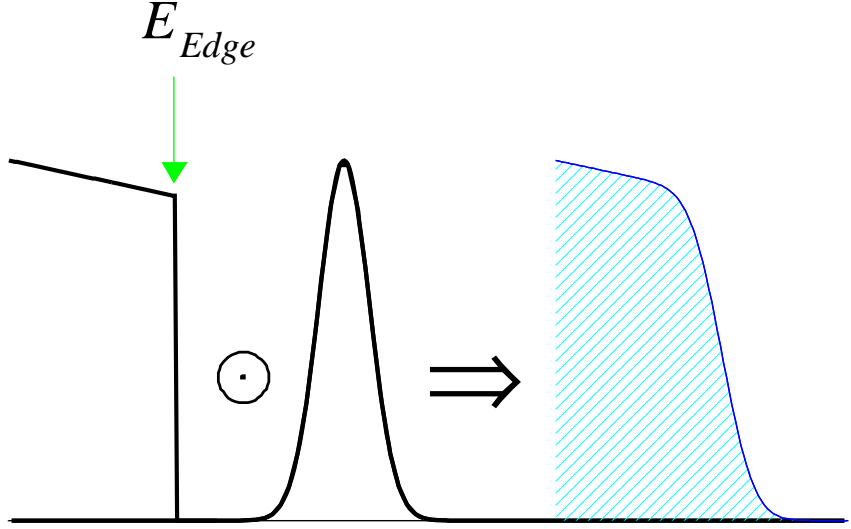}
\end{tabular}
\caption[End point model function]{\label{fig:end-point-sketch}
  A convolution of the cut-off linear function with a Gaussian gives
  the shape on the right, which is used to fit the end point of the
  reconstructed spectrum.
}
\end{figure}
%%================================================================
%

To overcome these issues, a fitting procedure that uses a global fit
to the 2-dimen\-sional momentum and $\costh$ reconstructed distribution
has been developed~\cite{andr-new-ec-posting}.  The model function is
constructed as following.  The momentum dependence is given by a
convolution of a ``slope and a step'' shape
\begin{equation}
  a\,(1+by)\,\Theta(y),\quad\text{where}\quad
  y = p-p_{\text{edge}}
\label{eq:slope-and-step}
\end{equation}
with a Gaussian, illustrated on Fig.~\ref{fig:end-point-sketch}.  Here
$\Theta(y)$ is the Heaviside step function.  There are 4 parameters
$a$, $b$, $p_{\text{edge}}$, which defines the position of the end
point, and the Gaussian $\sigma$.  The explicit form of the end point
model is
\begin{equation}
  \frac12 (a+by) \erfc(\frac{y}{\sqrt{2}\,\sigma)})
  - \frac{b\sigma}{\sqrt{2\pi}} \exp(-\frac{y^2}{2\sigma^2}).
\label{eq:convolution}
\end{equation}
This one-dimensional function is used to describe the momentum
dependence of the spectrum for a fixed angle $\theta$.  To obtain a
2-dimensional function, a dependence of the parameters $a$, $b$,
$p_{\text{edge}}$ and $\sigma$ on $\theta$ is introduced.  For
$p_{\text{edge}}$, the expected linear in $1/|\costh|$ behavior from
Eq.~\eqref{eq:ec-main} is used, with different momentum loss
parameters $\au$ and $\ad$ but common field scale $\beta$ in the
upstream and the downstream:
\begin{align}
  p^{\text{raw}}_{\text{edge}}(\theta) &=
  \left(1+\frac{{\beta}}{p_0}\right)
  \left(p_0 -\frac{\au\Theta(-\costh)+\ad\Theta(\costh)}{|\cos(\theta)|}\right).
\label{eq:ec-param-edge}
\\
\intertext{The muon decay spectrum is linear in
$\costh$, therefore a linear parameterization}
%\begin{equation}
a(\theta) &= a_0 + a_1\costh
\\
%\end{equation}
\intertext{is used for the normalization.  For $\sigma$ and $b$,
suitable parametrizations were found empirically by fitting
Eq.~\eqref{eq:convolution} to the data distribution independently for
different angular bins, and observing the behavior of the fitted
parameters as functions of angle.
%
% Another method was to use a global fit for all but one of the 4
% parameters, and let the parameter under study to vary 
%
That resulted into the following choice:
}
%\begin{align}
b(\theta)  &= b_0 + b_1\cos(\theta).
\\
\sigma(\theta) &= \frac{\sigma_0}{|\sin(\theta)|},
\label{eq:ec-param-sigma}
\end{align}
Equations \eqref{eq:convolution}--\eqref{eq:ec-param-sigma} completely
define the shape of the fitting function.  They contain 8 parameters:
$\beta$, $\au$, $\ad$, $a_0$, $a_1$, $b_0$, $b_1$, $\sigma_0$, that
are determined from a fit to a reconstructed spectrum.
Fig.~\ref{fig:ecfits} shows an energy calibration fit to data
for several angles.

The fit is done by minimizing $-2\ln\lambda$, where $\lambda$ is the
Poisson likelihood ratio~\cite{Baker:1983tu}:
\begin{equation}
  -2\ln\lambda=2\sum_{i=1}^{N}\left[f_i-n_i+n_i\ln\frac{n_i}{f_i}\right].
  \label{eq:twologlambda}
\end{equation}
In \eqref{eq:twologlambda} the summation runs over the bins of the
histogram in the fit range, $n_i$ is the number of entries in a bin,
and $f_i$ is the expectation value for $n_i$ computed using
Eqs.~\eqref{eq:convolution}--\eqref{eq:ec-param-sigma}.  The advantage
of the binned likelihood statistics $-2\ln\lambda$ is that it can be
used not only to perform the fitting, but also to easily estimate a
goodness of the resulting fit~\cite{Baker:1983tu}.  In the large
sample limit the minimum value of $-2\ln\lambda$ is distributed as
$\chi^2$.

As is mentioned above, for a fixed angle the same momentum fit
range \emph{relative to the end point} must be selected, so that no
biases between data and Monte-Carlo are introduced due to different
absolute positions of end points.  A series of tests, described
in Appendix~\ref{chapter:ectuning}, led to the following choice:
\begin{equation}
p_{\text{edge}}(\theta)-0.75\text{~MeV/c}<p<p_{\text{edge}}(\theta)+0.5\sigma(\theta).
\label{eq:ec-fit-range}
\end{equation}
The fit is done iteratively.  It uses its current estimate of
$p_{\text{edge}}(\theta)$ to fix the fit range for the next iteration.
There is no explicit limit on the number of iterations.  Instead each
new 2-dimensional fit region $\Omega_k$ is compared with all
previously seen regions $\Omega_1,\ldots,\Omega_{k-1}$.  The stopping
condition is $\Omega_k=\Omega_{k-l}$.  Because a binned fit is used, a
rounding to histogram bin boundaries ensures that the number of
possible fit regions is finite, so the fit always converges. In
practice it usually takes ${}\lesssim10$ iterations.  It is common to
have $l>1$, in that case 
%
%vb: the average values of fit parameters over the
%vb: regions $\Omega_{k-l},\ldots,\Omega_{k-1}$ are used as the best
%vb: estimate.  
%
\vc{the average value of $\beta$ over iterations $k-l+1,\ldots,k$
is computed, and the iteration with $\beta$ closest to the average is
chosen as the final result.  This is an arbitrary procedure, but}
the spread of the results over the regions of the
``convergence cycle'' is always much smaller than the statistical
error of the fit.

Table \ref{table:ec-fits} shows results of energy calibration fits for
data sets and corresponding Monte-Carlo sets used in the extraction of
$\delta$.  A systematic difference in the momentum resolution
parameter $\sigma_0$ at the end point is apparent from the numbers. 

%%----------------------------------------------------------------
%  beta =    -0.005445 +-  0.007823250   logs-data-5/log-set3an26_all_ecb.txt
%  beta =     0.003628 +-  0.005477426   logs-data-5/log-set2an18_all_ecb.txt
%  beta =     0.008995 +-  0.005434261   logs-data-5/log-set1an3_all_ecb.txt
%  beta =     0.008037 +-  0.006026336   logs-data-5/log-set22an2_all_ecb.txt
%
%  beta =     0.005274 +-  0.006161363   logs-data-5/log-set18an3_all_ecb.txt
%
\begin{table}[htbp]
\begin{center}
\begin{tabular}{ld{2.7}d{2.7}d{2.7}d{2.7}}
Spectrum  
& \multicolumn{1}{c}{$\beta$} 
& \multicolumn{1}{c}{$\au$}
& \multicolumn{1}{c}{$\ad$}
& \multicolumn{1}{c}{$\sigma_0$}
\\ \hline
% s3
Set A & -5.4\pm6.2 &  67.7\pm5.4 & 56.9\pm5.4 & 76.0\pm1.7\\
% s2
Set B & 3.6\pm5.5 & 74.0\pm3.8 & 60.2\pm3.8 & 74.1\pm1.2\\
% s1
1.96~T & 9.0\pm5.4 & 79.0\pm3.8 & 71.3\pm3.8 & 76.8\pm1.2\\
2.04~T & 8.0\pm6.0 & 75.9\pm4.2 & 64.8\pm4.2 & 73.3\pm1.3 \\
\hline
 \hline
% s3 => g152
MC A & -41.2\pm4.5 & 61.4\pm3.1& 57.8\pm3.1& 69.3\pm0.9 \\
% s2 => g148
MC B & -42.0\pm4.5& 60.3\pm3.1& 57.3\pm3.1& 69.1\pm1.0\\
% s1 => g154
MC 1.96~T & -7.1\pm4.5& 58.4\pm3.2& 56.0\pm3.1& 69.7\pm1.0\\
% s22 => g156
MC 2.04~T & -9.0\pm4.5 & 56.8\pm3.1 & 55.4\pm3.1 & 68.7\pm0.9 \\
\hline
\end{tabular}
\caption[Energy calibration results]{\label{table:ec-fits}Energy
  calibration results. All parameters are in keV/c.  The large
  deviations in $\beta$ for two of the Monte-Carlos are because of a
  mistake in setting the field scale in MC production.  The energy
  calibration procedure corrects for this mistake.}
\end{center}
\end{table}
%
% %%----------------------------------------------------------------
% \begin{table}[htbp]
% \begin{center}
% \begin{tabular}{ld{3.7}d{2.7}d{2.7}d{2.7}}
% %
% Data Set  
% & \multicolumn{1}{c}{$\beta$} 
% & \multicolumn{1}{c}{$\au$}
% & \multicolumn{1}{c}{$\ad$}
% & \multicolumn{1}{c}{$\sigma_0$}
% \\ \hline
% % s3 => g152
% MC A & -41.2\pm4.5 & 61.4\pm3.1& 57.8\pm3.1& 69.3\pm0.9 \\
% % s2 => g148
% MC B & -42.0\pm4.5& 60.3\pm3.1& 57.3\pm3.1& 69.1\pm1.0\\
% % s1 => g154
% MC 1.96~T & -7.1\pm4.5& 58.4\pm3.2& 56.0\pm3.1& 69.7\pm1.0\\
% % s22 => g156
% MC 2.04~T & -9.0\pm4.5 & 56.8\pm3.1 & 55.4\pm3.1 & 68.7\pm0.9 \\
% %
% \hline
% \end{tabular}
% \caption{\label{table:ec-fits-mc}Energy calibration results for
%   Monte-Carlo. All parameters are in keV/c.}
% \end{center}
% \end{table}
% %
%%----------------------------------------------------------------
\begin{figure}[tbp]
\begin{tabular}{cc}
\includegraphics[width=0.5\textwidth]{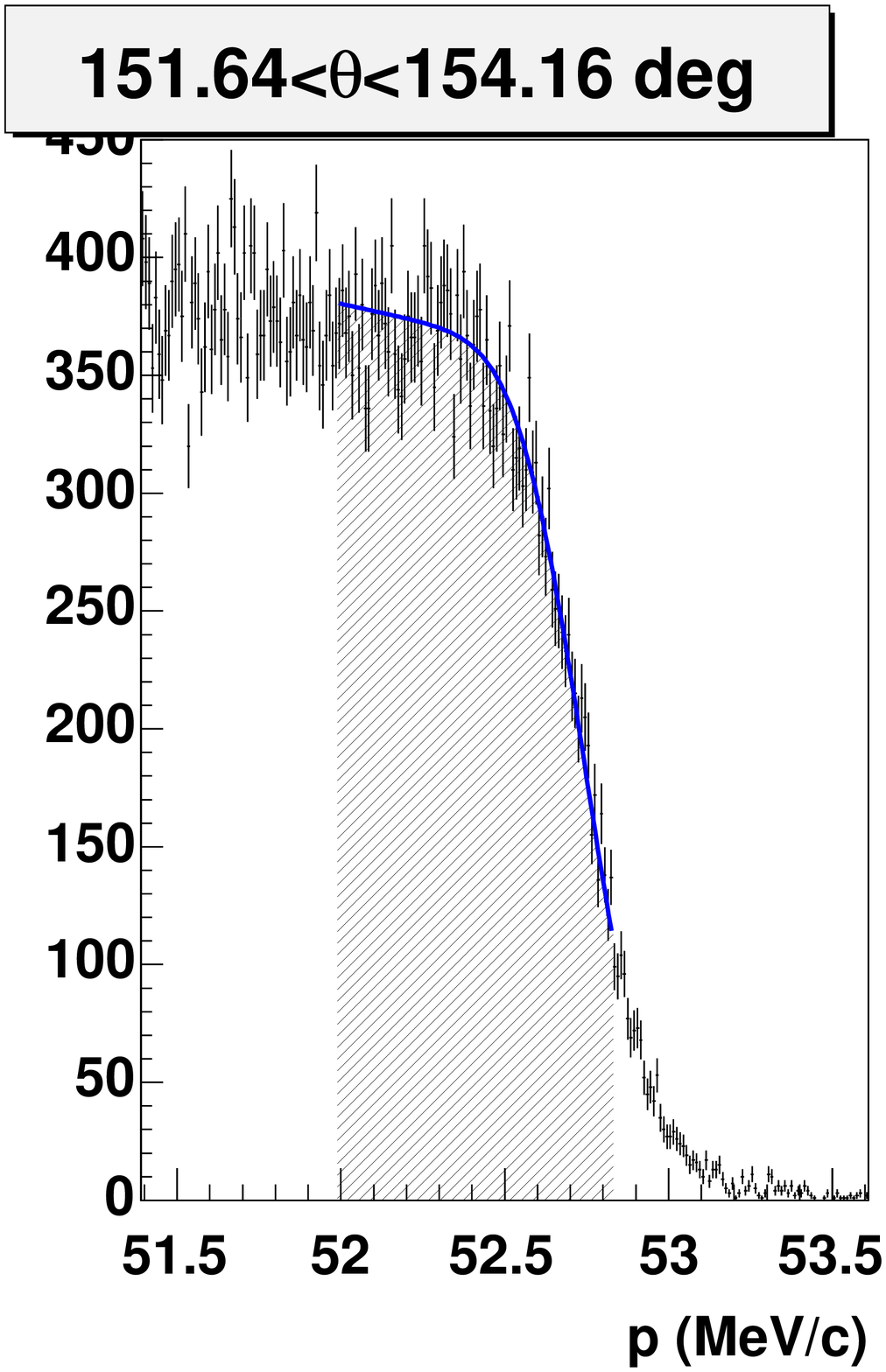}&
\includegraphics[width=0.5\textwidth]{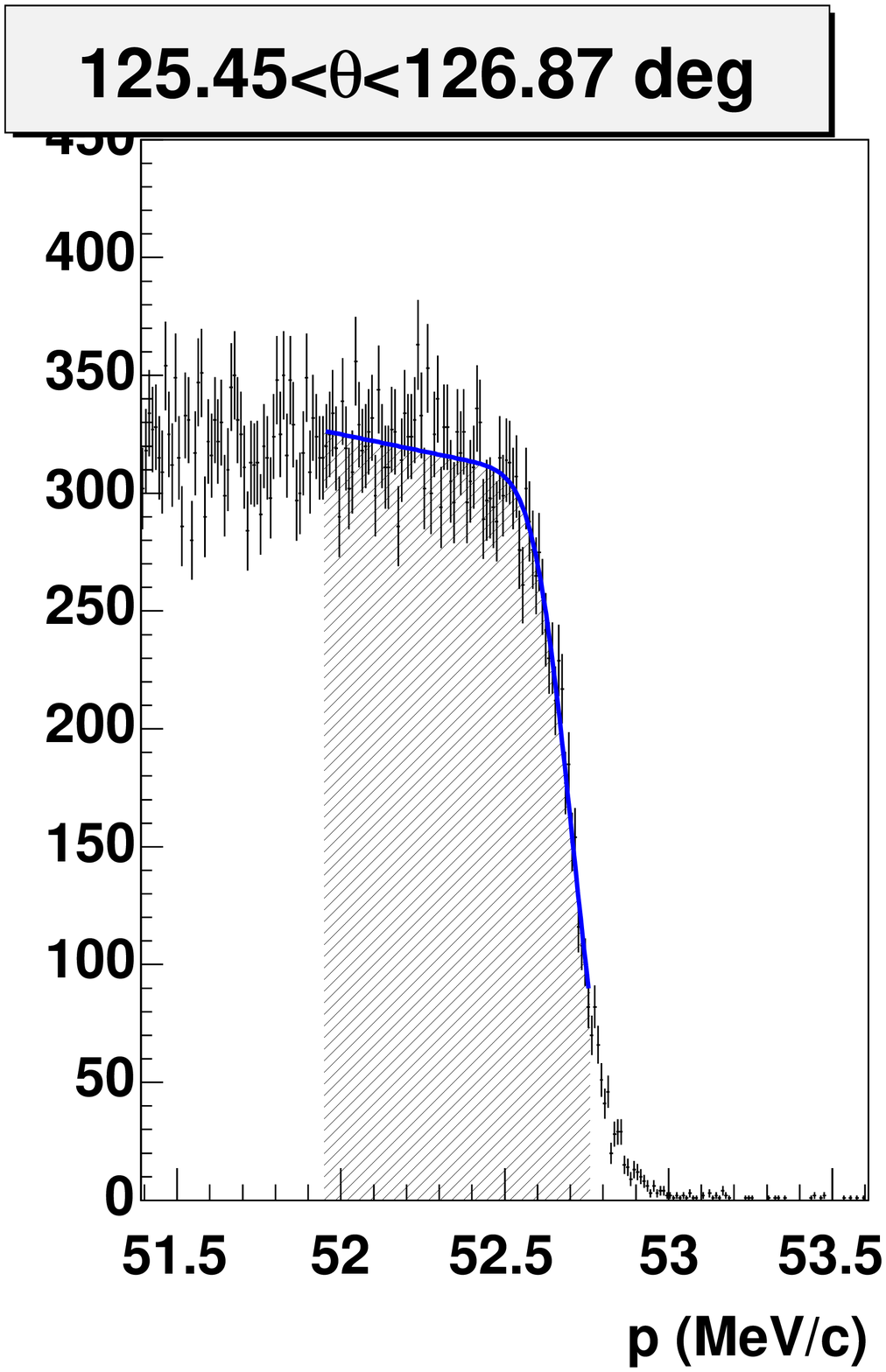}\\
\includegraphics[width=0.5\textwidth]{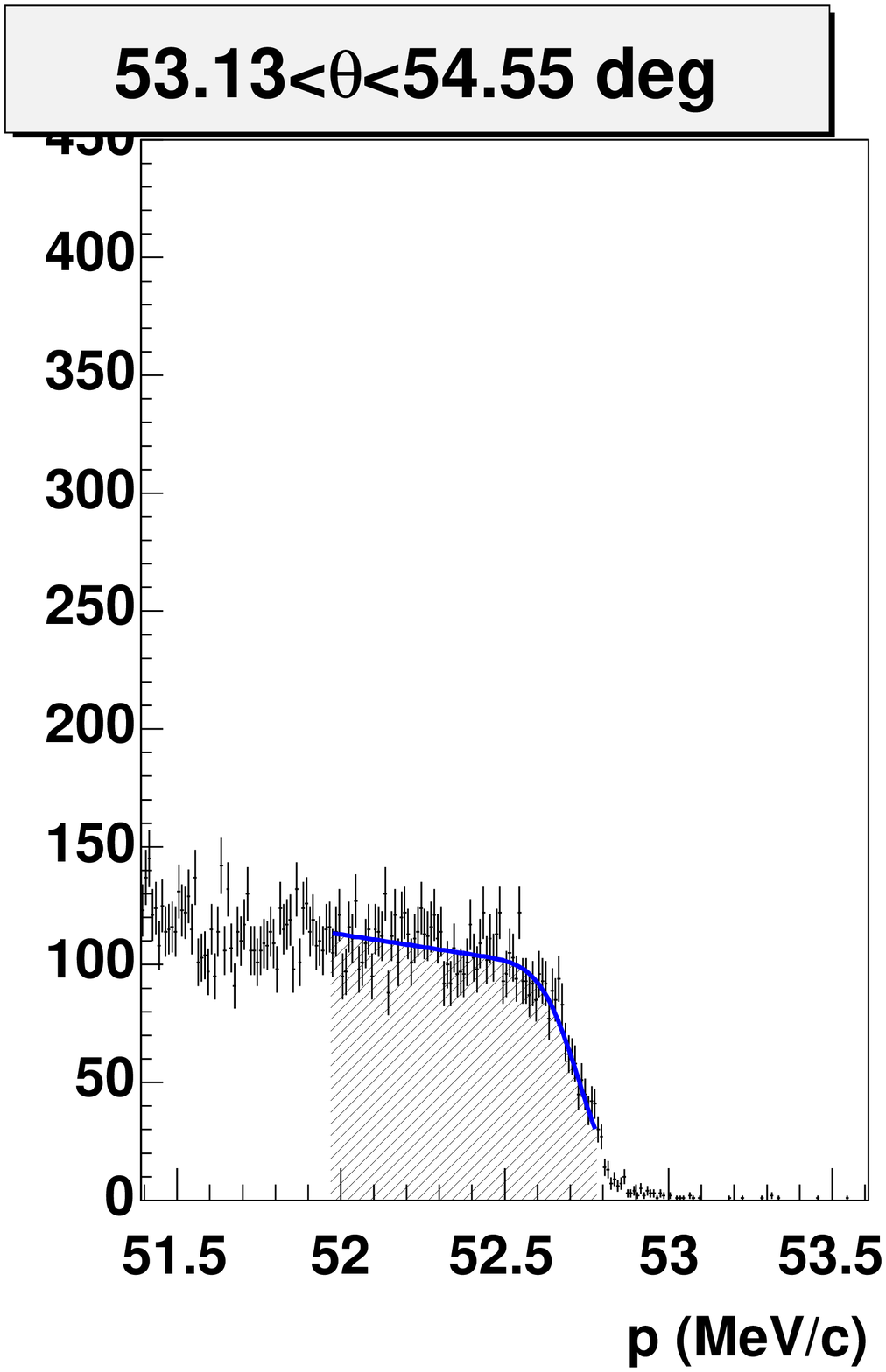}&
\includegraphics[width=0.5\textwidth]{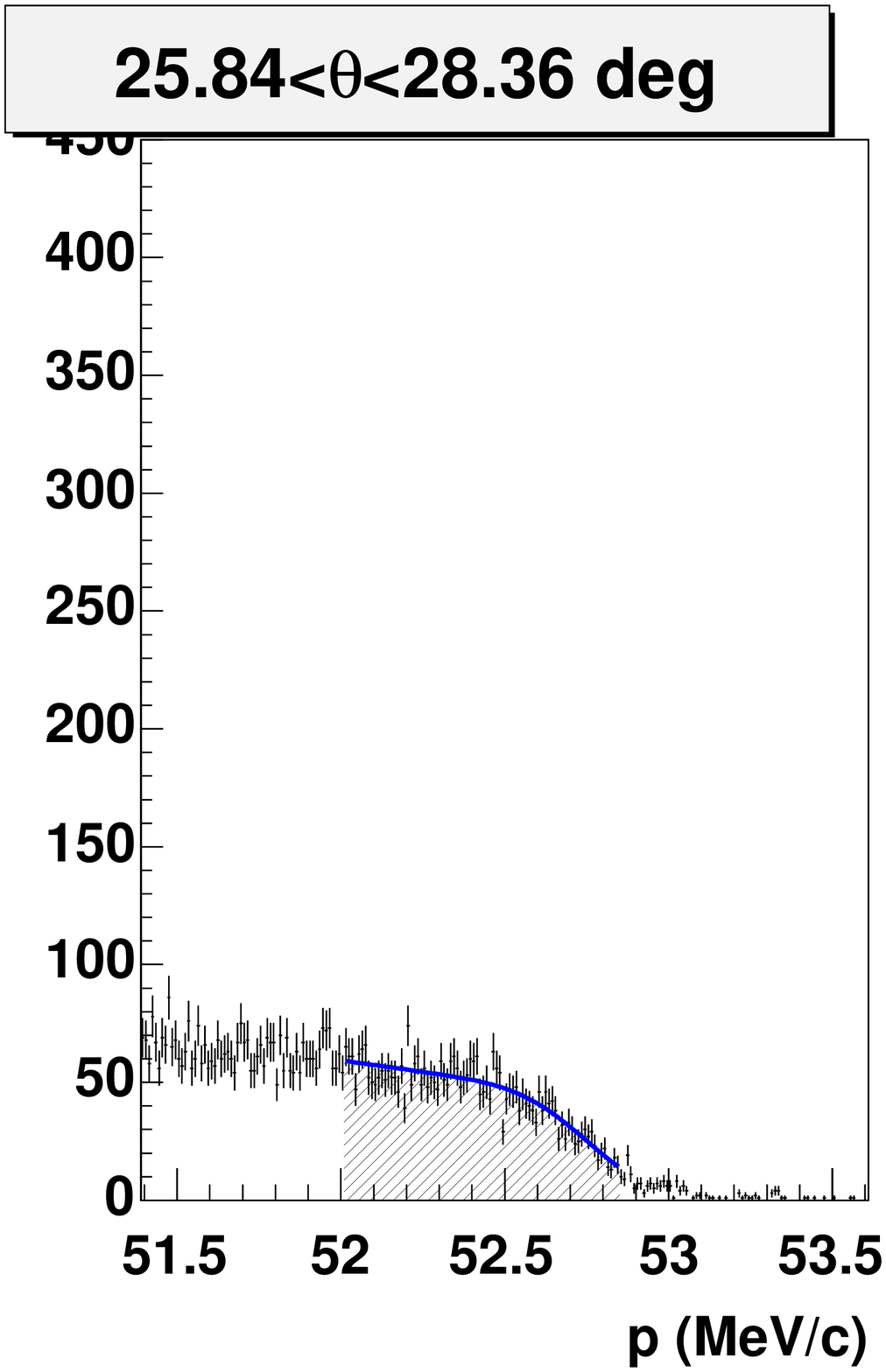}
\end{tabular}
\caption[End point fits]{\label{fig:ecfits}The end point of the muon
  decay spectrum, and sections of the 2-dimensional end point fit
  function for several angular slices.  
  The smallest and the largest angles in the upstream (top), and the
  downstream (bottom) are shown.  Data set~B.}
\end{figure}
%%----------------------------------------------------------------

This discrepancy may arise for multiple reasons.  For example, the
spectrum reconstruction uses the same drift chamber space-time
relations for data and Monte-Carlo.  These relations may be slightly
different in the real detector, but the simulation uses exactly the
same drift tables as the reconstruction to generate Monte-Carlo
events.  Therefore the reconstruction of Monte-Carlo can be expected
to perform better.  This holds true for all other calibrations, such
as alignment corrections, electronic timing offsets T0, etc.  The
GEANT handling of positron interactions in detector materials may also
contribute to the effect.  All the calibrations are counted as sources
of their corresponding systematic uncertainties, as well as
an imperfection of the GEANT simulation,
chapter~\ref{chapter:systematics}.
However 
%
% in the lack of convincing evidence that the resolution discrepancy can be
% fully explained by the known factors
%
an effect of the momentum resolution discrepancy on the final result,
regardless of its cause, has been estimated and included in the final
estimate of systematic uncertainty, chapter~\ref{chapter:systematics}.

Typical correlation coefficients among fit parameters are shown in
Table~\ref{table:ec-fit-correlations}.   Since the parameters of
interest $\beta$, $\au$ and $\ad$ are highly correlated, their
correlation must be taken into account while estimating a systematic
uncertainty of the result due to energy calibration.

%%----------------------------------------------------------------
% Correlation coefficients:       
% logs-data-5/log-set2an18_all_ecb.txt
% 
%                beta           au           ad       ResPar        norm0        norm1          sl0          sl1  
%   beta  1.000000000  0.978256053  0.926136762 -0.137491314 -0.271871138 -0.094945279 -0.242068520  0.137957998  
%     au  0.978256053  1.000000000  0.915509741 -0.057915735 -0.124834656 -0.144404054 -0.105036782  0.004390749  
%     ad  0.926136762  0.915509741  1.000000000 -0.058627502 -0.117064862  0.131698854 -0.100708177  0.155097317  
% ResPar -0.137491314 -0.057915735 -0.058627502  1.000000000  0.645883449 -0.006923008  0.583112021 -0.451838997  
%  norm0 -0.271871138 -0.124834656 -0.117064862  0.645883449  1.000000000  0.028671948  0.957182747 -0.718564194  
%  norm1 -0.094945279 -0.144404054  0.131698854 -0.006923008  0.028671948  1.000000000  0.024661470  0.592285263  
%
%    sl0 -0.242068520 -0.105036782 -0.100708177  0.583112021  0.957182747  0.024661470  1.000000000 -0.749863506  
%    sl1  0.137957998  0.004390749  0.155097317 -0.451838997 -0.718564194  0.592285263 -0.749863506  1.000000000  
%
\begin{table}[htbp]
$$
\begin{array}{l|d{2.2}d{2.2}d{2.2}d{2.2}d{2.2}d{2.2}d{2.2}}
{}       &  
\multicolumn{1}{c}{\au} & 
\multicolumn{1}{c}{\ad} & 
\multicolumn{1}{c}{\sigma_0} & 
\multicolumn{1}{c}{a_0}   & 
\multicolumn{1}{c}{a_1}   & 
\multicolumn{1}{c}{b_0}   & 
\multicolumn{1}{c}{b_1} 
\\ \hline
\beta    & 0.98 & 0.93 & -0.14   & -0.27 & -0.09 & -0.24 & 0.14 \\
\au      &      & 0.92 & -0.06   & -0.12 & -0.14 & -0.11 & 0.00 \\
\ad      &      &      & -0.06   & -0.12 & 0.13  & -0.10 & 0.16 \\
\sigma_0 &      &      &         & 0.65  & -0.01 & 0.58  & -0.45\\ 
a_0      &      &      &         &       & 0.03  & 0.96  & -0.72\\
a_1      &      &      &         &       &       & 0.02  & 0.59 \\
b_0      &      &      &         &       &       &       & -0.75
\end{array}
$$
\caption{\label{table:ec-fit-correlations}Correlation coefficients for
energy calibration fit to set~B.}
\end{table}
%%----------------------------------------------------------------

An energy calibrated decay spectrum is obtained from the corresponding
``raw'' spectrum by recomputing the reconstructed momentum for every
event:
%
%    return 
%      // Scale rec. momentum to account for the B field
%      ptot / (1 + beta/e0) 
%      // then add the momentum lost, which is already corrected for B
%      + ((costh<0) ? alpha_up : alpha_dn)/std::abs(costh);
%
\begin{equation}
p^{\text{ec}}= \frac{p^{\text{raw}}}{1+\beta/p_0}+\frac{\alpha}{|\costh|}
\label{eq:ec-application}
\end{equation}
The momentum is scaled by the magnetic field scale, and corrected for
the energy loss.  $\alpha=\au$ for upstream decays, $\ad$ for
downstream.  $p_0$ is defined after Eq.~\eqref{eq:ec-main}.  This
calibration procedure brings the end point of the calibrated spectrum
to its ``theoretical'' value, $p_{\text{edge}}(\theta)=p_0$.  Checks
by running energy calibration fits on calibrated (instead of raw)
spectra were done. As expected resulting $\beta$, $\au$, and $\ad$
were consistent with zero.

\endgroup
%

%%%%%%%%%%%%%%%%%%%%%%%%%%%%%%%%%%%%%%%%%%%%%%%%%%%%%%%%%%%%%%%%
% Method of extraction of decay parameters

\chapter{Method for extraction of the decay parameters\label{chapter:fitting-all}}
%\chapter{Method for extraction of the decay parameters}

%\section{Introduction}
\begingroup

\input fitting-macros.texinc

%%%%%%%%%%%%%%%%%%%%%%%%%%%%%%%%%%%%%%%%%%%%%%%%%%%%%%%%%%%%%%%%

A reconstructed spectrum differs from a theoretical one because of
finite experimental resolution and because the interaction of the decay
positrons with detector materials changes the energy and angle of the
particles.

Consider a parametrized theoretical probability distribution function
$\ft(x';\r)$, $x'\in\Omega_0$, where $\r$ are the parameters of the
theory, and the $x'$ are the kinematic variables of interest.  The
probability $K(x,x')\,dx$ of reconstructing an event that occurred at
$x'$ in a volume $dx$ 
around some point $x$ defines the response function $K$, which
describes the combined effect of the detector and the reconstruction.
The reconstructed spectrum $\fo(x;\r)$ can be written as
\begin{equation}
\fo(x;\r) = b(x) + \int_{\Omega_0} K(x,x')\, \ft(x';\r)\,dx',
\label{eq:rec-spectrum-C0}
\end{equation}
where $b(x)$ is the background term.  

Several approaches can be used to deduce the theory parameters $\r$
from a measured spectrum, $\fo$.  One can try to solve
\eqref{eq:rec-spectrum-C0} for $\ft$.  Deconvolution methods are
available \cite{cowan-deconvolution}, and were used by some
experiments (e.g. \cite{Barate:1997hv}). However they are not
practical for \TWIST{} where since $x=\{p,\costh)\}$, $K$ is a
function of 4 variables $p$, $\costh$, $p'$, $\cos(\theta')$.  It is
difficult to estimate a 4-dimensional function accurately from
Monte-Carlo.  In addition, a general feature of unfolding methods is a
need for a regularization parameter, which biases the result.

Another approach is to approximate \eqref{eq:rec-spectrum-C0} by an
analytic expression, and use the resulting $\fo(x;\r)$ to fit the
data.  Some terms in the approximate expression usually need to be
determined from simulations.  This method is used by
e.g. \cite{Balke:1988delta}.  It would be difficult to find a suitable
expression for the high precision representation of the spectrum
required by \TWIST{}.

The rest of this chapter describes yet another technique, which was
used by \TWIST{}.  The idea is to parametrize the \emph{reconstructed}
spectrum in terms of $\r$, and use that parameterization to fit the
data.  The method is similar to the one described in
\cite{fan_liu-phd}. The \TWIST{} method differs from
\cite{fan_liu-phd} in the fact that we fit only spectrum \emph{shape}
parameters, but not the absolute normalization.

%%%%%%%%%%%%%%%%%%%%%%%%%%%%%%%%%%%%%%%%%%%%%%%%%%%%%%%%%%%%%%%%
\section{\label{section:fitting-discussion}The fitting method}

Appendix \ref{chapter:spectrum-expansion} derives an expansion of a
reconstructed spectrum in the general case.  In \TWIST{} background
contamination is ${}\lesssim10^{-4}$~\cite{andr-background}.
Moreover, the primary background is simulated by \TWIST{}
Monte-Carlo. This means that there is a cancellation of the
corresponding spectrum distortion, and the effect is negligible at the
$10^{-3}$ level of precision.  Therefore Eq.~\eqref{eq:final2} can be
simplified to
\begin{equation}
  \ni(\r+\dr) = 
  \left[1-\sum_{\alpha=1}^{m}\dr_{\alpha}\,
    \Emc^{-1}\nu^{\alpha}
  \right] \ni(\r) 
%\\
  + \sum_{\alpha=1}^{m}\dr_\alpha\:\Emc^{-1}
  \nu_i^\alpha 
\label{eq:final3}
\end{equation}
where we also have omitted the $\OO(\dr^2)$ term.  Here $\ni(\r+\dr)$
are bin contents of a normalized data histogram, $\ni(\r)$ are
corresponding ``base'' Monte-Carlo values, $\nu_i^\alpha$ are bins of
a Monte-Carlo histogram for a ``derivative'' spectrum corresponding to
$\r^{\alpha}$, and $\Emc^{-1}$ is a constant that can be determined
from simulation.  See Appendix \ref{chapter:spectrum-expansion} for
details.

To extract values of the muon decay parameters, we minimize the
\begin{equation}
\chi^2 = \sum_{\Omega}
\frac{(\ni^{\text{Data}}-\ni^{\text{MC}})^2}{\sigma_i^2}
\label{fitter-chi2}
\end{equation}
where $\ni^{\text{Data}}$ are the normalized contents of a data
spectrum histogram, and $\ni^{\text{MC}}$ are calculated according to
\eqref{eq:final3} using Monte-Carlo ``base'' and ``derivative''
spectra. The errors are assumed to be Gaussian, since the available
statistics are sufficiently high%
\footnote{ The statistics are always high for ``data'' and ``base''
  spectra.  Some derivative functions cross zero in the fiducial, and
  around the crossings the count of events in a bin for the
  corresponding derivative may be small.  However the total error on
  the bin is dominated by ``data'' and ``base'' distributions, so the
  smallness of a derivative count has no importance.  }%
.
The error on the content of bin $i$ of an input
histogram is taken as the square root of the number of entries in
that bin, then the errors on different MC histograms are combined
following \eqref{eq:final3}, and $\sigma_i^2$ is calculated as 
$\sigma_i^2 = (\sigma^{\text{Data}}_i)^2 + (\sigma^{\text{MC}}_i)^2$.

%% Advantages

The technique of extracting muon decay parameters through expansion
\eqref{eq:final3} takes into account effects of the detector (such as
interactions of decay positrons in the detector materials) and of the
reconstruction (such as possible biases and inefficiencies of the
track fitting).  No further corrections to the fit result are
required.

An important advantage of the method is that effects of the
reconstruction cancel exactly, since the same software is used to
process real data and Monte-Carlo.  This fact allows for attributing
all systematic effects to deficiencies of the simulation.  (Indeed, a
perfect simulation would be reconstructed exactly as data by
\emph{any} reconstruction software.) On the other hand, the \TWIST{}
detector is very thin, so that the spectrum distortions it causes are
small in the first place.  Thus any discrepancy of the simulation is
multiplied by a small factor, and that helps to achieve a high
precision on the final result.

%%%%%%%%%%%%%%%%%%%%%%%%%%%%%%%%%%%%%%%%%%%%%%%%%%%%%%%%%%%%%%%%
%% Assumptions

The fitting technique assumes independently reconstructed decays (the
limit of zero beam intensity). If there is more than one muon decay in
an event, they are not reconstructed independently, and, strictly
speaking, the MC integration formulas become invalid.  Another
interpretation is that reconstruction of one decay from an event is
affected by the presence of the second, so the response function $K$
becomes dependent on $\r$.   A systematic error accounting for beam
intensity effects has been evaluated, see chapter~\ref{chapter:systematics}.
% 
% This effect can be investigated using
% data-to-data or MC-to-MC comparisons as is done for any other
% systematic.

It is beneficial to choose parameters $\r$ so that $F$ is linear in
$\r$.  Then $\frac{\p\Ni}{\p\r}$ does not depend on $\r$, and the same
Monte-Carlo derivative spectra can be used for any base $\r$.  
Here $\Ni$ is the number of entries in bin $i$ of the spectrum
histogram,
see Appendix~\ref{chapter:spectrum-expansion}.
This is
why instead of $\r=\{\rho,\Pmuxi,\delta\}$ \TWIST{} used
$\r=\{\rho,z,w\}$, where $z=\Pmuxi|_{\xidelta=\text{const}}$, and
$w=\xidelta$. However even with the linear parametrization
$\frac{\p^2\Ni}{\p\r^2}\equiv0$ the expansion \eqref{eq:final3} is still
an approximation, because $\frac{\p^2\ni}{\p\r^2}\ne0$ if the
normalization is affected by $\r$.  This bias could be overcome by
doing iterations: generate MC spectra using $\r=\r^{(n)}$, do the fit,
take the fit result as $\r^{(n+1)}$ and repeat.  In practice Michel
parameters $\rho$, $\xi$, $\delta$, were all already known to better
than $10^{-2}$ precisions before the \TWIST{} measurement, so that the
$\OO(\r^2)$ contribution could be brought down below the $10^{-4}$
level in the first fit, and no more iterations were required.

It is easy to show that the second order term in the expansion is
\begin{multline}
\frac{\dr_{\alpha}\dr_{\beta}}{2}\,\left\{
    \frac1{\Nr}\frac{\p^2\Ni}{\p\ra\p\rb}
    - \frac{\Ni}{\Nr^2}\frac{\p^2\Nr}{\p\ra\p\rb}
\right.
\\
\left.
    {} - \frac1{\Nr^2} \frac{\p\Nr}{\p\ra}\frac{\p\Ni}{\p\rb}
    - \frac1{\Nr^2} \frac{\p\Nr}{\p\rb}\frac{\p\Ni}{\p\ra}
    + \frac{2\Ni}{\Nr^3}\frac{\p\Nr}{\p\ra}\frac{\p\Nr}{\p\rb}
\right\}
\end{multline}
The first two terms in the braces contain a second derivative
of the rate and vanish in the case of a linear parameterization.  The
three remaining terms all contain a first derivative of the total
number of reconstructed events.  In the case of a symmetric response
function
$K(\energy,\costh,\energy',\costh')=K(\energy,-\costh,\energy',-\costh')$
derivatives $\p\Nr/\p(\xixidelta) = \p\Nr/\p(\xidelta) = 0$, so that in
fact the dependence \eqref{eq:final3} of $\ni$ on $\xixidelta$ and
$\xidelta$ is exact without the $\OO(\dr^2)$ term, and the only
deviation comes from $(\Delta\rho)^2$.

\section{\label{section:fitting-delta}Specifics of $\delta$}

Two ways of extracting $\delta$ from data were used by \TWIST{}.  A
2-dimensional spectrum \eqref{eq:theoretical-spectrum-2d} provides the
most detailed information, and can be fit to extract $\rho$, $\xi$,
and $\delta$ simultaneously.  (The $\eta$ parameter was fixed to the
world average, because \TWIST{} could not provide a competitive
constraint on it.) All systematics were evaluated, and the final
result extracted, using this approach.

From
\eqref{eq:theoretical-spectrum-2d}--\eqref{eq:theoretical-spectrum-fas} 
it is clear that an ``upstream minus downstream'' spectrum, 
%
% $\left.\frac{d^2\Gamma}{dx\,d\costh}\right|_{x,\costh}
% -\left.\frac{d^2\Gamma}{dx\,d\costh}\right|_{x,-\costh}
% $
\begin{equation}
f({p,\costh})-f(p,-\costh)\propto{}F_{\text{AS}}(p)\,\costh
\label{eq:up-minus-down}
\end{equation}
is manifestly independent of $\rho$ and $\eta$.  Of course, for a
reconstructed spectrum it is true only to the degree that the response
function is symmetric,
\begin{equation}
K(p,\costh,p',\cos(\theta'))=K(p,-\costh,p',-\cos(\theta')).
\end{equation}
Integrating out $\costh$ in \eqref{eq:up-minus-down} over a fiducial
region, we obtain a 1-dimensional momentum spectrum that can be fit
with only 2 parameters, $\lambda=\{\xixidelta,\xidelta\}$.  Such fits
were employed to cross-check the result,
chapter~\ref{chapter:data-fits}.

The previous measurement of $\delta$ \cite{Balke:1988delta} fitted
instead the momentum dependence of the asymmetry
$A(p)\propto{}F_{\text{AS}}(p)/F_{\text{IS}}(p)$.  However $A(p)$ is a
1-dimensional spectrum, which still depends on all the 4 muon decay
parameters.  A value of $\rho$ had to be assumed to extract $\delta$
in \cite{Balke:1988delta}.  \TWIST{} did not use asymmetry fits
because of the disadvantage of this method.

\label{page:pmu-for-delta}Obviously $\delta$ could not be measured with
unpolarized muons, because it only enters in the angle-dependent part
of Eq.~\eqref{eq:theoretical-spectrum-2d}.  However, the knowledge of
the absolute value of muon polarization $P_{\mu}$ was not required to
determine $\delta$.  Any inadequacies of simulating muon
depolarization in the beam line and the detector could be absorbed
into $\Pmu\xi$, which is a free fit parameter.  To achieve this,
we need to rewrite \eqref{eq:theoretical-spectrum-fas} replacing
\begin{equation}
F_{\text{AS}}^{\text{RC}}(x)\longrightarrow{}\xi{}F_{\text{AS}}^{\text{RC}}(x).
\label{eq:fas-rc}
\end{equation}
\TWIST{} uses radiative corrections computed within the Standard
Model, where $\xi=1$, therefore \eqref{eq:fas-rc} does not introduce
new assumptions.  However after this replacement
Eq.~\eqref{eq:theoretical-spectrum-2d} contains only a single
parameter $\Pmu\xi$, instead of separate $\Pmu$ and $\xi$.  

\endgroup

%%%%%%%%%%%%%%%%%%%%%%%%%%%%%%%%%%%%%%%%%%%%%%%%%%%%%%%%%%%%%%%%
%\section{Checks of the procedure}
\section{Tests of the fitter}

% ?? Add tests with random start, or a plot of chi2 vs \Delta\xidelta etc.

\begin{figure}
\begin{tabular}{c}
\includegraphics[width=\textwidth,height=\thirdpageh]{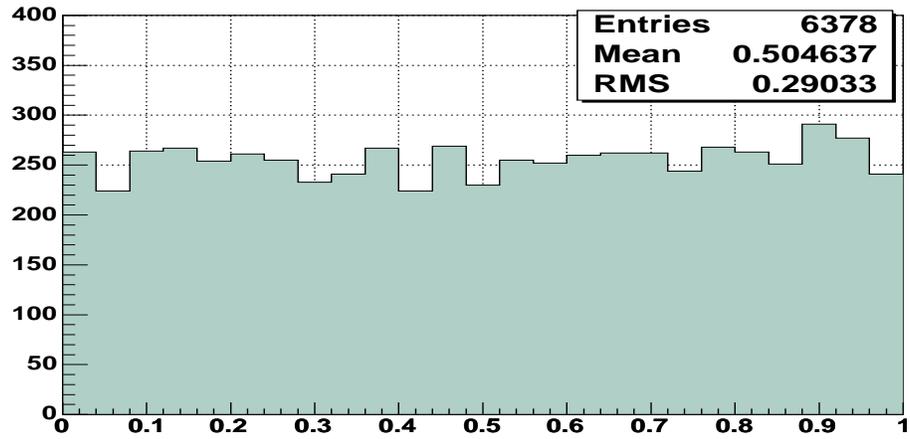}\\
\includegraphics[width=\textwidth,height=\thirdpageh]{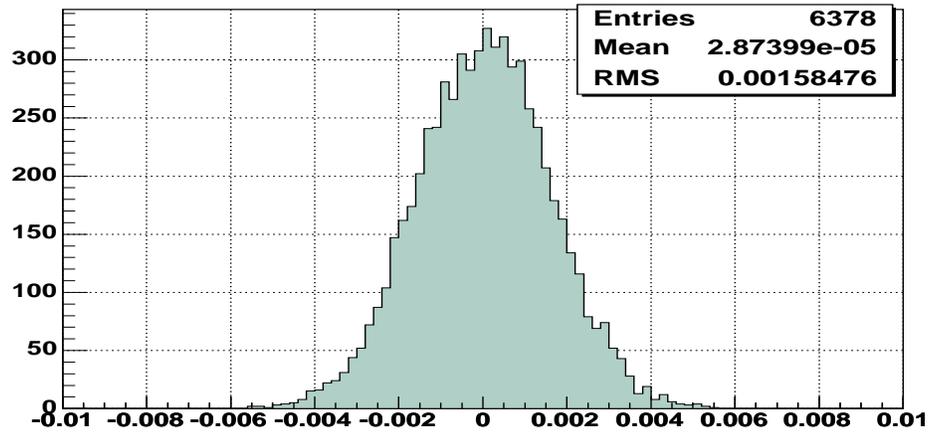}\\
\includegraphics[width=\textwidth,height=\thirdpageh]{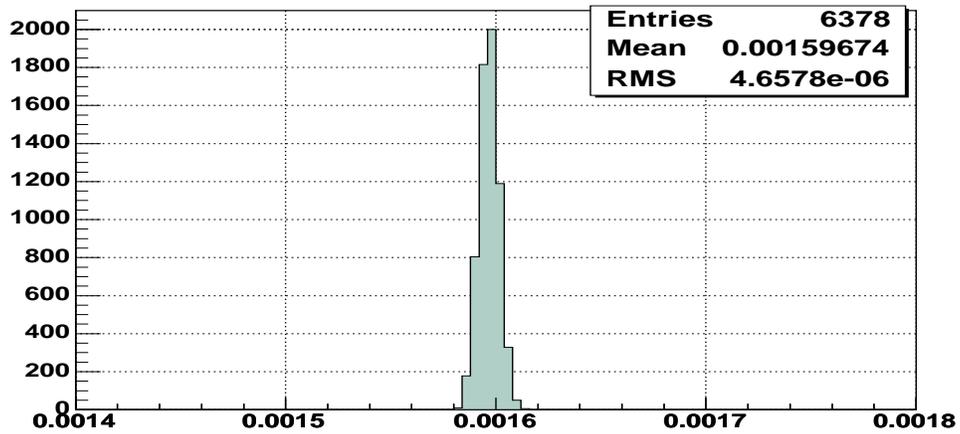}
\end{tabular}
\caption[Fixed statistics fitter tests]{\label{fig:mcfitter-e7}
Distributions of: (top) fit probability, (middle) deviation of the
parameter $\delta$ from the true value, (bottom) reported fit error on
$\delta$.  Each entry in the histograms comes from a statistically
independent fit using the same number of events.  }
\end{figure}

\TWIST{} implementation of the fitting technique
\eqref{eq:final3}--\eqref{fitter-chi2} is based on the \prog{ROOT}
rewrite \cite{root} of the \prog{MINUIT} package \cite{minuit}.

To test the program, about $3\times10^{11}$ muon decays were sampled
for ``data'' spectra using $\rho=0.76$, $\eta=0$, $\Pmuxi=1$,
$\delta=0.76$, the same amount for ``base'' spectra using $\rho=0.74$,
$\eta=0$, $\Pmuxi=0.97$, $\delta=0.74$, and 10\% of that amount for each
of the $\rho$, $\eta$, $\xixidelta$, $\xidelta$ derivatives.  (The
derivative spectra do not depend on Michel parameters.)  Then
different fits were done using these decays.  The fiducial region for
the tests was defined as
\begin{equation}
20\text{MeV/c}<\ptot<50\text{MeV/c},\qquad 0.54<|\costh|<0.80.
\label{eq:test-fiducial}
\end{equation}
%vc: The fiducial region used in the final analysis was further optimized
%vc: (see chapter~\ref{chapter:data-fits}), and was slightly different
%vc: from~\eqref{eq:test-fiducial}.
All histograms in the tests had 110 bins in momentum from 0 to 55MeV/c,
and 100 bins in $\costh$ from -1 to 1.  This is the same binning
as used in the actual data analysis.

For one test, the decays were split into samples of equal size.  The
size of the sample, $4.8\times10^7$ decays, was chosen as to obtain
approximately $10^{7}$ ``data'' events in the fiducial.  Each of the
``data'' samples was fit to a different ``base'' sample using a new
set of ``derivatives'', so that results of all the fits are
statistically independent.  
%
%v1: Both 4 and 3 (fixed $\eta$) parameter fits were tested.  
%
Tests were made in which all 4 parameters were fit as well as tests
in which $\eta$ was set to zero so that only 3 parameters were fit.
The fits were performed in the $\xixidelta$,
$\xidelta$ parametrization and the results were converted to 
the $\Pmuxi$, $\delta$ parametrization using formulas from 
Appendix~\ref{chapter:xideltaconv}.
%
% Fig.~\ref{fig:mcfitter-e7} shows
% distributions of some quantities produced in the fitting.  
%
The distribution of fit probability (computed from fit $\chi^2$
and the number of degrees of freedom) for the case of 
the 3 parameter fits is shown on the top plot of
Fig.~\ref{fig:mcfitter-e7}.  It is flat, as expected.  
The biases on
all fit parameters are consistent with zero, 
\vc{see table~\ref{table:fitter-biases},}
and the estimates of fit errors are close to the RMS of the
corresponding distribution.
\begin{table}
\begin{center}
\begin{tabular}{ccc}
Parameter & Mean bias${}\times10^4$ & Mean/(Error on mean)\\
\hline
$\rho$   & -0.026 & -0.24 \\
$\Pmuxi$    & -0.014 & -0.06 \\
$\delta$ & +0.287 & +1.45
\end{tabular}
\caption{\label{table:fitter-biases}\vc{Absolute and relative biases
    for different fit parameters.}}
\end{center}
\end{table}
As an example, the middle and the bottom
plots of Fig.~\ref{fig:mcfitter-e7} show distributions of
$\delta^{\text{fit}}-\delta^{\text{true}}$, and of the error
$\sigma_{\delta}$, respectively.

%%%%%%%%%%%%%%%%%%%%%%%%%%%%%%%%%%%%%%%%%%%%%%%%%%%%%%%%%%%%%%%%
\begin{figure}
\begin{tabular}{c}
\includegraphics[width=\textwidth,height=\quarterpageh]{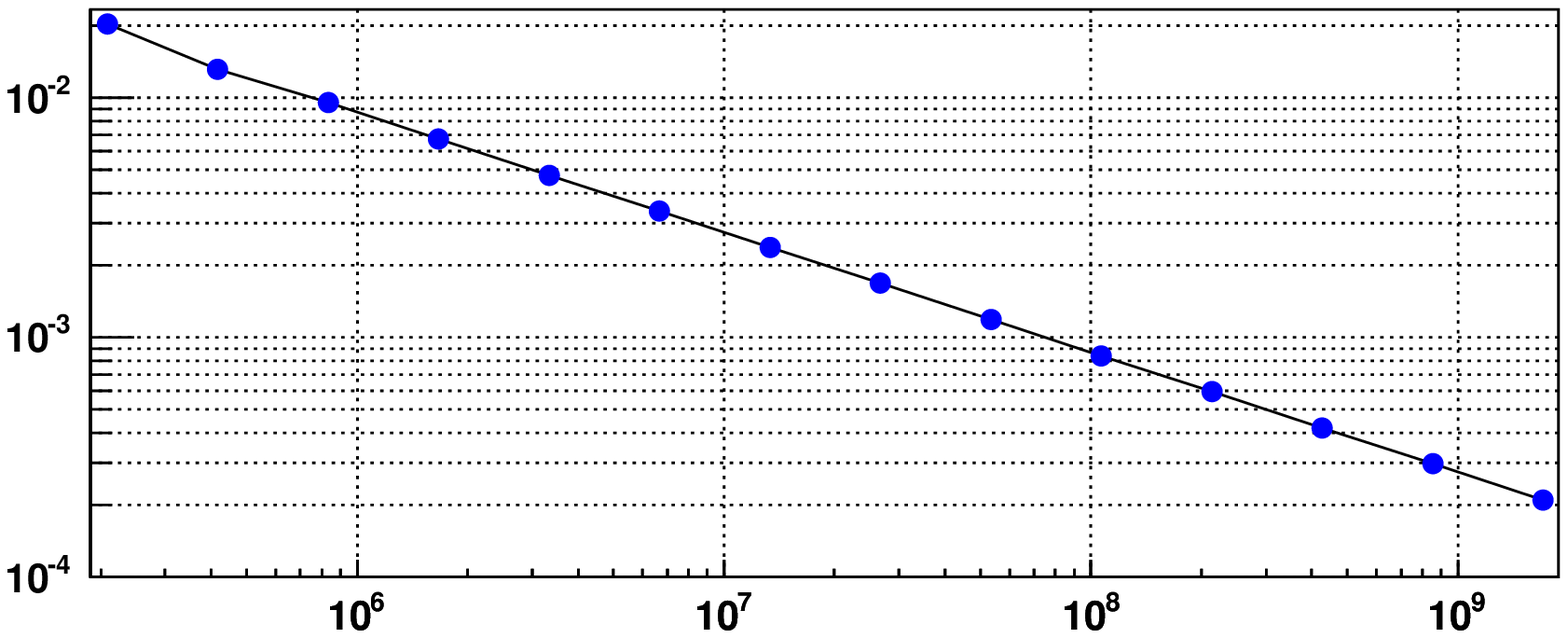}\\
\includegraphics[width=\textwidth,height=\quarterpageh]{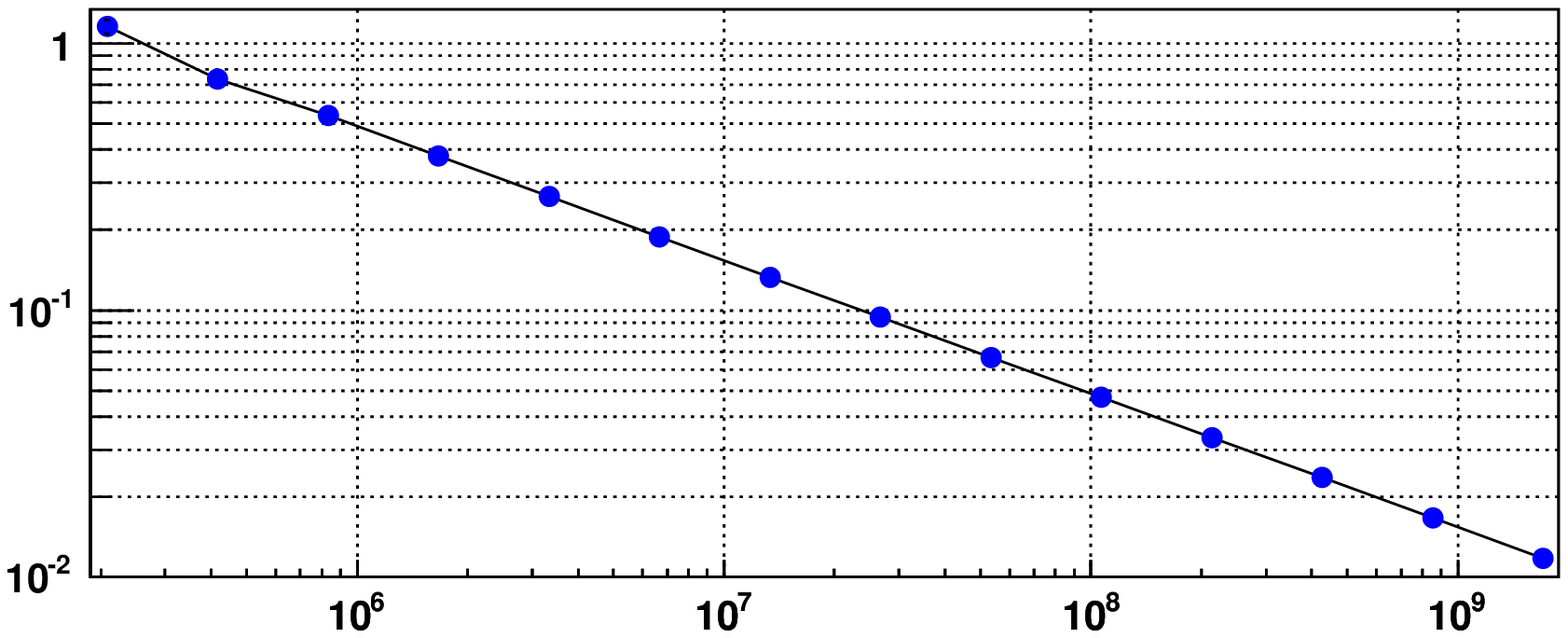}\\
\includegraphics[width=\textwidth,height=\quarterpageh]{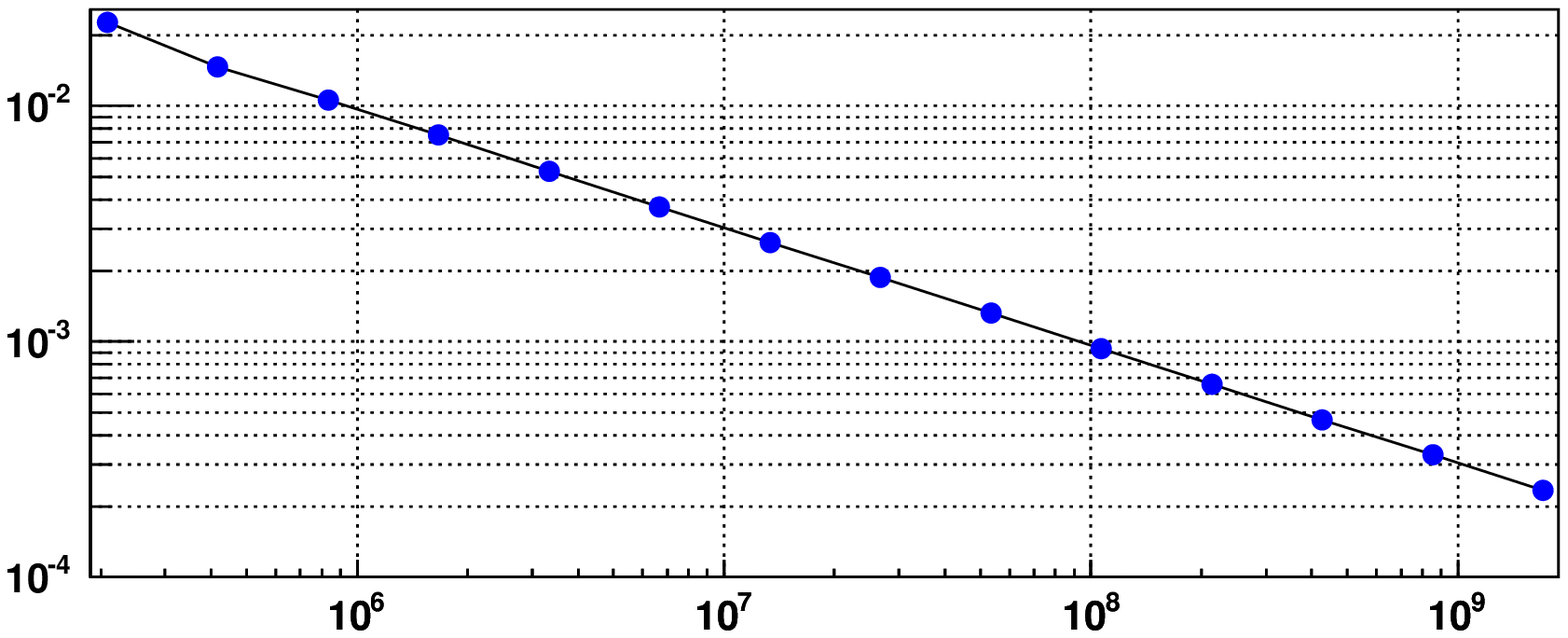}\\
\includegraphics[width=\textwidth,height=\quarterpageh]{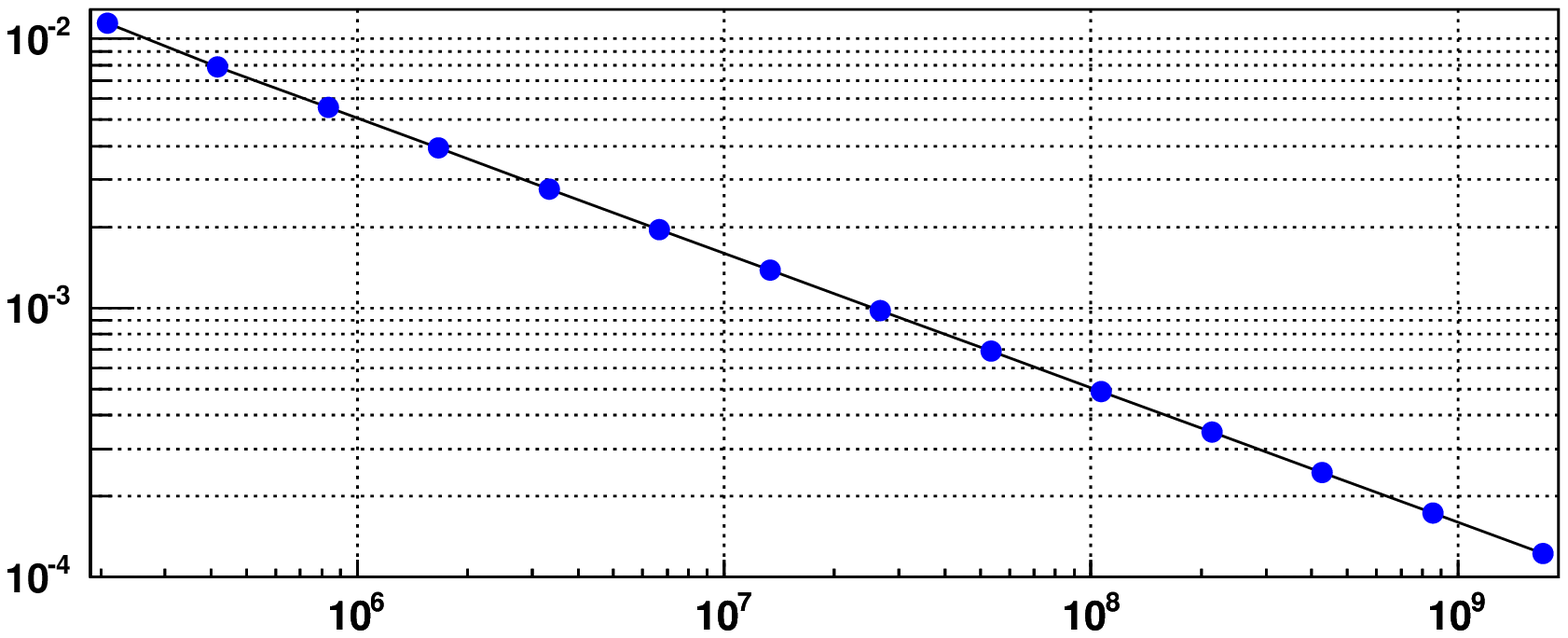}
\end{tabular}
\caption[Fitter errors vs statistics]{\label{fig:fit-error-vs-stat}
  Scaling of fit errors with statistics.  Horizontal axis: the number
  of data events in the fiducial region. Vertical axis: the mean value
  of reported fit error of 18 fits at the given statistics.  Top to
  bottom: $\rho$, $\eta$, $\Pmuxi$, $\delta$.  }
\end{figure}

\begin{figure}
\begin{tabular}{c}
\includegraphics[width=\textwidth,height=\quarterpageh]{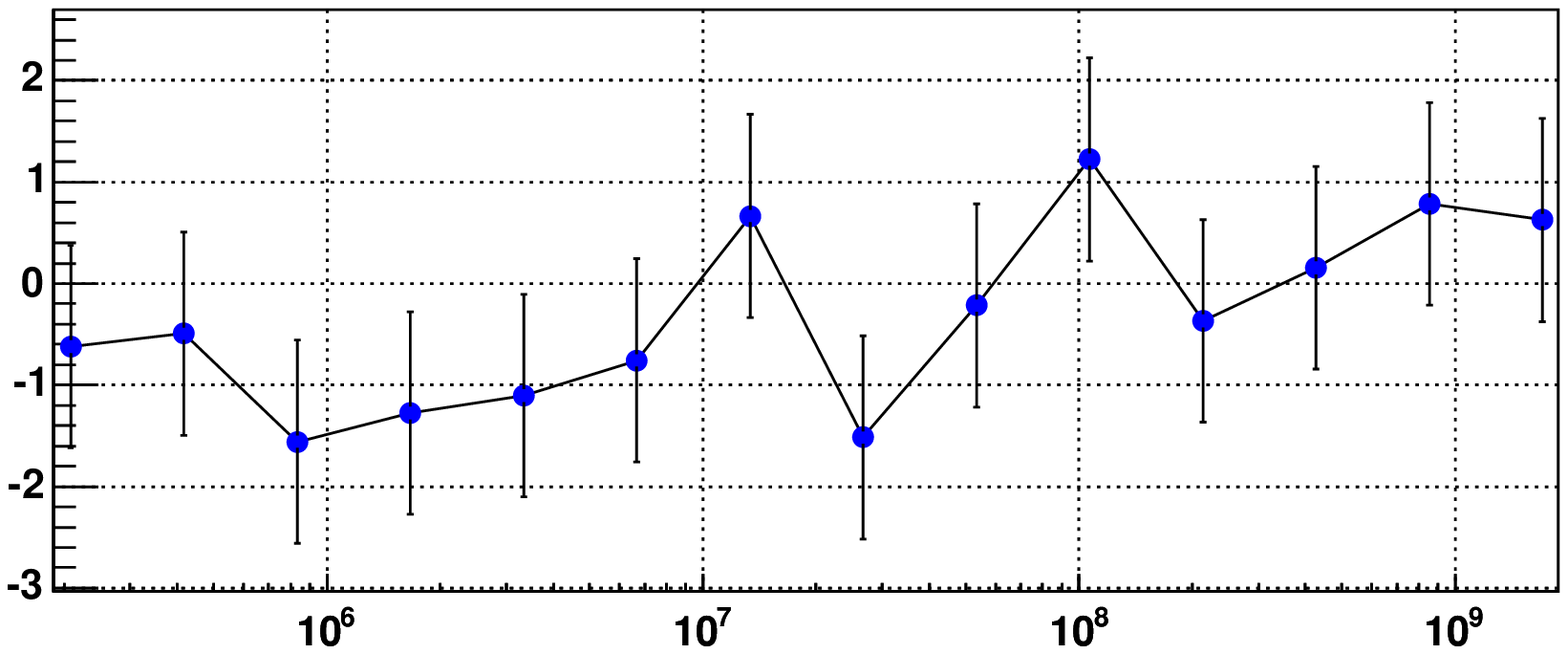}\\
\includegraphics[width=\textwidth,height=\quarterpageh]{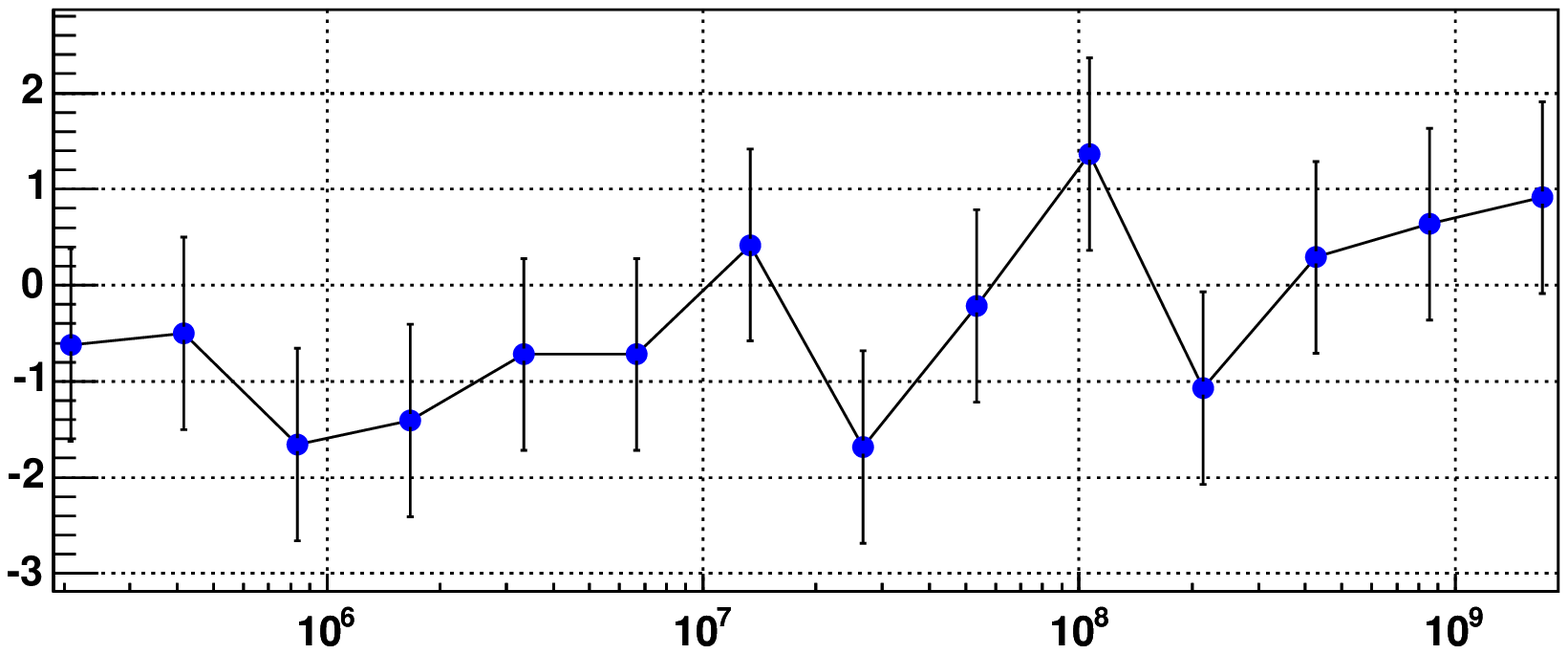}\\
\includegraphics[width=\textwidth,height=\quarterpageh]{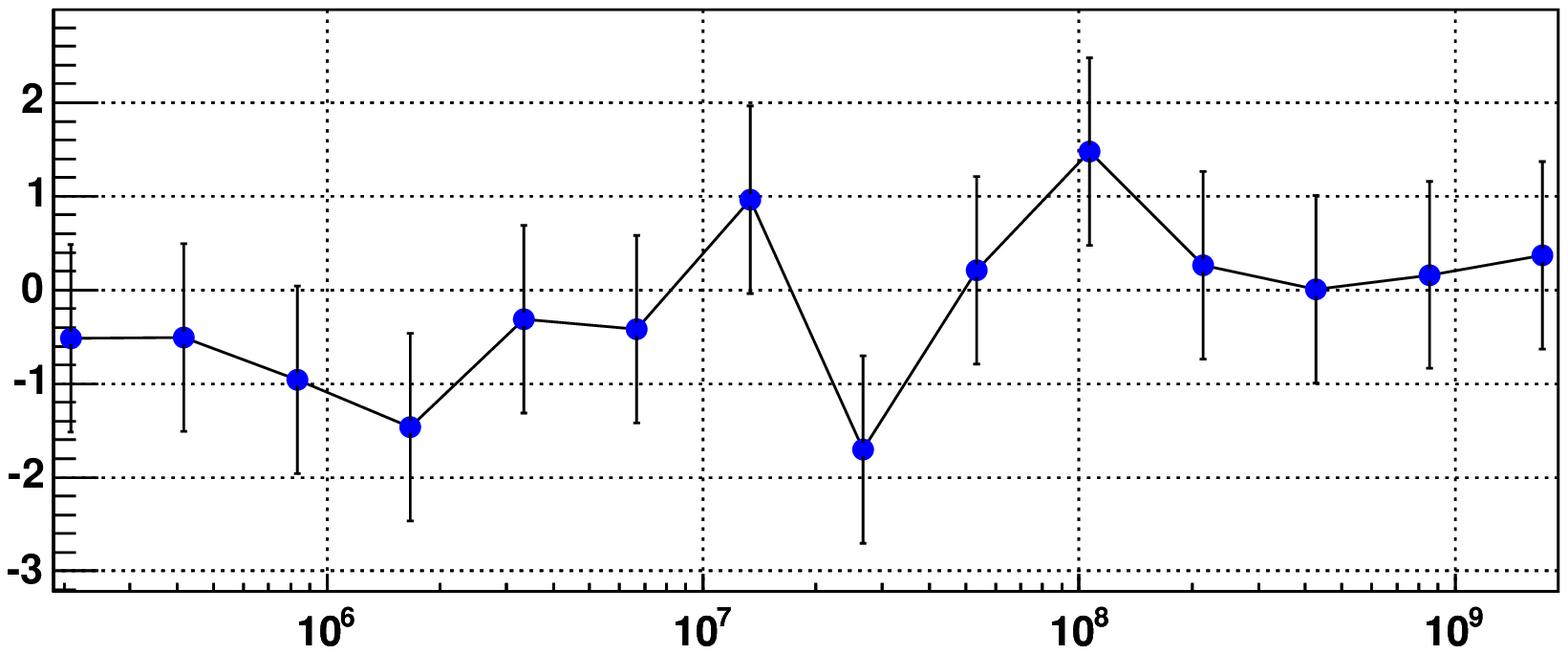}\\
\includegraphics[width=\textwidth,height=\quarterpageh]{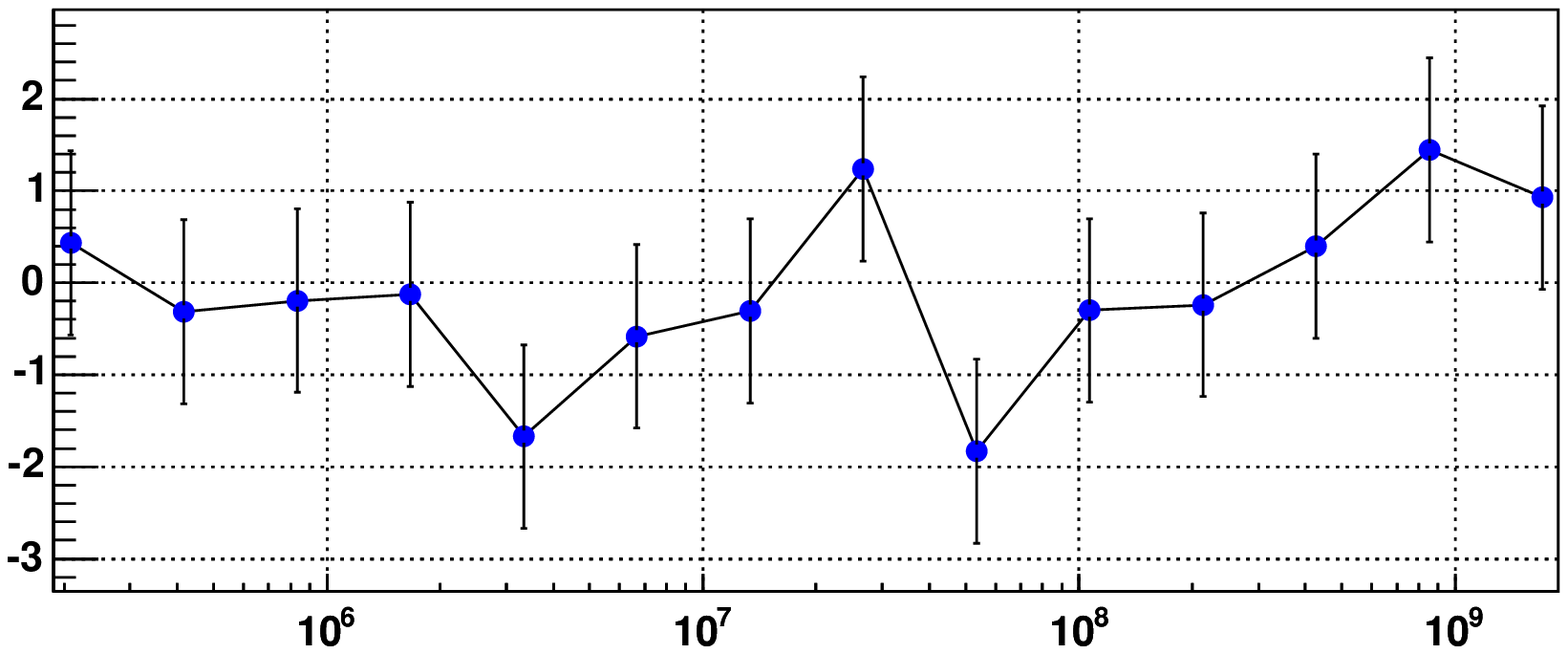}
\end{tabular}
\caption[Fitter biases vs statistics]{\label{fig:fit-bias-vs-stat}
  Normalized biases as functions of statistics.  Horizontal axis: the
  number of data events in the fiducial region. Vertical axis:
  (Mean)/(Error~on~mean) of the 18 fits at the given statistics. Top
  to bottom: $\rho$, $\eta$, $\Pmuxi$, $\delta$.  }
\end{figure}

\begin{figure}
\begin{tabular}{c}
\includegraphics[width=\textwidth,height=\quarterpageh]{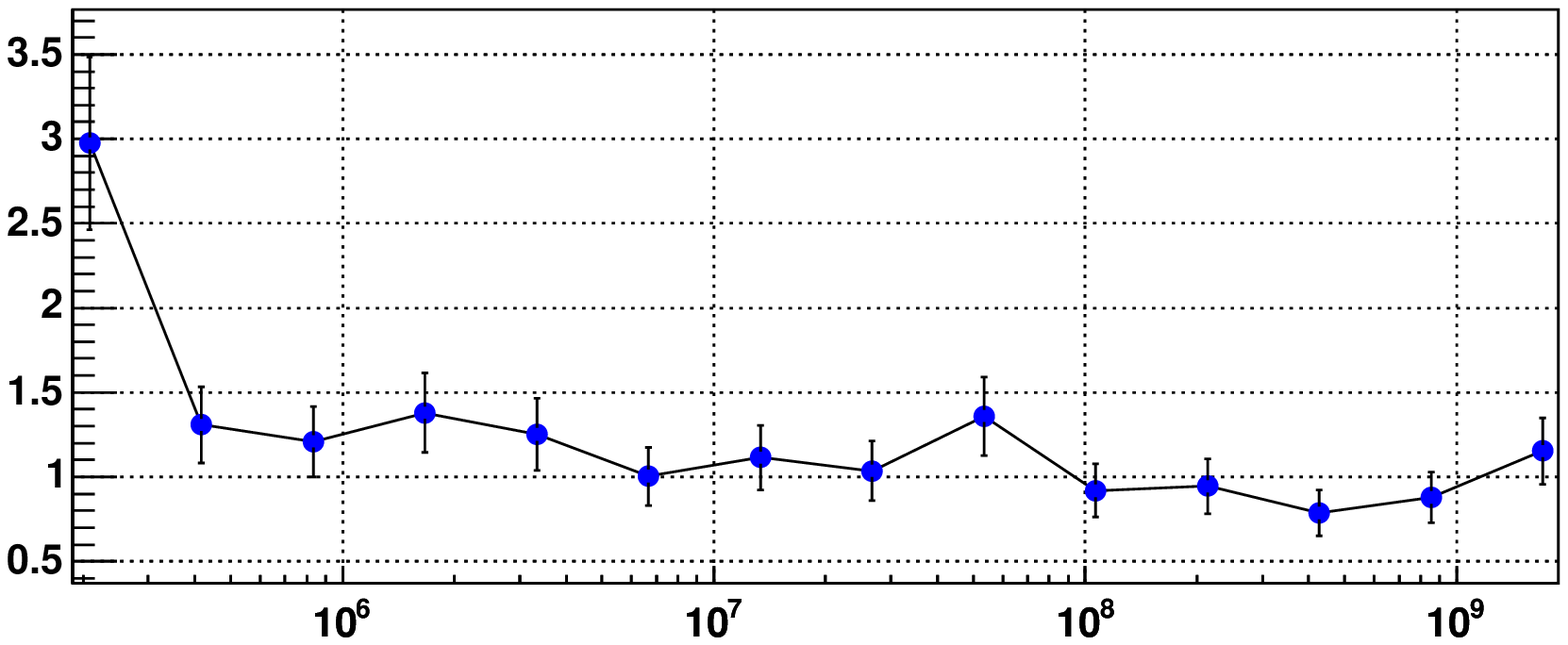}\\
\includegraphics[width=\textwidth,height=\quarterpageh]{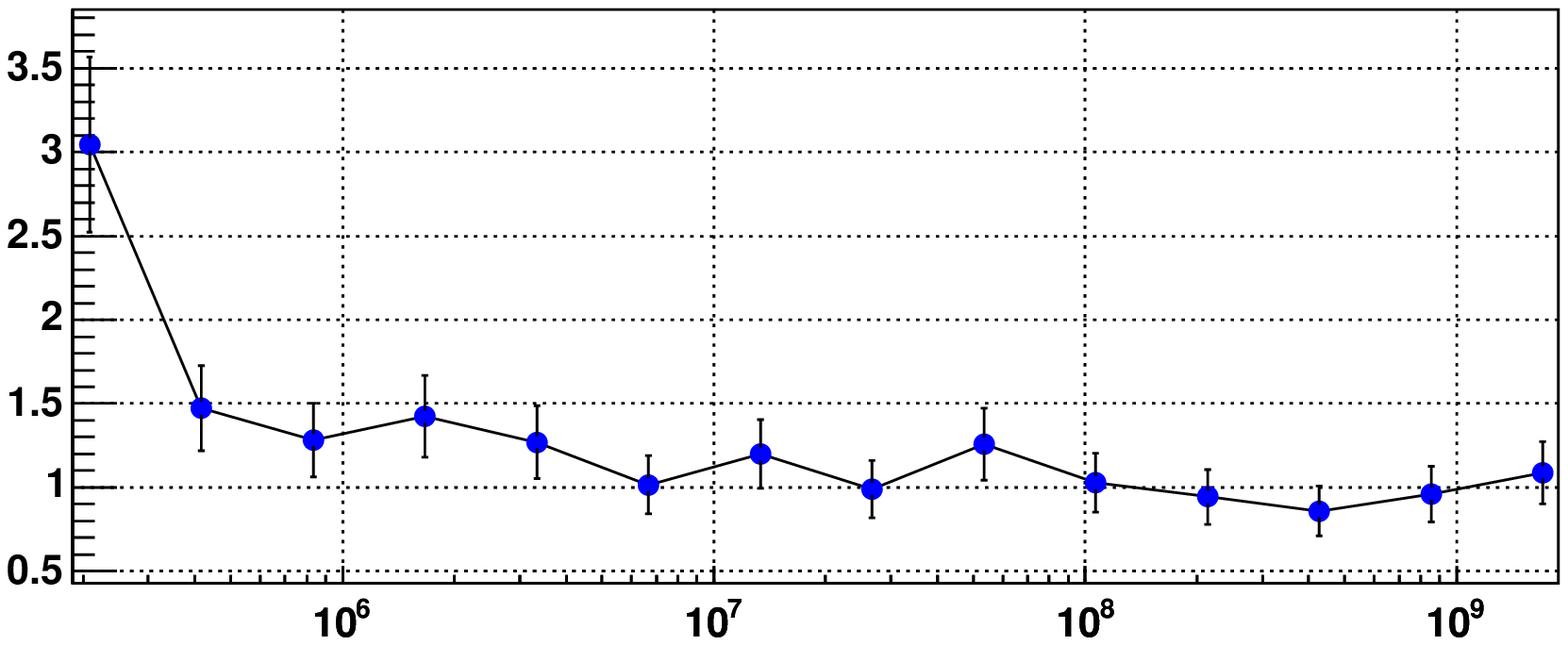}\\
\includegraphics[width=\textwidth,height=\quarterpageh]{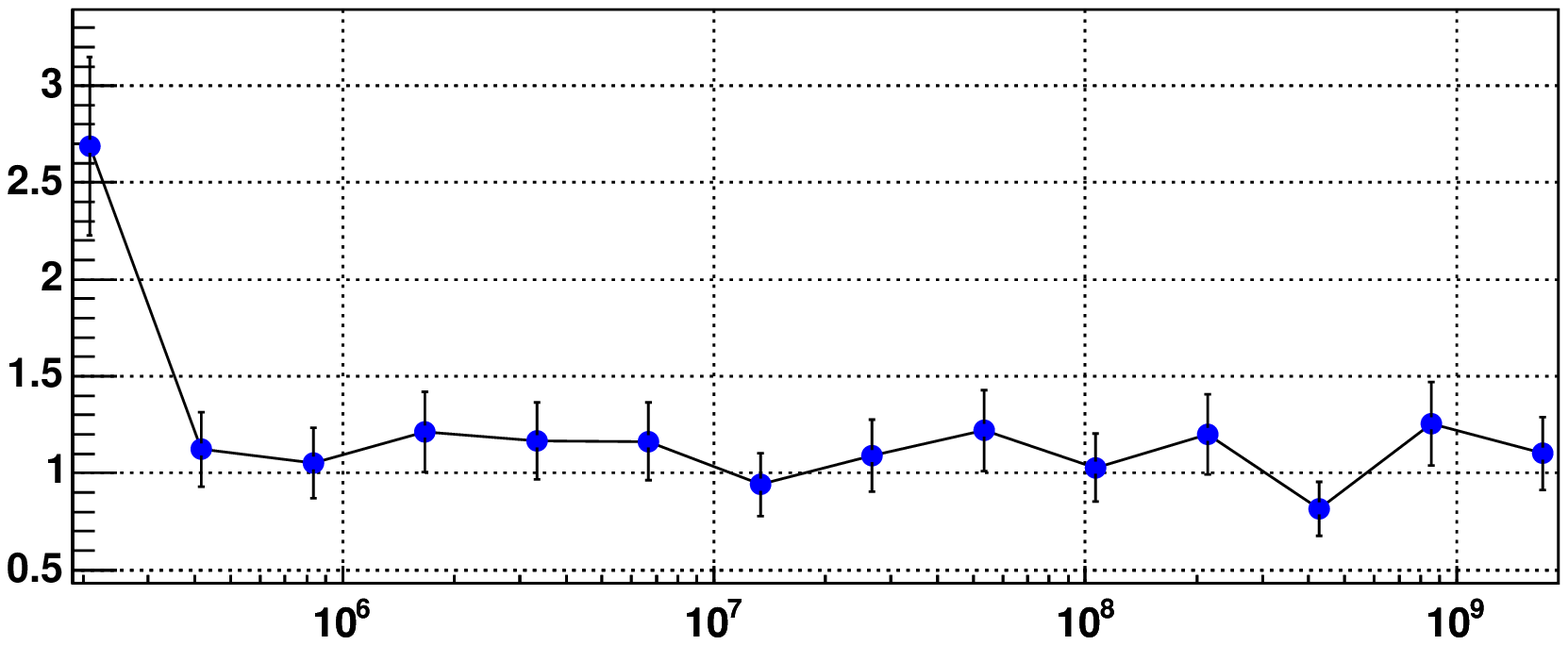}\\
\includegraphics[width=\textwidth,height=\quarterpageh]{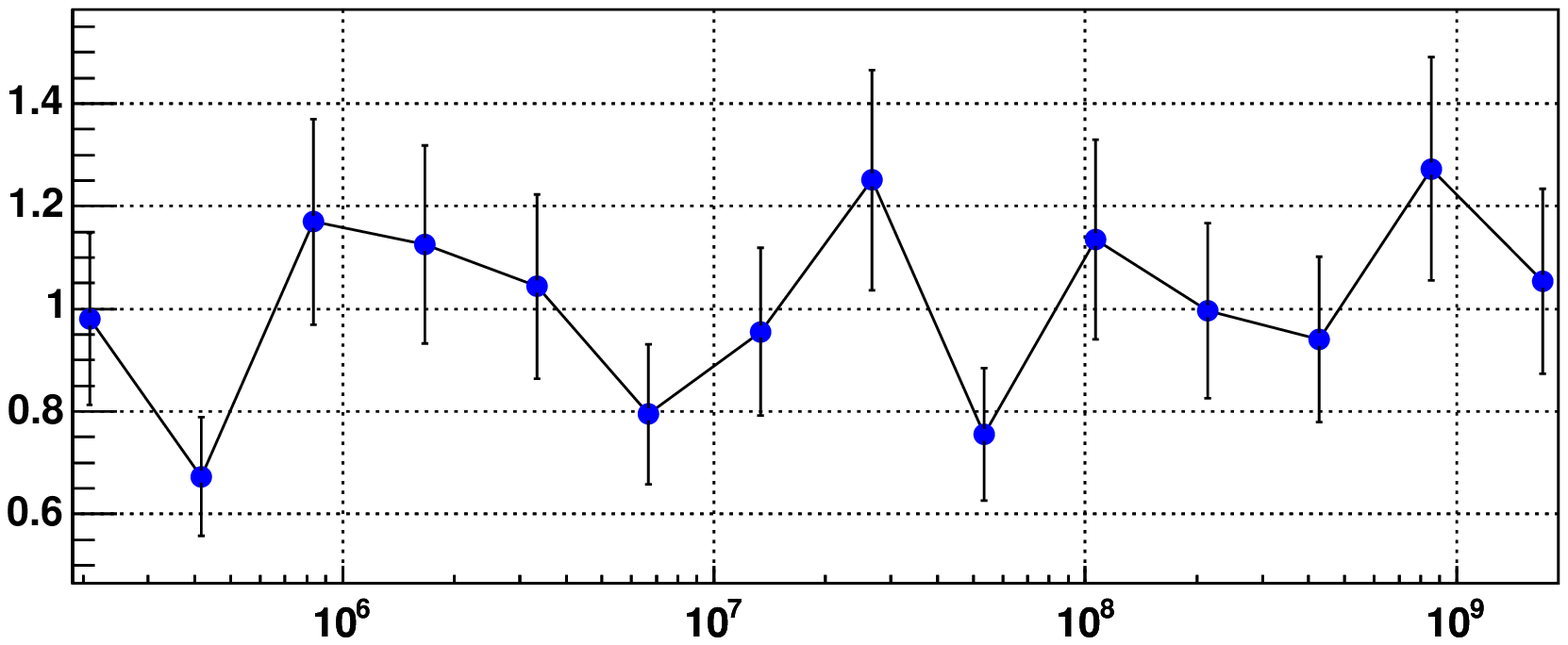}
\end{tabular}
\caption[Reliability of error estimates vs
  statistics]{\label{fig:fit-err_ratio-vs-stat} The reliability of fit
  error estimate as function of statistics.  Horizontal axis: the
  number of data events in the fiducial region. Vertical axis:
  RMS/(Mean~fit~error) of the 18 fits at the given statistics. Top to
  bottom: $\rho$, $\eta$, $\Pmuxi$, $\delta$.  }
\end{figure}
%%%%%%%%%%%%%%%%%%%%%%%%%%%%%%%%%%%%%%%%%%%%%%%%%%%%%%%%%%%%%%%%

Another test looked at the performance of the fitter as a function of
statistics.  The size of a ``data'' sample was varied from $10^6$ to
$2^{13}\times10^6$ events. (About $2\times10^5$ to $2\times10^9$
events in the fiducial.) For each sample size, 18 fits were
performed using the same size of the ``base'' sample, and 10\% of that
size for each of the four derivatives.  Each point on
Fig.~\ref{fig:fit-error-vs-stat}--\ref{fig:fit-err_ratio-vs-stat}
aggregates 18 fits.  Again, all of the fits in this test were
statistically independent.   
Fig.~\ref{fig:fit-error-vs-stat} demonstrates that fit errors, except
for the lowest tested statistics, scale as $1/\sqrt{N}$.
No statistically significant biases were observed in the test, see
Fig.~\ref{fig:fit-bias-vs-stat}.
It can be seen from Fig.~\ref{fig:fit-err_ratio-vs-stat} that fit
errors are underestimated when the statistics is low.  However they
are consistent with the spread of the fitted parameters when 
the statistics used is higher than about $10^6$ events in the fiducial
region.  Our measurement used more than $10^7$ data events
per typical fit, the lowest statistics fit had $0.79\times10^7$ 
events in its fiducial region.  Thus the fitting technique 
\eqref{eq:final3}--\eqref{fitter-chi2}, the conversion formulas from
Appendix~\ref{chapter:xideltaconv}, and the software implementation
of the fitter, have been completely validated for the measurement.

% \section{Blind analysis}
\section{Blind analysis\label{section:blind-analysis}}

\begin{figure}[tbp]
  \begin{center}
    \includegraphics[width=\textwidth]{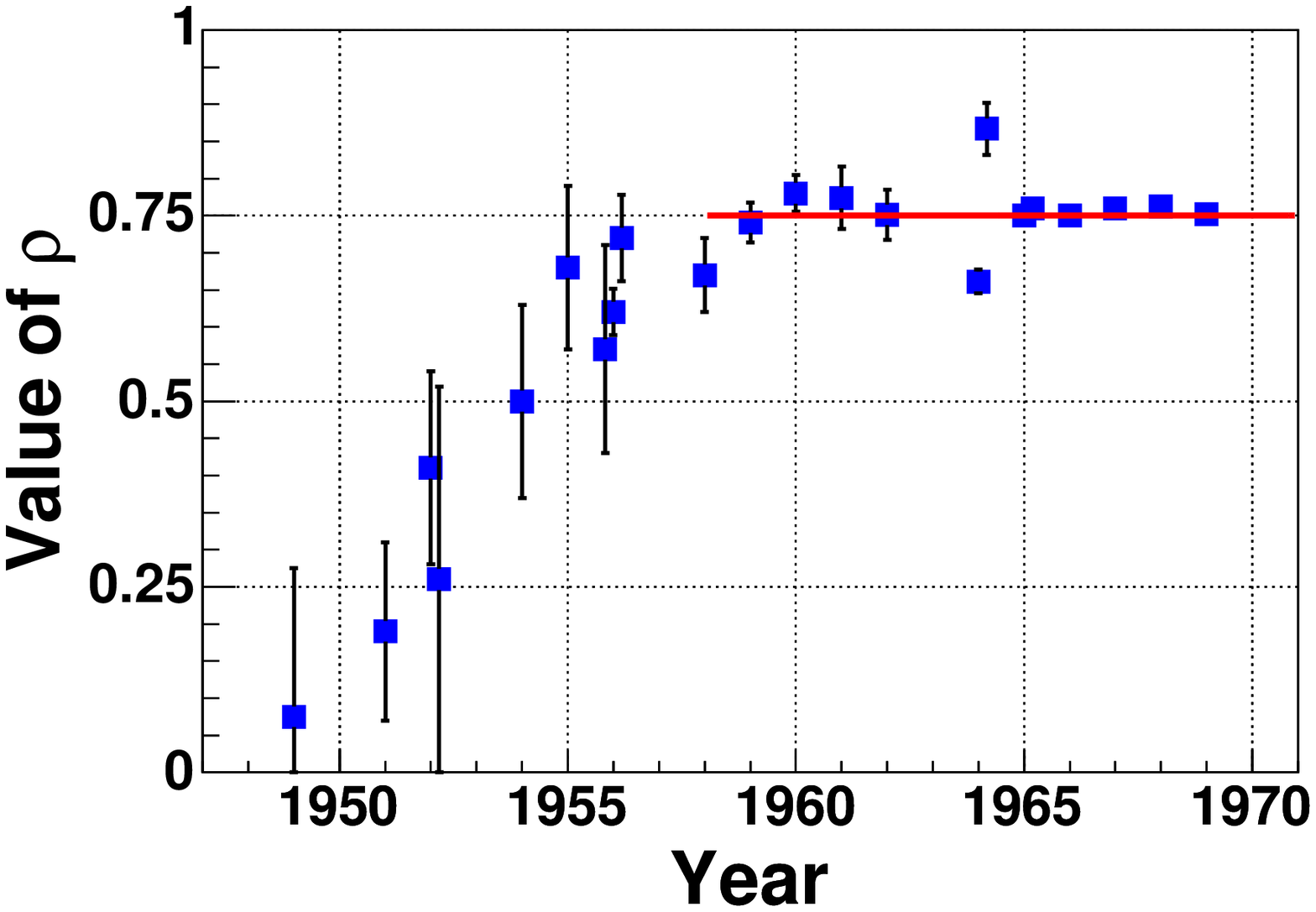} 
  \end{center}
  \caption[Experimental determination of the Michel parameter $\rho$
      since 1950.]{\label{fig:greiner-muller}Experimental
      determination of the Michel parameter $\rho$ since 1950.  The
      solid line represents the $V-A$ value $\rho=3/4$, }
\end{figure}

Blind analysis is an increasingly popular tool to avoid (subconscious)
experimenter's bias when doing a physics measurement.  There are
subjective decisions to be made in e.g. setting the cuts and rejecting
``bad'' data samples.  Several different choices may be equally valid
and what gets actually used may be affected by the knowledge of what
version gives a better agreement with the expected answer.  Another
possible source of a bias is looking for software bugs, or additional
sources of systematic uncertainty when a result does not agree with
the expectation, and not looking for them as hard otherwise.  
Often considerable judgment is involved in estimating the size of
systematic uncertainties.  Knowing how close a result is to an
expected answer may affect the quoted error.  A good discussion of
motivation for blind analysis, and more examples, can be found in
\cite{harrison-blind-analysis}.

There is evidence for such bias in some particle physics measurements.
For example, ``history plots'' in \cite{pdg-all:2004} show
non-statistical variations of several measured quantities with time.
In \cite{greiner-muller}, there is the following remark about history
of measurements of the Michel parameter $\rho$, which is shown on
Fig.~\ref{fig:greiner-muller}:
\begin{quote}
  The curve shows
  the improvement of the experiments, but perhaps also the
  prejudice of the experimentalists.
\end{quote}
% 
% A plot from \cite{greiner-muller} that is relevant for \TWIST{} is
% reproduced on Fig.~\ref{fig:greiner-muller}.  
%% 
The point is that human
bias may introduce an unquantifiable systematic uncertainty in the
result of a measurement.  It is desirable to avoid the possibility of
such a bias.  This can be accomplished by doing analysis in a
``blind'' fashion, i.e. keeping the final result hidden till the
analysis is essentially complete.  The value of a measurement does not
contain any information about its correctness and is of no use in
performing the analysis, therefore hiding the answer does not impede
the work.

%%%%%%%%%%%%%%%%%%%%%%%%%%%%%%%%%%%%%%%%%%%%%%%%%%%%%%%%%%%%%%%%
\subsection*{\TWIST{} implementation of a blind analysis}

Among our requirements for a blind analysis scheme were:
\begin{itemize}
\item Does not exclude any \TWIST{} member from doing any part of the
  analysis.
\item Convenient to use.
\item Minimal modifications to the existing software.
\item Hard to break.
\end{itemize}
\begin{figure}[tbp]
  \begin{center}
    \includegraphics[width=\textwidth]{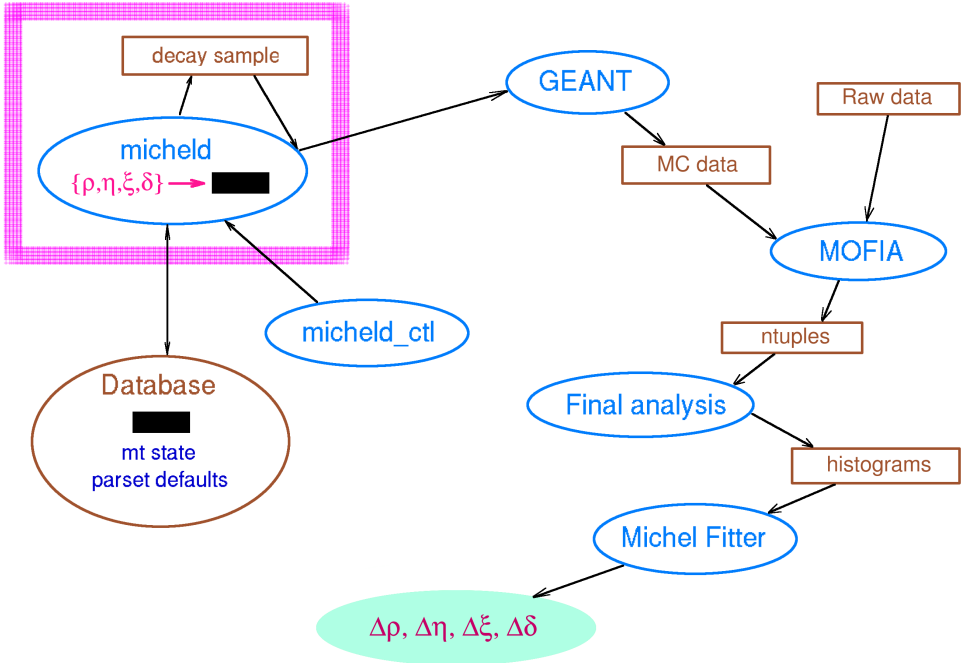} 
  \end{center}
  \caption{\label{fig:blind-analysis-sketch}
    \TWIST{} blind analysis scheme.
  }
\end{figure}
A scheme of implementation satisfying these criteria is shown on
Fig.~\ref{fig:blind-analysis-sketch}.  The idea of the method is to
blind the MC samples, not the fitter.  It is clear from \eqref{eq:final3}
that the fitting method gives only \emph{deviations} of the Michel
parameters in data from the values used to generate a base Monte-Carlo
spectrum.  Thus it was sufficient to hide the values of Michel parameters
used for MC production.  The secrecy was based on using an asymmetric
cryptography.  A private-public key pair was produced, and the
private key locked up in a place not accessible by \TWIST{} members.

The piece of software written specifically to make a blind analysis
possible is \prog{micheld}, which is essentially a muon decay spectrum
generator.  The spectrum it produces includes all radiative
corrections described in chapter \ref{chapter:monte-carlo}.  The
program runs on a computer which is not controlled by the \TWIST{}
group, and none of \TWIST{} members could login there during the
analysis period. (Symbolized by a ``wall'' around \prog{micheld} on
Fig.~\ref{fig:blind-analysis-sketch}.)  \prog{micheld} is a
multi-threaded TCP/IP server, accepting data and control requests from
a network.  An operator uses \prog{micheld\_ctl} to instruct
\prog{micheld} to produce a random set of Michel parameters.  They are
sampled uniformly within the following limits:
\begin{equation}
\rho=\frac34\pm0.02,\quad\eta=0,\quad\Pmuxi=1\pm0.03,\quad\delta=\frac34\pm0.02
\end{equation}
A candidate set of values is tested for being physically allowed (the
end point asymmetry $\Pmuxi\delta/\rho\le1$) before it is accepted.  An
accepted set of parameters is encrypted using the public key, and the
encryption result is stored in a database.  By another operator
request an accepted set of Michel parameters is used to generate a
series of muon decay samples, which are written to disk.

During a Monte-Carlo production run, a \prog{GEANT} process obtains a
sample of muon decays from the disk through \prog{micheld}. Every time
\prog{GEANT} needs to decay a muon, it uses the energy and angle (with
respect to the spin of the muon) of the next decay in the sample.
Since different muon decay samples were produced with the same
(unknown) Michel parameter values, we had the possibility to study
consistency between different data sets, and to estimate systematics
by fitting one MC sample to another, as explained in
chapter~\ref{chapter:systematics}.

After the analysis was complete, the ``black box'' was opened, and the
final values of parameters were computed using the results of the fits
and the revealed MC values.  A small complication arises from the fact
that the $(\xixidelta,\xidelta)\to(\Pmuxi,\delta)$ conversion
(Appendix~\ref{chapter:xideltaconv}) requires the knowledge of
``true'' MC parameters.  That was addressed by using the known
approximate numbers $\Pmuxi_0=1$, $\delta_0=0.75$, at the ``blind''
stage.  This approximation introduces an uncertainty of about 5\% on
the \emph{deviation} $\Delta\delta$ from the fit, which translates
into a 5\% uncertainty on sensitivities of $\delta$ to different
systematics (see chapter~\ref{chapter:systematics}).  This was
adequate for doing the analysis.  After opening the black box all data
and systematics fits were re-run using the revealed values of $\Pmuxi_0$
and $\delta_0$ to get rid of the additional uncertainty.  This was a
mechanical procedure not involving any judgment, so it did not violate
the philosophy of blind analysis.

%%%%%%%%%%%%%%%%%%%%%%%%%%%%%%%%%%%%%%%%%%%%%%%%%%%%%%%%%%%%%%%%

%%%%%%%%%%%%%%%%%%%%%%%%%%%%%%%%%%%%%%%%%%%%%%%%%%%%%%%%%%%%%%%%
\chapter{Determination of systematic uncertainties\label{chapter:systematics}}
%\chapter{Determination of systematic uncertainties\label{chapter:systematics}}

To estimate the systematic uncertainty of the result, it is essential
to account for all sources of systematic effects.  On the other hand,
it is important to avoid double counting, so that the same physical
cause of a bias is not included more than once in the estimate.  Also,
one has to distinguish evaluation of systematic uncertainties from
consistency checks \cite{Barlow:2002yb}.  The decision on what effects
to consider and exactly how to treat them involves judgment and is, to
some degree, arbitrary.  An important feature of the present
measurement is that it was done using a blind analysis scheme. (See
\ref{section:blind-analysis}.)  This means that a complete list of
systematic effects, along with a method to evaluate each of them, was
fixed before ``opening the black box'' and revealing the measured
value of $\delta$.  Such an approach reduces the possibility that the
obtained estimate of systematic uncertainty is subjectively biased.

In our approach (chapter \ref{chapter:fitting-all}) all systematic
effects can be attributed to imperfections of the Monte-Carlo.  A
simulation perfectly reproducing data would be reconstructed exactly
as data by \emph{any} reconstruction software, thus the result of a
fit of data to Monte-Carlo would be unbiased.  In other words, 
effects of reconstruction cancel in the comparison of data to
Monte-Carlo to the degree that the simulation reproduces the data.
Therefore the list of systematics does not include effects of the
reconstruction.  This of course does not mean that the quality of the
reconstruction software is irrelevant: the \emph{sensitivity} of the
result to a given imperfection of the simulation may be reduced by
improving the reconstruction program.

For \TWIST{} the possible sources of systematic uncertainties can be
classified into the following independent groups: positron
interactions, spectrometer alignment, chamber response, momentum
calibration, and muon beam stability.  Some of the specific effects
within these groups bias the result in the same way for all the data
sets, while other effects are time dependent and may contribute
differently to the different data sets.  The former effects only
contribute to the uncertainty of the result, while the time-dependent
uncertainties also should be used in computing the weighted average
value of $\delta$ from the four data
sets. (chapter~\ref{chapter:data-fits}.)

Most of the systematics effects were evaluated using the following
technique: evaluate the sensitivity $R'=d\delta/da$ of the result
$\delta$ to a systematic parameter $a$ known to a precision
$\pm\sigma_a$, and add $R'\sigma_a$ quadratically to the systematic
uncertainty \cite{Barlow:2002yb}.  To estimate $R'$, a data set set
was taken or a Monte-Carlo set was generated under a different condition
$a_{\text{test}}\ne{}a_{\text{nominal}}$.  The reconstructed ``test''
spectrum is then fit to the ``nominal'' one using
\eqref{eq:final3}--\eqref{fitter-chi2}.  This expresses the change in the
spectrum shape due to the systematic effect in terms of changes in the
Michel parameters.  The systematics estimate can therefore be written
as
\begin{equation}
R'\sigma_a=\frac{\Delta{R}}{\Delta{a}}\,\sigma_a
=(\delta_{\text{test}}-\delta_{\text{nominal}})\,
\frac{\sigma_a}{(a_{\text{test}}-a_{\text{nominal}})}
= \frac1S\,(\delta_{\text{test}}-\delta_{\text{nominal}})
\end{equation}
where we \vb{have} introduced the scaling factor
$S=(a_{\text{test}}-a_{\text{nominal}})/\sigma_a$.

\vb{The following subsections present in tabular format summaries of
  individual systematics for each group, followed by a short
  explanation of each entry in the table.  If all systematic uncertainties
  $R'\sigma_a$ are identical for all data sets used in the
  measurement, a single column is used to present them, as in
  Tables~\ref{table:sys-alignments}, \ref{table:sys-interactions},
  \ref{table:sys-mubeam}.  Otherwise individual numbers for data sets
  A, B, 1.96~T, and 2.04~T are shown, Tables
  \ref{table:sys-chamber-response}, \ref{table:sys-ecalib}.}

%%================================================================
\section{Spectrometer alignment\label{section:sys-alignments}}

Table \ref{table:sys-alignments} summarizes alignment related
systematic uncertainties, all of which are data set independent
because the \TWIST{} detector was mechanically stable and its parts
did not move during the data taking \cite{Henderson:2004zz}.
%
%
%================
\begin{table}[htbp]
%%\framebox[\width]{%
\begin{tabularx}{\textwidth}{|X|d{2}|d{1}|d{2.0}|}
\hline
Name 
& \multicolumn{1}{c|}{$10^3\times\Delta\delta$} 
& \multicolumn{1}{c|}{Scaling} 
& \multicolumn{1}{c|}{$10^3\times{}R'\sigma_a$}
\\ \hline\hline
%----------------
Translations &  0.39 & 28 & 0.01 
\\ \hline
%----------------
Rotations &  -4.33 & 39 & -0.11
\\ \hline
%----------------
Z (longitudinal) &  -1.07 & 10 & -0.11 
\\ \hline
%----------------
B field to detector axis & -1.86 & 3.1 & -0.60 
\\ \hline
%----------------
\hline
\multicolumn{3}{|l}{Total} & 0.62
\\ \hline
\end{tabularx}%
%
%% }%\framebox{}
\caption{Alignment systematics.\label{table:sys-alignments}}
\end{table}
\begingroup
%\ifagsdouble\onehalfspacing\else\relax\fi
\begin{description}
\item[The translational] alignment of the wire planes relative to each
  other is measured using straight tracks produced by 120~MeV pions
  with the solenoid off.  The accuracy of the \vb{resulting
  alignment}, $\sigma_a$, is 5~\mum{} \cite{Henderson:2004zz}.  The
  test spectrum was produced by analyzing the nominal data set B with
  a special alignment file, which was produced by applying random
  shifts to the nominal alignment corrections file.  The resulting
  translational spread of the wire planes from their nominal positions
  was 140~\mum{} (RMS), giving the scaling factor
  $S=140~\mum/5~\mum=28$.

\item[The rotational] alignment systematic was determined in a
  similar way, using a specially prepared rotational corrections
  file. The introduced angular spread of $0.39^\circ$ (RMS) yields a
  scaling factor of 39 compared to the $0.01^\circ$ precision of the
  nominal rotational alignment \cite{Henderson:2004zz}.

\item[The Z (longitudinal)] alignment of the wire planes is estimated
  to be accurate to 30\mum{} from mechanical precision of the detector
  construction \cite{Henderson:2004zz}.  A Monte-Carlo set was
  generated with Z positions of the planes offset by 300~\mum{} (RMS)
  and compared to a nominal MC set, thus producing a scaling factor of
  10.

\item[B field to detector axis.]  The nominal Monte-Carlo generation
  and data analysis assume a perfect alignment of the detector axis to
  the coordinate system of the magnetic field map.  To produce the
  test spectrum, the field map was rotated in GEANT by $0.25^\circ$.
  The actual misalignment is estimated from data by fitting (an
  approximation of) a helix \vb{that is} not aligned to the detector axis to
  \vb{the} positron tracks, with the alignment angles being two additional free
  parameters \vb{in the fit}.
%
% on a track by track basis.  
%
  The average misalignment \vb{found in this way} is 
  $0.08^\circ$, so the scaling factor is $0.25^\circ/0.08^\circ=3.1$.

\end{description}
\endgroup

%%================================================================
\section{Positron interactions}
%
%================
\begin{table}[htbp]
%%\framebox[\width]{%
\begin{tabularx}{\textwidth}{|X|d{2}|d{1}|d{2.0}|}
\hline
Name 
& \multicolumn{1}{c|}{$10^3\times\Delta\delta$} 
& \multicolumn{1}{c|}{Scaling} 
& \multicolumn{1}{c|}{$10^3\times{}R'\sigma_a$}
\\ \hline\hline
%----------------
Energy smearing & 0.58 & 4 & 0.15
% g148pt200/g148a1
% blind: delta =  0.000594 +- 0.001133867
% bb10:  delta =  0.000582 +- 0.001107264
% 
\\ \hline
%----------------
Multiple scattering & 0.10 & 20 & 0.00
% g148ag1/g148a1
% blind: delta =  0.000141 +- 0.001136320
% bb10:  delta =  0.000096 +- 0.001109592
% 
\\ \hline
%----------------
Hard and intermediate interactions &   &  & 0.53
% Carl's estimate
\\ \hline
%----------------
Detector materials & -0.73 & 2 & -0.36
% g143a1/g122a1  OK: 30um test vs 10um. GABS changed 0.42(std)->0.24(test)
% blind: delta = -0.000750 +- 0.001105399
% bb10:  delta = -0.000726 +- 0.001079533
% 
% bb10   drho  = -0.001010 +- 0.000595368
%
\\ \hline
%----------------
Outside materials & -2.09  & 70 & -0.03
% s7a2/s2a9
% blind: delta = -0.002134 +- 0.001638713
% bb10:  delta = -0.002087 +- 0.001600394
% 
\\ \hline
%----------------
\hline
% sqrt((0.000582/4)^2+(0.000096/20)^2+(0.00053)^2+(0.000726/2)^2+(0.002087/70)^2)
% = 0.00065935
\multicolumn{3}{|l}{Total} & 0.66
\\ \hline
\end{tabularx}%
%
%% }%\framebox{}
\caption{Positron interactions systematics.\label{table:sys-interactions}}
\end{table}
\begingroup
%\ifagsdouble\onehalfspacing\else\relax\fi
\begin{description}
\item[Energy smearing.] This systematic accounts for \vb{any} mismatch
  between data and MC in the momentum resolution.  This mismatch has
  been observed at the end point (chapter \ref{chapter:ecalib}).  A
  test spectrum was produced by applying a Gaussian smearing
  ($\sigma_{p_t\,\text{smear}}=200$~keV/c) to the transverse component of
  momentum, $p_t$, of reconstructed Monte-Carlo events.  The same
  Monte-Carlo data analyzed in the standard way, without any smearing,
  gave the reference spectrum.

  For $\sigma_{p_t\,\text{smear}}=50$~keV/c the width of the end
  point, as \vb{determined by the energy calibration procedure
  (chapter~\ref{chapter:ecalib})},
  agreed with data,
  $\sigma^{\text{smeared~MC}}_{\text{EC}}\approx\sigma^{\text{Data}}_{\text{EC}}$.
  Therefore the scaling factor used was 200/50=4.

%  \rrr{Conversation with Dick on 2005-01-25: smear $p_t$ and watch the
%  end point $\sigma$. For 50~keV smearing MC${}\approx{}$Data.  Test
%  done with 200~keV, thus the scaling is 4.}
%
%  \rrr{Dick's August 2004 talk refers to a Glen's posting, is it
%    bn=twist\_physics key=1087322081 ?  But that does not define the
%    scaling.}
%
%
% Dick's posting with the final scaling:
%
% https://twist.phys.ualberta.ca/forum/view.php?site=twist&bn=twist_physics&key=1094417426

\item[Multiple scattering.] The multiple scattering systematics
  addresses a potential deficiency in the simulation of multiple
  scattering (chapter \ref{chapter:monte-carlo}).  The angle $\theta$
  of reconstructed Monte-Carlo events was smeared with a Gaussian
  using
  \begin{equation}
  \sigma_{\theta\,\text{smear}}
  =\frac{K(\text{rad})}{p\text{(MeV/c)}\,\sqrt{\frac1{|\costh|}}}
  \label{eq:angle-smearing}
  \end{equation}
  to produce a test spectrum.  Here $K$ is a parameter, \vb{while} the
  functional dependence on momentum and angle comes from a simplified
  formula for \vb{the} multiple scattering angle of a relativistic particle in
  matter. (See e.g. \cite{pdg-all:2004}, page 245.)  The
  $\frac1{|\costh|}$ term is proportional to the amount of material
  traversed by a particle in the planar \TWIST{} geometry.  For
  $K=1\text{~rad}$, $p=30$~MeV/c, and $\costh=0.7$,
  Eq.~\eqref{eq:angle-smearing} gives
  $\sigma_{\theta\,\text{smear}}\approx29$~mrad.  The size of the
  discrepancy between data and Monte-Carlo was estimated as the bigger
  between the differences in mean and RMS between data and Monte-Carlo
  in the validation studies (chapter \ref{chapter:monte-carlo}).
  None of the differences exceeded 1.5~mrad \cite{dick-muls-posting},
  so a scaling factor of 20 was chosen\footnote{\vb{This} estimate of the
  scaling factor is not well justified.  However the typical smearing
  angle of 29~mrad is larger then the angular resolution (chapter
  \ref{chapter:reconstruction}), and the ``raw'' effect of the
  smearing is still small.  This is why the obtained estimate of this
  systematics was never refined.  }.

\item[Hard and intermediate interactions.]  

% The final 14% hard scatter systematics number:
% 
% https://twist.phys.ualberta.ca/forum/view.php?site=twist&bn=twist_physics&key=1094238154
%
% NB: the above uses Al target to get necessary statistics.
%
% Carl's calculation (1st round):
% https://twist.phys.ualberta.ca/forum/view.php?site=twist&bn=twist_physics&key=1095129648
%
% Carl-2nd:
% https://twist.phys.ualberta.ca/forum/view.php?site=twist&bn=twist_physics&key=1097289460

% from Carl-2nd:
% 
%      Fractional yield change due to hard interactions over ~25-50 MeV:  0.0068
%      Fractional yield change due to intermediate ints over ~25-50 MeV:  0.0066
% 
%      0.05*0.0066 + 0.14*0.0068 = 0.00128
%      Delta rho sensitivity factor for this range ~ 1.0/1.8 = 0.56
% 
%      Effect:  0.00128*0.56 = 0.00071
% 
% 
% Thus, the calculated systematic uncertainty for rho would be 0.00071.  For
% delta, we need to rescale this by the ratio of the target thickness
% uncertainties for delta/rho.  The most recent v2 target thickness sensitivities
% that I have are 0.00075/0.00101.  This may not be the final numbers.  But if
% they are:
% 
% 
% This implies the systematic uncertainty for delta due to GEANT treatment of
% intermediate and hard interactions is:  0.00053.

\begin{figure}[tbp]
\begin{tabular}{cc}
\includegraphics[width=0.5\textwidth]{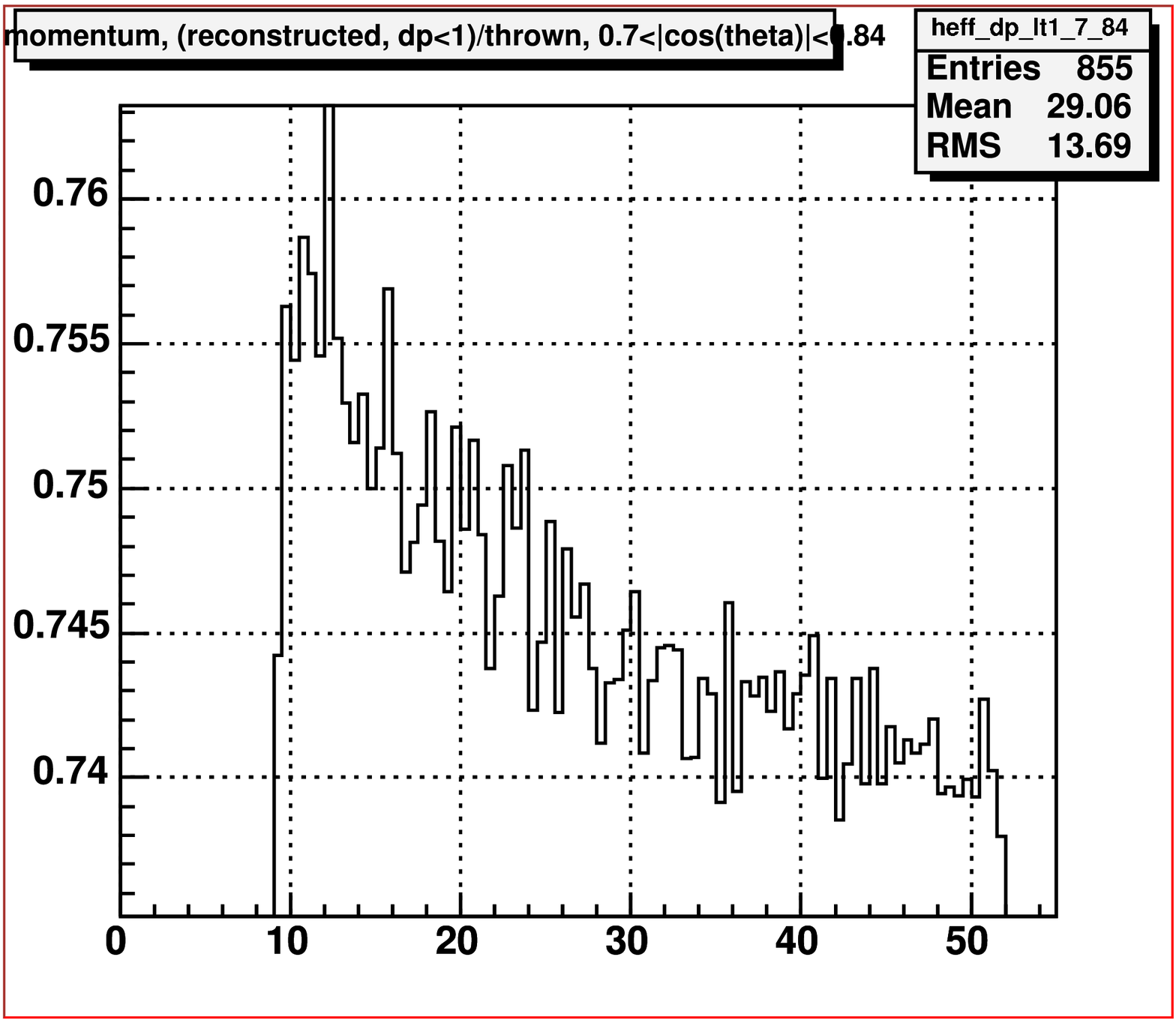}&
\includegraphics[width=0.5\textwidth]{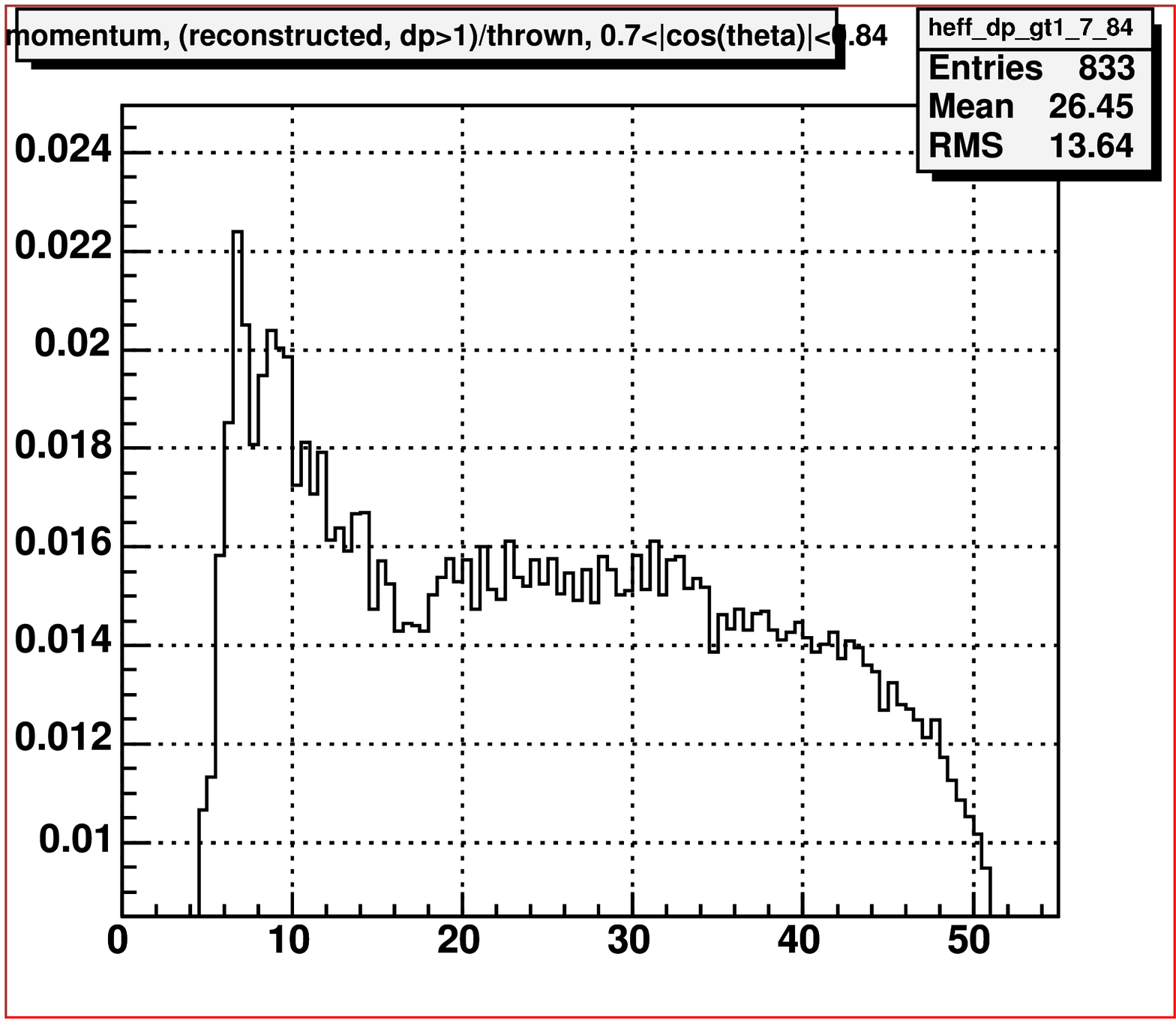}
\end{tabular}
\caption[Reconstructed/thrown momentum
  spectra]{\label{fig:hard-scatters}Ratios of (reconstructed)/(thrown)
  momentum spectra for: $\Delta{p_{\text{MC}}}\le1\text{~MeV}/c$
  (left), $\Delta{p_{\text{MC}}}>1\text{~MeV}/c$ (right).  Only events
  with $0.7<|\costh|<0.84$ are included in the spectra. (Plots from
  \cite{carl-hard-scatter-v2}.)  }
\end{figure}

  The systematic uncertainty due to the imperfect simulation of hard
  and intermediate interactions was estimated in the following way
  \cite{carl-hard-scatter-v2}.  \vb{A spectrum} of reconstructed and thrown
  positron momenta \vb{was} prepared for events that lost \vb{less than
  1~MeV/c in the detector} (according to Monte-Carlo information).
  \vb{A similar spectrum was generated for positrons that lost greater
  than 1~MeV/c.}
  The 1~MeV/c number is an
  arbitrary boundary between ``intermediate'' and ``hard''
  interactions, consistent with the boundary used in the Monte-Carlo
  validations (chapter~\ref{chapter:monte-carlo}). Ratios of the
  (reconstructed)/(thrown) distributions, presented on
  Fig.~\ref{fig:hard-scatters}, show the distortions of the momentum
  spectra for the two classes of events.

  The fractional yield changes over the range 25--50~MeV/c are
  \begin{align}
    y_s&=(s_1-s_2)/\text{norm}\approx0.0067 && \text{(left plot, intermediate interactions)},\\
    y_h&=(h_1-h_2)/\text{norm}\approx0.0070 && \text{(right plot, hard interactions)}.
  \end{align}
  Here $h_1\approx0.7458$, $h_2\approx0.7405$, $s_1\approx0.0153$,
  $s_2\approx0.0102$ are the readings at 25~MeV/c and 50~MeV/c from
  the plots, and $\text{norm}=\frac12(h_1+h_2+s_1+s_2)$.  The change
  of \vb{the Michel parameter} $\rho$ by $0.0010$ leads to a fractional yield change of $0.0018$
  over the same range.  GEANT has been validated to 14\% for hard
  interactions and to 5\% for intermediate interactions
  (chapter~\ref{chapter:monte-carlo}).  Therefore an uncertainty on
  $\rho$ can be estimated as
  $\frac{0.0010}{0.0018}\times(0.05\times{}y_s+0.14\times{}y_h)$.  To
  \vb{determine the} effect on $\delta$, we scale the uncertainty on $\rho$ by the
  ratio of $\Delta\delta/\Delta\rho\approx0.719$ obtained in the
  ``Detector materials'' systematic below.  The final number is
  \begin{equation}
    0.719\times\frac{0.0010}{0.0018}\times(0.05\times{}y_s+0.14\times{}y_h)
      \approx 0.00053
  \end{equation}

% 
% h1=0.7458; h2=0.7405; s1=0.0153; s2=0.0102; norm=(h1+h2+s1+s2)/2; (0.000726/0.001010) * (1.0/1.8) * (0.05 *(s1-s2)/norm + 0.14*(h1-h2)/norm)
% .00052671241459375464
% 
% (s1-s2)/norm
% .00674692419632226484
% 
% (h1-h2)/norm
% .00701150945892313798
% 
% norm
% .75590000000000000000
% 

\item[Detector materials.]  The nominal thickness of the graphite
  coating on each side of the stopping target is 10~\mum{}.  A test
  Monte-Carlo set was generated with 30~\mum{} graphite coating and
  fit to a nominal 10~\mum{} MC set.  The setting of the gas absorber
  in the test MC run was tuned to keep the muon stopping distribution
  centered in the target, because the systematic effect related to
  shifts of this distribution is accounted for separately.
  Since the thickness of the graphite coating on
  each side of the \vb{stopping} target is known to be between 5~\mum{} and
  20~\mum{} \cite{Henderson:2004zz}, a scaling factor of
  $(30-10)/(20-10)=2$ has been applied.

  Another output from this systematic study is the relative size of
  effects on $\delta$ and $\rho$ due to interactions of the positrons
  in detector materials,
  $\Delta\delta/\Delta\rho\approx0.000726/0.001010\approx0.719$, used
  above in the estimation of the uncertainty due to hard and
  intermediate interactions.

%  $10\genfrac{}{}{0pt}{}{+10}{-5}~\mum$ 

%  
%  \rrr{gen143: 30~\mum{} graphite on each side
%  and 24\% CO2.  Nominal: 10~\mum{} (and 42\%). E.g. the fit shows an
%  effect of 20~$\mu$m of graphite. (The mustop is compensated for
%  by GABS?) The nominal thickness (g122) is 10, but the target can be
%  as thick as 20~\mum{} (NIM paper, p.14).  Thus the uncertainty on the
%  thickness is 10, and the scaling factor is 20/10=2.}

% g122 
% 
%   beta =    -0.037537 +-  0.004413725
%     au =     0.062607 +-  0.003075724
%     ad =     0.061245 +-  0.003078801
% ResPar =     0.067807 +-  0.000923308
%  norm0 = 14020.354737 +- 70.658902863
%  norm1 =    -0.866459 +-  0.003612010
%    sl0 = -1737.972646 +- 136.117299290
%    sl1 =   754.006895 +- 157.083764470
% 
% g143
% 
%   beta =     0.000402 +-  0.004275357
%     au =     0.071107 +-  0.002984310
%     ad =     0.063249 +-  0.002970111
% ResPar =     0.068477 +-  0.000889946
%  norm0 = 15234.271952 +- 73.856730411
%  norm1 =    -0.875224 +-  0.003476995
%    sl0 = -1979.783224 +- 142.103337612
%    sl1 =   599.218875 +- 164.601897641

%  \rrr{Only $\alpha_{\text{up}}$ is changed in EC of gen143,
%  $\alpha_{\text{dn}}$ is the same as for gen122.  Did we add the
%  graphite on both sides? }

\item[Outside materials.]  Decay positrons after leaving the tracking
  volume may scatter off the outside structures of the detector,
  re-enter the tracking volume, and produce more hits, \vb{and consequently}
  confuse track reconstruction.  The biggest source of backscatters is
  the \vb{upstream} beam package, which holds the trigger scintillator and the
  degraders (Fig.~\ref{fig:twist-drawing}). There were no materials
  other than air at the downstream end during the normal data taking. 

  To estimate the effect of an imperfect Monte-Carlo simulation of the
  positron backscattering process, a special data set was taken with
  an aluminum plate mounted outside of the downstream end of the
  detector.  A fit of this data set to a nominal data set produced the
  shift $\Delta\delta=-2.09\times10^{-3}$, shown in
  Table~\ref{table:sys-interactions}.

  \vb{Backscatters and} beam particle\vb{s} overlapping in time with
  \vb{a} decay positron may not be distinguishable on an event by
  event basis.  A study of backscatter rates in data and Monte-Carlo
  \cite{rob-backscatter-scaling} used the ``PC time of flight''
  variable, $T_{\text{PC}}=t_u-t_d$, where $t_u$ and $t_d$ are the
  average times of hits in the 4 most upstream and most downstream
  PCs.  Accidental overlaps\vb{, such as those with beam positrons,}
  produce a flat background, \vb{while} backscatters \vb{produce} a
  peak in \vb{the} $T_{\text{PC}}$ distribution. This \vb{difference}
  provides a way to measure the rate of backscatters.  
  \vc{The rate of backscatters from the downstream direction under 
    nominal conditions was demonstrated to be ${}\approx0$, and}
  the scaling
  factor was estimated as \vc{the ratio of the backscatter rate from
  the downstream aluminum to the difference of the backscatter rates
  between the nominal data and simulation \cite{rob-backscatter-scaling}:
  \begin{equation}
    S = \frac{N_{DdD}-0}{|N_{GuS}-N_{DuS}|}\approx70.
  \end{equation}
  Here} $N_{GuS}$ is the number of backscatters seen in GEANT from the
  upstream material under the ``standard'' conditions, $N_{DuS}$ is
  the number of backscatters seen in data from the upstream material
  under the ``standard'' conditions, and $N_{DdD}$ is the number of
  backscatters seen in data from the downstream aluminum.

\end{description}
\endgroup

%%================================================================
\section{Chamber response}
%
%================
\begin{table}[htbp]
%%\framebox[\width]{%
\begin{tabularx}{\textwidth}{|X|d{2}|c|d{2}|d{2}|d{2}|d{2}|}
\hline
Name 
& \multicolumn{1}{c|}{$10^3\times\Delta\delta$} 
& Scaling
& \multicolumn{4}{c|}{$10^3\times{}R'\sigma_a$}
\\ \cline{4-7}
{}&{}&{}
& \multicolumn{1}{c|}{A}
& \multicolumn{1}{c|}{B}
& \multicolumn{1}{c|}{1.96}
& \multicolumn{1}{c|}{2.04}
\\ \hline\hline
%----------------
DC efficiency & 0.27 & 50 & 0.01 & 0.01 & 0.01 & 0.01
% s2a12/s2a9 OK eff=95%, time=700ns
% blind: delta =  0.000277 +- 0.001328827
% bb10:  delta =  0.000271 +- 0.001297649
% 0.271/50 = 0.00542
\\ \hline
%----------------
PC efficiency & 0.07 & 50 & 0.00 & 0.00 & 0.00 & 0.00
% s2a17/s2a9  OK, eff=95%, time=50ns
% blind: delta =  0.000074 +- 0.001348529
% bb10:  delta =  0.000065 +- 0.001316881
% 0.065/50 = 0.0013
\\ \hline
%----------------
Dead zone PC & 0.46 & 6 & 0.08 &0.08 &0.08 &0.08
% g158/g148  OK correlated
% blind: delta =  0.000488 +- 0.001138143
% bb10:  delta =  0.000455 +- 0.001111401
% 0.455/6 = 0.0758
\\ \hline
%----------------
Dead zone DC & 1.38 & 15 &0.09 &0.09 &0.09 &0.09
% g159/g148  OK  correlated
% blind: delta =  0.001445 +- 0.001139098
% bb10:  delta =  0.001379 +- 0.001112277
% 1.379/15 = 0.0919
\\ \hline
%----------------
Up-down differences & -0.19 & 4 & -0.05 & -0.05 & -0.05 & -0.05
%
% s2a31/a9 - g122a5/a1? Must be s2a31/s2a18-...???
% 
% s2a31/a9
% blind: delta =  0.000339 +- 0.001325095
% bb10:  delta =  0.000338 +- 0.001294016
% 
% s2a31/a18
% blind: delta =  0.000009 +- 0.001327220
% bb10:  delta =  0.000000 +- 0.001296071
%
% g122a5/a1
% blind: delta =  0.000198 +- 0.001130542
% bb10:  delta =  0.000193 +- 0.001104037
%
% bb10 s2a31/a9-g122a5/a1  = 0.338 - 0.193 =  0.145
% bb10 s2a31/a18-g122a5/a1 = 0.000 - 0.193 = -0.193
%
% -0.193/4 = -0.04825
\\ \hline
%----------------
HV variations & 0.08 & 20 &0.00 &0.00 &0.00 &0.00
% s2a15/s2a9  OK 
% blind: delta =  0.000088 +- 0.001325784
% bb10:  delta =  0.000081 +- 0.001294673
% 0.081/20 = 0.00405
\\ \hline
%----------------
Temp, Pressure & -2.66 && -0.35 & -0.35 & -0.22 & -0.35
% g103a1/g122a1-mustop
% blind: delta = (-0.002734 +- 0.001146997) - mustop
% bb10:  delta = (-0.002662 +- 0.001120242) - mustop
% 
% % bb10 mustop:  delta =  0.000521, from s4a2/s3a25
%
% g103a1: au =     0.053266 +-  0.003432298
%         ad =     0.065286 +-  0.003428722
% ==> (au-ad)/(au+ad) = -0.101
%
% The slighly upstream set:
% s4a25: au =     0.060022 +-  0.005583861
%        ad =     0.075653 +-  0.005593784
% ==> (au-ad)/(au+ad) = -0.115
%
% On the 4 data sets: (au-ad)/(au+ad) mean=0.080042,  rms=0.0216077
% On the 5 data:                      mean=0.0723665, rms=0.0253916       
%
% On 5 MC:                            mean=0.0040844, rms=0.040957 
%
% awk -f extract_ecfit_labeled.awk logs-data-4/log-set* |grep -E 'au|ad' | paste --serial -d'\t\n' | awk '{au=$3; ad=$9; print (au-ad)/(au+ad)"\t"$0}' | awk '{sw+=$1; sw2+=$1*$1; n+=1; mean=sw/n; print mean"\t"sqrt((sw2-n*mean*mean)/(n>1?(n-1):1))"\t"$0; }'
%
\\ \hline
%----------------
Foil bulges & -1.3 && -0.52 &-0.26 &-0.52 &-0.26
% g145a1/g122a1
% blind: delta = -0.001352 +- 0.001124913
% bb10:  delta = -0.001302 +- 0.001098635
% 
\\ \hline
%----------------
Crosstalk & 0.01 & 10 & 0.00& 0.00& 0.00& 0.00
% s2a16/s2a9  OK
% blind: delta = -0.000005 +- 0.001327393
% bb10:  delta = -0.000005 +- 0.001296260
% 
\\ \hline
%----------------
T0 variations & -1.83 & 10 & -0.18 & -0.18 & -0.18 & -0.18 
% s2a11/s2a9  OK
% blind: delta = -0.001918 +- 0.001315805
% bb10:  delta = -0.001826 +- 0.001285135
% -1.826/10 = -0.1826
\\ \hline
%----------------
\hline
\multicolumn{3}{|l|}{Total} &0.67 &0.49 &0.61 &0.49
\\ \hline
\end{tabularx}%
%
%% }%\framebox{}
\caption{Chamber response systematics.\label{table:sys-chamber-response}}
\end{table}
\begingroup
%\ifagsdouble\onehalfspacing\else\relax\fi
\begin{description}

\item[DC efficiency.]  A special analysis of the nominal data set B
  deleted 5\% of DC hits before passing events through the standard
  reconstruction chain.  Once a hit was marked for deletion, all hits
  on the same wire within 700~ns were also deleted, since they were
  likely to come from the same track.   The obtained Michel spectrum
  was fit against the standard analysis of the same set.
  The 5\% inefficiency that was introduced for this test corresponds
  to an exaggeration factor of 50, because the actual efficiency of
  the DCs is about 99.9\% \cite{Henderson:2004zz}.
  
\item[PC efficiency.] The effect of inefficiencies in the PCs was
  estimated in a similar way.  The test spectrum was produced from set
  B with a 5\% artificial inefficiency and a hit removal time interval
  of 50~ns.  This systematics becomes negligible after applying the
  scaling factor of 50.  Note that even the ``raw'' effect of 5\% PC
  inefficiency is small, because PCs are not used in the track
  fitting.

\item[Dead zone PC.]  Muons slow down in the detector and become
  highly ionizing before they stop.  The large space charge they
  create in wire chambers may ``deaden'' a section of wire for some
  time after a passage of a muon.  The effect is largest in PC~6, the
  chamber which is closest to the muon stopping target on the upstream
  side, and is also observable in PC~5.  The nominal Monte-Carlo 
  did not simulate this effect.

  In a special study \cite{Brynle-report,art-deadzone}, the dead zone
  effect was simulated by introducing a 100\% inefficiency along the
  wire around \vb{the point of the} muon hit.  The rectangular in
  space inefficiency zone exponentially shrank in time.  The initial
  length of the zone was computed as the length of the projection of
  the muon created space charge on the wire plus $\Delta{L}$.  A
  ``realistic'' simulation with the parameters $\Delta{L}/2=0.24$~cm
  and $\tau_{\text{heal}}=2444$~ns reproduced the \vb{``dip'' in PC~6
  efficiency around a muon hit that was observed in data} reasonably
  well.  An ``exaggerated'' simulation used $\Delta{L}/2=5.00$~cm and
  $\tau_{\text{heal}}=3500$~ns.  Both simulations used the same dead
  zone parameters for all PC planes.

  To measure the sensitivity $R'$, the ``exaggerated'' MC was fit
  against a nominal (no dead zone) simulation.  The \vb{scaling}
  factor was estimated using the ratio of the number of PC positron
  hits lost due to the dead zone in the ``exaggerated'' MC to the
  number lost in the ``realistic'' MC.

\item[Dead zone DC.]  A similar ``dead zone'' effect is also expected
  in the DCs.  Its magnitude is smaller then in PC~6, because muons
  are less ionizing farther from their stopping position.  A DC dead
  zone is also seen at a smaller solid angle from the stopping target
  than PC~5 or 6 dead zone, making the effect harder to observe.

  With no estimate of DC dead zone parameters available from data, the
  same $\Delta{L}/2$ and $\tau_{\text{heal}}$ parameters as used in
  the PC study were applied to the DCs.  The scaling factor was
  estimated using the ratio of the number of DC positron hits lost due to
  the dead zone in the ``exaggerated'' MC to the number lost in the
  ``realistic'' MC.

\item[Upstream-downstream differences.] 

% On Wed, 26 Jan 2005, Carl Gagliardi wrote:
% 
% > Hi Andrei:
% > 
% > Dick indicated that he was having a hard time finding the supporting 
% > documentation for the drift time clip scale factors.  I dug a bit and found 
% > the corresponding forum postings.  They are:
% >   https://twist.phys.ualberta.ca/forum/view.php?site=twist&bn=twist_physics&key=1094832158
% > and
% >   https://twist.phys.ualberta.ca/forum/view.php?site=twist&bn=twist_physics&key=1094834856
% > 
% > The former as an initial discussion of this issue near the end.  The latter 
% > has a .pdf attached from the <ndof> spreadsheet that I put together.  It 
% > also discusses the <ndof> issues as they impact rho and delta.  My 
% > conclusion was that rho should have a scale factor of 1 and delta should 
% > have a larger scale factor, but smaller than 15.
% > 
% >  From there, the final choice of 4 was made by a "committee of the whole" 
% > during our pre-box-opening phone call later that day.
% > 
% > I've attached a copy of the .xls file.  It doesn't contain anything that 
% > isn't in the .pdf forum attachment, but it will make it easier for you to 
% > play with the numbers if you wish.
% > 
% > I hope this helps!
% > Carl

  \begingroup  
  \newcommand\ndf{\left<\text{ndof}\right>}
  \newcommand\tmax{t_{\text{max}}}
  \newcommand\Andf{A_{\text{ndof}}}
  %\rrr{Long drift time cut.}
  
  \vb{The average number} of degrees of freedom of
  the track fits, $\ndf$, related to the number of drift chamber hits
  used in the tracking, is different between data and Monte-Carlo.
  Among many possible explanations (like the cathode foil bulges, see
  below) is an inaccuracy of simulation of the drift chamber response
  in the corners of a drift cell.  Special analyses were run, which
  excluded ``\vb{cell corner} hits'' with $t>\tmax$ from the final track
  fitting.  For a special analysis of Monte-Carlo, $\tmax=400$~ns was
  used.  In the special data analysis, the cut was tuned to match the
  average $\ndf$ of the data track fitting to that of the 400~ns
  Monte-Carlo analysis.  The tuning resulted in $\tmax=522$~ns.

  The data spectrum produced in the special analysis was fit to a
  nominal analysis of the same set, giving
  $\Delta\delta=0.000\times10^{-3}$.  A similar fit \vb{of special to
  nominal Monte-Carlo spectra}
  gave $\Delta\delta=0.193\times10^{-3}$.
  The ``raw'' effect was taken as the difference between the data and
  the Monte-Carlo fits: $\Delta\delta=-0.193\times10^{-3}$.

  The effect of corner cell inefficiencies on the spectrum could be
  taken with a scaling factor of 1.  However while tuning the average
  $\ndf=\frac12\,(\ndf_{up}+\ndf_{dn})$, the long drift time cuts
  exaggerated a difference in the asymmetry
  $\Andf=(\ndf_{up}-\ndf_{dn})/(\ndf_{up}+\ndf_{dn})$ between data and
  Monte-Carlo by about a factor of 15, from
  $\Andf^{\text{Data}}-\Andf^{\text{MC}}\sim-0.16$\% for the nominal
  analyses to 2--3\% for the long drift time cut analyses.
  Since $\delta$ is an asymmetry parameter, a scaling factor of 15 was 
  another \vb{possible} choice.  It has been decided to use a factor of 4 (the
  geometric mean) for this systematic.
  
  \endgroup

\item[HV variations.] This systematic represents the effect of our
  \vb{imperfect} knowledge of high voltage on the wire chambers.
  The test spectrum was produced by analyzing the nominal data set B 
  using drift tables corresponding to 1850~V, and comparing it to the
%
%  nominal spectrum (1950~V).  
% 
  \vb{same data set analyzed with the nominal, 1950~V, drift tables.}
  The scaling factor is 20 because the 
  accuracy of the high voltage if 5~V.

\item[Temperature and pressure.]  This systematic uncertainty
  represents effects of varying gas density in the \TWIST{} drift
  chambers, caused by variations of the atmospheric pressure and
  outside temperature.  

%  primarily through changing the space-time relationship of the drift
%  chambers.

  A special Monte-Carlo set was generated with settings corresponding
  to the temperature of $-10^\circ$~C, instead of the nominal
  $+20^\circ$~C, and fit to a nominal Monte-Carlo set, yielding
  $\Delta\delta=-2.66\times10^{-3}$.  Scale factors were determined
  individually for each data set by comparing variations in the gas
  density that occurred during data taking (available from the Slow
  Control DAQ records) to the ${}\approx10$\% density change in the
  $-10^\circ$~C test MC set.  The scaling factors are: 10 (set A), 10
  (set B), 20 (1.96~T), 10 (2.04~T).

  A change in the gas density also affects the muon stopping
  distribution. \vb{To avoid double counting the stopping distribution
  effect, the latter systematic number, 0.09 (discussed below), was
  linearly subtracted from scaled per-set temperature and pressure
  estimates.}

%v0:  (The stopping distribution displacement
%v0:  $\alphadiff/\alphasum\approx-0.101$ for the $-10^\circ$~C MC set is
%v0:  close to the value $-0.115$ for the ``slightly upstream'' data set
%v0:  used to measure the stopping distribution systematic, thus no
%v0:  additional scaling was necessary. The $\alphadiff$ and $\alphasum$
%v0:  symbols are the energy calibration variables from
%v0:  chapter~\ref{chapter:ecalib}. The stopping distribution systematic
%v0:  is explained below.)
  
%  \rrr{Set-dependent. Talk: set1=20,
%  10, 10, 10.  gen103: -10 deg C, gen122: +20 deg C.  Another change
%  is the ``surface'' cut for e+ in MSOR removed in gen103, it was ON
%  by mistake in gen122.}
%
%  KO's final posting containing density data:
%  \verb|bn=twist_physics| \verb|key=1093556369|

% https://twist.phys.ualberta.ca/forum/view.php?site=twist&bn=twist_physics&key=1093556369

%  \rrr{How to subtract mustop?}

\item[Foil bulges.]  During the data taking, the differential pressure
  between the drift chambers and the enclosing He/N2 volume
  was not always stable.  That led to a movement of the cathode foils,
  affecting both the space-time relationship of the drift chambers,
  and the average number of wires in a plane hit by a track at a given
  angle.  For example, in the nominal geometry with square
  $4\text{~mm}\times4\text{~mm}$ drift cells a track at
  $\theta<45^\circ$ can never hit more than 2 cells in a plane.  But
  if the cathode foils bulge out, extending the cell size along
  the detector axis, the same track might produce more hits per plane.

  To estimate the systematic uncertainty due to the foil movements, a
  special Monte-Carlo set with the foils moved out by 500~\mum{} was
  produced and fit to a nominal MC set, giving $\Delta\delta=-0.00130$.

  Several measurement of the differential pressure, done during the
  running period, were correlated to an analysis variable derived from
  the \vb{most probable} value of $\chi^2/\text{ndof}$ in the tracking and corrected
  for gas density effects \cite{olchansk-density-plots}.  That
  variable was then used to set the following scaling factors for
  different data sets: 2.5 (set A), 5 (set B), 2.5 (1.96~T), 5
  (2.04~T).

%  \rrr{Set-dependent. SFU: set18 scale=5, refers to a meeting on Sept
%    6. Talk: set18 scale=0.4} See KO posting above (in Temp and
%    pressure).  The change in the scaling is probably related to a
%    different foil bulge in the MC study (?), and for set18 to the use
%    of the ``bulged'' base MC.

\item[Crosstalk.] 

% See \cite{lalancette-crosstalk}, and 
% Dick's talk on October 17, 2003.

  The test spectrum was produced by turning off the
  crosstalk removal algorithm (page
  \pageref{page:crosstalk-algorithm}) and re-analyzing nominal set B.
  As expected, the effect is very small since decay positrons are
  weakly ionizing and do not produce large signals in the chambers,
  which could induce crosstalk.  The removal algorithm is estimated to
  be correct at least 90\% of time, thus the scaling factor of 10.

\item[T0 variations.]  Time offsets T0 for different wires were
  calibrated using 120~MeV pion tracks with solenoid off.  Results
  from a calibration run taken at the beginning of the data taking
  period were used in the nominal data analysis.
  
  To estimate the systematic uncertainty associated with the
  calibrations, another T0 run was taken at the end of the data taking
  period.  The per-channel differences in the extracted offsets
  between the two runs were multiplied by 10 (the scaling factor) and
  added to the nominal T0 values.  The obtained T0 calibration file was
  used to analyze data set B to produce the test spectrum, which was
  fit against the nominal set B spectrum, \vb{resulting in $\Delta\delta=-0.00183$}.

%
% dc_t0.00012: runs 5627-29:   Jun 19 2002 
% dc_t0.00013: runs 13972-76:  Dec  9 2002
%
% t0.scaled time calibrations = t0.00012 + 10*(t0.00013 - t0.00012)
%  
% \rrr{Are the ends are ``extreme'' or ``typical'' for T0??? Different
%   assumptions lead to different scaling factors.}

\end{description}
\endgroup

%%================================================================
\section{Momentum calibration}

%
%================
\begin{table}[htbp]
%%\framebox[\width]{%
\begin{tabularx}{\textwidth}{|X|d{2}|d{2}|d{2}|d{2}|d{2}|}
\hline
Name 
& \multicolumn{1}{c|}{$10^3\times\Delta\delta$} 
& \multicolumn{4}{c|}{$10^3\times{}R'\sigma_a$}
\\ \cline{3-6}
{}&{}
& \multicolumn{1}{c|}{A}
& \multicolumn{1}{c|}{B}
& \multicolumn{1}{c|}{1.96}
& \multicolumn{1}{c|}{2.04}
\\ \hline\hline
%----------------
% ~/analysis/twt.20041001/datav.v2/fit3.20050301-ecsys/, using input
% file from fit3.20041013-bb10open, and today's version of ecsys:
% 
% sigma_ec(delta) = 2.095559944e-04    fit3_set1an3_to_gen154an1.out
% sigma_ec(delta) = 2.353787159e-04    fit3_set22an2_to_gen156an1.out
% sigma_ec(delta) = 2.106584187e-04    fit3_set2an18_to_gen148an1.out
% sigma_ec(delta) = 2.721976993e-04    fit3_set3an26_to_gen152an1.out
% 
End point fits &  & 0.27 & 0.21 & 0.21 & 0.24
\\ \hline
%----------------
Field map & 0.68 & 0.07 & 0.07 & 0.17 & 0.34
% blind:delta =  0.000706 +- 0.001326671
% bb10: delta =  0.000683 +- 0.001295510
% 
\\ \hline
%----------------
\hline
\multicolumn{2}{|l|}{Total} &0.28 &0.22 &0.27 &0.42
\\ \hline
\end{tabularx}%
%
%% }%\framebox{}
\caption{\label{table:sys-ecalib}Momentum calibration systematics.}
\end{table}
\begingroup
%\ifagsdouble\onehalfspacing\else\relax\fi
\begin{description}
\item[End point fits.] Sensitivities of $\delta$ to 
  \vb{the energy calibration parameters} $\beta$, $\au$, and
  $\ad$ (chapter~\ref{chapter:ecalib}) were determined by fitting a
  nominal spectrum to a spectrum with the appropriate energy
  calibration constant offset by 100~keV/c.  Covariance sub-matrices
  $V$ for $(\beta,\au,\ad)$ from energy calibration fits to data and
  to the corresponding MC spectrum were added, and the resulting
  matrix multiplied with the vector of sensitivities $A$ in the usual
  way~\cite{eadie:statmethods}:
  \begin{equation}
    \sigma^2 = A^T (V_{\text{Data}}+V_{\text{MC}}) A
  \end{equation}

% Carl's posting
%
% https://twist.phys.ualberta.ca/forum/view.php?site=twist&bn=twist_physics&key=1092271777

\item[Field map.]  A mismatch between the measured $B_z$ component of
  the nominal 2~T spectrometer field and the OPERA field map was fit
  using the expression $\Delta{B_z}=c_2z^2+c_3z^3+c_rr$
  \cite{carl-field-map}.  This simple function describes
  the residuals to within 1~G throughout the entire tracking region,
  and to within 0.5~G over most of the tracking region.  Nominal data
  set B was re-analyzed using a test field map, which was prepared by
  adding $10\times\Delta{B_z}$ to the nominal OPERA map, and then fit
  against the standard analysis of the same data.  Thus the scaling
  factors are 10 for the 2~T sets A and B.  Nominal analyses of the
  1.96~T and 2.04~T data sets were done using a scaled 2~T field map.
  Comparisons of the scaled versions to the actual $B_z$ measurements
  at those fields gave the scaling factors of 2 for the 2.04~T data
  set, and -4 for the 1.96~T data set.

% where delta B_z is in Gauss and z and r are in cm.  The coefficients
% were:
%
% C2 = -6 * 10^{-4}
% C3 = -4 * 10^{-6}
% CR = -0.125

\end{description}
\endgroup

%%================================================================
\section{Muon beam stability\label{section:sys-beam-stability}}

Systematics related to the stability of muon beam parameters are
``data set dependent'' by their nature.  Since none
contributed significantly to the final result, a common estimate for
each of the scaling factors based on the worst case data set was used,
instead of assigning individual scalings to different data sets.
%
%================
\begin{table}[htbp]
%%\framebox[\width]{%
\begin{tabularx}{\textwidth}{|X|d{2}|d{1}|d{2.0}|}
\hline
Name 
& \multicolumn{1}{c|}{$10^3\times\Delta\delta$} 
& \multicolumn{1}{c|}{Scaling} 
& \multicolumn{1}{c|}{$10^3\times{}R'\sigma_a$}
\\ \hline\hline
%----------------
Stopping location & 0.52 & 6 & 0.09 
% s4a2/s3a25
% blind: delta =  0.000168 +- 0.001859391
% bb10:  delta =  0.000521 +- 0.001816935
% 0.521/6 = 0.08683
\\ \hline
%----------------
Beam intensity &  0.26 & 6 & 0.04
% blind: delta =  0.000300 +- 0.001334865
% bb10:  delta =  0.000255 +- 0.001303463
% 0.255/6 = 0.0425
\\ \hline
%----------------
Channel magnets &  -1.29 & 50 & -0.03
% blind: delta = -0.001189 +- 0.001640579
% bb10:  delta = -0.001290 +- 0.001601848
%1.29/50 = 0.0258
\\ \hline
%----------------
\hline
\multicolumn{3}{|l}{Total} & 0.10
% rounded => 0.103
% exact   => 0.100
\\ \hline
\end{tabularx}%
%
%% }%\framebox{}
\caption{\label{table:sys-mubeam}Muon beam systematics.}
\end{table}
\begingroup
%\ifagsdouble\onehalfspacing\else\relax\fi
\begin{description}
\item[Stopping location.] The average position of muon stops in the
  stopping target affects the amount of the target material seen by
  decay positrons.  E.g. if muons stop before reaching the center of
  the target, positrons going upstream will be less affected by energy
  loss and multiple scattering than those going downstream.  The
  energy calibration procedure (chapter \ref{chapter:ecalib})
  compensates, to first order, for differences in energy loss.
  The ``stopping location'' systematics covers any remaining effects. 

  A special data set was taken with the muon stopping position
  displaced slightly upstream (by introducing more $CO_2$ in the gas
  degrader).  That set was fit to a nominal data set to measure the
  effect.  The scaling factor was obtained by comparing the ratio
  $\alphadiff/\alphasum$ for the special set\vb{, $-0.12$,} to the spread of that
  ratio for other data sets\vb{, ${}\sim0.02$}.  Here $\alphadiff$ and $\alphasum$ are
  the energy calibration variables, see chapter~\ref{chapter:ecalib}.

  % \rrr{Dick's plot.}

\item[Beam intensity.]  The signal rate on the \TWIST{} trigger
  counter was recorded during the data taking.  After rejection of bad
  runs, the spread of the average trigger rate for different runs
  within the four nominal data sets was found to be smaller than
  $0.6\times10^3$~\persec.  To measure sensitivities of the Michel
  parameters to beam intensity, a low rate ($1.1\times10^3$~\persec)
  and a high rate ($4.7\times10^3$~\persec) data set were taken and
  fit against each other.  The exaggeration factor $S$ in this
  measurement is $(4.7-1.1)/0.6=6$.

\item[Channel magnets.]  This systematic accounts primarily for
  instabilities in the B2 beamline dipole
  (Fig.~\ref{fig:m13beamline}), which directly affects the position of
  the muon beam as it enters the \TWIST{} spectrometer.  The other
  dipole magnet, B1, defines the momentum of the muons accepted by the
  channel, and its instabilities are included in the ``Stopping
  location'' systematic.  

  The strength of the magnetic field in B2 was continuously monitored
  and was stable to 0.2~G.  A test data set was taken with B2
  intentionally offset from the nominal value by 10~G, giving the
  exaggeration factor of 50.

  The deflection of beam particles by M13 quadrupole magnets is small
  compared to the $60^\circ$ bend by B2. Since the systematic effect
  of B2 is already small, contributions of the quadrupoles to the
  systematic were neglected.

\end{description}
\endgroup

\section{\label{section:sys-summary}Summary of systematics}

Table~\ref{table:sys-summary} shows a summary of the systematic
uncertainties.  \vb{The ``stopping target thickness'' is shown
separately from the rest of ``positron interactions'' uncertainties,
and ``Upstream/Downstream differences'' separately from the rest of
``Chamber response'', following~\cite{Gaponenko:2004mi}.}

An entry not discussed above is the uncertainty from
theoretical radiative corrections.  Theoretical uncertainty on
$\delta$ is estimated as $1\times10^{-4}$ \cite{aaa-leadinglog:2002},
if terms of up to $\OO(\alpha^2L^2)$ are included in the
spectrum. Here $L=\ln(m^2_\mu/m^2_e)\approx10.66$, and $\alpha$ is
the fine structure constant.  \TWIST{} uses \vb{an} even more precise spectrum
description (chapter~\ref{chapter:monte-carlo}), therefore this
estimate provides a safe upper bound on the uncertainty.

% from delta PRD
%
\begin{table}
\begin{center}
\begin{tabular}{lc}
  Effect   & Uncertainty \\ \hline
  Spectrometer alignment & $\pm$0.00062 \\
  Chamber response(ave)  & $\pm$0.00056 \\
  Positron interactions  & $\pm$0.00055 \\
  Stopping target thickness & $\pm$0.00037 \\
  Momentum calibration(ave) & $\pm$0.00030 \\
  Muon beam stability(ave)  & $\pm$0.00010 \\
  Theoretical radiative corrections & $\pm$0.00010 \\
  Upstream/Downstream \vb{differences(ave)} & $\pm$0.00005 \\ 
\hline Total & $\pm$0.00112
\end{tabular}
\caption[Summary of systematics]{\label{table:sys-summary}
Contributions to the systematic uncertainty for $\delta$.  For
set-dependent systematics the average values, denoted by (ave), are
shown.  }
\end{center}
\end{table}

%
% sqrt((0.00061)^2+(0.00056)^2+(0.00055)^2+(0.00037)^2+(0.00029)^2+(0.00010)^2+(0.00010)^2+(0.00004)^2)
% 

%%%%%%%%%%%%%%%%%%%%%%%%%%%%%%%%%%%%%%%%%%%%%%%%%%%%%%%%%%%%%%%%

\chapter{Determination of decay parameters\label{chapter:data-fits}}
%\chapter{Determination of decay parameters\label{chapter:data-fits}}

% values for the delta paper:
%
% https://twist.phys.ualberta.ca/forum/view.php?bn=twist_physics&key=1097719859

According to the philosophy of a blind analysis, the decision on which
data sets to use for the extraction of a final result had been made
prior to ``opening the box'' and revealing the physics result.
Within each of the chosen data sets, an identification and exclusion
of bad runs was also done at the blind stage of analysis.

All fits used for the extraction of $\delta$ were done using
2-dimensional histograms of reconstructed data and Monte-Carlo spectra
in momentum and $\costh$, which is the complete information available
from the detector.  The fits were done in the linear parametrization
$\{\Delta\rho,\Delta{z},\Delta{w}\}$, where
$z=\Pmuxi|_{\xidelta=\text{const}}$, and $w=\xidelta$
(Section~\ref{section:fitting-discussion}), then the results were
converted to the usual $\{\rho,\Pmuxi,\delta\}$ parametrization and
the covariance matrices recomputed.  The 3-parameter fits, unlike the
fits to the spectrum asymmetry
$A(p)=F_{\text{AS}}(p)/F_{\text{IS}}(p)$ (see
Eqs.~\ref{eq:theoretical-spectrum-2d}--\ref{eq:theoretical-spectrum-fas})
used in the previous measurement \cite{Balke:1988delta}, do not
require making any assumption regarding the value of $\rho$ in order
to find $\delta$.  Sensitivity of $\delta$ to the value of
$\eta=-0.007\pm0.013$~\cite{pdg-all:2004} assumed in MC production was
checked and found negligible.

The measurement of the decay parameter $\delta$ uses the following
four data sets: set~A, set~B, 1.96~T, and 2.04~T. (See
Table~\ref{table:twist-data}.)  Table~\ref{table:data-fits} shows
results of fits to the chosen data sets, computed using black box
offset values $\rho_0=0.74766$, $\Pmuxi_0=1.0148$, $\delta_0=0.73645$.
%\vc{Fit results for $\rho$ }

%================
\begin{table}[tbp]
\begingroup
\vc{%
\begin{center}
\begin{tabular}{lcccc}
Data Set  & $\delta$                      & $\rho$            & $\chi^2$ & Probability\\ \hline
%					  			      
% s3					  			      
Set A     & $0.75087\pm0.00156\pm0.00073$ &$0.75083\pm0.00083$& 1924 & 0.27\\
%					  			      
% s2					  			      
Set B     & $0.74979\pm0.00124\pm0.00055$ &$0.74911\pm0.00066$& 1880 & 0.54\\
%					  			      
% s1					  			      
1.96~T    & $0.74918\pm0.00124\pm0.00067$ &$0.74956\pm0.00066$& 1987 & 0.05\\
%					  			      
% s22					  			      
2.04~T    & $0.74908\pm0.00132\pm0.00065$ &$0.75203\pm0.00071$& 1947 & 0.16\\
\hline
\end{tabular}
\caption[Fit results]{\label{table:data-fits}Fit results.
  Set-dependent systematic uncertainties from Tables
  \ref{table:sys-chamber-response}, \ref{table:sys-ecalib},
  \ref{section:sys-beam-stability} are shown for $\delta$ after the
  statistical errors.  Results for $\rho$ are consistent with our
  previous measurement \cite{Musser:2004zw}. Large depolarizing
  effects in the graphite coated Mylar target (chapter
  \ref{chapter:monte-carlo}) made the present data unsuitable for an
  improved measurement of $\Pmuxi$, therefore this parameter is not
  shown.  Each fit has 1887 degrees of freedom (NDOF).  The last
  column is the fit probability computed from $\chi^2$ and NDOF. }
\end{center}%
}
\endgroup
\end{table}

Correlation coefficients for set B are shown in
Table.~\ref{table:correlations}.  Correlations for other surface muon sets
are very similar.  The small (less than 10\%) correlation between
$\rho$ and $\delta$ confirms that in our approach the two parameters
are independent.

%%================
% =====> fit3_set2an18_to_gen148an1.out <=====
% Correlation coefficients:
%                  rho  xixidelta    xidelta
%       rho   1.000000   0.156623   0.262293
% xixidelta   0.156623   1.000000   0.422372
%   xidelta   0.262293   0.422372   1.000000
% 
% =====> fit3_set2an18_to_gen148an1.out <=====
% Correlation coefficients:
%              rho         xi      delta
%   rho   1.000000   0.156623   0.097221
%    xi   0.156623   1.000000  -0.541111
% delta   0.097221  -0.541111   1.000000
%
\begin{table}
$$
\begin{array}{l|c|c}
{}         &  \Delta\xixidelta & \Delta\xidelta \\ \hline
\Delta\rho       & 0.157    & 0.262 \\ %\hline
\Delta\xixidelta &          & 0.422
\end{array}
\qquad\qquad
\begin{array}{l|c|c}
{}               &  \Delta\Pmuxi & \Delta\delta \\ \hline
\Delta\rho       & 0.157      & 0.097        \\ %\hline
\Delta\Pmuxi        &            & -0.541
\end{array}
$$
\caption[Correlation coefficients]{\label{table:correlations}Correlation
  coefficients.  Left: in the original fit parametrization. Right:
  converted to the usual muon decay parameters, according to formulas
  from Appendix~\ref{chapter:xideltaconv}.  }
\end{table}
%%================

\begin{figure}[tbp]
\begin{tabular}{cc}
\includegraphics[width=0.48\textwidth]{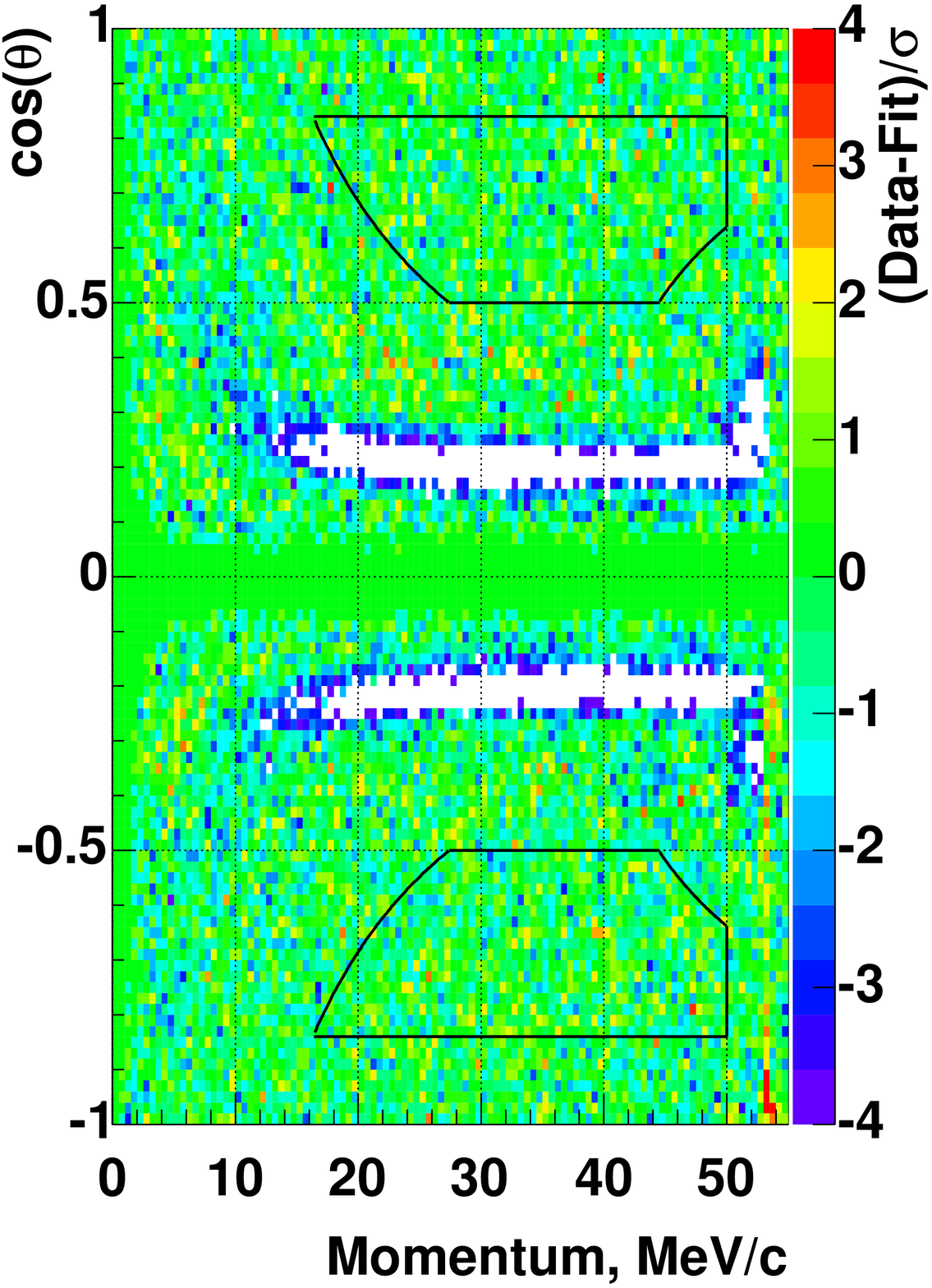} 
&
\includegraphics[width=0.48\textwidth]{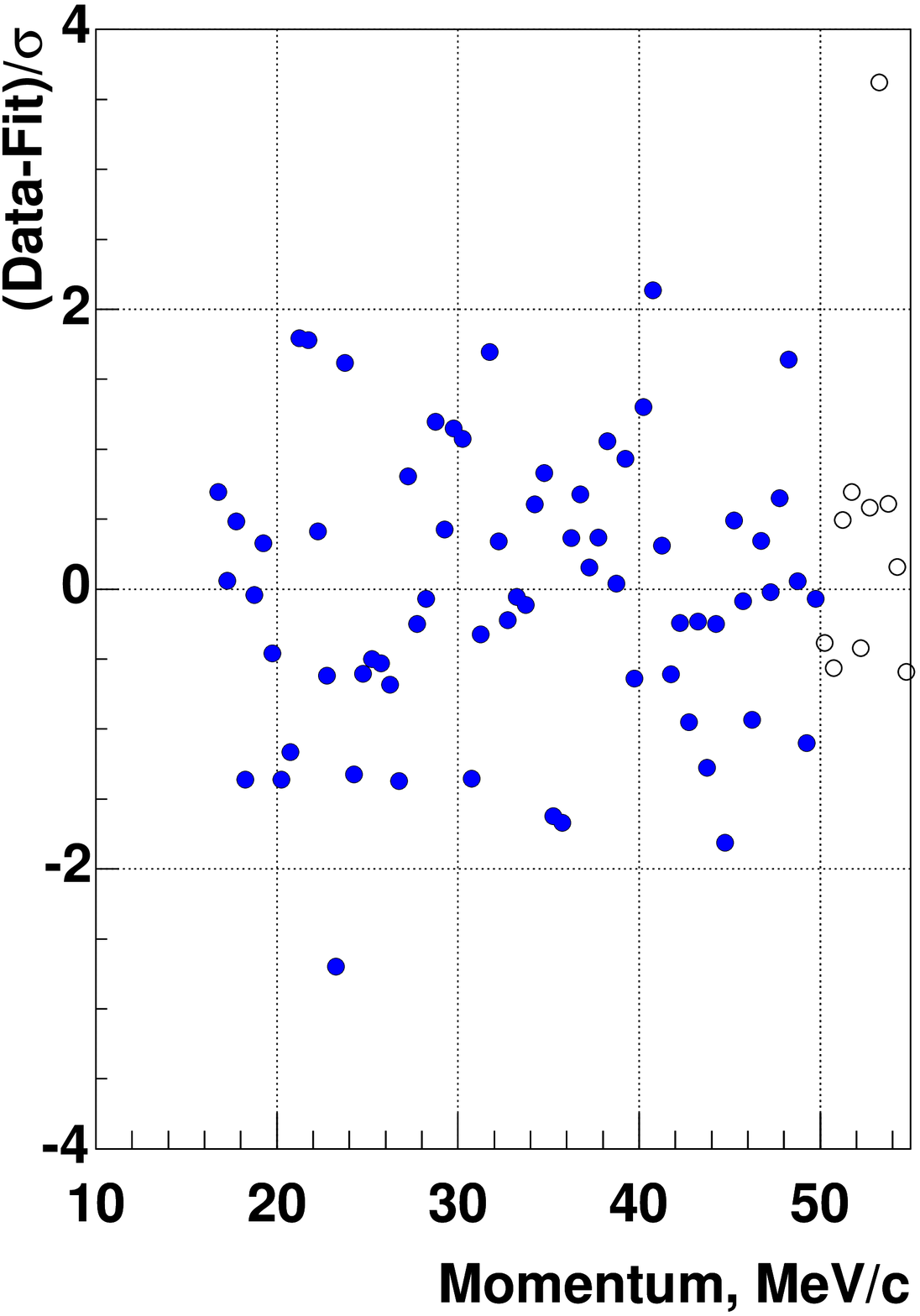}
\\
\includegraphics[width=0.48\textwidth]{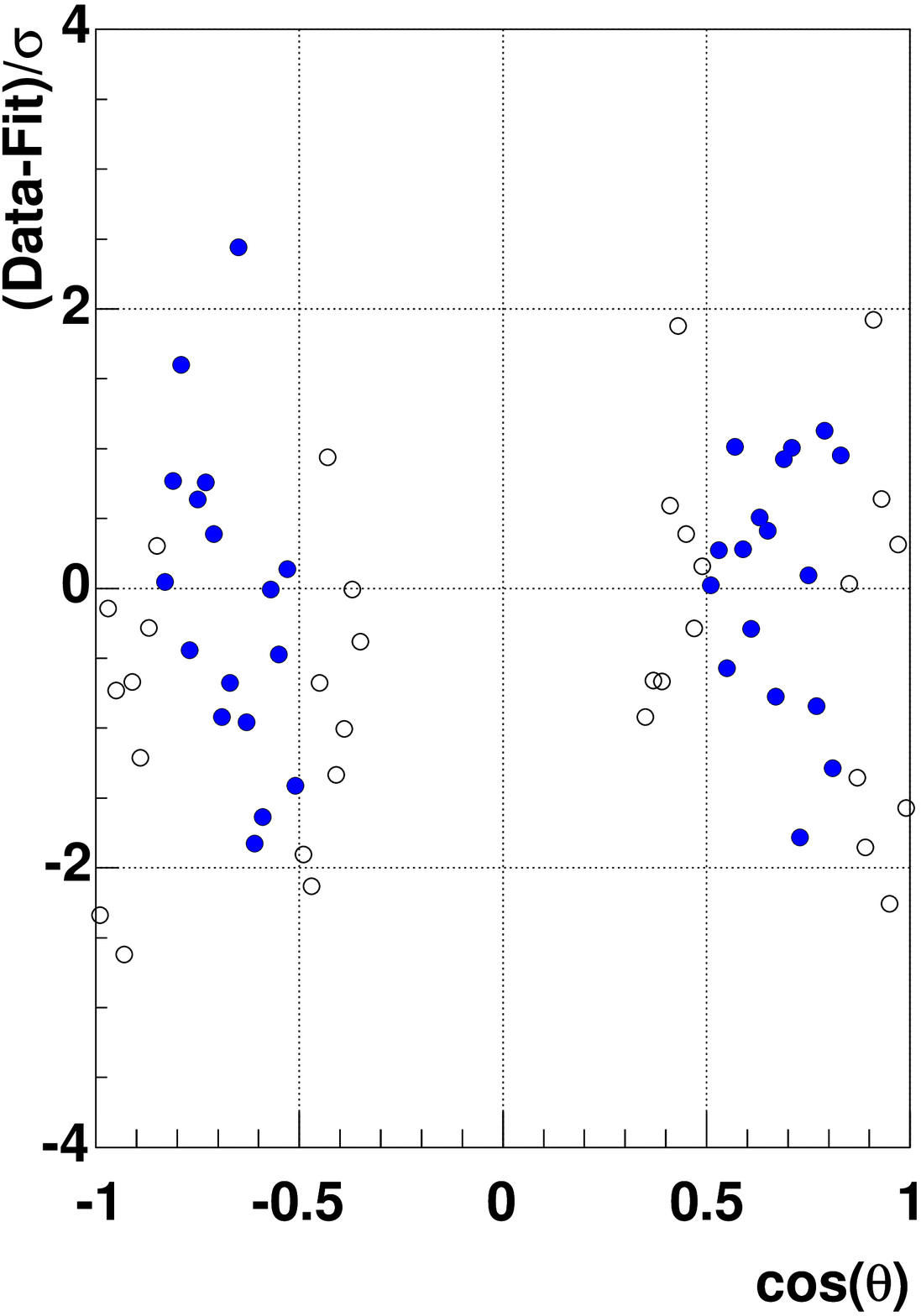} 
&
\includegraphics[width=0.48\textwidth]{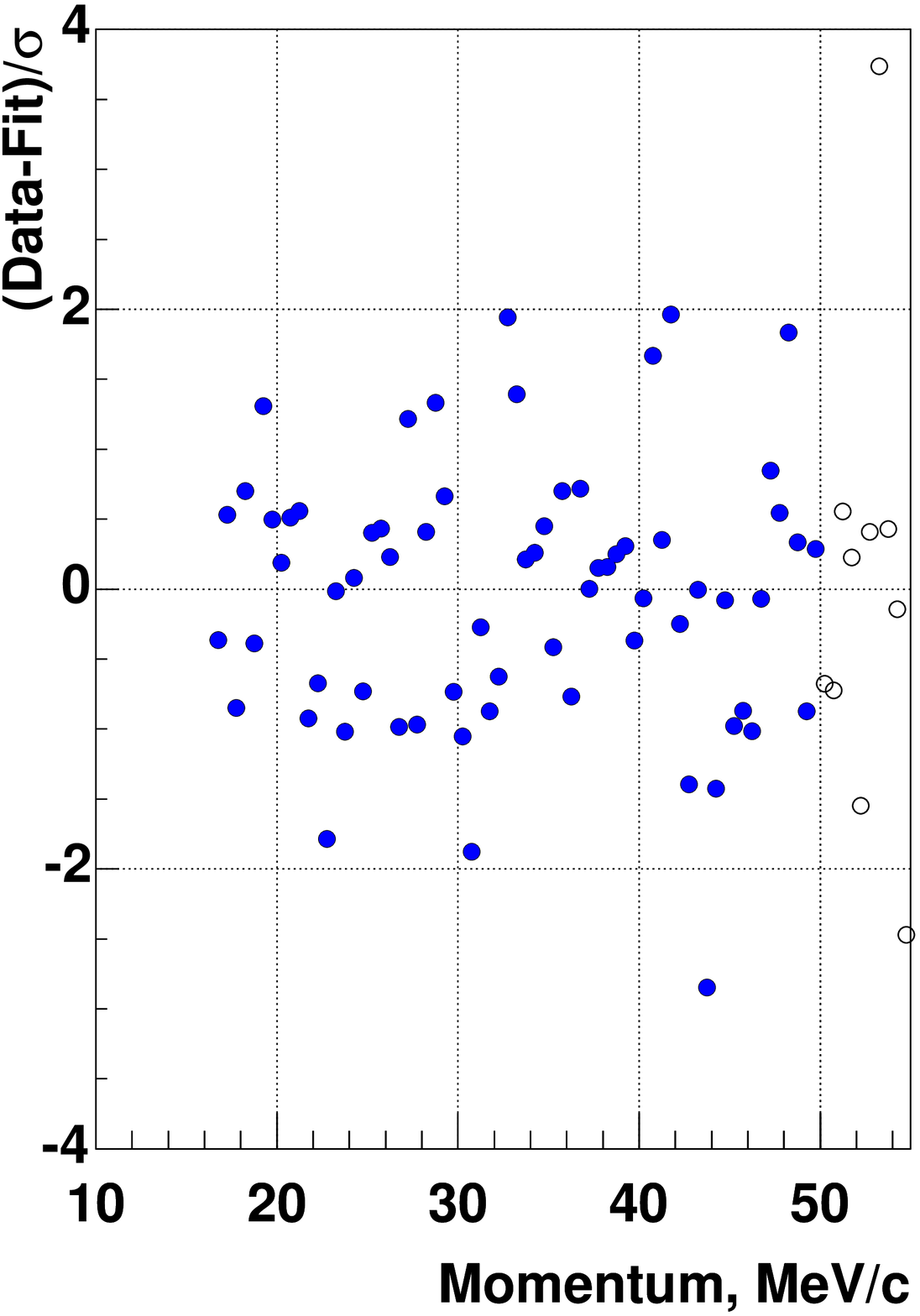}
\end{tabular}
\caption[Fit residuals for set A.]{\label{fig:residuals-setA}Fit
  residuals for set A.  Top left: color-coded residuals in the
  $\costh$ vs momentum plane.  Top right: projection on the momentum
  axis.  Bottom left: projection on the $\costh$ axis.  Bottom right:
  projection of the ``upstream minus downstream'' distribution. Solid
  markers are for points inside the fiducial region, empty markers for
  outside. The contours delimit the fiducial region. See text.  }
\end{figure}

\begin{figure}[tbp]
\begin{tabular}{cc}
\includegraphics[width=0.48\textwidth]{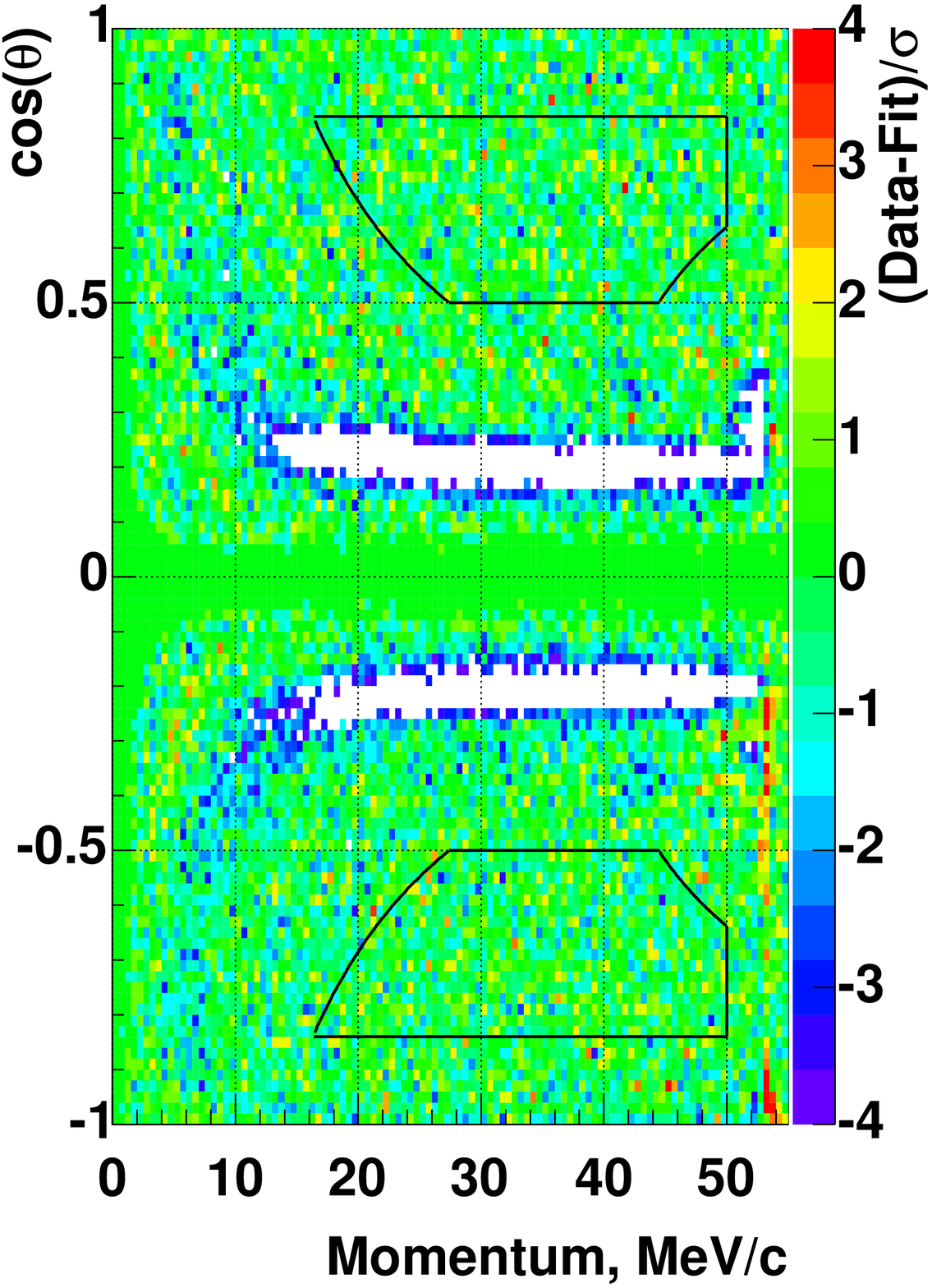} 
&
\includegraphics[width=0.48\textwidth]{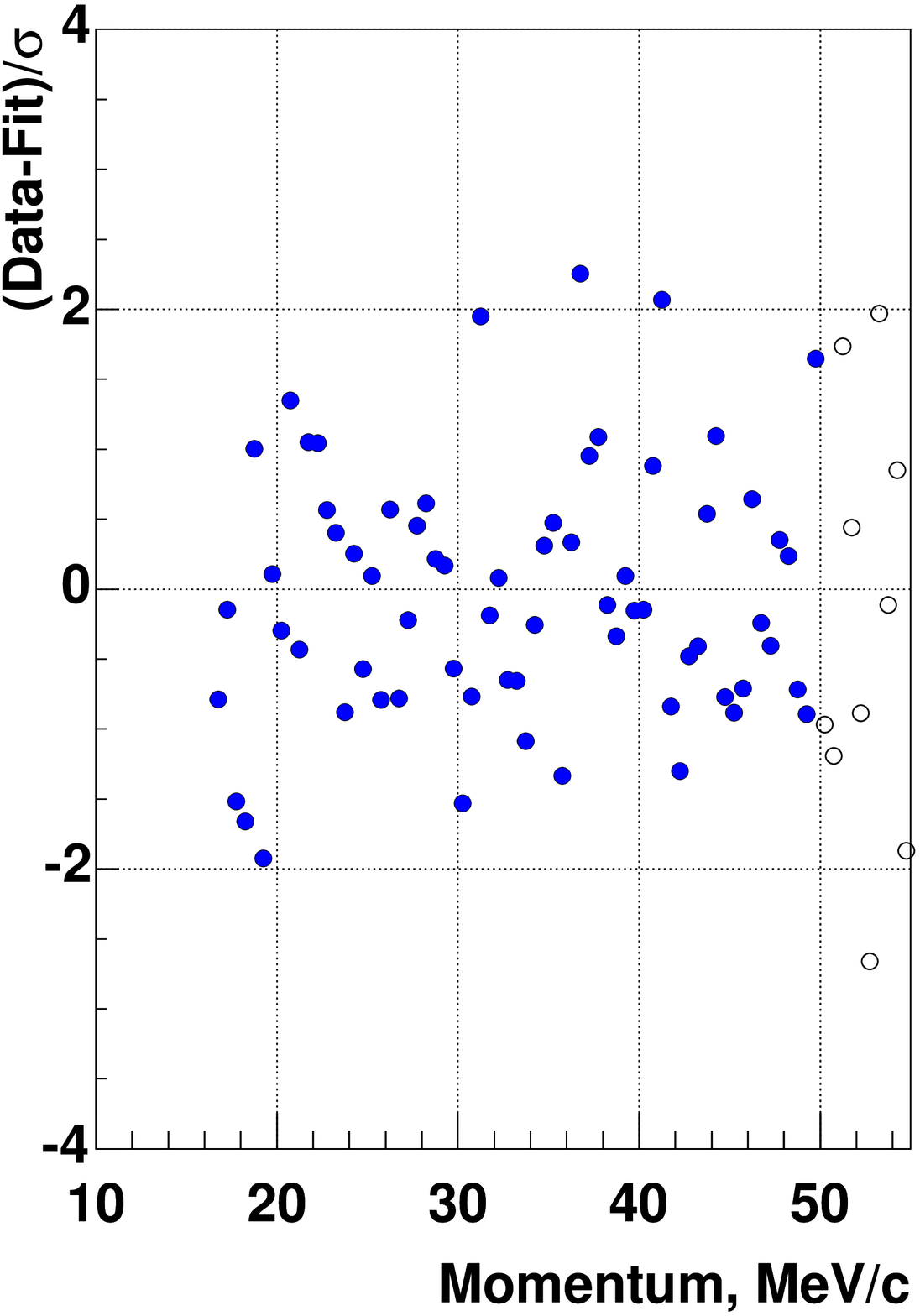}
\\
\includegraphics[width=0.48\textwidth]{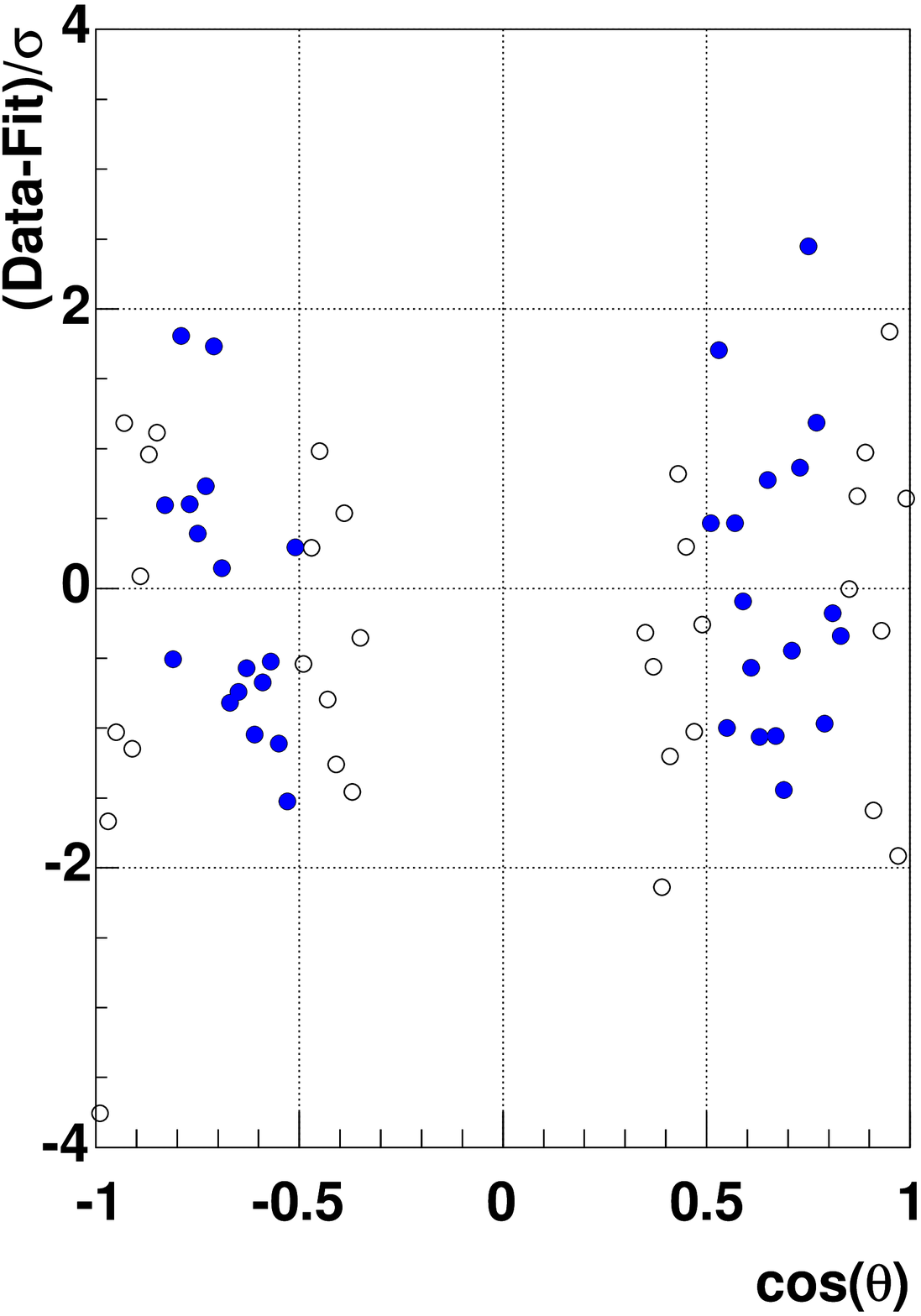} 
&
\includegraphics[width=0.48\textwidth]{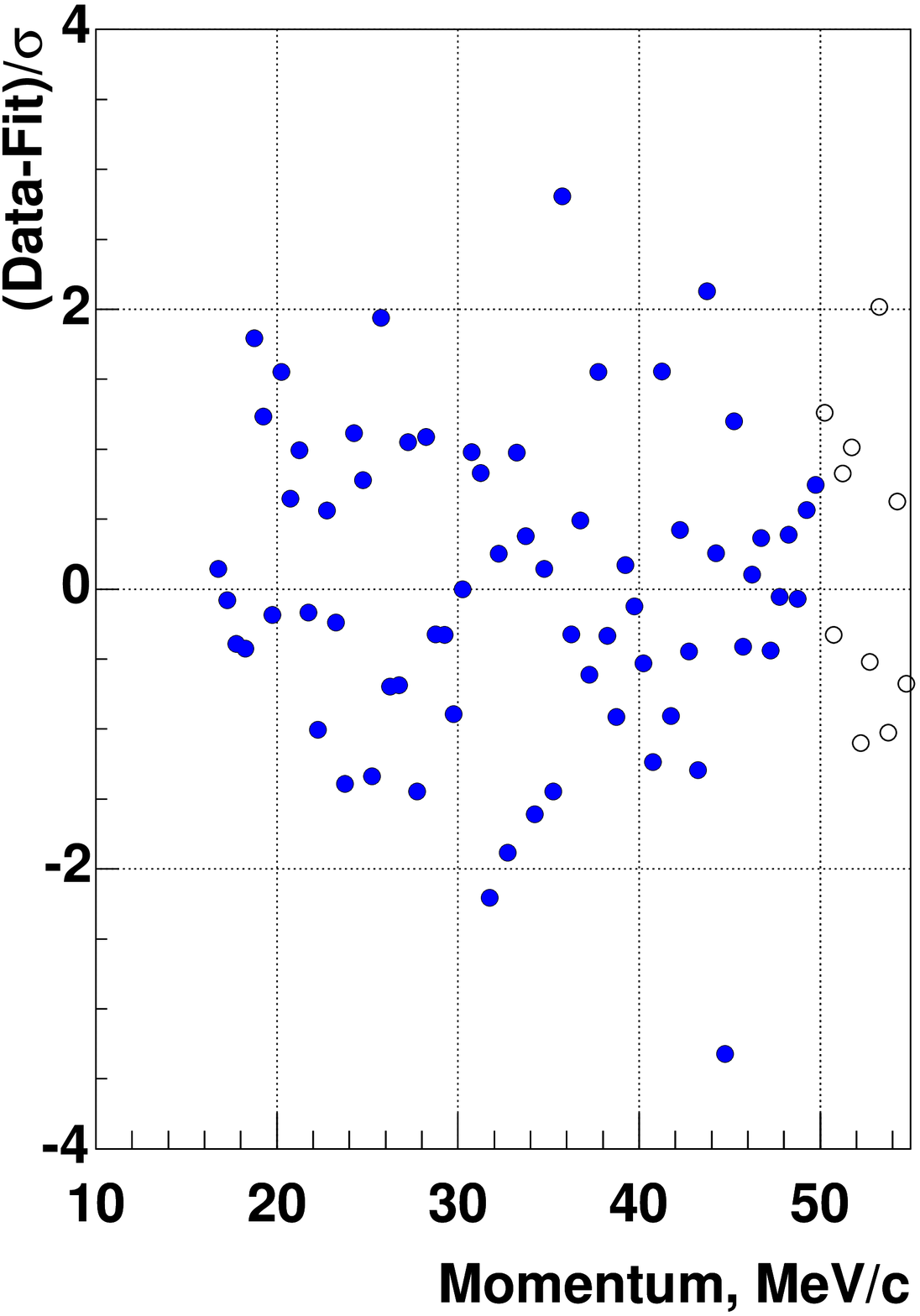}
\end{tabular}
\caption[Fit residuals for set B.]{\label{fig:residuals-setB}Fit
  residuals for set B.  Top left: color-coded residuals in the
  $\costh$ vs momentum plane.  Top right: projection on the momentum
  axis.  Bottom left: projection on the $\costh$ axis.  Bottom right:
  projection of the ``upstream minus downstream'' distribution. Solid
  markers are for points inside the fiducial region, empty markers for
  outside. The contours delimit the fiducial region. See text.  }
\end{figure}

\begin{figure}[tbp]
\begin{tabular}{cc}
\includegraphics[width=0.48\textwidth]{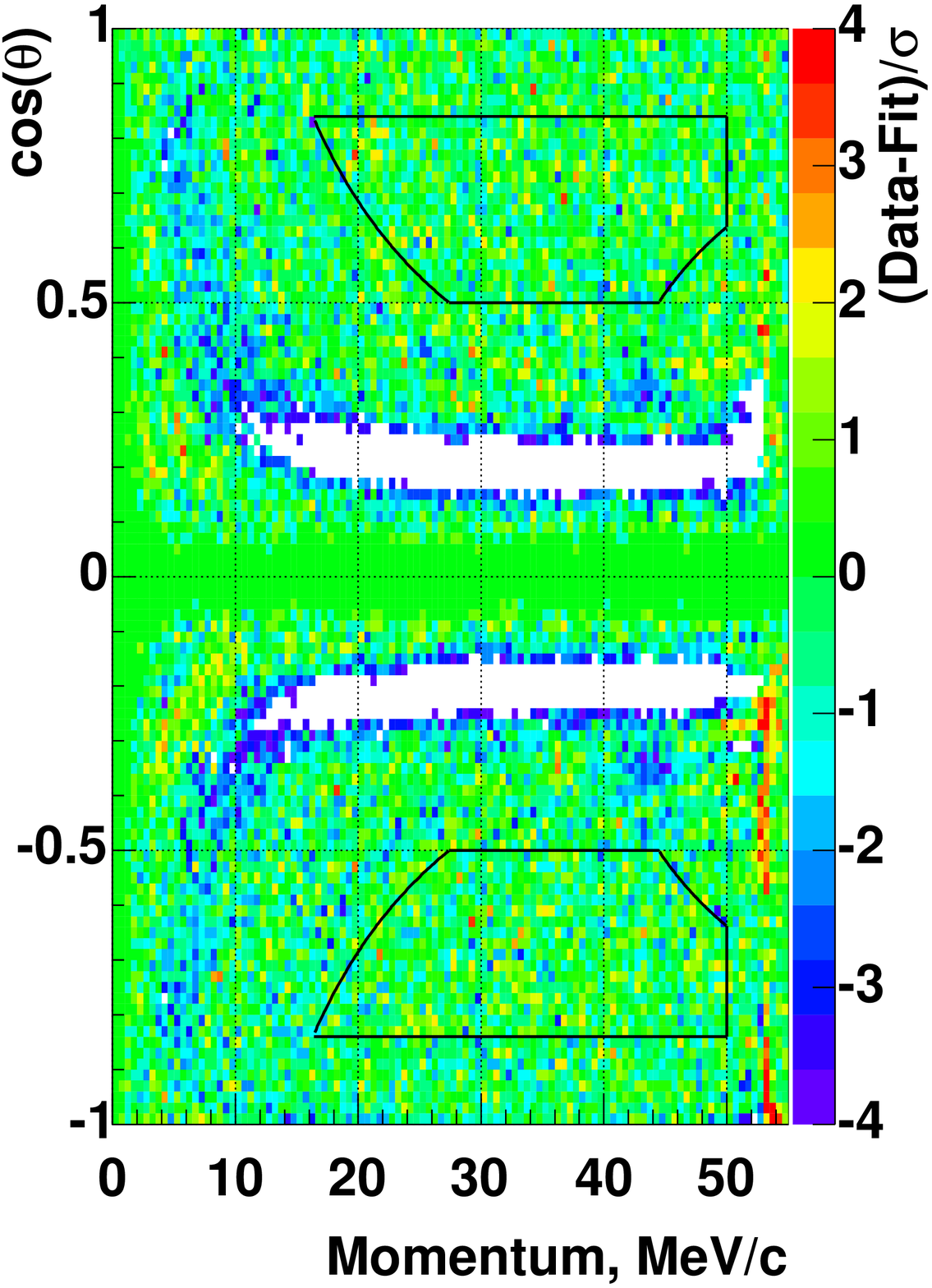} 
&
\includegraphics[width=0.48\textwidth]{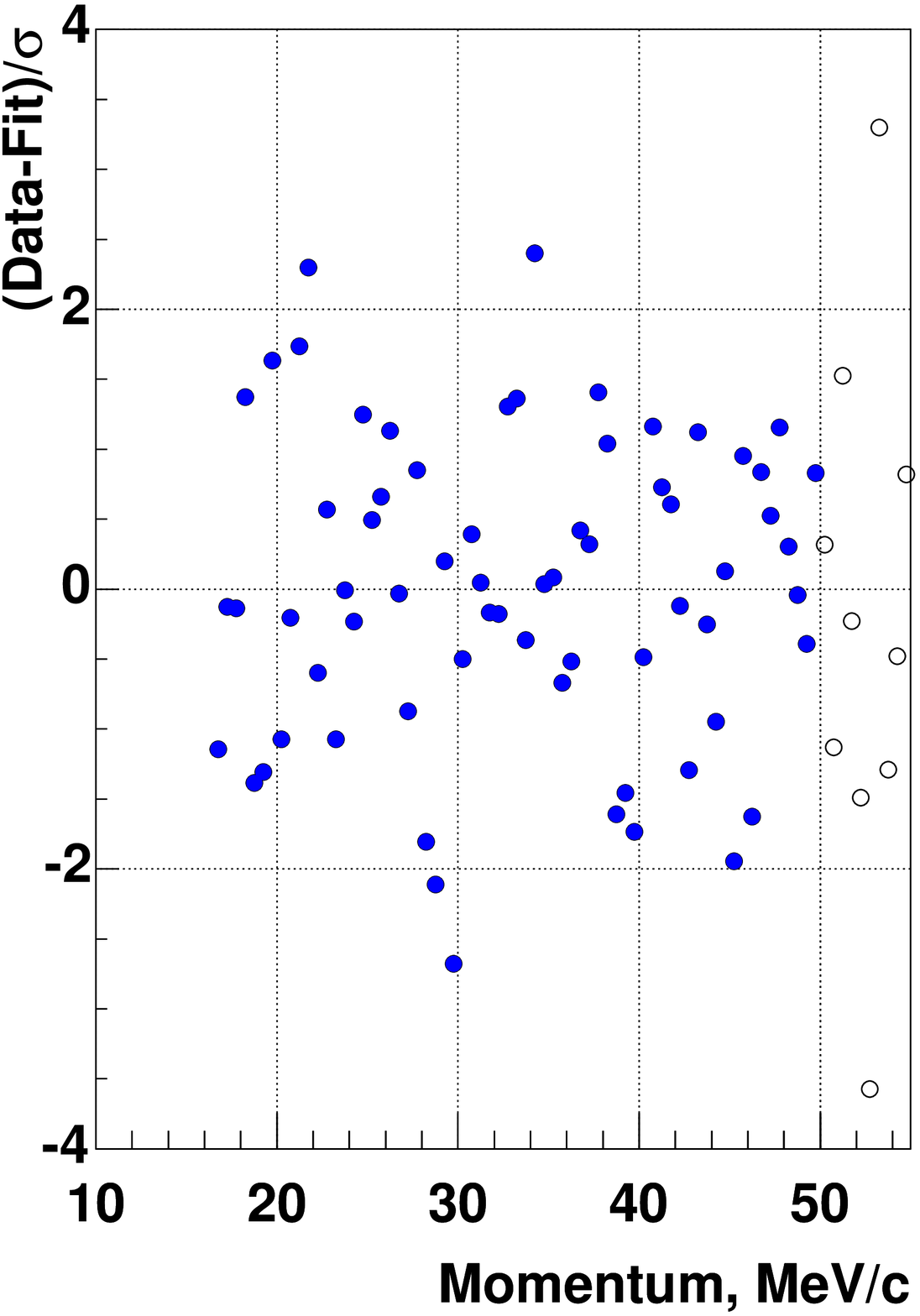}
\\
\includegraphics[width=0.48\textwidth]{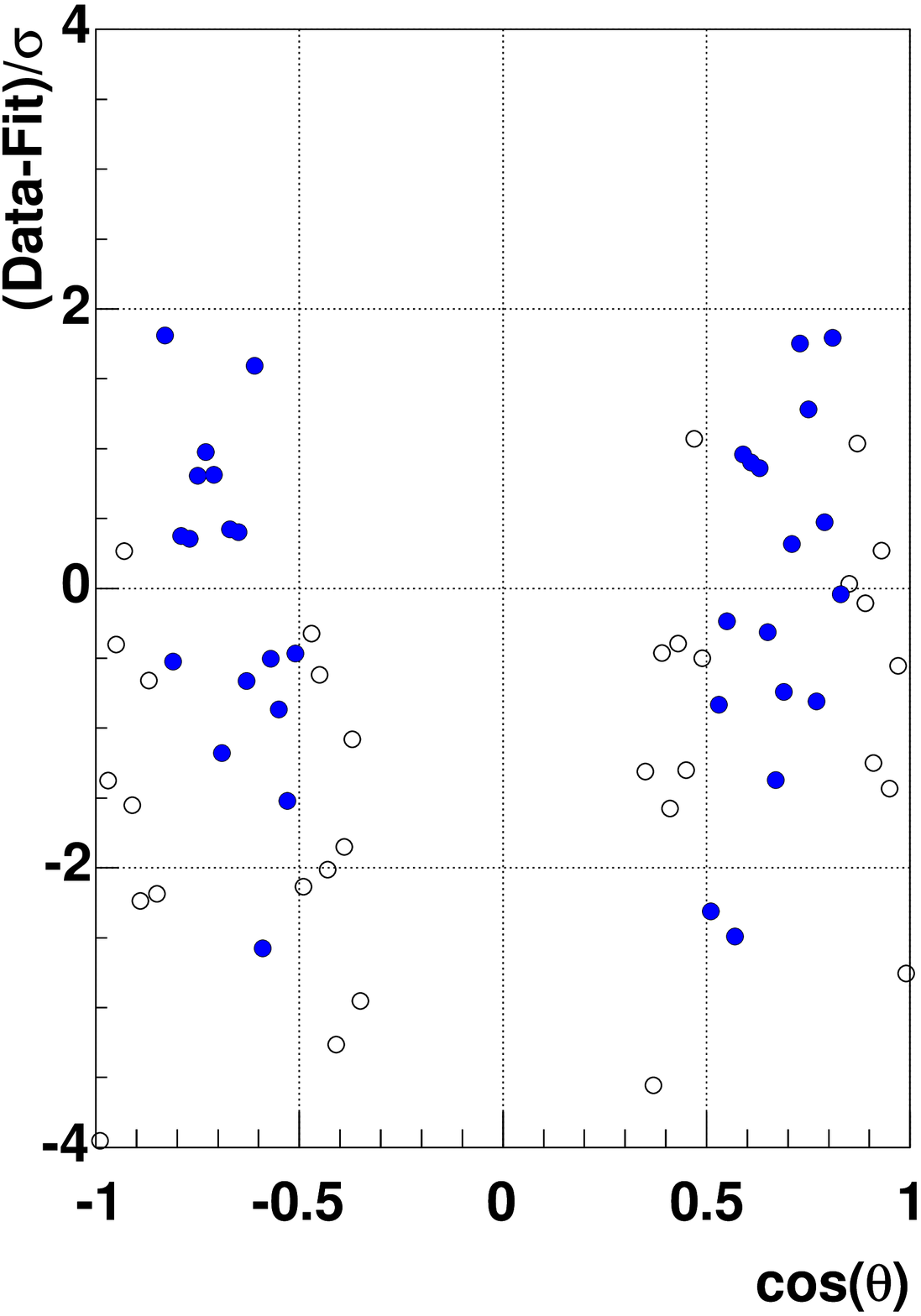} 
&
\includegraphics[width=0.48\textwidth]{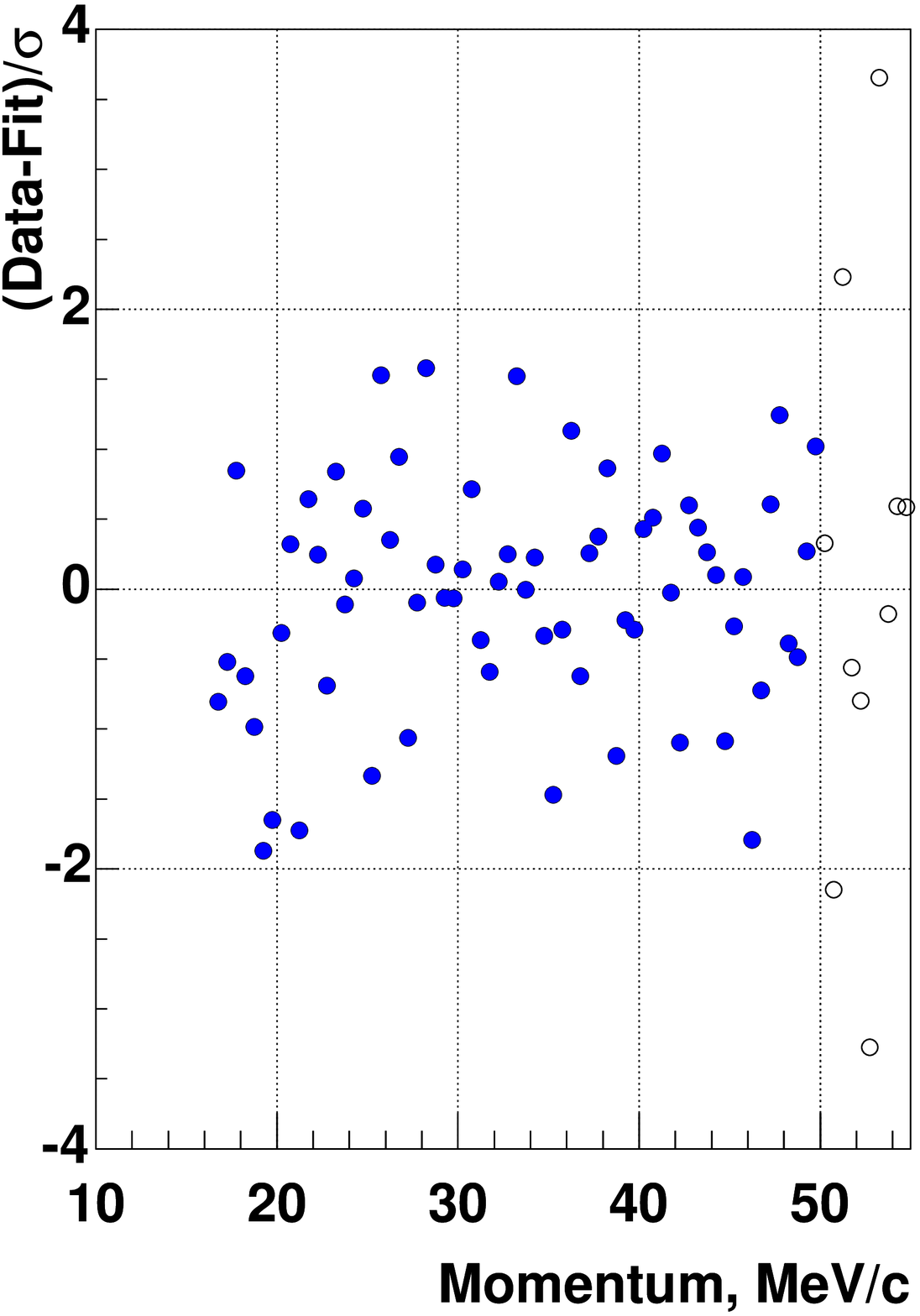}
\end{tabular}
\caption[Fit residuals for set 1.96~T.]{\label{fig:residuals-set196}Fit
  residuals for set 1.96~T.  Top left: color-coded residuals in the
  $\costh$ vs momentum plane.  Top right: projection on the momentum
  axis.  Bottom left: projection on the $\costh$ axis.  Bottom right:
  projection of the ``upstream minus downstream'' distribution. Solid
  markers are for points inside the fiducial region, empty markers for
  outside. The contours delimit the fiducial region. See text.  }
\end{figure}

\begin{figure}[tbp]
\begin{tabular}{cc}
\includegraphics[width=0.48\textwidth]{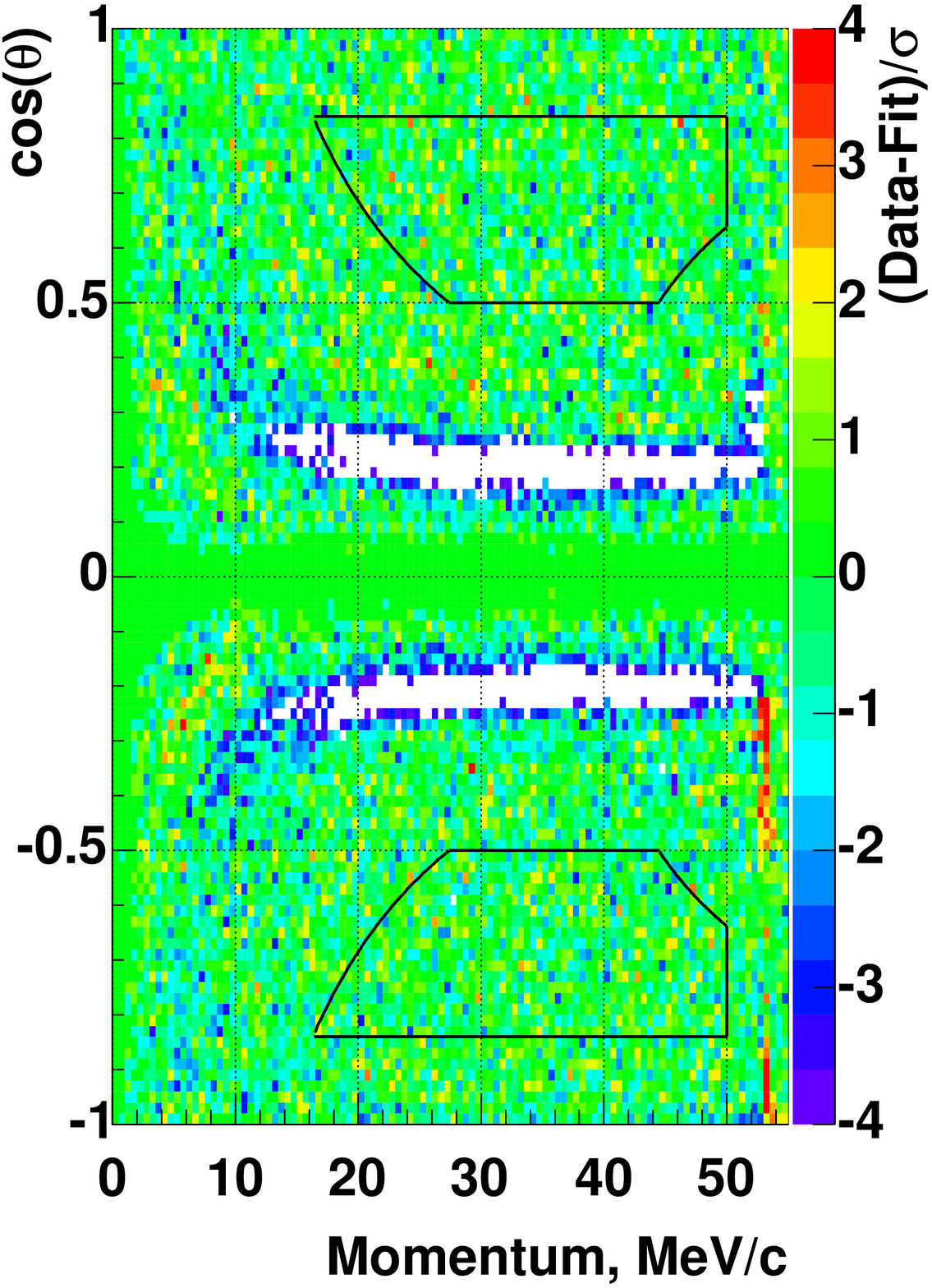} 
&
\includegraphics[width=0.48\textwidth]{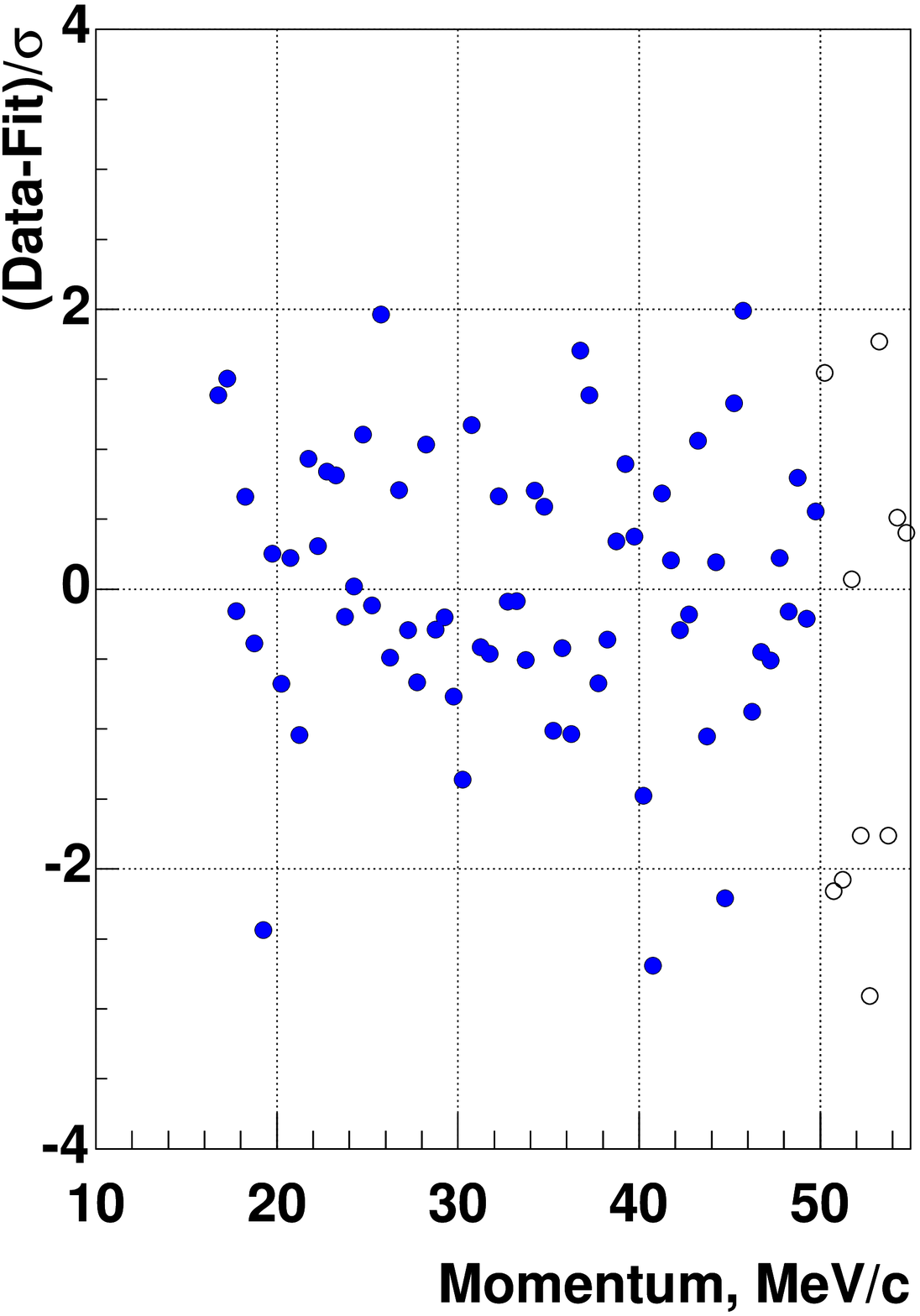}
\\
\includegraphics[width=0.48\textwidth]{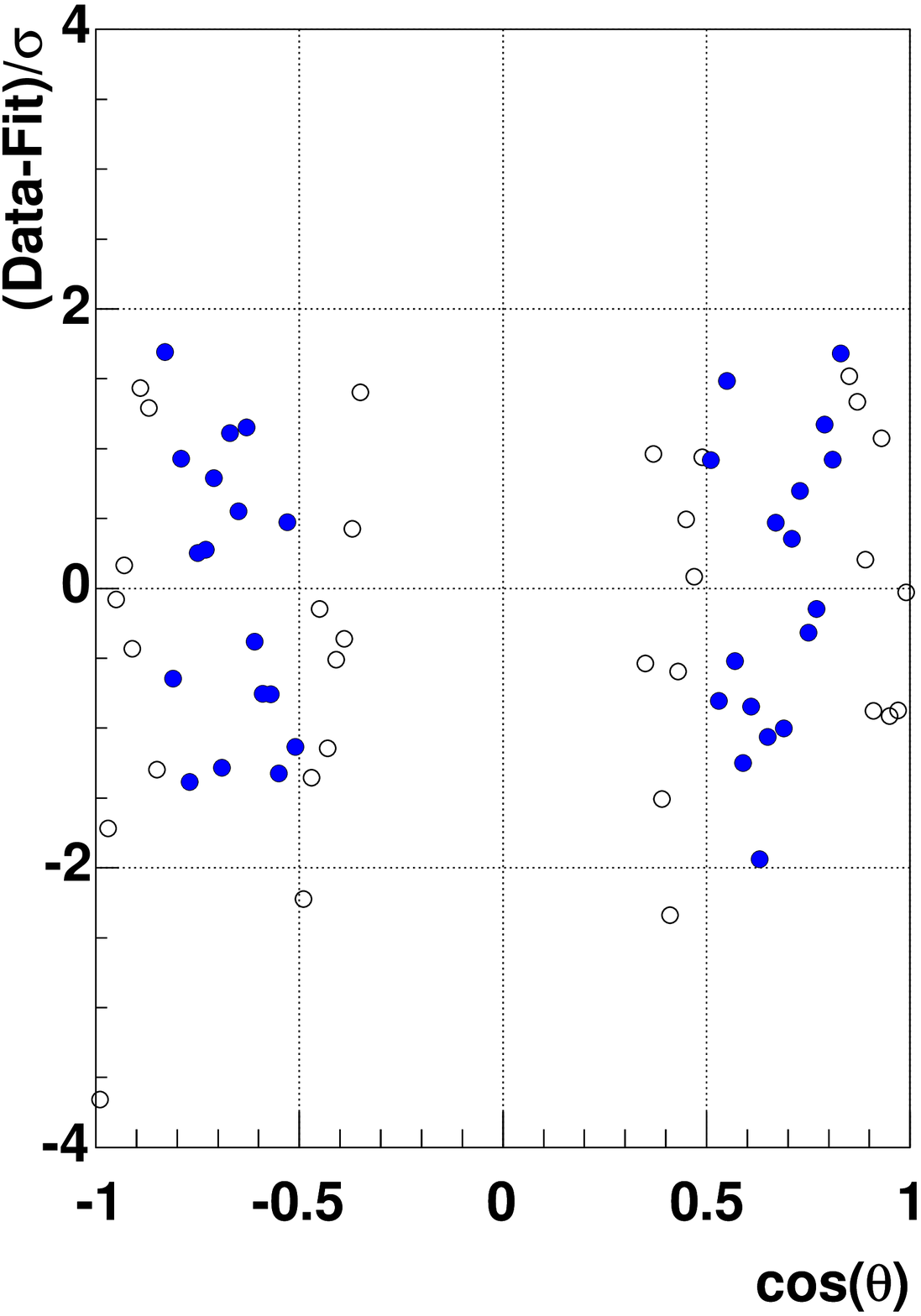} 
&
\includegraphics[width=0.48\textwidth]{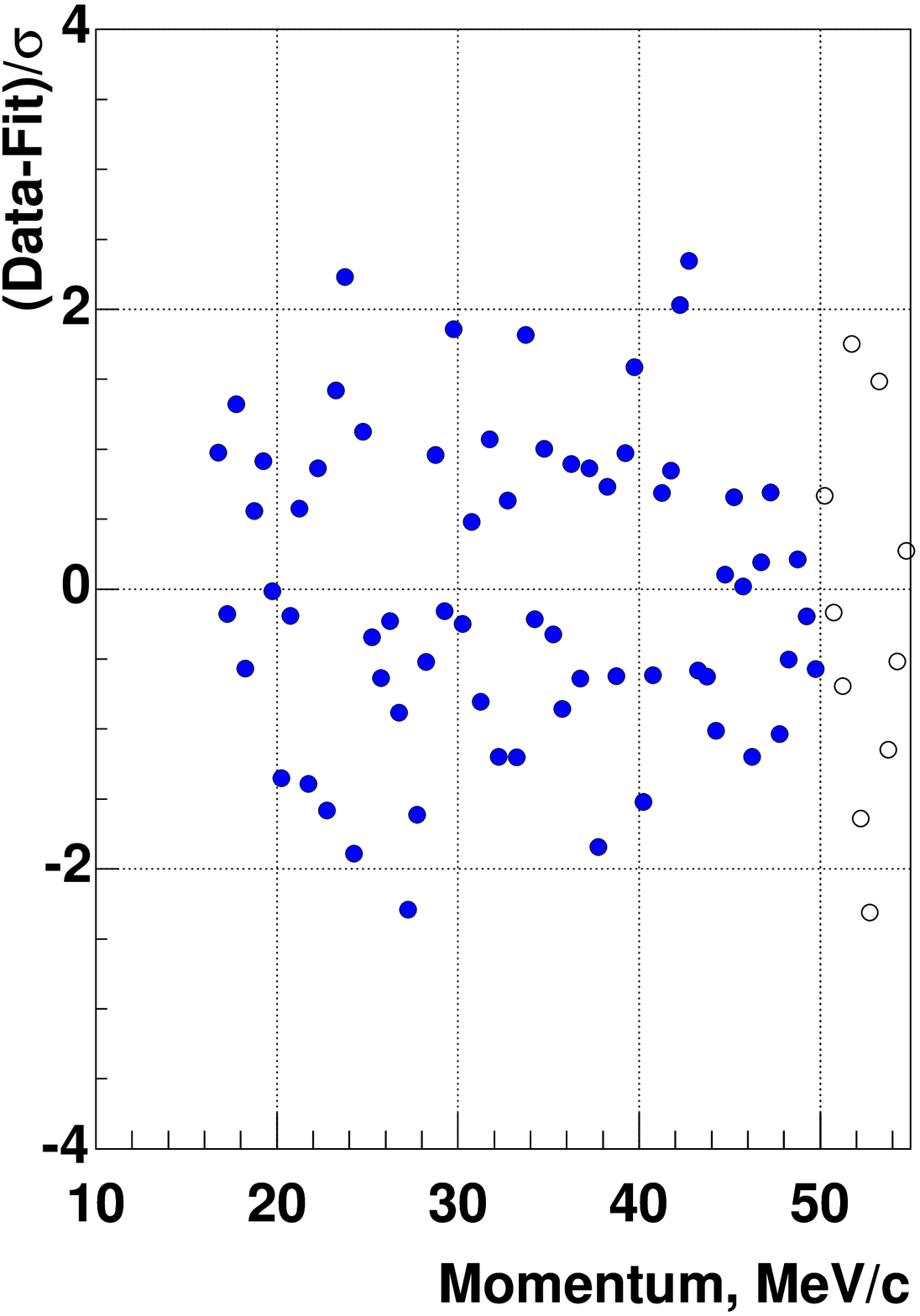}
\end{tabular}
\caption[Fit residuals for set 2.04~T.]{\label{fig:residuals-set204}Fit
  residuals for set 2.04~T.  Top left: color-coded residuals in the
  $\costh$ vs momentum plane.  Top right: projection on the momentum
  axis.  Bottom left: projection on the $\costh$ axis.  Bottom right:
  projection of the ``upstream minus downstream'' distribution. Solid
  markers are for points inside the fiducial region, empty markers for
  outside. The contours delimit the fiducial region. See text.  }
\end{figure}

\begin{figure}[tbp]
\begin{tabular}{cc}
\includegraphics[width=0.48\textwidth]{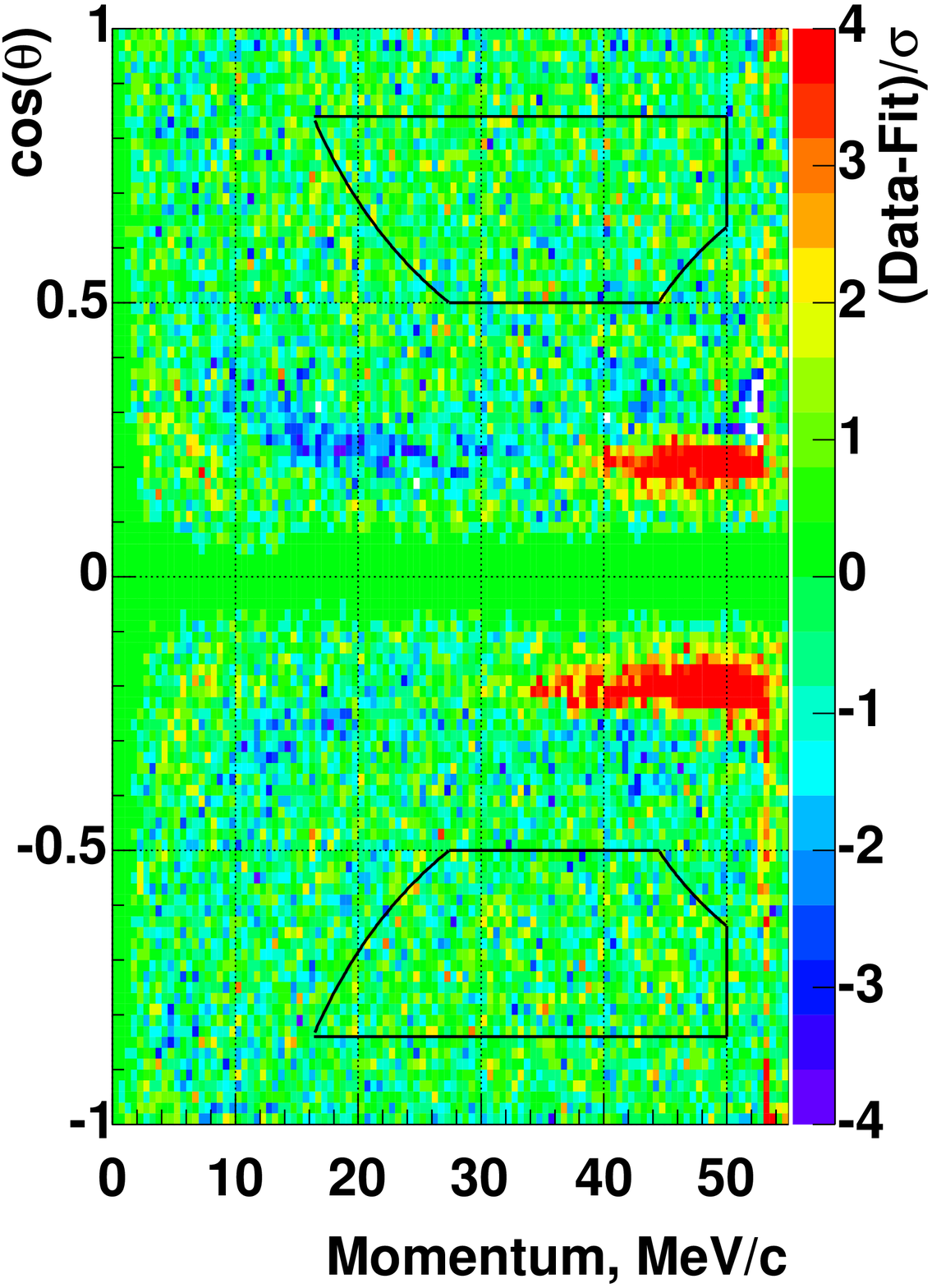} 
&
\includegraphics[width=0.48\textwidth]{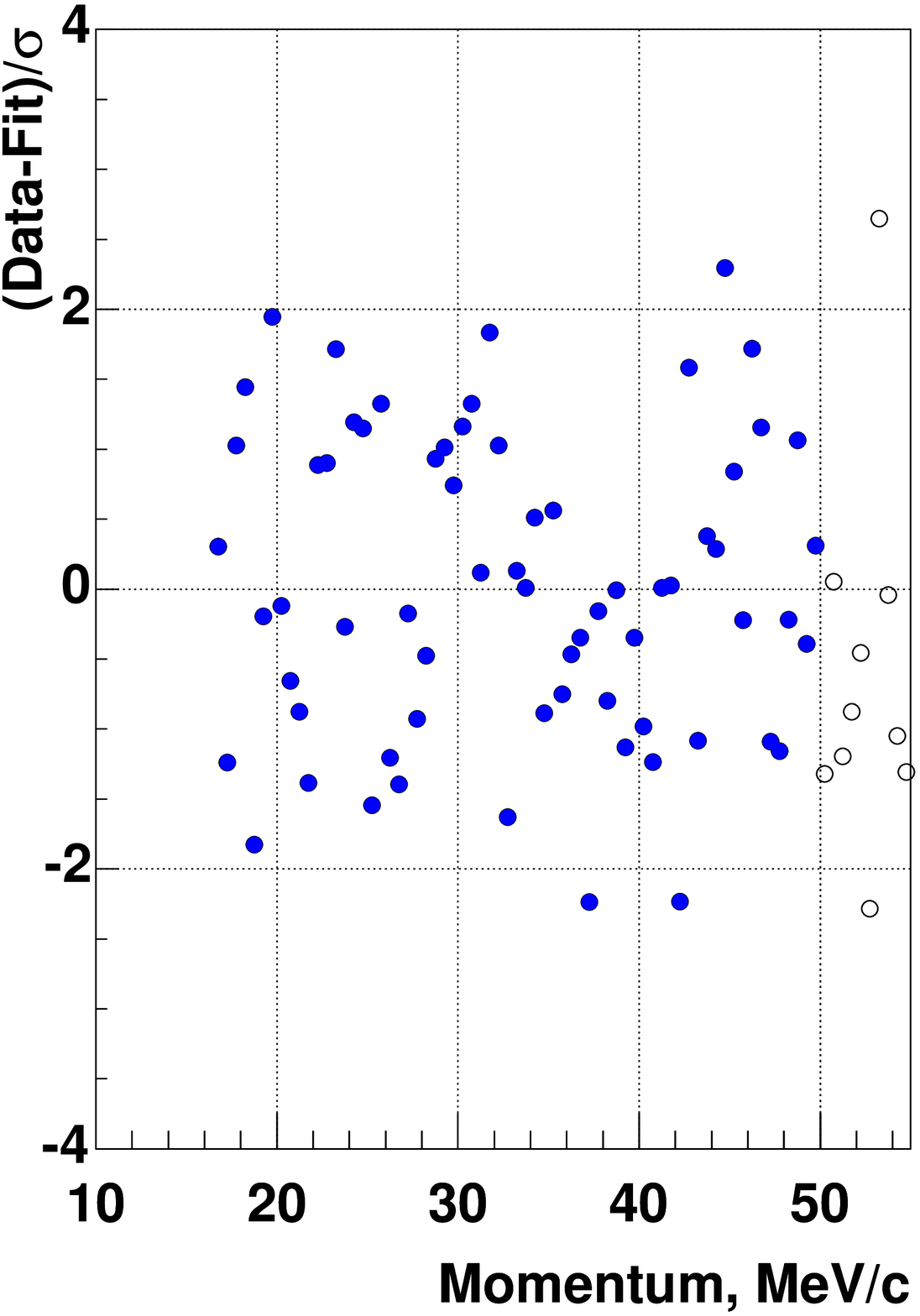}
\\
\includegraphics[width=0.48\textwidth]{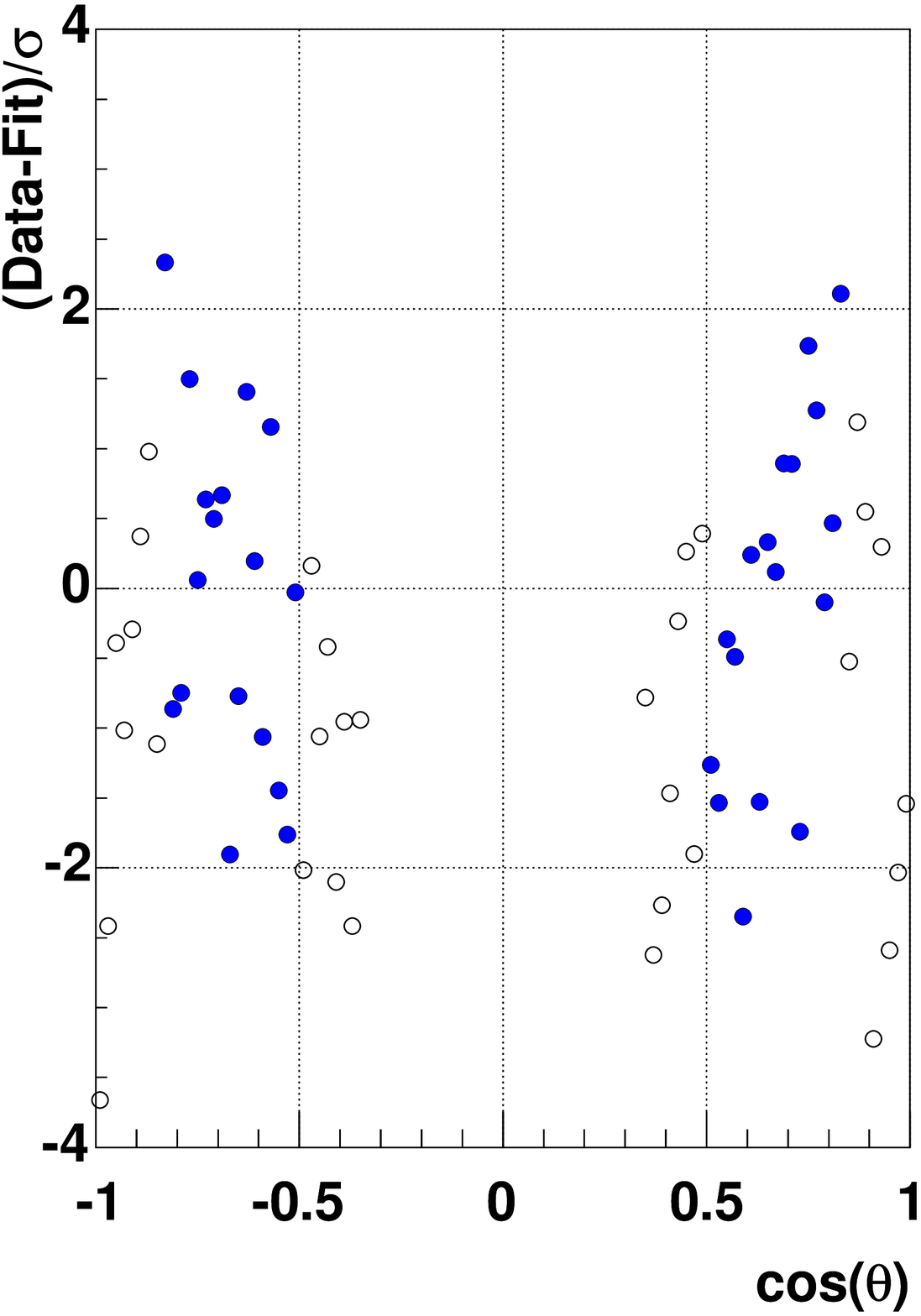} 
&
\includegraphics[width=0.48\textwidth]{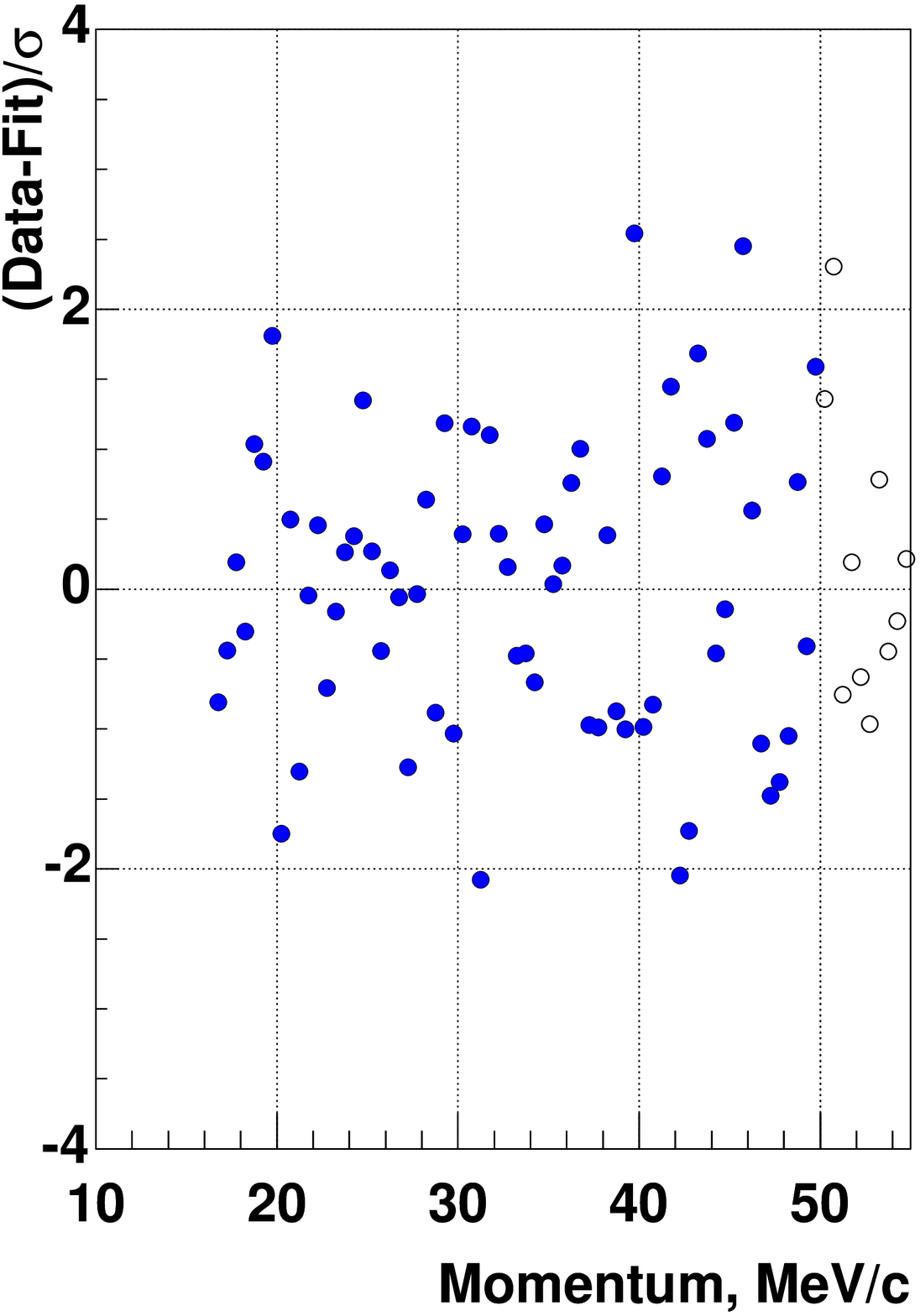}
\end{tabular}
\caption[Fit residuals for cloud muon set.]{\label{fig:residuals-cloud}Fit
  residuals for cloud muon set.  Top left: color-coded residuals in the
  $\costh$ vs momentum plane.  Top right: projection on the momentum
  axis.  Bottom left: projection on the $\costh$ axis.  Bottom right:
  projection of the ``upstream minus downstream'' distribution. Solid
  markers are for points inside the fiducial region, empty markers for
  outside. The contours delimit the fiducial region. See text.  }
\end{figure}

The simulation describes the data well, as it is demonstrated by
$\chi^2/\text{NDOF}\approx1$ and the reasonable fit probabilities
shown in Table~\ref{table:data-fits}.  Figures
\ref{fig:residuals-setA}--\ref{fig:residuals-cloud} show residuals of
the fits, providing more details on fit quality.  On each of the
figures, the top left panel shows the normalized deviation of the best
fit, $(\text{Data}-\text{Fit})/\sigma$, for each bin of the
2-dimensional fit histogram.   The deviation of a
bin is normalized to its statistical error $\sigma$, and shown on a
color scale.  
The two solid contours delimit the
fiducial region (page~\pageref{page:fiducial}). 
It can be seen that most bins within the fiducial agree
to 1--2~$\sigma$.

The 2-dimensional fit and data histograms were 
independently projected onto the
momentum (top right), and $\costh$ (bottom left) axes, ignoring bins
outside of the fiducial region.  Deviations between the obtained data
and fit projections are shown using the solid marker.  The empty
marker points on the top right plot were obtained by removing the
$p<50$~MeV/c fiducial cut, and projecting the 2-dimensional
distributions from the extended region.  Similarly the
$0.50<|\costh|<0.84$ cut was removed to obtain points outside of the
fiducial on the bottom left panel.  The bottom right panel shows a
match between the data and the fit for the $F_{AS}(p)$ distribution,
which is the most relevant for $\delta$.  It was obtained by
angle-integrating the spectra separately in the upstream ($\costh<0$)
and the downstream ($\costh>0$) parts of the fiducial region, and
computing the difference
$F_{AS}(p)\propto\text{Upstream}-\text{Downstream}$.  Most of the
residuals for bins inside the fiducial region are within 2~$\sigma$
off zero for all the views, and no pattern indicates a systematic 
difference between the data and simulation.

A series of consistency checks were done.  A fit to the
Cloud set gave $\delta=0.75245\pm0.00526\text{~(stat)}$.  The fact
that $\delta$ extracted from a data set with an \emph{opposite} (and
small) muon polarization, $\Pmu^{\text{cloud}}\sim+0.25$, is
consistent with the value extracted from surface muon data, with
$\Pmu^{\text{surface}}\sim-1$, demonstrates the absence of detector
asymmetries that would lead to different biases on $\delta$ for the
two cases.

Another check involved generating a Monte-Carlo set with values of the
muon decay parameters determined from set~B, and using the produced
spectrum to do another fit to set~B.  The fit yielded all deviations
of the decay parameters consistent with zero, as
expected\footnote{\label{footnote-reanalysis}That 
test failed for $\delta$ in the first round of
\TWIST{} analysis.  The problem was traced to a failure of
implementing Eq.~\eqref{eq:fas-rc} in the spectrum generator.  Since
$\rho$ is decoupled from the asymmetry parameters, this flaw had no
impact on $\rho$, and its value was published \cite{Musser:2004zw}.
On the other hand, it introduced an additional systematic uncertainty
on $\delta$, dependent on the difference of the average muon
depolarization in data and Monte-Carlo.  This systematics was
estimated to be ${}\lesssim0.001$.  This large value necessitated
a reanalysis, with the generator fixed.  A new black box was created,
and a second round of analysis performed, which is presented in this
work.  Because the effect of the mistake on $\delta$ was large, we did
not know the value of $\delta$, so the analysis was still blind. 
\vc{Other changes between \cite{Musser:2004zw} and this analysis 
are improvements in track selection (chapter~\ref{chapter:reconstruction})
and rotational alignment of the drift chambers.}
}.

Yet another check used a modified fitting procedure.  Instead of doing
a 3-parameter fit to the 2-dimensional spectrum, a 1-dimensional
distribution proportional to $F_{AS}(p)$ 
was extracted from it.  
(See \eqref{eq:theoretical-spectrum-2d}--\eqref{eq:theoretical-spectrum-fas}.) 
The
shape of that distribution is manifestly independent of $\rho$ and
$\eta$ (provided the detector response function is symmetric).  This
1-dimensional distribution was fit using the $\xixidelta$ and
$\xidelta$ parameters, and an alternative value of $\delta$ was
extracted from that fit.  A comparison of the alternative values,
$\delta_{\text{ud}}$, with the values extracted from fits to the
2-dimensional $(p,\costh)$ distribution, $\delta_{2D}$, is shown in
Table~\ref{table:fits-ud}.  The expected variance of the difference
$\sigma^2_{\text{diff}}$ for correlated data was computed as
$\sigma^2_{\text{diff}}={|\sigma^2_{\text{ud}}-\sigma^2_{\text{2D}}|}$,
assuming that one of the estimators saturates the Minimum Variance
Bound~\cite{Barlow:2002yb,kendall-stuart}.  It can be seen from
Table~\ref{table:fits-ud} that results of the alternative technique
are highly consistent with those given by the 3-parameter fits.
%
%fit3_set18an3_to_gen170an1.out  delta = 0.752580 +- 0.005309252 delta = 0.752449 +- 0.005261453,        diff = 0.000131 +- 0.000710823 (0.184293 sigma)
%fit3_set1an3_to_gen154an1.out   delta = 0.749350 +- 0.001262019 delta = 0.749182 +- 0.001241271,        diff = 0.000168 +- 0.0002279 (0.737166 sigma)
%fit3_set22an2_to_gen156an1.out  delta = 0.749073 +- 0.001337525 delta = 0.749080 +- 0.001315914,        diff = -7e-06 +- 0.000239465 (-0.0292318 sigma)
%fit3_set2an18_to_gen148an1.out  delta = 0.749788 +- 0.001264934 delta = 0.749794 +- 0.001243876,        diff = -6e-06 +- 0.000229849 (-0.0261041 sigma)
%fit3_set3an26_to_gen152an1.out  delta = 0.751083 +- 0.001589785 delta = 0.750865 +- 0.001562475,        diff = 0.000218 +- 0.000293408 (0.742993 sigma)
%
\begin{table}[hbtp]
\begin{center}
\begin{tabular}{ld{1.6}d{1.6}d{1.2}}
Data Set  
& \multicolumn{1}{c}{$\delta_{\text{ud}}-\delta_{\text{2D}}$} 
& \multicolumn{1}{c}{$\sigma_{\text{diff}}$}
& \multicolumn{1}{c}{$(\delta_{\text{ud}}-\delta_{\text{2D}})/\sigma_{\text{diff}}$}
\\ \hline
Set A  &  0.000218& 0.000293 &0.74 \\
Set B  & -0.000006& 0.000230 & -0.03 \\
1.96~T &  0.000168& 0.000228 & 0.74 \\
2.04~T & -0.000007& 0.000239 & -0.03 \\
Cloud  &  0.000131& 0.000711 & 0.18 \\
\hline
\end{tabular}
\caption[Differences between two and three parameter
  fits]{\label{table:fits-ud} Per-set differences between
  $\delta_{\text{ud}}$ extracted from a fit to the ``upstream minus
  downstream'' distribution $F_{AS}(p)$, and $\delta_{2D}$ from a
  3-parameter fit to the $(p,\costh)$ spectrum.  }
\end{center}
\end{table}

%%%%%%%%%%%%%%%%%%%%%%%%%%%%%%%%%%%%%%%%%%%%%%%%%%%%%%%%%%%%%%%%
%
%  w_i = 1/\sigma^2_i
%  
%  <x> = (\sum w_i x_i ) / (\sum w_i)   +-  1/sqrt(\sum w_i)
%

We compute the central value of $\delta$ as a weighted average
\cite{pdg-all:2004}, using for the weights a quadratic sum of 
statistical and set-dependent systematic uncertainties from
Table~\ref{table:data-fits}.  In the calculation of the final
systematic uncertainty we do not assume that it shrinks in the
combined measurement, and quadratically add set-independent and
average values of set-dependent systematics as shown in
Table~\ref{table:sys-summary}.  The final result is
\begin{equation}
\delta = 0.74964\pm0.00066\,\text{(stat.)}\pm0.00112\,\text{(syst.)}
\label{eq:delta-result}
\end{equation}

%%================================================================
%% Set A

%%================================================================
%% Set B

%%================================================================
%% Set 1.96

%%================================================================
%% Set 2.04

%%================================================================
%% Cloud set

%%================================================================

%%%%%%%%%%%%%%%%%%%%%%%%%%%%%%%%%%%%%%%%%%%%%%%%%%%%%%%%%%%%%%%%
\chapter{\label{chapter:summary}Conclusion}
%\chapter{\label{chapter:summary}Summary and discussion}

The \TWIST{} result for $\delta$, Eq.~\eqref{eq:delta-result}, is
consistent with the Standard Model prediction $\delta=\frac34$.  It
is also consistent with the previous best measurement
\cite{Balke:1988delta,pdg-all:2004}
$\delta=0.7486\pm0.0026\,\text{(stat.)}\pm0.0028\,\text{(syst.)}$.
\TWIST{} result Eq.~\eqref{eq:delta-result} can be rewritten with the
errors combined:
\begin{equation}
\delta = 0.74964\pm0.00130.
\label{eq:delta-combined}
\end{equation}
Compared to the combined error of \cite{Balke:1988delta}, the \TWIST{}
result is an improvement of a factor of $2.9$.  Because the measured
value is consistent with the Standard Model, it places new limits on
possible deviations from the theory.

\section{\label{section:qmur-limits}Model-independent limits \\
  on right-handed muon interactions}

Model-independent limits on right-handed couplings of the muon can be
obtained using Eq.~\eqref{eq:delta-combined}, the \TWIST{}
result~\cite{Musser:2004zw}
\begin{equation}
\rho=0.75080\pm0.00105,
\label{eq:rho-value}
\end{equation}
and a value of $P_{\mu}\xi\delta/\rho$.  Using
\cite{Jodidio:1986mz,Stoker:1985sj,Stoker-thesis} we get%
\footnote{No erratum correcting a mistake in $\mu-e$ scattering
\cite{Jodidio:1986mz} has been published for \cite{Stoker:1985sj}.  In
\cite{Stoker-thesis}, page~103, the value of
$P_{\mu}\xi\delta/\rho=0.9984\pm0.0016\pm0.0016$ is quoted.  Removing
an upward correction factor of $1.0007$ (page~86) for depolarization
in $\mu-e$ scattering, we obtain
$P_{\mu}\xi\delta/\rho=0.9977\pm0.0016\pm0.0016$.  The latter number,
combined with $P_{\mu}\xi\delta/\rho=0.99790\pm0.00046\pm0.00075$ from
\cite{Jodidio:1986mz}, gives the value quoted in the text.  
\label{page:limits-footnote}Because
\cite{Jodidio:1986mz,Stoker:1985sj,Stoker-thesis} quote their final
results not as values but only as lower limits on
$P_{\mu}\xi\delta/\rho$, Eq.~\eqref{eq:pmuxideltarho-value} can only
be used to produce limits on, but not values of, other parameters.
}
\begin{equation}
P_{\mu}{\xi\delta}/{\rho} = 0.99787 \pm 0.00082.
\label{eq:pmuxideltarho-value}
\end{equation}

We can transform Eq.~\eqref{eq:qmur} 
\begin{align}
%\begin{split}
Q^{\mu}_{R}&=\frac12\,\left\{1 + \frac13\,\xi-\frac{16}{9}\,\xi\delta\right\}
\\
&=\frac12\,\left\{1 - 
\xi\left(\frac{16}{9}\,\delta-\frac13\right)\right\}.
\\
\intertext{Since $\delta\approx3/4$, $16\,\delta/9-1/3>0$. Also
  $P_{\mu}\le1$, therefore}
Q^{\mu}_{R}&\le
\frac12\,\left\{1 - P_{\mu}\xi
\left(\frac{16}{9}\,\delta-\frac13\right)\right\}
\\
&=
\frac12\,\left\{1 - \left(P_{\mu}\xi\delta/\rho\right)\,\rho
\left(\frac{16}{9}-\frac{1}{3\delta}\right)\right\}.
\\
\intertext{Substituting the values of $P_{\mu}\xi\delta/\rho$, $\rho$,
and $\delta$}
Q^{\mu}_{R}& = 0.00061 \pm 0.00086.
\label{eq:qmur-value}
%\end{split}
\end{align}
Because mathematically $Q^{\mu}_{R}\ge0$, we convert
Eq.~\eqref{eq:qmur-value} to a 1-sided limit:
\begin{equation}
Q^{\mu}_{R}<0.00184,\quad\text{90\% confidence level.}
\label{eq:qmur-limit}
\end{equation}
This is our new limit on the fraction of muons decaying through
right-handed interactions.

Using Eqs.~\eqref{eq:qrr}--\eqref{eq:qll} and \eqref{eq:qmur-limit},
we can put new limits on interactions that couple right-handed muons
to left-handed electrons. These limits are summarized in Table~\ref{table:glr}

%: $|g^S_{LR}|< 0.086$, $|g^V_{LR}|<0.043$, and $|g^T_{LR}|<0.025$.

\begin{table}[htbp]
\begin{center}
\begin{tabular}{ccc}
Coupling & \TWIST{} limit & Previous limit from \cite{pdg-all:2004}\\
\hline
$|g^S_{LR}|\vphantom{\Big|}$ & 0.086 & 0.125 \\
$|g^V_{LR}|\vphantom{\Big|}$ &0.043  & 0.060 \\
$|g^T_{LR}|\vphantom{\Big|}$ &0.025  & 0.036 \\
\hline
\end{tabular}
\end{center}
\caption[Limits on right-handed muon
  interactions]{\label{table:glr}90\% confidence level upper limits on
  couplings between right-handed muons and left-handed electrons.}
\end{table}

%%================================================================
\section{Limits on $P_{\mu}\xi$}

% Lower limit:  (Strovink lower limit for Pmu*xi*delta/rho)*(our rho)/(our 
% delta).  This is not a "value" since Strovink et al only quote their result 
% as a limit.  This is a lower limit for Pmu*xi.  But Pmu*xi <= xi, so it 
% also becomes a lower limit for xi.
% 
% Upper limit:  Q^mu_R >= 0, combined with our value of delta.  This is an 
% upper limit for xi.  But Pmu*xi <= xi, so it also becomes an upper limit 
% for Pmu*xi.
%
% Formally, each of these is a one-way limit.  Hence, one is tempted to take 
% 1.28 sigma values.  However, that isn't fair, since it would assume the 
% other direction is unconstrained, which it isn't.  I don't know of an 
% "exact" way to handle a situation like this.  What I did was to require the 
% SAME sigma multiplier for both limits, such that each included essentially 
% 90% of the total probability within the JOINTLY accepted region.  You can 
% see that in the .xls spreadsheet that I've attached.

From the same inputs of $\delta$, $\rho$, and $P_{\mu}\xi\delta/\rho$,
as in the previous section, it is possible to place new limits on
$P_{\mu}\xi$.  
Using \eqref{eq:pmuxideltarho-value}, \eqref{eq:rho-value}, and
\eqref{eq:delta-combined}, we obtain an intermediate result in the
form $P_{\mu}\xi=v_L\pm\sigma_L$, which is not a ``value'', but can be
used to set a lower limit on $P_{\mu}\xi$ (see footnote on
page~\pageref{page:limits-footnote}).
%
%A lower limit on $P_{\mu}\xi$ can be extracted from
%\eqref{eq:pmuxideltarho-value}, \eqref{eq:rho-value}, and
%\eqref{eq:delta-combined}.  
%
Upper limits on $\xi$ are imposed by $\xi\delta/\rho\le1$ and
$Q^{\mu}_{R}\ge0$, with the latter being the strongest. From
$Q^{\mu}_{R}=0$ we get $\xi=v_U\pm\sigma_U$, and can use it to compute
a limit.
Because $P_{\mu}\xi\le\xi$, the final result should be in the form
$L<P_{\mu}\xi\le\xi<U$.  This makes $L$ and $U$ weaker than the
corresponding one-sided limits would be.
We defined the lower, $L$, and the upper, $U$, bounds, as
$L=v_L-k\sigma_L$, $U=v_U+k\sigma_U$, imposing the same sigma
multipliers at both ends.  Demanding that the sum of integrals of
the normal distributions $\text{Gauss}(v_L,\sigma_L)$ and
$\text{Gauss}(v_U,\sigma_U)$ between $L$ and $U$ is $2\times0.9$, we
obtain a ``90\% confidence interval''
\begin{equation}
0.9960<P_{\mu}\xi\le\xi<1.0040.
\label{eq:pmuxi-limits}
\end{equation}
%
% Integral(L) = 0.8975
% Integral(U) = 0.9025
%
(In fact the two integrals are $0.8975$ (L), and $0.9025$ (U), so that
each of the limits is very close to~90\%.)

\section{Limits on left-right symmetric models}

The lower limit on $P_{\mu}\xi$ \eqref{eq:pmuxi-limits} can be used to
put limits on mass of the second charged gauge boson and its mixing
angle with the Standard Model $W$ in left-right symmetric theories,
see \eqref{eq:rls-xi}--\eqref{eq:rls-pmu}.  

Manifest left-right symmetric models assume that $g_L=g_R$, the
right-handed CKM matrix coincides with the known left-handed CKM
matrix, and there is no CP violation in the mixing: $\omega=0$.
(Notation from section~\ref{section:lrs}.)  Pseudo-manifest models 
allow CP violation, but require $V^L=(V^R)^*$ and $g_L=g_R$. 
See pp.~377--379 in \cite{pdg-all:2004} for a recent review.

\begin{figure}[tbp]
%\begin{tabular}{c}
\vbox to 0pt{%
\hbox to\textwidth{\hspace*{2mm}\hfill\includegraphics[width=0.97\textwidth]{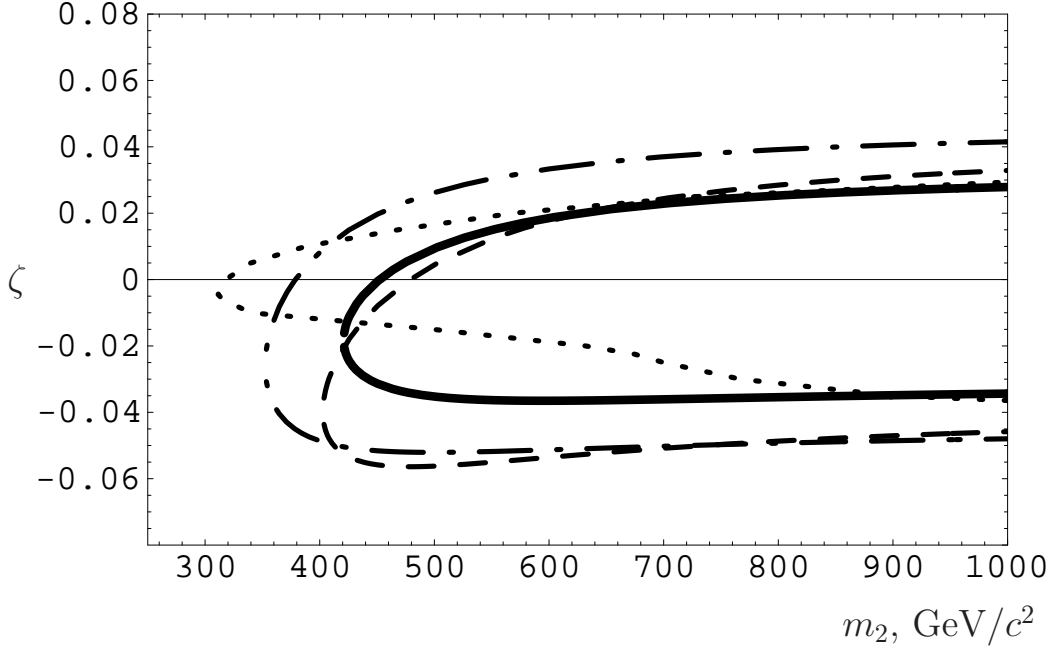}\hfill}\vss}%
% 
% Axis labels:
\vbox to 0pt{\vspace*{31mm}\hbox to \textwidth{\hspace*{0mm}{\Large$\zeta$}\hfill}\vss}%
\vbox to 80mm{\vspace*{76mm}\hbox to
  \textwidth{\hspace*{111mm}{\Large$m_2$,~GeV/$c^2$}\hfill}\vss}
%
%\end{tabular}
\caption[Constraints on manifest left-right symmetric models]{%
\label{fig:lrs-manifest}90\%~CL constraints 
on manifest left-right symmetric models.
%
%vd: Horizontal scale: $m_2$, GeV/c${}^2$. Vertical scale: $\zeta$. 
%
The excluded region is low mass and large $|\zeta|$.  Bold line:
\TWIST{} lower limit on $P_{\mu}\xi$.  Dashed line: dedicated search
for right-handed currents in muon decay
\cite{Jodidio:1986mz}. Dash-dotted line: one-sided limit
$P_{\mu}\xi>0.991867$ (90\%~CL) from a direct measurement of
$P_{\mu}\xi$ \cite{Beltrami:1987ne}.  Dotted line: combined nuclear
beta decay data~\cite{Thomas:2001}.  }
\end{figure}

Some of the existing constraints in the mass---mixing angle plane for
the manifest case are shown on Fig.~\ref{fig:lrs-manifest}.  It can be
seen that \TWIST{} constraints indeed provide an improvement over
previous muon decay data, including a dedicated search
\cite{Jodidio:1986mz}.  The new limit on $W_R$ mass is
$m_2>420$~GeV/c${}^2$, compared with the previous limit of
406~GeV/c${}^2$ \cite{Jodidio:1986mz} (402~GeV/c${}^2$ with the modern
value $m_1=80.423$~GeV/c${}^2$).  This new limit 
is also significantly stronger than the combined
limit from many nuclear beta decay experiments summarized
in~\cite{Thomas:2001}.

%%================================================================
%% LRS exclusion plot
%%================================================================

\vc{A measurement of the Michel parameter $\rho$ provides a constraint
on the mixing angle, which does not depend on the mass.  The best
limit, established by \TWIST{}, is
$|\zeta|<0.030$ at 90\% confidence level~\cite{Musser:2004zw}.}
There is \vc{also} a very tight constraint on mixing angle from superallowed
nuclear beta decays \cite{Hardy:2004dm}, that is dependent on nuclear
theory and other inputs.  When PDG recommended values \cite{pdg-all:2004}
are used for elements of the CKM matrix, it yields a non-zero result
$\zeta=0.00176\pm0.00074$.

% From Carl's Limits_02b.xls, E14=1.7646 gives 1.9000 in G17.
%
%\TWIST{} $0.99523<P_{\mu}\xi\le\xi<1.00472$~(95\%~CL).

% Hi Glen,
% 
% I got the following limits on left-right symmetric models:
% 
% from 
% 
% (1-Pmu*xi) < a^2
% 
% follows
% 
% (g_L/g_R) * M_2 > K * M_1 / sqrt(a)
% 
% where K=2^(1/4) for the general case, K=3^(1/4) for the pseudomanifest
% case, M_1 ~ 80.425 GeV.  (Of course, g_L=g_R for the pseudomanifest
% case.)

%
% masslimitG[a2_] := (2^(1/4))*80.425/a2^(1/4);
% masslimitM[a2_] := (3^(1/4)) * 80.425/a2^(1/4);
% masslimitG[1 - 0.9960]
% masslimitM[1 - 0.9960]
% masslimitG[1 - 0.99523]
% masslimitM[1 - 0.99523]
% 
% 380.306
% 420.878
% 363.931
% 402.756

Limits from direct searches at colliders do not constrain $\zeta$, but
provide better constraints on the mass.  The strongest combined result
is $m_2>786$~GeV/c${}^2$ at 95\% confidence level
\cite{Affolder:2001gr}.  (At 95\%~CL,
$0.99523<P_{\mu}\xi\le\xi<1.00472$, and $m_2>402$~GeV/c${}^2$ from
\TWIST{}.)  Collider results need less restrictive assumptions about
mass of right-handed neutrino than low-energy tests, but depend on the
assumed decay channels of the right-handed boson.

Under the assumption of manifest left-right symmetry a yet stronger
limit of $m_2>1.6$~TeV/c${}^2$ can be extracted from the $K_L-K_S$
mass difference $\Delta{m_K}$ \cite{Beall:1981ze}.

%%================================================================
%% LRS exclusion plot
\begin{figure}[tbp]
%\begin{tabular}{c}
\vbox to 0pt{%
\hbox to\textwidth{\hspace*{8mm}\includegraphics[width=0.97\textwidth]{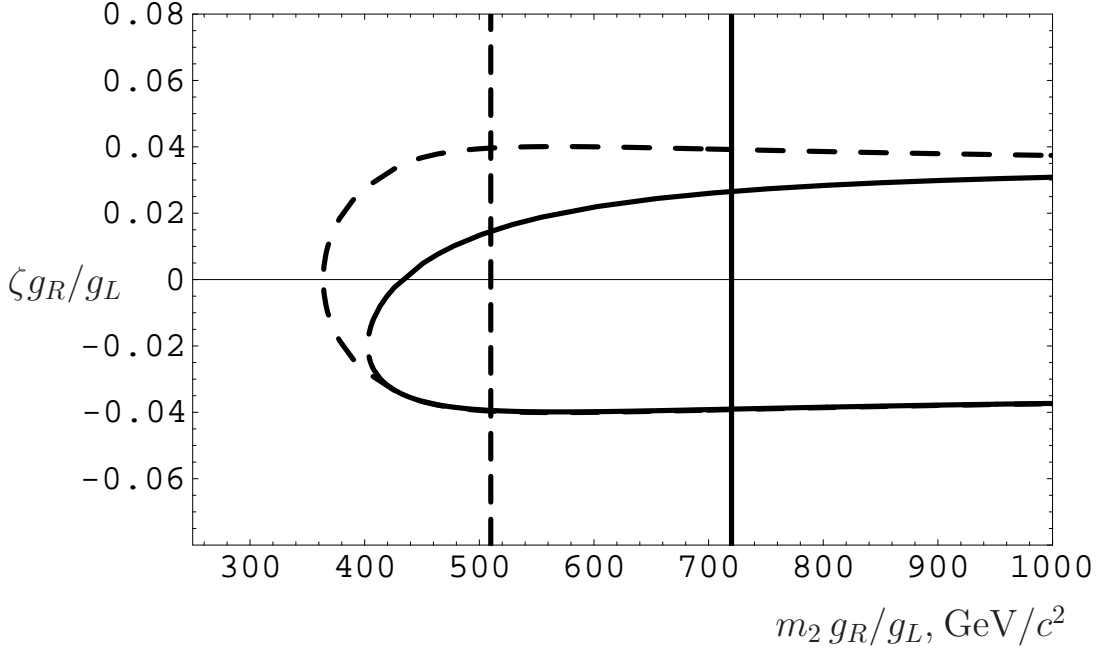}\hfill}\vss}%
% 
% Axis labels:
\vbox to 0pt{\vspace*{31mm}\hbox to \textwidth{\hspace*{-1mm}{\Large$\zeta{}{g_{R}}/{g_L}$}\hfill}\vss}%
\vbox to 80mm{\vspace*{76mm}\hbox to
  \textwidth{\hspace*{101mm}{\Large$m_{2}\,g_{R}/g_L$,~GeV/$c^2$}\hfill}\vss}
%
%\end{tabular}
\caption[Model dependence of LRS constraints]{%
\label{fig:lrs-general}A comparison of model dependence of muon decay
and collider limits on left-right symmetric models.  
%
%vd: Horizontal scale:
%vd: $m_{2}g_{R}/g_L$, GeV/c${}^2$. Vertical scale: $\zeta{}g_{R}/g_L$.
%
Solid lines: 95\%~CL limits on manifest left-right symmetry from
\TWIST{} (curved line) and D0~\cite{Abachi:1995yi} (vertical).  Dashed
lines: \TWIST{} exclusion for arbitrary CP violation and right-handed
CKM matrices (curved), D0 mass limit for a specific set of parameters
of non-manifest model (vertical).  }
\end{figure}
%%================================================================

Manifest and pseudo-manifest left-right symmetric models have severe
difficulties explaining experimental data (see
e.g. \cite{Langacker:1989xa}), therefore a more general case has to be
considered.  The number of parameters become much larger in
generalized models, since no statement about the right-handed CKM
matrix can be made.  Many limits become significantly weaker for
generalized left-right symmetric models, and may cease to be useful if
fine-tuning is allowed.  This is true for constraints from
$\Delta{m_K}$ \cite{Olness:1984xb,Langacker:1989xa}, nuclear beta
decays (Fig.~12 in \cite{Thomas:2001}), and collider results
\cite{Rizzo:1993fe}.  Muon decay data, on the other hand, have very
little sensitivity to assumptions about the unknown right-handed
sector.  An illustration of this statement is given on
Fig.~\ref{fig:lrs-general}.  The two vertical lines represent 95\% D0
limit under assumption of manifest symmetry (solid), and a
significantly weaker one obtained assuming a different specific set of
model parameters (dashed) \cite{Abachi:1995yi}.  To make a direct
comparison to the quoted D0 results, the \TWIST{} limit from
Fig.~\ref{fig:lrs-manifest} was converted to 95\% CL, and shown as the
solid curved line.  The dashed curved line on
Fig.~\ref{fig:lrs-general} shows 95\%~CL excluded region from \TWIST{}
for the most general case, with an arbitrary fine tuning allowed.  If
a point is outside of the dashed curve, it is excluded for \emph{any}
possible combination of the parameters with at least 95\% confidence
level.  (Of course, right-handed neutrinos still have to be light.)

% On the other hand, for each point inside the curve, there exist a set
% of parameters of generalized left-right symmetric model such that the
% point is withing a 5\%~CL allowed region.

A very strong limit on the mass of the second $W$ boson, of the order
of 3~TeV/c${}^2$, can be obtained from big bang
nucleosynthesis~\cite{Cyburt:2004yc}.  They require right-handed
neutrinos to be lighter than about $1$~MeV/c${}^2$, and depend on
assumptions about cosmological models that are outside of the scope of
particle physics.

Among tests of left-right symmetric models done within the context of
particle physics, muon decay provides better mass limits than nuclear
(and neutron) decay data.  These limits are still weaker than limits
from direct collider searches.  Unlike direct collider searches, muon
decay constrains the mixing angle as well as the mass parameter.  But
muon decay constraints are conditional on the lightness of
right-handed neutrinos.  On the other hand, collider results are
obtained using assumptions of manifest or pseudo-manifest symmetry.
An important advantage of muon decay results is their weak dependence
on unknown parameters of more general left-right symmetric models,
making them complementary to data from other sources.  A discussion on
complementarity of different observables for generalized left-right
symmetric models can be found in \cite{Thomas:2001}.  Muon decay
results are also not subject to complications of QCD and nuclear
theory.

%%%%%%%%%%%%%%%%%%%%%%%%%%%%%%%%%%%%%%%%%%%%%%%%%%%%%%%%%%%%%%%%
\section{Limit on non-local tensor interactions}

Using \eqref{eq:kappa-effect} and a 90\% confidence level lower limit
for $\delta$ from  \eqref{eq:delta-combined}, we obtain
\begin{equation}
|g^T_{RR}|<0.024,\quad\text{90\% confidence level}.
\label{eq:kappa-limit}
\end{equation}
Since the proposed value is $g^T_{RR}\approx0.013$
(Section~\ref{section:kappa}), this limit does not constrain the
model significantly.

%%================================================================
%

%%%%%%%%%%%%%%%%%%%%%%%%%%%%%%%%%%%%%%%%%%%%%%%%%%%%%%%%%%%%%%%%

\clearpage\addcontentsline{toc}{chapter}{References}
     %add the above line to get "Bibliography" in the table of contents.
%

%\singlespacing % optional;  Bibliography is better in single spacing
               %            but you may choose different
               %            Don't use \singlespacing if your thesis
               %            is already in single spacing

\ifagsdouble\onehalfspacing\else\relax\fi

%
%\bibliographystyle{plain} 
%\bibliographystyle{abbrv} 
%\bibliographystyle{unsrt} % Or which ever you wish. Plain is good
                          % for long bibs.  This style will list
                          % references in the order you cited them.

% From SPIRES, handles the additional eprint, collaboration (and may
% be other?) fields.

\bibliographystyle{utphys}

\bibliography{main,gaponenko}

%
%\printgloss{default}

%%%%%%%%%%%%%%%%%%%%%%%%%%%%%%%%%%%%%%%%%%%%%%%%%%%%%%%%%%%%%%%%
\appendix  %  If you have any appendicies
            % Use standard Latex sectioning commands
            % like \chapter etc.
\begingroup
\ifagsdouble\doublespacing\else\relax\fi
\newcommand\mychapter[2]{%
  \stepcounter{chapter}\protect\chapter*{\label{#1}{\normalfont\huge\bfseries
      Appendix~\thechapter}\\[20pt]#2}\protect\addcontentsline{toc}{chapter}{Appendix~\thechapter. ~#2}}

%%%%%%%%%%%%%%%%%%%%%%%%%%%%%%%%%%%%%%%%%%%%%%%%%%%%%%%%%%%%%%%%
\mychapter{chapter:twist-coordinates}{\TWIST{} coordinate system\\
  and kinematic variables}

%\chapter{\label{chapter:twist-coordinates}\TWIST{} coordinate system\\
%  and kinematic variables}

The origin of the \TWIST{} coordinate system is in the center of the
muon stopping target.  The $Z$ axis is along the detector stack, in
the direction of the muon beam (see Fig.~\ref{fig:stack}).  The $Y$
axis points upwards, and the $X$ axis completes a right-handed $XYZ$
coordinate system.  The $U$ and $V$ axes are obtained from the $X$ and
$Y$ axes, correspondingly, by a $+45^\circ$ rotation about the $Z$
axis.  \TWIST{} wire chambers measure $U$ and $V$ coordinates, not $X$
and $Y$, see chapter~\ref{chapter:experiment}.

The angle $\theta$ of a track is defined by $\costh=p_z/p$, where
$p=|\vec{p}|$ is the momentum of the particle, and $p_z$ is the
projection of the momentum on the $Z$ axes.  The transverse momentum,
$p_t$, is defined as $p_t^2=p^2-p_z^2$.

The ``upstream'' and ``downstream'' directions are defined relative to
the muon beam, that is $\costh<0$ for an upstream decay, $\costh>0$ for
a downstream decay.  Upstream part of the detector is the one seen by
an incoming muon before it comes to rest in the central stopping
target.

%%%%%%%%%%%%%%%%%%%%%%%%%%%%%%%%%%%%%%%%%%%%%%%%%%%%%%%%%%%%%%%%
\mychapter{chapter:ectuning}{Optimization of fit range \\
  for the energy calibration procedure}

%\chapter{\label{chapter:ectuning}Optimization of fit range \\
%  for the energy calibration procedure}

The energy calibration procedure attempts to compensate for differences
between data and Monte-Carlo that affect the position of the end point.
To accomplish that task, the calibration results should not be sensitive
to these same differences.  This can be re-stated as a requirement
that the fit should be able to recoup a \emph{change} in $\beta$,
$\au$, $\ad$.  So the optimization criteria is not the minimization of a
fit bias, but the minimization of any dependence of a bias on the shape of
the end point region.

To quantify the ability of the fit to recoup a shape change, a data
spectrum was distorted by applying the energy calibration
transformation \eqref{eq:ec-application} with e.g.
$\beta^{\text{shift}}=25$~keV/c.  The energy calibration procedure was
run on both the original and the distorted spectrum, and the changes
\begin{align}
\Delta\beta&=\beta^{\text{shifted}}-\beta^{\text{raw}}-\beta^{\text{shift}},\\
\Delta\au&=\au^{\text{shifted}}-\au^{\text{raw}},\\
\Delta\ad&=\ad^{\text{shifted}}-\ad^{\text{raw}},\\
\Delta\sigma_0&=\sigma_0^{\text{shifted}}-\sigma_0^{\text{raw}},
\end{align}
were computed.  These changes characterize stabilities of different
parameters to the given shape change.  To obtain more reliable
estimates, several values of $\beta^{\text{shift}}$, from $-75$~keV/c
to $+75$~keV/c in steps of $25$~keV/c, were used, and RMSes of
$\Delta\beta$, $\Delta\au$, $\Delta\ad$, $\Delta\sigma$ computed.
RMSes of $\Delta\alphasum$ and $\Delta\alphadiff$ were also obtained
from the same data.  Similar scans were done for $\au$ and $\ad$, to
quantify stabilities of fit results under different shape changes.  A
``variation'' of $\sigma_0$ was accomplished by smearing the
reconstructed momentum with a Gaussian, whose width was defined following
\eqref{eq:ec-param-sigma} as $\sigma_0^{\text{shift}}/|\sin(\theta)|$.
For that scan, $\Delta\sigma_0$ was defined through a quadratic
difference
$\Delta\sigma_0=\sqrt{(\sigma_0^{\text{shifted}})^2-(\sigma_0^{\text{shift}})^2}-\sigma_0^{\text{raw}}$.
Therefore, for a fixed choice of fit range, we had 24 numbers,
characterizing stabilities of 6 fit results ($\beta$, $\au$, $\ad$,
$\sigma_0$, $\alphasum$, $\alphadiff$) under 4 different shape changes
(the scans of $\beta^{\text{shift}}$, $\au^{\text{shift}}$,
$\ad^{\text{shift}}$, $\sigma_0^{\text{shift}}$).

Three versions of choosing the momentum range were tested: 
\begin{align}
p_{\text{edge}}(\theta)-c_1&<p<p_{\text{edge}}(\theta)+c_2,
\label{eq:fit-range-cc}
\\
p_{\text{edge}}(\theta)-c_1&<p<p_{\text{edge}}(\theta)+s_2\sigma(\theta),
\label{eq:fit-range-cs}
\\
p_{\text{edge}}(\theta)-s_1\sigma(\theta)&<p<p_{\text{edge}}(\theta)+s_2\sigma(\theta).
\label{eq:fit-range-ss}
\end{align}
where $c_i$ are constant momentum intervals and $s_i$ are constant
multipliers.  

For each of the schemes
\eqref{eq:fit-range-cc}--\eqref{eq:fit-range-ss} a 2-dimensional scan
of the parameters $c_i$ and/or $s_i$ was performed, with
$c_1=0\ldots2.5$~MeV/c in 0.25~MeV/c steps, $c_2=0\ldots0.5$~MeV/c in
0.05~MeV/c steps, $s_1=0\ldots5$ in steps of 0.5, and $s_2=0\ldots5$
in steps of 0.5.  At each scan point, the 24 ``stability'' numbers
were computed, so that a scan yielded 24 2-dimensional ``maps'' of the
fit range parameter space.

These 24 maps were examined by eye, and a ``compromise'' region,
approximately minimizing all of the stability parameters, was
identified.  Then the best regions found for
\eqref{eq:fit-range-cc}--\eqref{eq:fit-range-ss} were compared to each
other.  The best results were given by the scheme
\eqref{eq:fit-range-cs}, with $c_1=0.75$~MeV/c and $s_2=0.5$.  So
these were the settings used for the energy calibration during production
fitting.

% Because the resolution at the
% end point may also be different for data and Monte-Carlo, a definition
% of ``the same'' fit range may involve $\sigma(\theta)$.

%
% const/sigma versions tried: cc, ss, cs.
%
% sc1=const_minus=10,0.25,2.5
% sc1=const_minus=11,0.,2.5
% sc1=const_minus=4,0.25,1
% sc1=sigma_minus=11,0,5
% 
% sc2=const_plus=11,0.,0.5
% sc2=sigma_plus=11,0,5
% sc2=sigma_plus=5,0,2
% 

% Iterate.  Finite num bins means always converges.  $ncycle>1$.

%%%%%%%%%%%%%%%%%%%%%%%%%%%%%%%%%%%%%%%%%%%%%%%%%%%%%%%%%%%%%%%%

\mychapter{chapter:spectrum-expansion}{The spectrum expansion}
%\chapter{\label{chapter:spectrum-expansion}The spectrum expansion}
%\chapter{\label{chapter:spectrum-expansion}The spectrum expansion}

\begingroup

%%%%%%%%%%%%%%%%%%%%%%%%%%%%%%%%%%%%%%%%%%%%%%
\input fitting-macros.texinc

%%%%%%%%%%%%%%%%%%%%%%%%%%%%%%%%%%%%%%%%%%%%%%%%%%%%%%%%%%%%%%%%

Integrating the response function $K$ from \eqref{eq:rec-spectrum-C0}
over $x$ in bin $i$ of the spectrum histogram, we get a ``binned''
response function $K_i(x')$, that is, the probability to get the
reconstructed event in bin $i$.  For a dataset with $\Nd$ true decays,
the expected number of events reconstructed in bin $i$ is:
\begin{equation}
\Ni(\r) = B_i + \Nd\int_{\Omega_0} K_i(x')\, f(x';\r)\,dx'  
\label{Nexpected}
\end{equation}
where $B_i$ is the background, $f(x'; \r)$ is the true distribution of
decays, and $\Omega_0$ is the whole kinematically allowed phase space.
Often we have an analytical expression for the theoretical
distribution representing $f$ only up to a normalization factor:
\begin{equation}
f(x';\r) = A(\r)\, F(x';\r), \qquad
\int_{\Omega_0} f(x';\r)\,dx'\equiv1,
\label{Adef}
\end{equation}
we know $F$ but do not know $A$. (Of course $A$ can be
calculated numerically.)  This is true for muon decay:
\eqref{eq:theoretical-spectrum-2d} gives the differential decay \emph{rate}
$F$, but not the \emph{probability distribution} $f$.

\TWIST{} measures the shape of the spectrum.  We can get rid of
the absolute count by normalizing $\Ni$ to the total number of
reconstructed events in fiducial volume $\Omega$:
\begin{equation}
\ni(\r) = \Ni(\r)/\Nr(\r), \qquad\text{where}\quad
\Nr(\r) = \sum_{\Omega}\Ni(\r).
\label{ndef}
\end{equation}

A change in the parameters $\r$ modifies the spectrum shape as
\begin{equation}
\ni(\r+\dr)-\ni(\r) = \dr\, \dNfrac
- \ni(\r)\, \dr \sum_{\Omega}\dNfrac
+ \OO({\dr}^2)
\label{dn1}
\end{equation}
Using \eqref{Nexpected}--\eqref{ndef} we can express the derivative on
the right hand side as:
\begin{align}
\begin{split}
\dr\dNfrac & = \frac{\Nd}{\Nr} A(\r) \,\dr
\int_{\Omega_0} K_i(x')\,\frac{\p F(x';\r)}{\p\r}\,dx'\\
&\qquad + \frac{\Ni}{\Nr} \,\dr\,\dAfrac - \frac{B_i}{\Nr}\,\dr\,\dAfrac
+ \OO(\dr^2).
\end{split}
\label{deriv1} \\
\intertext{Also,}
\dr\,\dAfrac & = -A(\r) \,\dr\int_{\Omega_0} \frac{\p F(x';\r)}{\p\r}\,dx'.
\label{Ader}\\
\intertext{and}
A^{-1}(\r) & = \int_{\Omega_0} F(x';\r)\,dx'.
\label{Aval}
\end{align}

Analytical expressions for $F$ and $\p F/\p\r$ are known, so there
are available many ways to calculate integrals
(\ref{Ader})--(\ref{Aval}).  The integral in
(\ref{deriv1}) contains an unknown response function $K$, and 
therefore must 
be evaluated with Monte-Carlo.  It is convenient to calculate 
(\ref{Ader})--(\ref{Aval}) by Monte-Carlo integration as well.

%\subsection{Doing the integrals}
\subsection*{Doing the integrals}

A definite integral of a bounded non-negative function $g(x')$
can be evaluated using the acceptance-rejection method: choose 
%
%vb: a $\Ym \ge \sup_{x'\in\Omega_0}g(x')$ and sample $\Nt$ points 
%
a $\Ym \ge \vc{\max_{x'\in\Omega_0}{}}g(x')$ and sample $\Nt$ points 
$\{x',y\}$ from a uniform distribution on $\Omega_0\times[0,\Ym]$.  
Call a point accepted if $y<g(x')$. Then
\begin{equation}
  \int_{\Omega_0} g(x')\,dx' \approx \frac{\Na}{\Nt}\,\Ym\int_{\Omega_0}dx'.
  \label{acc-rej1}
\end{equation}
where $\Na$ is the count of accepted points.
This recipe is applicable for evaluating~(\ref{Aval}). 
An integral for a more general $g(x')$ can be written
\begin{equation}
  \int_{\Omega_0} g(x')\,dx' = \int_{\Omega_0} g^{+}(x')\,dx' 
  -\int_{\Omega_0} g^{-}(x')\,dx'
\end{equation}
where $g^{+}$ and $g^{-}$ are non-negative functions:
\begin{equation}
g^{+}(x') 
= \begin{cases}
  g(x')& \text{if $g(x')\ge0$},\\
  0&\text{otherwise},
  \end{cases}
\qquad
g^{-}(x') 
= \begin{cases}
  |g(x')|& \text{if $g(x')<0$},\\
  0&\text{otherwise}.
  \end{cases}
\end{equation}
Then from (\ref{acc-rej1})
\begin{align}
  \int_{\Omega_0} g(x')\,dx' &\approx
  \frac{\Na^{+}-\Na^{-}}{\Nt}\,\Ym\int_{\Omega_0}dx',
  \label{Ymaxdef}
\\
\intertext{where}
%\Ym&\ge\sup_{x'\in\Omega_0}|g(x')|,\\
\Ym&\ge\vc{\max_{x'\in\Omega_0}{}}|g(x')|,\\
y &< g^{+}(x') \quad \text{for}\quad  \Na^{+}, \label{accplus}\\
%\quad \text{and}
y &< g^{-}(x') \quad \text{for}\quad  \Na^{-}. \label{accminus}
\end{align}
The integral in (\ref{Ader}) can be calculated in this way.

To evaluate the integral in (\ref{deriv1}) we can use the following
procedure: with the ``generating function'' $g(x')={\p F(x';\r)}/{\p\r}$
sample $\{x',y\}$ from a uniform distribution as before. For every
point accepted according to (\ref{accplus}) or (\ref{accminus}) use
$x'$ to define the kinematics of an event in GEANT, simulate and
reconstruct the event.  If the event passed the whole analysis chain
and landed in bin $i$, count it as $\Ni^{+}[\pF]$ or $\Ni^{-}[\pF]$.
Then
\begin{equation}
  \int_{\Omega_0} K_i(x')\,\frac{\p F(x';\r)}{\p\r}\,dx'
  = \frac{\Ni^+[\pF]-\Ni^-[\pF]}{\Nt[\pF]}\:\Ym[\pF]\,\int_{\Omega_0}\,dx'
\end{equation}
and
\begin{multline}
\dr\,\dNfrac = \dr\,\frac{\Ym[\pF]}{\Ym[F]}\,
\frac{\Nt[F]}{\Nr[F]}
\\
{}\times\left\{
\frac{\Nr^{+}_i[\pF] - \Nr^{-}_i[\pF]}{\Nt[\pF]} 
- \frac{\Ni[F]-B_i}{\Nd[F]}\, \frac{\Nd^{+}[\pF]-\Nd^{-}[\pF]}{\Nt[\pF]}
\right\}
\\ 
+ \OO(\dr^2).
\label{dNfracvalue}
\end{multline}
$\Nd$ here, as before, denotes the number of ``true'' decays, that is,
those accepted in generation but not necessarily passed through the
analysis.  The argument in the square brackets indicates which
generating function was involved.  For example, $\Ni[F]\equiv\Ni$,
and $\Nd[F]\equiv\Nd$ from (\ref{Nexpected}).  The numbers $\Nr$,
$\Ni$, $\Ni^{\pm}$, $\Nd$ , $\Nd^{\pm}$ are to be understood
as statistical expectations.

Substituting (\ref{dNfracvalue}) into (\ref{dn1}) 
and introducing 
\begin{equation}
\Ni[\pF] = \Ni^+[\pF]-\Ni^-[\pF], 
\qquad
\Nd[\pF] = \Nd^+[\pF]-\Nd^-[\pF],
\end{equation}
we finally get
\begin{multline}
\ni(\r+\dr)-\ni(\r) = 
\dr\,\frac{\Ym[\pF]}{\Ym[F]}\,\frac{\Nt[F]}{\Nr[F]}
\\
{}\times\left\{
\frac{\Ni[\pF]}{\Nt[\pF]} 
+ \frac{B_i}{\Nd[F]}\,\frac{\Nd[\pF]}{\Nt[\pF]}
- \frac{\Ni[F]}{\Nr[F]}\,\frac{\Nr[\pF]}{\Nt[\pF]} 
\right.
\\
\left.
- \frac{\Ni[F]}{\Nr[F]}\,\frac{B}{\Nd[F]}\,\frac{\Nd[\pF]}{\Nt[\pF]}
\right\} 
+\OO(\dr^2),
\label{final1}
\end{multline}
where $B=\sum_{i\in\Omega} B_i$.
%
%  We can choose $\Ym[\pF]=\Ym[F]$ and process the ``base'' spectrum $F$
%  and derivative spectra $\pF$ in a single pass $\Nt[\pF]\equiv\Nt[F]$.
%  Then from (\ref{final1}) one can see that the reconstructed derivative
%  spectra could be normalized to the reconstructed base spectrum from
%  the same pass, and we would not need to know $\Nt$.  However
%  computationally it's more efficient to produce a base spectrum and
%  each derivative in separate runs.
%
Introducing the efficiency
\begin{equation}
\Emc(\r) = \Ym[F]\:\frac{\Nr[F]}{\Nt[F]}
\end{equation}
of the mapping of $\Omega_0\times[0,\Ym]$ into $\Omega$,
the integral of $\p F/\p\r$
\begin{equation}
\D(\r) = \Ym[\pF]\,\frac{\Nd[\pF]}{\Nt[\pF]},
\end{equation}
and normalized spectra
\begin{equation}
\nu_i(\r)=\Ym[{\textstyle \pF}]\:\frac{\Ni[\pF]}{\Nt[\pF]},
\qquad
\beta_i = \frac{B_i}{\Nd[F]},
\end{equation}
we can re-write (\ref{final1}) as
\begin{multline}
  \ni(\r+\dr) = 
  \left[1-\sum_{\alpha=1}^{m}\dr_{\alpha}\,
    \Emc^{-1}(\nu^{\alpha} +\beta\D^{\alpha})
  \right] \ni(\r) 
\\
  + \sum_{\alpha=1}^{m}\dr_\alpha\:\Emc^{-1}
  (\nu_i^\alpha + \beta_i\D^{\alpha})
+\OO(\dr^2).
\label{eq:final2}
\end{multline}
Here the index $\alpha=1,\ldots,m$ numbering the components of
$\r$ is shown explicitly, $\beta = \sum_{i\in\Omega}\beta_i$, 
and $\nu^{\alpha}=\sum_{i\in\Omega}\nu_i^{\alpha}$.

Equation \eqref{eq:final2} shows that a reconstructed spectrum for
parameter values $\r+\dr$ can be represented as a linear combination
of reconstructed spectra with different values of parameters $\r$.
The coefficients in front of the ``derivative'' terms in the linear
combination are proportional to the deviations of parameters $\dr$.
% 
% So, a reconstructed spectrum for ``shifted'' values of parameters
% $\ni(\r+\dr)$ can be represented as a linear combination of a ``base''
% spectrum $\ni(\r)$ and $m$ ``derivative'' spectra $\nu_i(\r)$.
% 
%
% So, a ``shifted'' spectrum $\ni(\r+\dr)$ can be represented as a
% linear combination of a ``base'' spectrum $\ni(\r)$ and $m$
% ``derivative'' spectra.  
%

%%%%%%%%%%%%%%%%%%%%%%%%%%%%%%%%%%%%%%%%%%%%%%%%%%%%%%%%%%%%%%%%

\endgroup

%inc: \end{document}

%%%%%%%%%%%%%%%%%%%%%%%%%%%%%%%%%%%%%%%%%%%%%%%%%%%%%%%%%%%%%%%%
\mychapter{chapter:xideltaconv}{Conversion formulas for the $\xidelta$ parametrization}

Here are the formulas for conversion between the $\Pmuxi,\delta$,
covariance matrix $V$, and $z=\xixidelta,w=\xidelta$,
covariance matrix $U$, parametrizations. \\[3pt]

1) $\Pmuxi,\delta\longrightarrow\xidelta$
\begin{eqnarray*}
\Delta z & = &\Delta\Pmuxi\\
\Delta w & = &(\Pmuxi_0 + \Delta\Pmuxi)(\delta_0+\Delta\delta) - \Pmuxi_0\delta_0\\
U_{ij} &= &V_{ij},\qquad i,j\ne w\\
U_{iw} & = & (\delta_0 + \Delta\delta)\,V_{i\xi} +
	(\Pmuxi_0+\Delta\Pmuxi)\,V_{i\delta},\qquad i\ne w\\
U_{ww} & = & (\delta_0+\Delta\delta)^2\,V_{\xi\xi} 
	+ 2(\Pmuxi_0+\Delta\Pmuxi)(\delta_0+\Delta\delta)\,V_{\xi\delta} 
	+ (\Pmuxi_0+\Delta\Pmuxi)^2\,V_{\delta\delta}
\end{eqnarray*}

2) $\xidelta\longrightarrow\Pmuxi,\delta$
\begin{eqnarray*}
\Delta\Pmuxi & = & \Delta z\\
\Delta\delta & = & (\Delta w - \delta_0\Delta z)/(\Pmuxi_0+\Delta z)\\
V_{ij} &= &U_{ij},\qquad i,j\ne\delta\\
V_{i\delta} & = & -\frac{\Pmuxi_0\delta_0+\Delta w}{(\Pmuxi_0+\Delta z)^2}\,U_{iz} 
	+ \frac{1}{\Pmuxi_0 + \Delta z}\,U_{iw},\qquad i\ne\delta\\
V_{\delta\delta} & = & 
	\frac{(\Pmuxi_0\delta_0+\Delta w)^2}{(\Pmuxi_0 +\Delta z)^4}\,U_{zz} 
	- 2\,\frac{\Pmuxi_0\delta_0 + \Delta w}{(\Pmuxi_0+\Delta z)^3}\,U_{zw}
	+\frac1{(\Pmuxi_0+\Delta z)^2}\,U_{ww}
\end{eqnarray*}

The covariance matrix conversion is approximate, see
e.g.~\cite{eadie:statmethods} for details.

% Note that the value of $\Delta\delta$ depends on the reference values
% $\xi_0$, $\delta_0$ used for the fit.  In the blind analysis technique
% those values are not known exactly; however we know that $\xi_0 = 1
% \pm 0.03$, $\delta_0 = 0.75 \pm 0.02$, that gives about 5\% precision
% on the {\em deviation} $\Delta\delta$.  This is sufficient for
% systematic studies: it gives a 5\% uncertainty on sensitivities to
% different systematics and a 5\% uncertainty on the estimated
% systematic error on $\delta$.  Other parameters, including $\xi$, are
% not affected by this uncertainty.  In the fitter, conversion function
% is called with arguments $\xi_0=1$, $\delta_0=0.75$ by default.  After
% the ``black box'' was open the fits were rerun with the revealed
% $\xi_0$ and $\delta_0$ to find the true value of $\Delta\delta$.

%%%%%%%%%%%%%%%%%%%%%%%%%%%%%%%%%%%%%%%%%%%%%%%%%%%%%%%%%%%%%%%%

\endgroup

%% % AG: add an empty page at the end
%% \pagestyle{empty}
%% \cleardoublepage
%% \hbox to 0pt{}

%

%AG: \printindex
\end{document}